# Modeling for the Growth of Unorganized Retailing in the Presence of Organized and E-Retailing in Indian Pharmaceutical Industry

Koushik Mondal[a*], Balagopal G Menon[b], Sunil Sahadev[c]

[a, b] Department of Industrial and Systems Engineering,
Indian Institute of Technology Kharagpur, Kharagpur 721 302, West Bengal, India.

[c] Stoddart Building, Sheffield Hallam University,
S11WA Sheffield, United Kingdom.

**Abstract**

The present study considers the rural pharmaceutical retail sector in India, where the arrival of organized retailers and e-retailers is testing the survival strategies of unorganized retailers. Grounded in a field investigation of the Indian pharmaceutical retail sector, this study integrates primary data collection, consumer conjoint analysis and design of experiments to develop an empirically grounded agent-based simulation of multi-channel competition among unorganized, organized and e-pharmaceutical retailers. The results of the conjoint analysis reveal that 'store attributes' of price discount, quality of products offered, variety of assortment, and degree of personalized service, and 'customer attributes' of distance, degree of mobility, and degree of emergency are key determinants of optimal store choice strategies. The primary insight obtained from the agent-based modeling is that the attribute levels of each individual retailer have some effect on other retailers' performance. The field-calibrated simulation also evidenced counterintuitive behavior that an increase in unorganized price discounts initially leads to an increase in average footprint at unorganized retailers, but eventually leads to these retailers moving out of the market. Hence, the unorganized retailers should not increase the price discount offered beyond a 'tipping point' or it will be detrimental to them. Another counterintuitive behavior found was that high emergency customers give less importance to 'variety of assortment' than low emergency customers. This study aids in understanding the levers for policy design towards improving the competition dynamics among retail channels in the pharmaceutical retail sector in India.

## 1. Introduction

The retail landscape in emerging economies like India is changing at a fast pace. Along with e-retailers, organized retail chains are also penetrating deep into small towns and villages. Customers, even in the less urbanized parts of the country, are now offered a range of retail choices. Unorganized retailers, especially those who were operating in rural areas and small towns, are now facing a difficult future. While large retail chains and e-retailers have professionally managed logistics systems and significant capital to invest in their inventories, unorganized retailers have to often rely entirely on their personal relationships with customers as a strategy to survive. However, fluctuations in demand and lack of loyalty among customers test the limits of such strategies. The unorganized retail sector is therefore gradually shrinking in market share.

The retail industry constitutes a substantial proportion of the gross domestic product of a country, as evidenced by the retail sales figures of over $6.5 trillion in the United States (U.S. Census Bureau 2022), and approximately $850 billion in a developing country like India which constitutes about 30% of the India's GDP (Technopak 2020). In general, the retail sector in any country can be categorized into organized and unorganized sectors. The term "organized retailing" refers to the business activity carried out by registered or licensed shops (for example, supermarkets, privately operated large retail chains, hypermarkets, etc.). On the contrary, "unorganized retailing" refers to the traditional small neighbourhood shops such as mom-and-pop shops, grocery shops, general stores, etc. In literature, these types of stores are known by several names like "nanostores" (Blanco and Fransoo 2013), "small unorganized retailers" (Dugar and Chamola 2021), and "micro-retailers" (Zhang et al. 2017). In the context of developing countries, these unorganized retail outlets are also known by various region-specific native names, such as "kirana" in India, "changarro" in Mexico, "sari sari" in the Philippines, "bodega" in Peru, "warung" in Indonesia, or "spaza" in South Africa (Escamilla et al. 2021). These independent shops are mainly run by the owner or family of the owner or managed by them, have cheap real estate and labor costs, and low taxes. The major benefits of unorganized retail stores are proximity and customer familiarity, which may be passed down through generations. These unorganized retail stores operate on a small scale, serving only a few hundred consumers around their vicinity, having storage and cash constraints (Fransoo et al. 2017). They also offer relationship-based credits, personalized services, and even delivery services to become competitive in the market (Child et al. 2015). In most developed countries, organized retailing dominates the market, while unorganized retailing, often in the informal sector, is still dominant in emerging economies (Jerath et al. 2016). Blanco and Fransoo (2013) estimate around 50 million nanostores in emerging economies. There are about twelve million nanostores in India (Kohli and Bhagwati, 2012), six million in China (Ge 2017), and millions more across other emerging economies like Mexico, Brazil, Nigeria, etc. In the majority of developing economies, traditional or unorganized retail is the major distribution channel for consumer-packaged goods (CPG), with a market share of more than 85% in South Asia and Africa and approximately 50% in Latin America (Fransoo et al. 2017). Despite the significance of this phenomenon, the effect of organized retailers and e-retailers on the growth of unorganized retailers in developing countries is still not fully understood (Jerath et al. 2016). Moreover, the research in this line is sparse in general, especially in the pharmaceutical retail sector in a developing country like India.

Moreover, pharmaceutical is deemed to be a critical national supply chain in the United States (White House 2021) as it has a crucial part in the health and well-being of millions of people around the world (OECD 2025). Thus, the above facts motivate the authors to conduct the present research on the pharmaceutical retail sector in India.

### 1.1. Pharmaceutical Retail Channel Competition in India

A pharmaceutical retail (also referred to as "community pharmacy" in literature) delivers medications to patients as prescribed by their doctors. Professional and qualified healthcare pharmacists work in a community pharmacy. Pharmacy accessibility is essential for the emerging role of community pharmacies as providers of patient-centred, medication management services in addition to their traditional dispensing responsibilities (Berenbrok et al. 2022). The responsibilities of a community pharmacist encompass processing of prescriptions, verifying of drug interactions, dispensing and disposing of medication, providing advice, and promoting a healthy lifestyle. Pioneering work in this field has been led by Yadav (2015) along with Larson et al. (2013), through the documentation of supply chains and the exploration of innovative distribution and financing models. The authors analysed health product supply chains of developing countries, highlighting key issues and needed reforms to improve health outcomes. The type of pharmaceutical retail ranges from small, individually owned unorganized pharmacies in isolated rural towns and urban areas to large organized chain pharmacies and online e-pharmacies in the country.

According to the Indian Pharmaceutical Alliance (IPA 2024), the Indian pharmaceutical retail industry is projected to grow from INR 2,42,000 crores (USD 29 billion) in 2023 to INR 4,60,000 crores (USD 55 billion) by 2030, expanding at a compound annual growth rate (CAGR) of 9.6%. The e-pharmacy industry in India has been growing rapidly and is expected to grow from INR 38 billion in 2019 to INR 240 billion by 2030 (IPA 2024; Market Research 2022). Organized pharmacies are also growing substantially, with a projected growth of INR 370 billion by 2026 from INR 134 billion in 2020. Hence, the market shares of both organized and e-pharmacies are growing rapidly (IBEF 2024; Market Research 2022). The unorganized pharmaceutical retail in India still holds the largest market share with an estimated growth from INR 1548 billion in 2020 to INR 2390 billion by 2026 (IBEF 2024; Market Research 2022), which makes this retail channel very significant for this study. Organized, unorganized, and e-pharmaceutical retailers have been fiercely competing for market share in emerging countries for the last two decades (IPA 2024; Market Research 2022). Surprisingly, unorganized retailers have demonstrated remarkable resilience despite the competition (Ge 2017). A study by Joseph et al. (2008) investigated the impact of organized retailing on the unorganized retail sector in India, where traditional unorganized retail dominated 96% of the market. The authors surveyed a sample of 2020 small unorganized retailers across 10 major cities and observed a decline in turnover and profits for those unorganized retailers located in close proximity to organized retailers. As per their analysis, the entrance of an organized retailer results in the annual closure of approximately 1.7% of local small unorganized businesses. However, this adverse effect slowly becomes less over time. In contrast, unorganized retailers situated further from organized retailers experienced a negligible impact. The unorganized pharmaceutical retail industry, which constitutes approximately 54% of the Indian pharmaceutical retail market, faces significant challenges despite the growth in the market (IPA 2024). Notably, traditional unorganized retail continues to grow at a steady annual rate of 10% despite the competition with supermarkets in the overall market. Large organized chain pharmacies have entered the market with predatory pricing, like a minimum of 20% off on all medicines, customer loyalty programs, and extensive product variety that the unorganized pharmaceutical retail sector is struggling to compete with. The unorganized pharmaceutical retail sector has seen a drastic decline in market share, falling from 62.5% in 2019 to 54.2% in 2023, due to high competition from organized and e-pharmacies (IPA 2024; IBEF 2024; Market Research 2022).

### 1.2. Prior Studies and Research Gaps

The literature on competition among different retail formats in emerging economies is limited and mostly empirical in nature (Ge 2017). Most of the studies are survey or data-based and analyze the retail market dynamics from a model-free perspective (Kohli and Bhagwati 2012; Joseph et al. 2008). When a new organized retailer enters the market, it causes significant alterations in consumer purchasing behavior, how infrastructure performs, and how the government makes policy changes, especially in emerging economies. Empirical studies can capture these short-term effects of retail competition (Ge 2017). It is pertinent and intriguing to analyse how the retail landscape will eventually converge and become stable in the long run and achieve an equilibrium state (Ge 2017). From this perspective, Jerath et al. (2016) presented a game-theoretical model that encompasses retail pricing and retail store location to depict the competition between unorganized retailers and a single organized retailer. Their findings indicate that the entry of an organized retailer reduces the number of unorganized retailers, but the prices charged by the surviving unorganized retailers may increase. This study represents a pioneering effort to explore the competition between unorganized and organized retail through an analytical lens. Notably, this study does not construct a dynamic model of the entry of the organized retailer in an unorganized retail market; rather, it estimates the long-term implications of the scenarios without and with an organized retailer separately and then compares the two outcomes (Jerath et al. 2016). In the existing literature on the competition between organized chain pharmacies and small unorganized independent pharmacies (Miller and Goodman 2017), the importance of independently owned pharmacies providing access to community pharmacies in rural areas (Berenbrok et al. 2022) and, the role of healthcare provider from a retailer in chronic disease management (Mossialos et al. 2015) are discussed in great details. However, these discussions, especially in the pharma marketing studies, have been considered exclusively from a pharmacist's angle rather than a management angle.

The consumer purchasing behavior plays a key role in the competition dynamics among unorganized, organized, and e-pharmaceutical retailers. Consumer has a key trade-off, whether to buy from the local unorganized shops multiple times whenever the demand occurs, or to purchase larger quantities from an organized retailer while making a few trips with higher transaction costs, or to purchase from an online retailer from the ease of their home and tolerate the waiting time. Bulk purchasing at lower cost from organized retailers often leads to either wastage or consumption with reduced utility because the future demand for the purchased product may not arise (Jerath et al. 2016). In their model, Jerath et al. (2016) showed that after the entry of an organized retail in an unorganized retail landscape, overall consumer surplus may decrease, which may further decrease the social surplus. Thus, the study concluded that policy makers should not take for granted that the introduction of an efficient organized retailer in the emerging market is always good for the retailing environment. The limitation of this study is that the authors assume the unorganized retailers to be located symmetrically in the market. Moreover, the study considers exactly one organized retailer in the market, considers a two-period setting, and doesn't consider the product assortments in the model.

Ge and Tomlin (2025) explored the competition between the supermarket and nanostores and examined the manufacturer's influence on the retail market structure for its product through wholesale price contracts, utilizing a multi-party supply-chain game. The study established that aggregate nanostore demand and profit can decrease in nanostore density, and the presence of a high-power supermarket can increase consumer surplus. This study shares a common limitation with that of Jerath et al. (2016) in assuming the unorganized retailers to be located symmetrically in the market, thereby adopting the 'circle model'. This study also doesn't consider the exit/entry decisions of the nanostore in the model.

The unorganized retail supply chain literature is fragmented due to the fact that the scope of the research is broad and diverse. Therefore, it is difficult to properly consolidate existing knowledge to advance the study of unorganized retailing. Research questions such as why the customer will prefer one channel over the other, under what circumstances the customer will switch channels, and how the market share for the competing retail channel is affected by these require a more integrated research approach. Empirical-simulation type or field-analytical type study can provide deeper insights into these. Building on these insights, the present research is an attempt to study the impact of organized and e-pharmaceutical retailing on unorganized pharmaceutical retailing in the Indian pharmaceutical retail landscape from a competitive dynamic perspective. In this backdrop, the present study models the influence of organized chain pharmacy and e-pharmacy on the growth of independent small unorganized pharmacies in terms of customer footprints and market share in the Indian context. Using the competitive dynamics perspective, we simultaneously utilise customer switching dynamics captured through conjoint analysis to develop the agent-based model. To summarize, the research questions guiding our empirical investigation into the impact of presence of organized pharmacy and e-pharmacy retails on the growth of unorganized pharmacy retail are as follows: (1) What are the factors influencing the customers purchasing decision (i.e. customer preferences) among the organized, unorganized and e-pharmaceutical retailer channels, (2) How customer's choice of a particular pharmaceutical retail distribution channel effects the customer footprints and market share of these three pharmaceutical retail distribution channels, and (3) How will the market dynamics change over time. The study considers the organized, unorganized, and e-pharmaceutical retailers, and thus it is confined only within the pharmaceutical retail sector in India. Moreover, this study is the need of the hour to specifically understand the behavioural dynamics of the Indian pharmaceutical retail sector emerging out of the co-existence of organized, unorganized, and e-pharmaceutical retailers in the country.

A block in the state of West Bengal, India, has been selected as the study area for this research. Several field studies, including interviews of unorganized, organized, and e-pharmaceutical retailers, customers, experts, and different stakeholders of the pharmaceutical supply chain, are carried out to identify the factors behind the consumers' purchasing decisions among the organized, unorganized, and e-pharmaceutical retailer channels. After a GIS-based mapping of all the retailers and customers from a field study and open-source satellite data, an agent-based model is developed in NetLogo 6.4 (Wilensky 1999). Using conjoint analysis, we attempt to calibrate the agent-based model (ABM) with Sawtooth software Lighthouse Studio 9.16.8.

The study makes several contributions to the existing body of knowledge in the operations management domain within the pharmaceutical retail sector. First, unlike in existing studies in retail sector which adopted a 'circle model' that considers unorganized retailers to be located symmetrically in the market (Ge and Tomlin 2025; Jerath et al. 2016), the present study has considered the spatial mapping of unorganized and organized pharmaceutical retailers as spatial characteristics plays a vital role in the customers' purchasing decisions. As there may be geographical constraints that prevent retailers to locate perfectly symmetrically in reality (Jerath et al. 2016), our approach of spatial mapping makes the model closer to the real market environment. Moreover, the actual spatial data of each household in the study area was captured utilizing a novel methodology by the authors (see sections 2.1.3 and 2.1.4). Second, unlike Jerath et al. (2016) study that considers exactly one organized retailer in the market, our approach considers multiple organized, unorganized, and e-pharmaceutical retailers in the market to induce competition in the market dynamics. Third, Jerath et al. (2016) have excluded product assortments in their model. The present study considers product assortment as it considerably influences the customers' purchasing behavior. Finally, the organized, unorganized, and e-pharmaceutical retailer channels serve customers in completely different ways (Fransoo et al. 2017). Hence, the present study is among the early contributions to the field by considering spatial attributes of customers and three types of retailers in a single framework, along with customer switching dynamics in the unorganized pharmaceutical retail sector in India.

From a methodological standpoint, the present study utilizes an empirically grounded agent-based modeling approach as it can simulate macro systems through micro-level individual behaviors involved in retail activities (Li and Liu 2024), thereby allowing for the modeling of individual heterogeneity as well as accumulated memory (Gilbert 2019; Sahadev et al. 2024). Unlike existing retail sector studies that mainly rely on a game theoretic approach (Ge and Tomlin 2025; Jerath et al. 2016), the present study utilizes field-calibrated agent-based modeling as the latter have better emergent properties when compared with game theoretic models. Recently, a study by Li and Liu (2024) has utilized agent-based modeling for analyzing and optimizing omnichannel retailing operation decisions. The present study utilizes choice-based conjoint analysis to capture the "voice of the customer" in order to understand the needs and preferences of customers regarding pharmaceutical products, which can in turn shed light on the customers' decision-making process. Moreover, the final experimentations with the base-run agent-based model are conducted based on a full factorial statistical experimental design scheme, thereby comparing all possible scenarios. Hence, all the above contributions make the present research important.

### 1.3. The Study Area

To model the competitive dynamics in a more focused manner, we, therefore, consider a smaller geographic area, predominantly rural, where unorganized pharmaceutical retailers hitherto dominated for decades but are now being threatened by the arrival of several large organized and e-pharmaceutical retail players. Our field study focused on the "Bishnupur II" block of the district "South 24 Parganas" of an eastern Indian state, West Bengal (Figure 1). As per Census 2011, this block has an area of 81.71 square km, a population of 214531 (around which 7pprox.. 65 % lives in villages), 53 villages, 11 census towns, a literacy rate of 81.37 % and a sex ratio of 957 out of 1000. As per our field study, there are 159 unorganized, 7 organized, and 4 e-pharmaceutical retailers present in the study area. The first organized chain pharmaceutical retailer opened its first store in 2010 and started to give a 10 % discount on all medicines. Then the Bengal Chemist and Druggist Association raised an objection to huge discounts and kept a level playing field for everyone. The first online e-pharmaceutical retailer started its franchisee operation in 2017. But in recent years, more organized chain pharmacies and e-pharmacies have entered the market with a minimum of 20% off on all medicines and created a market standard. Small unorganized retailers usually have a profit margin of around 18 to 25 % from distributors (Panda and Sahadev 2019; IBEF 2024). So, these small unorganized stores have started losing customers at a high rate and eventually several stores have closed down or switched their business. From our field study and interview with the experts, it is found that around 5 to 10 % of unorganized pharmaceutical retail stores are closing down every year. These dynamics among three retail players and customers make the "Bishnupur II" block suitable for our case study.

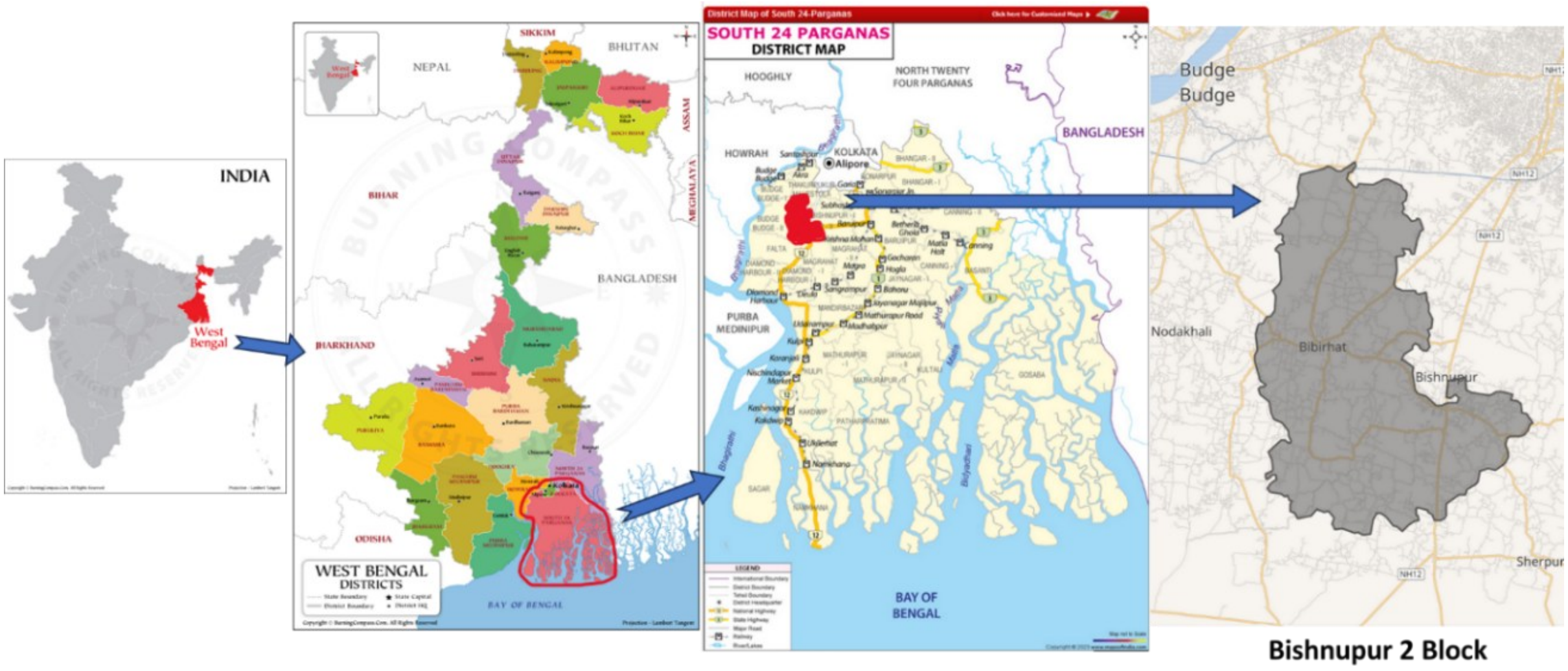

FIGURE 1 The study area
(Sources: OpenStreetMap, Wikimedia, burningcompass.com, mapsofindia.com)

## 2. Methodology

Modelling the purchasing behaviour of consumers needs to include all connecting elements, namely the demand side, the supply side, the specific products, and the market where they interact physically. The methodological framework adopted in the present study is depicted in Figure 2. To identify the systemic variables, authors have carried out both literature review and field study exploring the environment under investigation. Initial observations aided to validate assumptions that the unorganized retailers face a decline in customer footprint and market share after the organized and e-pharmaceutical retailers have entered the market. Initial observations also aided in observing unique characteristics of the pharmaceutical retail sector in India that defined the research questions posed in this work. The field study is divided into three parts. The first part is the interview of the unorganized, organized, and e-pharmaceutical retailers to identify the different service output levels offered by each of these three retail distribution channels. The different operational characteristics (like price discount offered, variety of assortment, quality of products served, degree of personalized service, etc.) of these three pharmaceutical retail channels are identified. The second part is the interview of customers to identify the customer desired service output levels, especially in pharmaceutical products. Perhaps, these are the factors that customers give prime importance to while making purchasing decisions. Thus, four retail-specific attributes and two customer-specific attributes are identified for the present study. The third part of the field study is spatial mapping of retailers and customers from real-world GIS data. This GIS mapping helps to simulate real-world scenarios in an ABM environment. From the attributes identified in the field study, a choice-based conjoint (CBC) analysis is carried out. CBC analysis provides the partial utilities of the attribute levels and the relative importance of each attribute for the customers. These partial utilities are used in the customer's utility calculation in ABM. After the robustness check, the base-run agent-based model is further utilized to perform statistically designed experiments (DOE) with different levels of retailers' attributes. The detailed description of these steps is as follows.

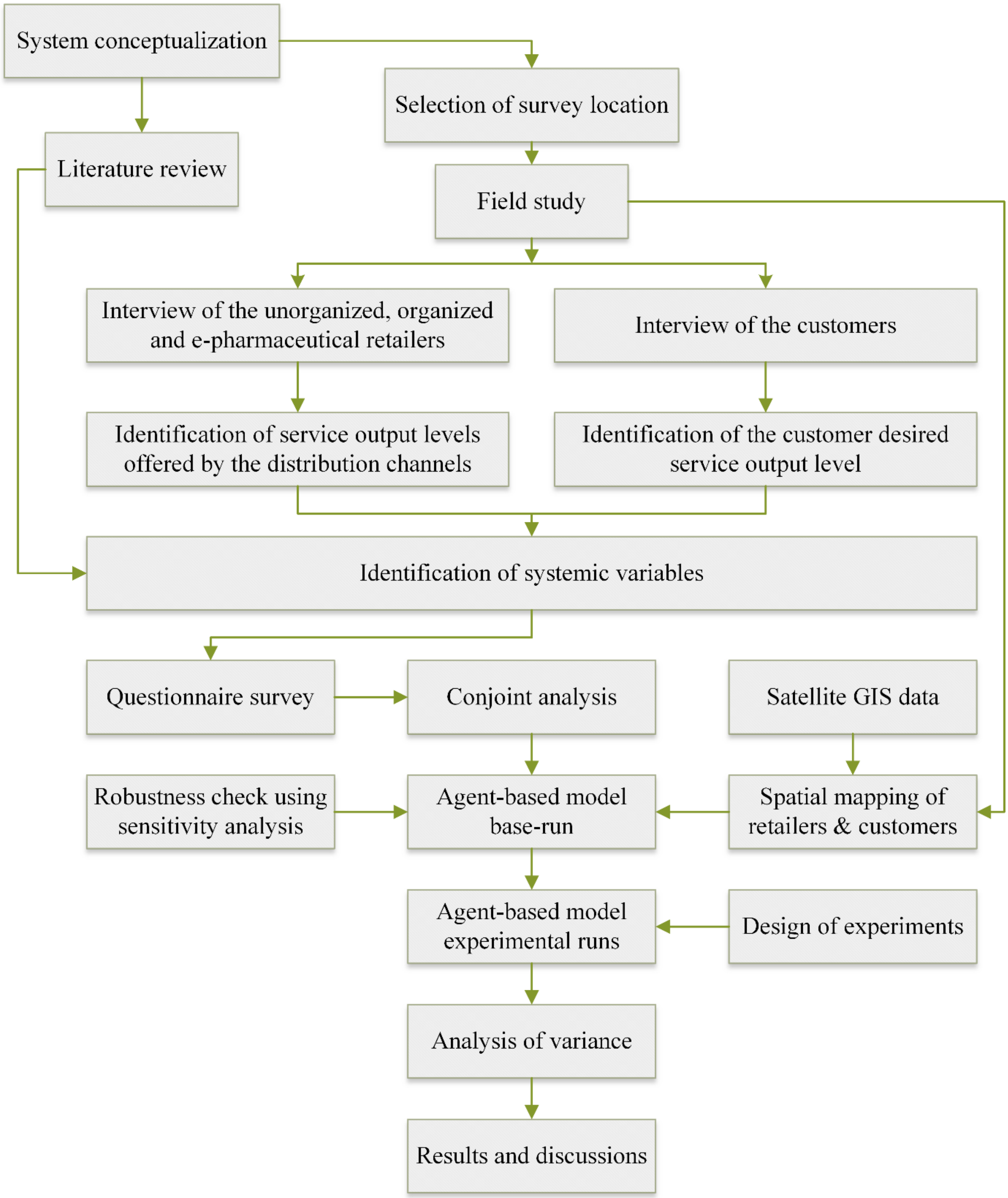

FIGURE 2 Methodological framework

## 2.1 Field Study

The consumer attributes include income, sex, age, household size, distance from the nearest retailer, and degree of emergency, which form their purchasing habits. For the present study, only distance from the nearest retailer and degree of emergency have been considered in ABM. It is assumed that only one family member will visit the shop when the demand occurs, in spite of individual visits. So, each household is assumed as one customer in the modelling.

The store attributes can be classified into tangible and intangible ones. Tangible factors include store location, price discount offered, variety of assortment, store format, and quality of the product (Krishnan et al. 2002; Dukes et al. 2009; Yang et al. 2025). The public perception about the store or the owner and the degree of personalized service offered are the intangible factors. This is a salient feature of unorganized retailers over organized and e-pharmacies (Child et al. 2015; Fransoo et al. 2017). A small unorganized retailer knows its clientele personally along with their preferences, and thus can guide them about how to use certain products to better match their personal tastes and needs (Jerath et al. 2016). Hence the utility of the product increases.

*The pharmaceutical products are specific in nature*. The consumption of pharmaceutical drugs or products is limited and depends upon the disease or medical conditions of the customer or the members of the family (Naumov et al. 2025). Usually, doctors prescribe medications in generic names according to the patient's condition. So, the customers can switch between different drug brands as long as the generic composition is the same. This makes pharmaceutical drugs a completely substitute product. From the field study and literature review, it is found that customers heterogeneously value the drugs purchased from unorganized, organized, and online retail channels. Recently, the Drugs Controller General of India (DCGI) issued show cause notices to illegal e-pharmacies that sell medications on the internet (Livemint 2023; The Print 2023). This influenced authors to take perceived quality or value of the drugs purchased from retail channels as a store attribute, which customers take into consideration while rating stores. The specialty of pharmaceutical drugs is that it is neither a regular consumable product like groceries nor a hedonic purchase like fashion products. The budget of a customer also does not necessarily dictate the degree of purchase (Naumov et al. 2025). In case of emergency, a patient is bound to purchase the specific medicine despite its high price or their budget constraint. It is assumed that customer utility could be negative, but they will compare the utilities derived from the three respective retail channels and choose to purchase certainly from the retailer with the relatively highest utility.

The market is where the customers meet retailers physically. It is observed from the review of the literature and our field study that those retailers are more prone to the adverse effect of organized chain pharmacies if they fall inside the catchment area of the organized pharmacy retail outlets (Joseph et al. 2008). 'Catchment area' is defined as the area (radius of distance in km) from where the organized retailers are expected to draw potential customers.

### 2.1.1 Deriving the Demand Side Dimensions of the Framework

To get the demand side data, the authors conducted a series of field studies that involved interviewing the customers, unorganized, organized, and e-pharmaceutical retailers, and several other stakeholders in the pharmaceutical supply chain. The stakeholders interviewed are doctors from private and government nursing homes, Auxiliary Nurse and Midwife (ANM) of block primary health centre (BPHC), medical representatives, pharmacists, quality control officer of a reputed pharmaceutical manufacturing company and two inspectors and assistant director of the Directorate of Drugs Control, Govt. of West Bengal, India.

The demand for pharmaceutical products directly depends on how many people are affected by a disease in the study area in a specific time period of the year. To capture the specificity of pharmaceutical drug demand, one should understand the fluctuating nature of diseases specific to the study area. For that purpose, the International Statistical Classification of Diseases and Related Health Problems (ICD) by World Health Organization (WHO) is adopted and relevant disease data is collected from the field study and government health reports from the Department of Health & Family Welfare, Government of West Bengal, India (refer to Appendix A). The Indian pharmaceutical sector is expected to have a compound annual growth rate of 9.6% between 2025 and 2030 (IPA 2024). For each time period, the total number of registered patients (in %) with the disease is calculated and fed to the ABM model. To maintain the generalizability of the model, the customers are randomly chosen from the populations in each iteration. Customer demographic data are collected from the last census in 2011. One of the challenging tasks is to collect customer spatial data in India, especially in rural and semi-urban areas. The authors have proposed a novel methodology to map household locations from real-world satellite data into an ABM model. The detailed step-by-step framework is discussed in subsection 2.1.4.

The present study models the individual heterogeneous consumer preferences for the store attributes using utility theory. Each customer has different individual desirability or worth (part-worth) of the specific store attribute levels, as well as their own customer-specific attribute levels. The part-worth is a numerical score that is used to estimate the total utility or desirability of a retailer by summing the part-worths of its individual attributes. They rank the stores based on the total utility derived from purchasing and select the store that gives them maximum total utility. To obtain these preference probabilities or part-worths, a consumer survey is carried out in the study area, followed by a choice-based conjoint analysis. These preference probabilities are used in the agent's decision-making process in the ABM model.

### 2.1.2. Deriving the Supply Side Dimensions of the Framework

Organized, unorganized, and e-pharmaceutical retailer channels serve customers differently (Fransoo et al. 2017). The sales, profit, and market share of these three pharmaceutical retail channels depend on the customer demand for each of these three channels (Jerath et al. 2016). For identification of the attributes of the retailers, the authors have conducted interviews with the three types of retailers to find out average sales volume, price, offers, average customer footprint, and whether they provide any personalized services like home delivery, relationship-based credit, etc. With the help of literature and field study, the following attributes are considered in the present study, which influence consumer purchasing decisions (refer to Table 1).

*Price discount* (*P*): Usually, the price charged by unorganized retailers is higher than that of organized and e-pharmaceutical retailers due to lower economies of scale and smaller bargaining power of the unorganized stores with storage and cash constraints. In the field study, it is found that organized and e-pharmaceutical retailers are providing a minimum 20% discount on medicine purchases, while most of the unorganized retailers are able to give a maximum 10% to 15% price discount. This makes price discount a key attribute for the customers while making a purchasing decision.

*Quality or value* (*v*) of a pharmaceutical drug depends upon whether the drug is generic or branded, whether it is a new medicine (which has more usage time before expiry) or near to the expiry date, and perceived quality by the customer. In recent news, illegal e-pharmacies selling medicines on the internet (Livemint 2023; The Print 2023) create a negative word-of-mouth (WOM) which may decrease the perceived value for the customer while making a purchasing decision from an online channel. Also, the absence of touch and feel, compared with other brands, and manual checking of expiry date may affect the customer's perceived quality.

*Variety of assortment* (*A*) refers to the range and diversity of products a business offers, encompassing the breadth (different product categories), depth (variations within a category), length (total number of items), and consistency (how related the products are). Consumers prioritize variety of assortment just after location and price when identifying reasons for patronizing their preferred stores (Hoch et al. 1999). Unorganized retail stores have relatively less variety of assortments with respect to organized and e-pharmaceutical retailers because of the store size and cash constraints (Fransoo et al. 2017).

*Distance between customer and retailer* (*d*) plays a vital role in deciding which store the customer will visit. Usually, the unorganized retailers are larger in number than the organized and e-pharmaceutical retailers in a market. Hence, the locational advantage of the unorganized retailers is relatively greater. For e-pharmaceutical retailers, the main concern for a customer is lead time. For the present study area, the minimum lead time is three days, as it falls under a remote service area. For main cities, this lead time could be less than a day. To capture this phenomenon and also to make it comparable with the other retailers, distance is taken as a proxy for the lead time while purchasing from e-pharmaceutical retailers. While unorganized and organized retailers are on average within 5-10 km from customers, the e-pharmaceutical retailers are assumed to be at least 10 km away from the customers.

Pharmaceutical drugs fall under the category of necessary goods. So, the *Degree of emergency* I will influence the purchasing decision of the customers. The degree of emergency for the medicine may be high, medium, or low, based on the medical condition faced by an individual or their family members. A customer with a higher emergency would not tolerate a higher waiting time. They will prefer immediate possession within a shorter timeframe. In these cases, customers may prefer the nearest pharmacy with a higher assortment so that they can surely get their required medicine within the least time, without thinking much of the discounts offered. On the contrary, if the degree of emergency for medicine is low, one can wait for a day or two to get the medicine. In these cases, customers can look for better price discounts without worrying about the lead time to wait.

Many unorganized retailers provide a certain Degree of personalized service (S), engaging directly with customers to better understand their tastes and preferences, potentially offering added value (Child et al. 2015). For example, an unorganized retailer might provide guidance on how to use a product in a way that aligns more closely with the customer's needs, thereby enhancing the product's overall utility. In contrast, customers purchasing from organized and e-pharmaceutical retailers typically do not receive this level of personalized service and utility (Jerath et al. 2017; Fransoo et al. 2024).

### 2.1.3. Spatial Mapping of Retailers

The spatial mapping of 159 unorganized, 7 organized, and 4 e-pharmaceutical retailers in the survey area is carried out in the third stage of the field study using "GPS Map Camera" to collect latitude and longitude data of every pharmaceutical retail shop visited (Figure 3). Those data were imported into a GIS software "QGIS 3.32 Lima" along with base maps from 'Maps of India', Google satellite maps, and 'OpenStreetMap'. A buffer of 1 kilometer is used beyond the actual block boundary to accommodate the customers in the catchment area (Figure 4a and 4b). The spatial mapping of the unorganized (green dots) and organized (⊕ dots) pharmaceutical retailers in Bishnupur II block is shown in Figure 4c. The e-pharmaceutical retailers are not shown in the spatial mapping. It is taken into ABM directly in Figure 7b as hypothetical locations from where pharmaceutical drugs are delivered directly to customers.

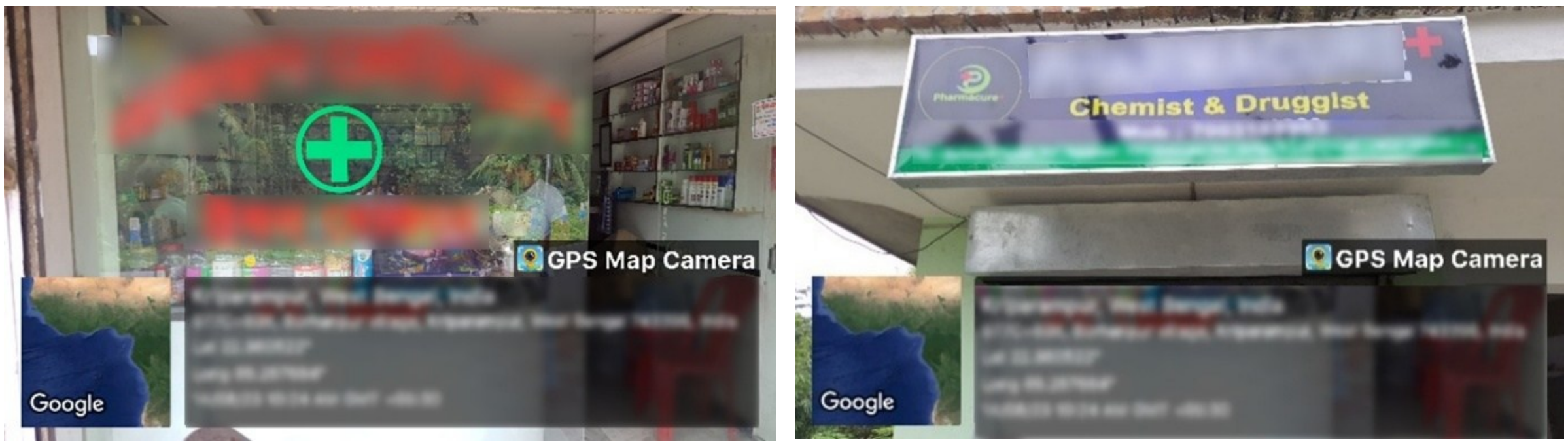

FIGURE 3 Spatial data collection of the retailers in GPS Map Camera

#### 2.1.4. Spatial Mapping of Customers

The spatial characteristics of customers play a vital role in their purchasing decisions in the advent of emergency as well as convenience. To obtain the actual spatial data of each household in a given area is an arduous task. A novel methodology is followed in this study to find out the spatial household data of any given area in the world. Firstly, create the base map of the required area under study in any GIS software. Then, from the “Open Buildings” dataset provided by Google, the specific region needs to be downloaded and imported into the GIS software as point data (Figure 4d). Open Buildings (Sirko et al. 2021) is a large-scale open dataset that contains the outlines of buildings derived from high-resolution satellite imagery by training a deep learning model. The next step is to clean the data by setting an appropriate confidence level and manually checking the map to ensure buildings are detected correctly (Figure 4e). Lastly, the required area is selected from the filtered dataset as per the area under study. The spatial mapping of the customer households and pharmaceutical retailers in Bishnupur II block is shown in Figure 4f. The data from the GIS model is exported as a shape (.shp) file and imported into the NetLogo environment for Agent-Based Modeling (Figure 7b).

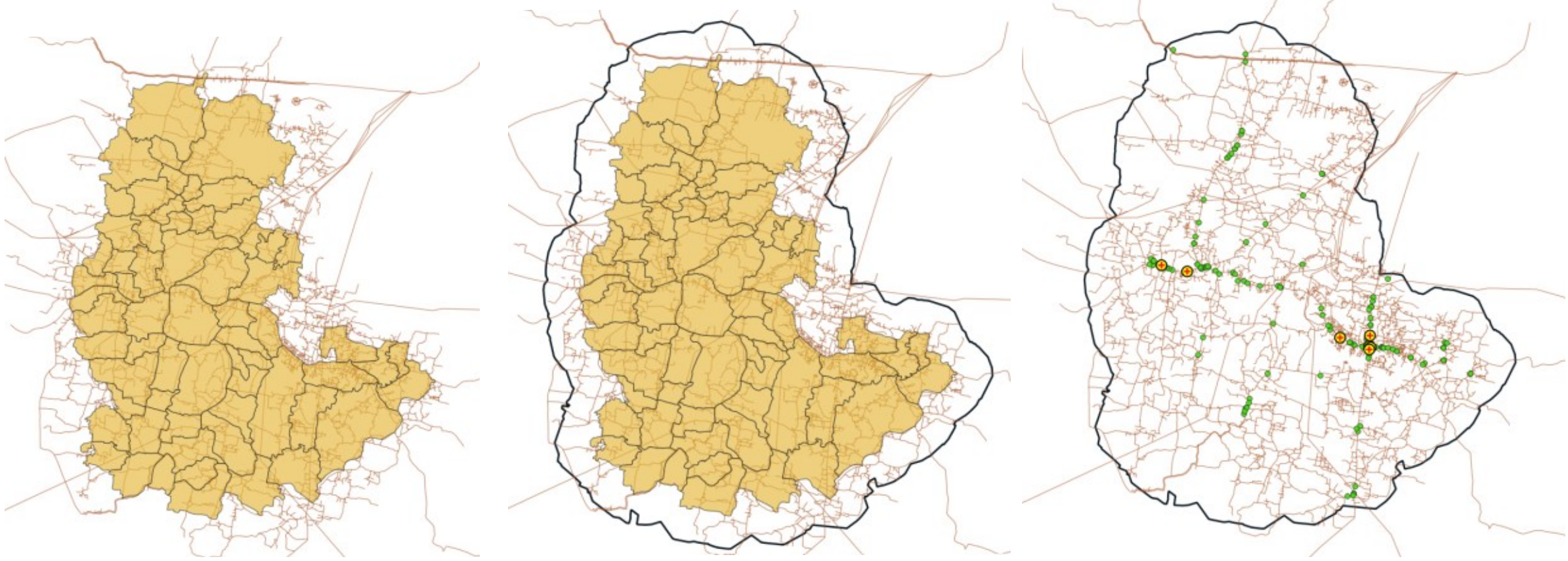

(a) (b) (c)

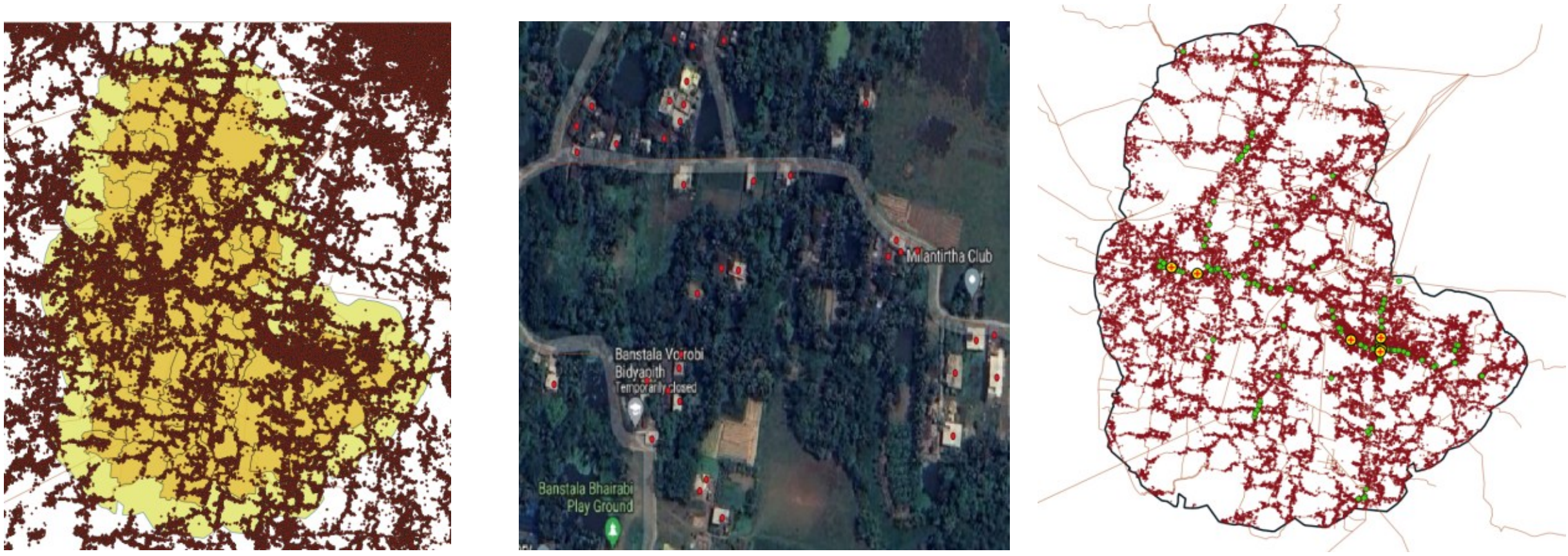

(d) (e) (f)

FIGURE 4 Spatial mapping in QGIS: (a) Bishnupur 2 block boundary (b) block boundary with 1 km buffer I spatial mapping of unorganized (green dots) and organized (⊕ dots) pharmaceutical retailers (d) spatial mapping of consumers (red dots) from google-earth raw dataset I manual data cleaning (f) final spatial mapping of consumers and retailers in QGIS

## 2.2. The Conjoint Analysis

The study employs choice-based conjoint (CBC) analysis to determine the utility of individual customers from purchasing a pharmaceutical product from a particular retail distribution channel. To conduct CBC analysis, a set of profiles with different combinations of retail store attributes and levels has to be generated and presented to the respondents. Previous studies have identified price discount, quality of product, variety of assortment, degree of personalized service, and distance between customer and retailer as key factors that may influence customers' retail channel choice behavior (Jerath et al. 2016; Chiang et al. 2003; Hoch et al. 1999). Together with a field study and literature review, we selected these five factors as attributes in the estimation of retail channel choice behavior for customers in the study area. Three separate CBC analyses are conducted for three different levels of the customer's degree of emergency. For the available set of choices from 5 attributes, the full factorial design for each of the CBCs comes out to be $3^5$ = 243 choice sets, which may reduce the response rate by burdening the respondents. To avoid choice overloading for such big choice sets, the authors have implemented an orthogonal design and have reduced the 243 choice sets to 16 choice sets for each of the high, medium, and low degree of emergency profiles. Each choice task consisted of four options. The respondents are instructed to choose an option they prefer the most out of four choice sets given at a time. So, a total of 4 times they have to repeat the task, and subsequently, the utility scores are derived.

TABLE 1 Attributes with levels

| | Attributes | Levels 1 | 2 | 3 | References |
|---|---|---|---|---|---|
| Retail specific | Price discount ($P$) [Rupees] | Less than 10% off [0-10) | [10-20] % off [10-20] | More than 20% off (20-100] | Jerath et al. (2016) |
| | Quality or value ($v$) [Unitless] | Low [0-0.4) | Medium [0.4-0.7) | High [0.7-1] | Chiang et al. (2003) |
| | Variety of assortment ($A$) [Unitless] | Small [0-0.4) | Medium [0.4-0.7) | Large [0.7-1] | Hoch et al. (1999) |
| | Degree of personalized service ($S$) [Unitless] | Low [0-0.4) | Medium [0.4-0.7) | High [0.7-1] | Jerath et al. (2016) |
| Customer specific | Distance between customer and retailer ($d$) [km] | Less than or equal to 2 km | (2-10] km | More than 10 km | Jerath et al. (2016) |
| | Degree of emergency ($\beta$) [Unitless] | Low [0-0.4) | Medium [0.4-0.7) | High [0.7-1] | Present study |

*Note:* Square bracket indicates that the value is included, and round bracket indicates not included

### 2.2.1. Sample Size

Numerous studies have proposed guidelines for determining the minimum sample size required for stated preference choice surveys. Orme (1998) introduced a widely adopted rule of thumb, recommending the following equation for estimating the sample size necessary for the reliable estimation of main effects:

$$N \geq \frac{500 \times l}{J \times T} \tag{1}$$

Here, $l$ is the largest number of attribute levels, $T$ is the number of choice tasks, and $J$ is the number of alternatives. Following this rule of thumb, the sample size would need to be greater than the calculated minimum $N$ ((500 × 3) ÷ (4 × 4) ≈ 94). Lancsar and Louviere (2008) recommended a sample size of 20 respondents per choice task as a bare minimum, while 100 was the sample size threshold for choice modelling analysis according to Pearmain and Kroes (1990). It is notable that the present survey was well above the minimum sample size requirement (138) of all these guidelines.

### 2.2.2. Data Collection

The questionnaire was made available online and circulated to 300 randomly selected customers in the study area. To get proper and clear answers, only those who had previously purchased pharmaceutical products from the three retail channels were considered. The "terms" used in the questionnaire were explained beforehand to the respondents. Respondents were informed that there were no right or wrong answers in order to relieve any pressure and enable honest feedback. This was carried out with the purpose of reducing bias and increasing the validity and reliability of the responses and thus the quality of the study findings. A total of 138 completed questionnaires were received, and the respondents' profiles are shown in Table B1 in Appendix B.

### 2.3. Theoretical Model Development

The Agent-Based Model (hereafter ABM) is considered suitable to model complex processes that do not come under the analytic framework of modeling. In the present study, an agent-based model for three retail channels in an oligopolistic market environment is developed and utilized to investigate the effects of consumer choice dynamics on sales, profits, and market shares of these retailers in the country.

#### 2.3.1. Simulation Model Architecture

The customers and pharmaceutical retailers in the Indian market are modelled as agents in the simulation. Both the agent sets have specified attributes associated with their respective functions. The different types of agents used in ABM are shown in Figure 5. The customer agents (Figure 5d and 5e) in ABM refer to the customers in the Indian market who carry out transactions when a demand for pharmaceutical drugs arises. Customers are assumed to be rational. They want to maximize their utility by combining their individual preferences with the store attributes and customer attributes before making the purchasing decision. In a multi-channel market, customers can buy pharmaceutical products either from the independent unorganized stores, or from the organized chain stores, or from the online e-pharmaceutical retailer. In order to capture the dynamic nature of perceived inconvenience due to travel, another environmental factor considered in the model is the degree of mobility (*n*). It is defined as the ease of transportation or movement allowed for the customers. A low degree of mobility indicates a restricted movement of the mass (e.g., bad weather like snow, storm, etc, or compulsory restrictions like COVID-19 lockdown, etc). Depending on the degree of mobility, the store's locational advantages can change. The discrete choice utility model proposed in the present model is as follows.

$$U_{i,s} = \frac{1}{d_{i,s}^{n}} \left[ \sum_{j} w_{i,j} A_{s,j} + \sum_{k} w_{i,k} A_{i,k} \right] \tag{2}$$

Where, for $i^{th}$ customer and $s^{th}$ retailer, $j$ = respective store attribute level, $k$ = respective customer attribute level, $d_{i,s}$ = distance between $i^{th}$ customer and the $s^{th}$ store, $w_{i,j}$ = $i^{th}$ individual's preference for the store attribute level $j$, $A_{s,j}$ = 1 if 'yes' or 0 if 'no' for the store attribute level $j$, $w_{i,k}$ = $i^{th}$ individual preference for the customer attribute level $k$, $A_{s,k}$ = 1 if 'yes' or 0 if 'no' for the customer attribute level $k$, $n$ = degree of mobility. The utility is assumed to be a linear additive function of the attribute's score.

Individual preferences for attribute levels are estimated from the conjoint analysis and utilized for ABM calibration. Utilities derived from the unorganized, organized, and e-pharmaceutical retailers are $U_u$, $U_o$, and $U_e$, respectively. When customers can choose between unorganized, organized, and e-pharmaceutical retailers, they compare utilities to decide which channel to purchase from. If $U_u>U_o>U_e$, then the unorganized retailer is preferred over the organized retailer and then the e-pharmaceutical retailer. If $U_o>U_e>U_u$, then the organized retailer is preferred over the e-pharmaceutical retailer and then the unorganized retailer. If $U_e>U_o>U_u$, then the e-pharmaceutical retailer is preferred over the organized retailer and then the unorganized retailer. If $U_u=U_o$ and $U_u=U_e$, the customer is indifferent between purchasing from unorganized versus organized and e-pharmaceutical retailers. Figure 6 shows the flowchart of the customer's decision-making process.

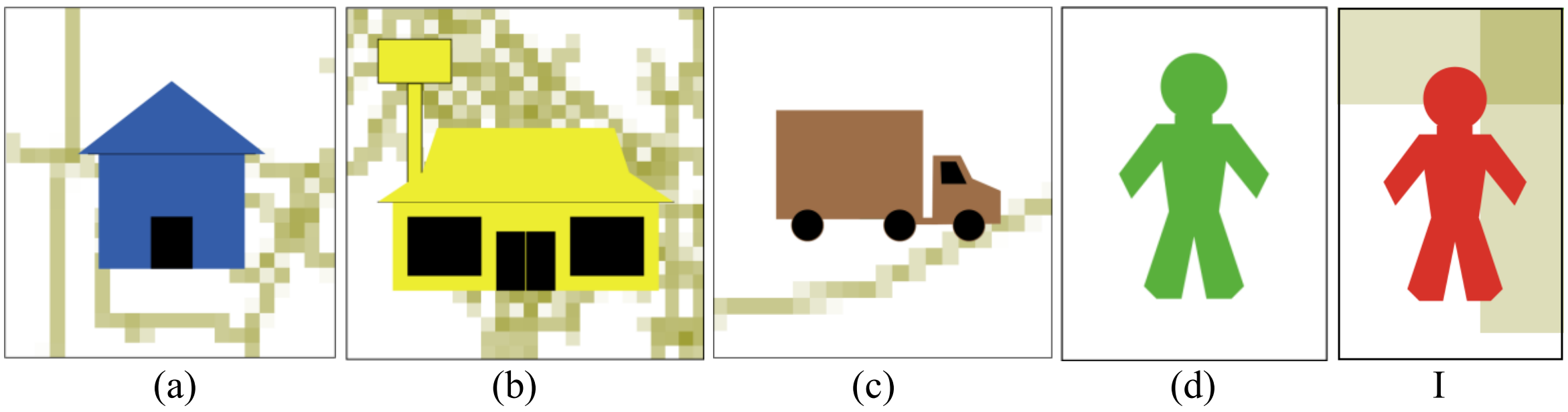

FIGURE 5 Types of agents used in ABM: (a) unorganized retailer (b) organized retailer I e-pharmaceutical retailer (d) customer without demand I customer with demand

From the GIS mapping of retailers and customers, the retailer and customer agents are imported into an agent-based model environment using the NetLogo GIS extension. The retailer agent variables include price discount, quality, variety of assortment, degree of personalized service, position, weekly customer footprint, average customer order size, gross margin, last six months sales and profits, weekly minimum profits to survive, weekly total cost including rent, operations, overhead charges, and salary of employees and pharmacists. The customer agent variables include degree of emergency, distance from the retailers, distance traveled, preferences for each of the store attributes, preferences for customer attributes, utilities of purchasing from the three retail channels, and degree of mobility.

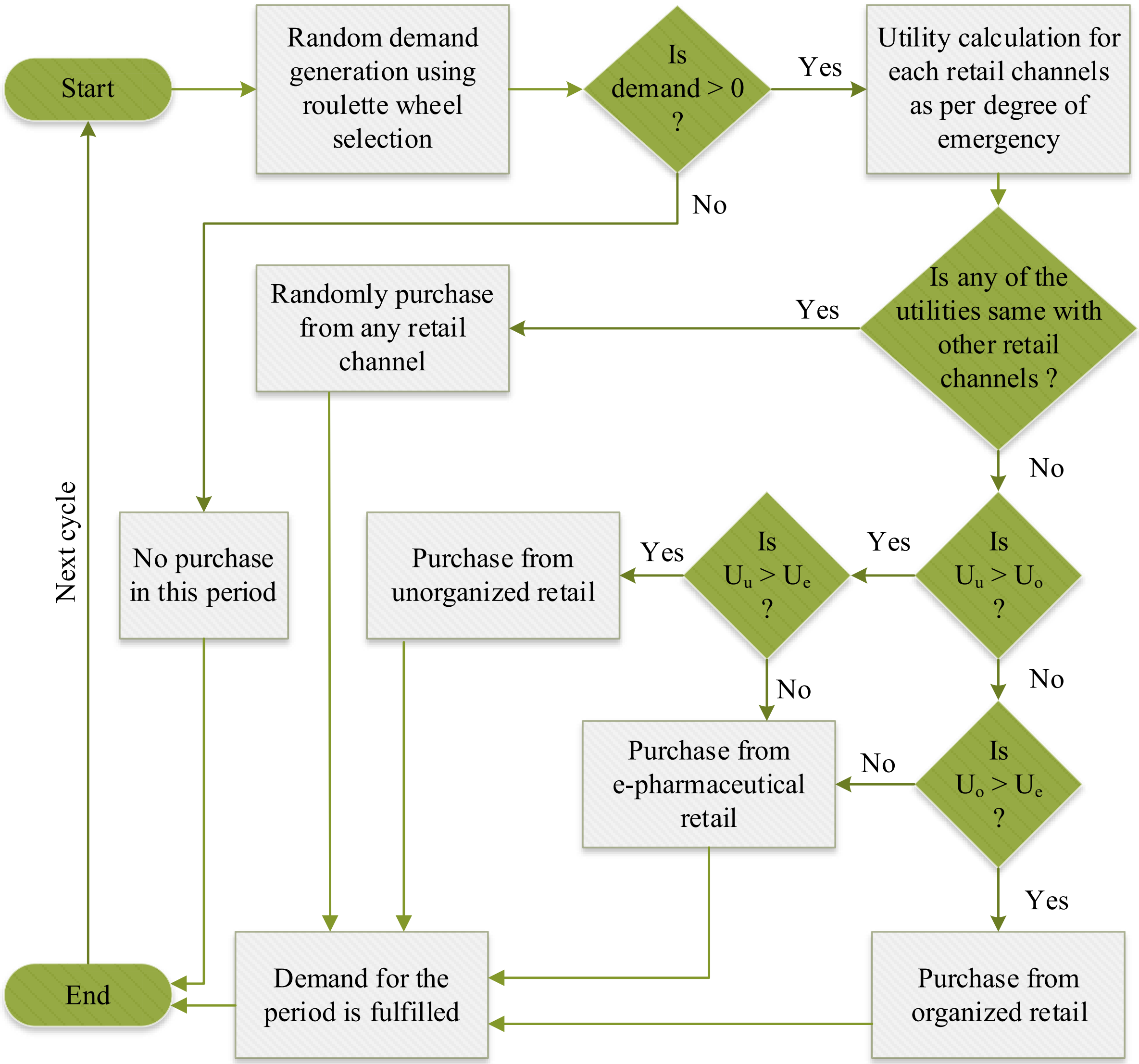

FIGURE 6 Flowchart of the customer's decision-making process

### 2.3.2 Agent-Based Model (ABM) Setup, Base-Run, and Robustness Check

The setup procedure for the ABM environment is designed to integrate GIS data and establish four distinct agent types, which are unorganized retailers, organized retailers, e-pharmaceutical retailers, and customers. The demand data is extracted from the disease data, which serves as the foundation for initializing the simulation. Retailer attribute values are assigned according to the base case specifications (Table 5) and are assumed to remain constant throughout all iterations. Every household is represented as a customer, and customer attributes are initialized accordingly. At the start of the simulation, all customers are marked as "green," indicating an absence of demand (Figure 5d). At the beginning of each iteration, a roulette wheel mechanism (Chen et al. 2016; Pham et al. 2023; Smith 2012) is employed to randomly designate a proportion of customers as "customers with demand", changing their status from green to red (Figure 5e). The proportion of red customers is determined based on the disease data for each iteration. Only these red customers, referred to as ‘potential customers’, actively participate in the simulation. The remaining green customers, termed ‘dormant customers’, remain inactive during that iteration.

The potential buyers proceed with searching for the nearest unorganized, organized, and e-pharmaceutical retailers. As per their degree of emergency, each potential customer will have different partial utilities from Table 4, or Tables C1 or C2 (in Appendix C). Next, they will calculate the utilities derived from purchasing from the three nearest retail channels separately using equation 2. After comparison of the computed utilities, each potential buyer selects the retailer with the highest utility and transacts with that retailer. The transaction is captured in the ABM as a link between the buyer and the selected retailer. The type of retail is distinguished by link color with yellow for organized retailers, blue for unorganized retailers, and magenta for e-pharmaceutical retailers (Figure 7c). The customer footprint of each retailer is calculated by counting the total links customers have made with that retailer. In the present model, one iteration is equivalent to one week. Retailers are assumed to have sufficient inventory to satisfy the demand of the customers who have chosen them. Based on observed retail practices it is assumed that unorganized retailers calculate the profits at six-month intervals to see if the business is viable to continue (using equations 3, 4, and 5). If an unorganized retailer finds that his/her last six months’ profit is less than the last six months’ total cost and minimum profit to survive, then they will shut down the business and will not participate in the subsequent iterations. These closed-down stores will turn into ‘red’ (Figure 7d). It is also assumed that organized and e-pharmaceutical retailers will not shut down during the simulation because these players follow a “growth first, profit later” model and are heavily backed by investors.

$$Net\ Profit = Profit\ from\ sales - Total\ cost\ of\ the\ retailer \tag{3}$$

$$Profit\ from\ sales = Average\ customer\ order\ size\ \times Customer\ footprint\ \times (Gross\ margin -\ Price\ discount)/100 \tag{4}$$

$$Total\ cost\ of\ the\ retailer = Shop\ rent + Pharmacist\ salary + Employee\ salary + Overhead\ charges \tag{5}$$

The system is reset after each iteration by deleting all existing links and resetting all the customers to their initial green state. In the next iteration, the process is repeated with the roulette wheel mechanism in order to determine the new set of ‘potential customers’, and then the identification of the type of emergency is carried out, similar to the setup process. The iterative process is repeated during the simulation length. The simulation length is six years from 2024 to 2030. Once the model is developed and after the base run, a sensitivity analysis is conducted in order to assess the robustness of the agent-based simulation model. Sensitivity analysis investigates if the results differ significantly with respect to the objective when assumptions are changed within an acceptable range of variability (Menon and Mahanty 2015; Sterman 2000). This study conducts the best-case and worst-case sensitivity analysis.

### 2.3.3 Performance Analysis of Pharmaceutical Retail Sector Using Design of Experiments

After the base model robustness check, to analyze the performance of the pharmaceutical retailer, analysis of variance (ANOVA) is adopted to quantify the effects of input variables (price discount and product quality) on the output performance measures (customer footprints, market share, and retail shutdown). The input variables selected for this study are price discount and quality of the three different retail channels. The performance measures are as follows;

1. Average weekly customer footprints per retail channel (customers per week)
2. Average market share per retail channel (%)
3. Unorganized retails shut down during the simulation time (no. of stores shut down)

To examine the influence of discount and quality on the output performance measures, different simulation experiments are designed to simulate the ABM model based on a full factorial $L_{27}(3^3)$ Orthogonal Array (OA) design. In the experimental settings to generate simulation scenarios, each variable is considered to have three levels: low (L), medium (M), and high (H). In this study, two separate scenarios are being investigated: one with varying discount and the other with varying quality. At first, ANOVA analysis is carried out for the 'discount' offered, and the results are discussed. Then, a similar analysis is done for 'quality', and a comparison of results is made. Table 2 presents the input variables for the price discount with their levels. Three factors with three levels, i.e., a total $3^3$ = 27 experimental runs, are generated in Minitab software based on a full factorial design. The ABM model is simulated with these variables for 312 weeks (6 years), and simulation results are summarized in Table D1 in the online Appendix D.

TABLE 2 Input variables with their levels for price discount

| Sl no. | Input variables | Levels | | |
|---|---|---|---|---|
| | | Low (L) | Medium (M) | High (H) |
| 1 | Unorganized discount | Less than 10% off | (10-20) % off | More than 20% off |
| 2 | Organized discount | Less than 10% off | (10-20) % off | More than 20% off |
| 3 | E-Pharmacy discount | Less than 10% off | (10-20) % off | More than 20% off |

Table 3 presents the input variables for quality with their levels. Three factors with three levels, i.e., a total $3^3$ = 27 experimental runs, are generated in Minitab software based on a full factorial design. The simulation results are summarized in Table D9 in the online Appendix D.

TABLE 3 Input variables with their levels for quality

| Sl no. | Input variables | Levels | | |
|---|---|---|---|---|
| | | Low (L) | Medium (M) | High (H) |
| 1 | Unorganized quality | Low [0-0.4) | Medium [0.4-0.7) | High [0.7-1] |
| 2 | Organized quality | Low [0-0.4) | Medium [0.4-0.7) | High [0.7-1] |
| 3 | E-Pharmacy quality | Low [0-0.4) | Medium [0.4-0.7) | High [0.7-1] |

*Note:* Square bracket indicates that the value is included, and round bracket indicates not included

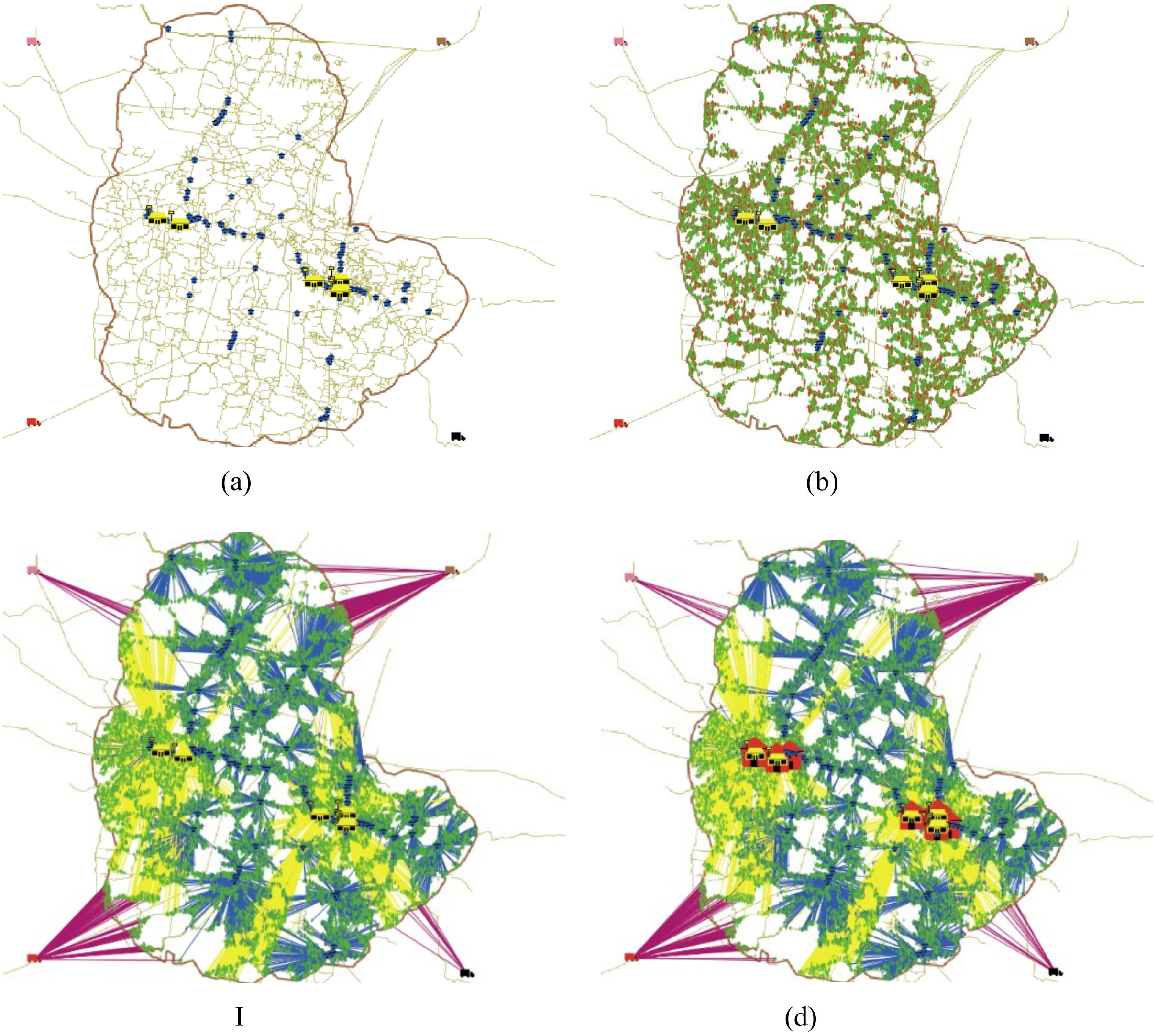

FIGURE 7 Pictorial representation of agent-based model in NetLogo: (a) simulation environment imported from GIS to ABM (b) setup procedure (c) after each iteration (d) unorganized retailers shut down during base case simulation shown in “red”.

## 3. Results

### 3.1. Conjoint Analysis Results

In order to validate the practical applicability of the proposed simulation model, interviews were carried out with 138 customers possessing prior purchase experience across all three pharmaceutical retail channels. The conjoint utilities or part-worths in Table 4 are estimated by conditional logit in Sawtooth Lighthouse Studio 9 (Orme and Chrzan 2017; Sawtooth Software 2019). The main effects are found to be all significant with $p < 0.01$.

TABLE 4 Partial utility and relative importance for customers with high degree of emergency (HE)

| Attributes | Levels | Partial Utility | Std. Error | *p*-value | Relative Importance (in %) |
|---|---|---|---|---|---|
| Price discount | Less than 10% off | -0.17793 | 0.06932 | 0.000 | 14.51 |
| | [10-20] % off | 0.05226 | 0.06805 | 0.000 | |
| | More than 20% off | 0.12566 | 0.06783 | 0.000 | |
| Quality or value | Low | -0.68490 | 0.08076 | 0.000 | 33.27 |
| | Medium | 0.00292 | 0.06937 | 0.000 | |
| | High | 0.68198 | 0.06535 | 0.000 | |
| Variety of assortment | Small | -0.15470 | 0.06985 | 0.000 | 12.82 |
| | Medium | 0.06973 | 0.06779 | 0.000 | |
| | Large | 0.08497 | 0.06709 | 0.000 | |
| Degree of personalized service | Low | -0.04340 | 0.06838 | 0.000 | 8.66 |
| | Medium | 0.03511 | 0.06796 | 0.000 | |
| | High | 0.00829 | 0.06798 | 0.000 | |
| Distance between customer and retailer | Less than or equal to 2 km | 0.66468 | 0.06577 | 0.000 | 30.74 |
| | (2-10] km | 0.07211 | 0.06948 | 0.000 | |
| | More than 10 km | -0.73678 | 0.08252 | 0.000 | |

The partial utility of each attribute level for high degree of emergency (HE) customers is shown in Table 4. The rest of the partial utilities of attribute level for medium and low degree of emergency (ME and LE, respectively) customers are shown in Tables C1 and C2, respectively, in Appendix C. In the case of partial utilities in HE, ME and LE, the customers tend to derive maximum benefit from 'more than 20% off' price discount along with 'high' quality of medicine, 'high' variety of assortment, 'medium' degree of personalized service and 'less than or equal to 2 km' distance between customer and retailer. However, the relative values of partial utilities among HE, ME, and LE are different, as shown in Figure 8. The results signify that customers value different attribute levels differently depending on the degree of emergency. Negative utility signifies that those customers are not interested (or rather hesitant) in those levels of attributes. Figure 8a shows LE customers value higher price discounts while HE customers value them the least. All the customer types do not compromise on quality, as shown in Figure 8b. Variety of assortment and personalized service are valued relatively less by all three customer types. However, the distance between customer and retailer has been highly valued by the HE customers and least by the LE customers (Figure 8e).

The relative importance of these attributes is computed from the partial utilities, as shown in Tables 4, C1, C2, and Figure 9. For HE customers, ‘quality’ is the most important attribute, followed by ‘distance between customer and retailer’, and then ‘price discount’. A similar trend can be seen in the case of ME customers. However, LE customers show a different trend where ‘quality’ is the most important attribute, followed by ‘price discount’ and then ‘distance between customer and retailer’.

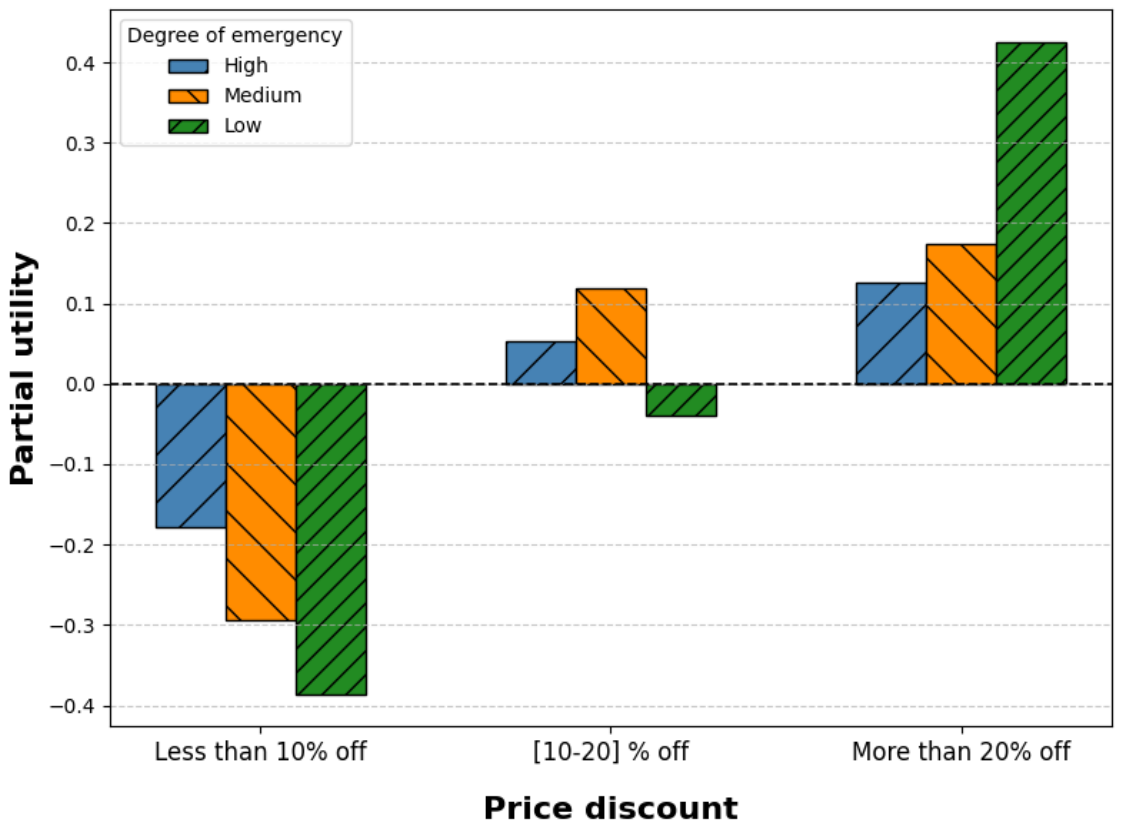

(a)

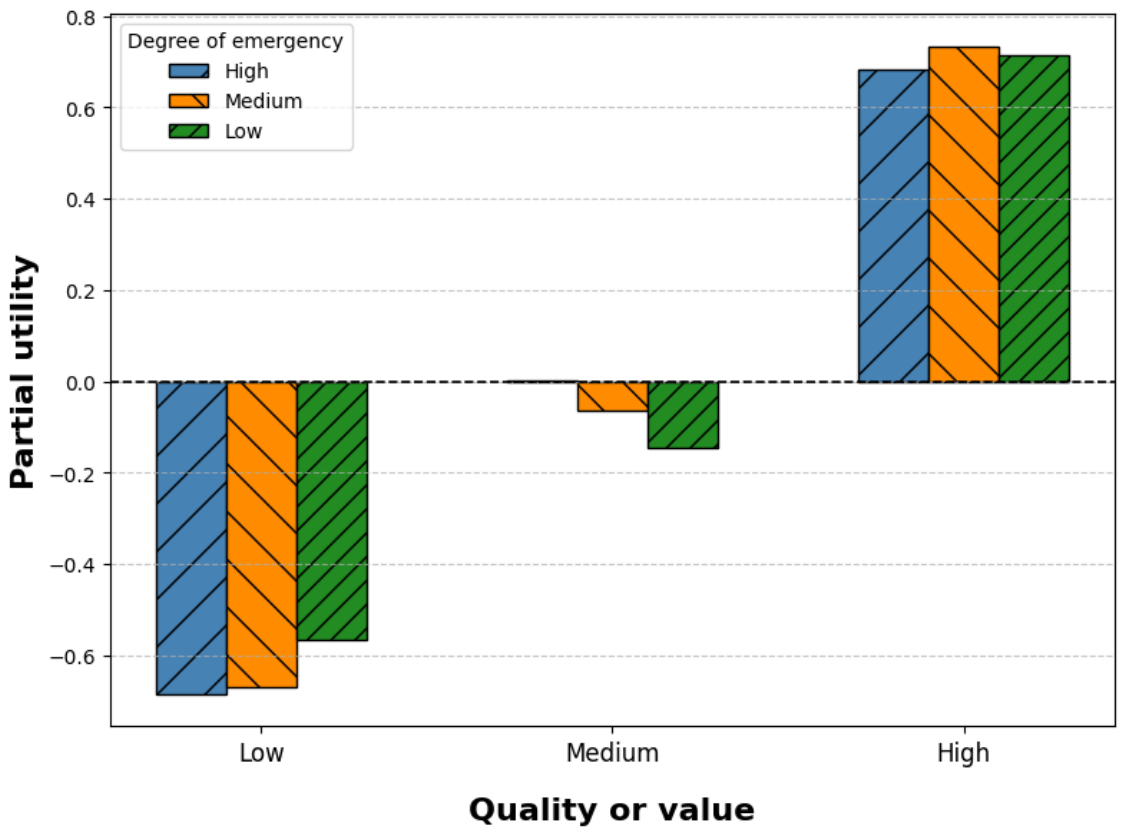

(b)

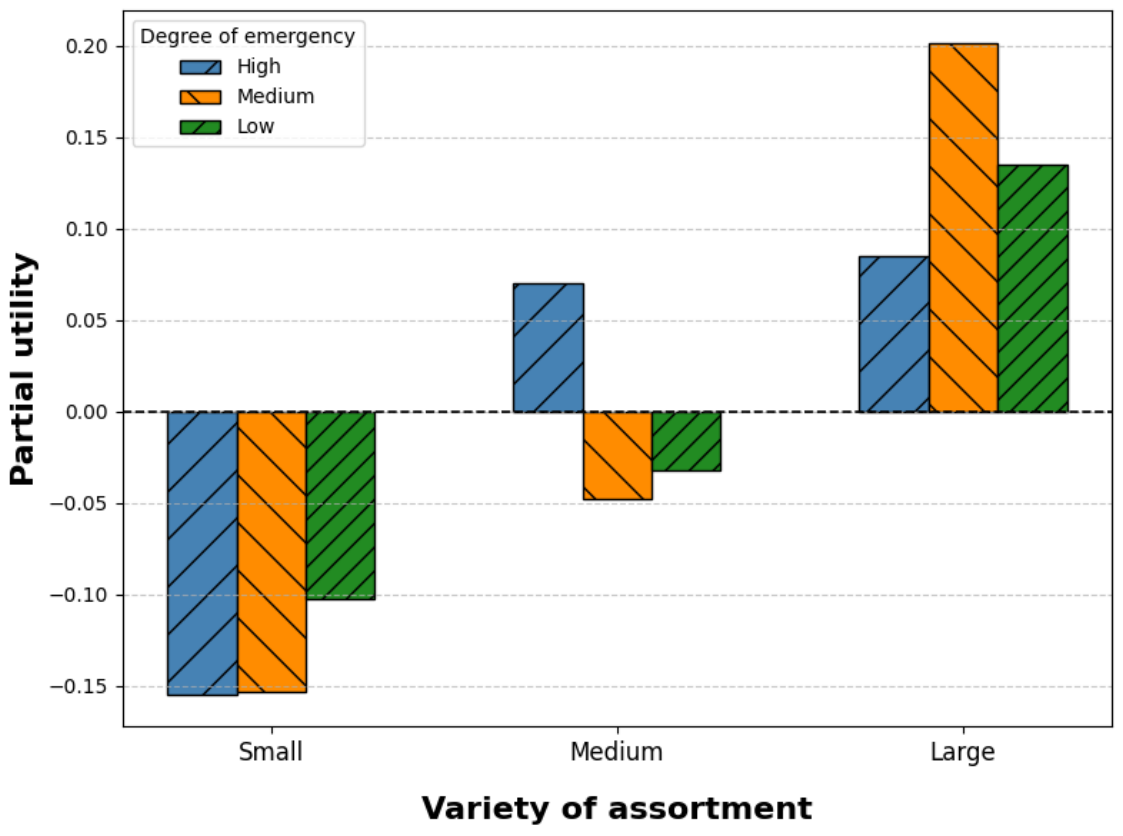

I

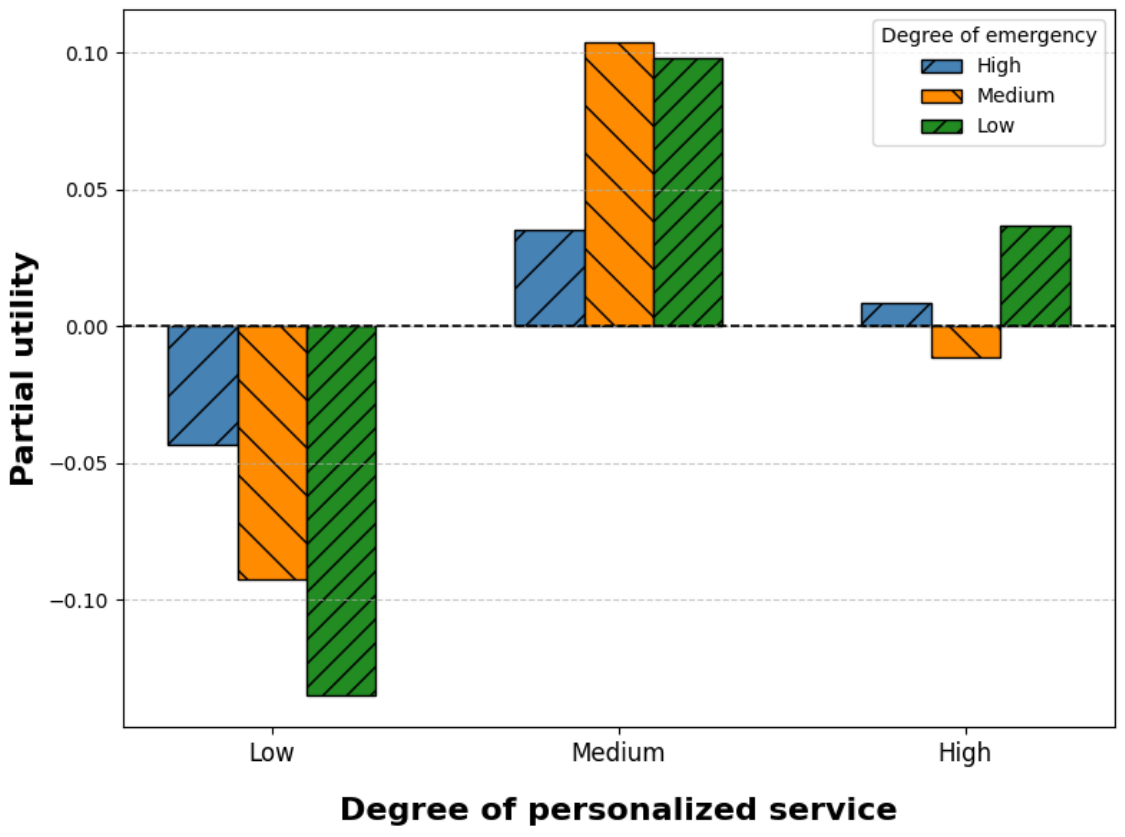

(d)

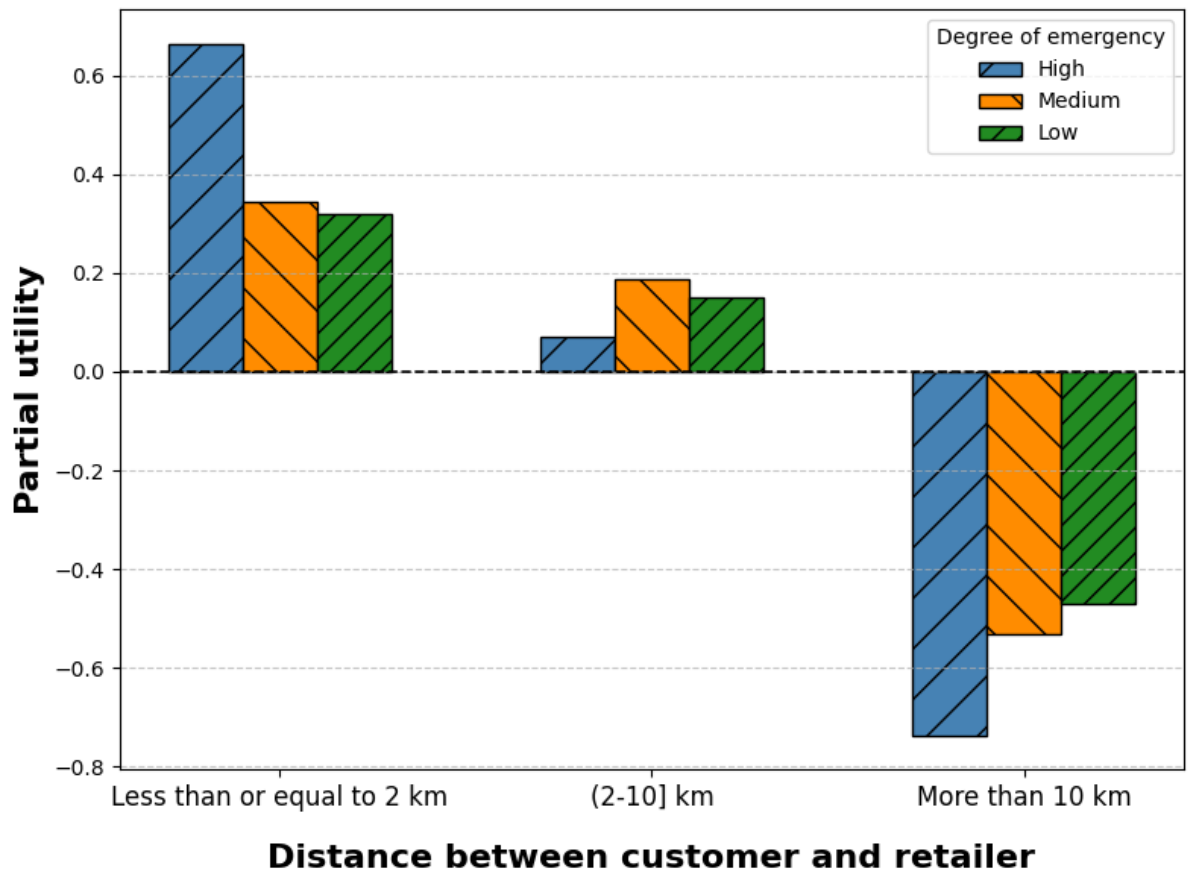

I

FIGURE 8 Partial utility of the attribute levels: (a) price discount, (b) quality or value, (c) variety of assortment, (d) degree of personalized service, I distance between customer and retailer

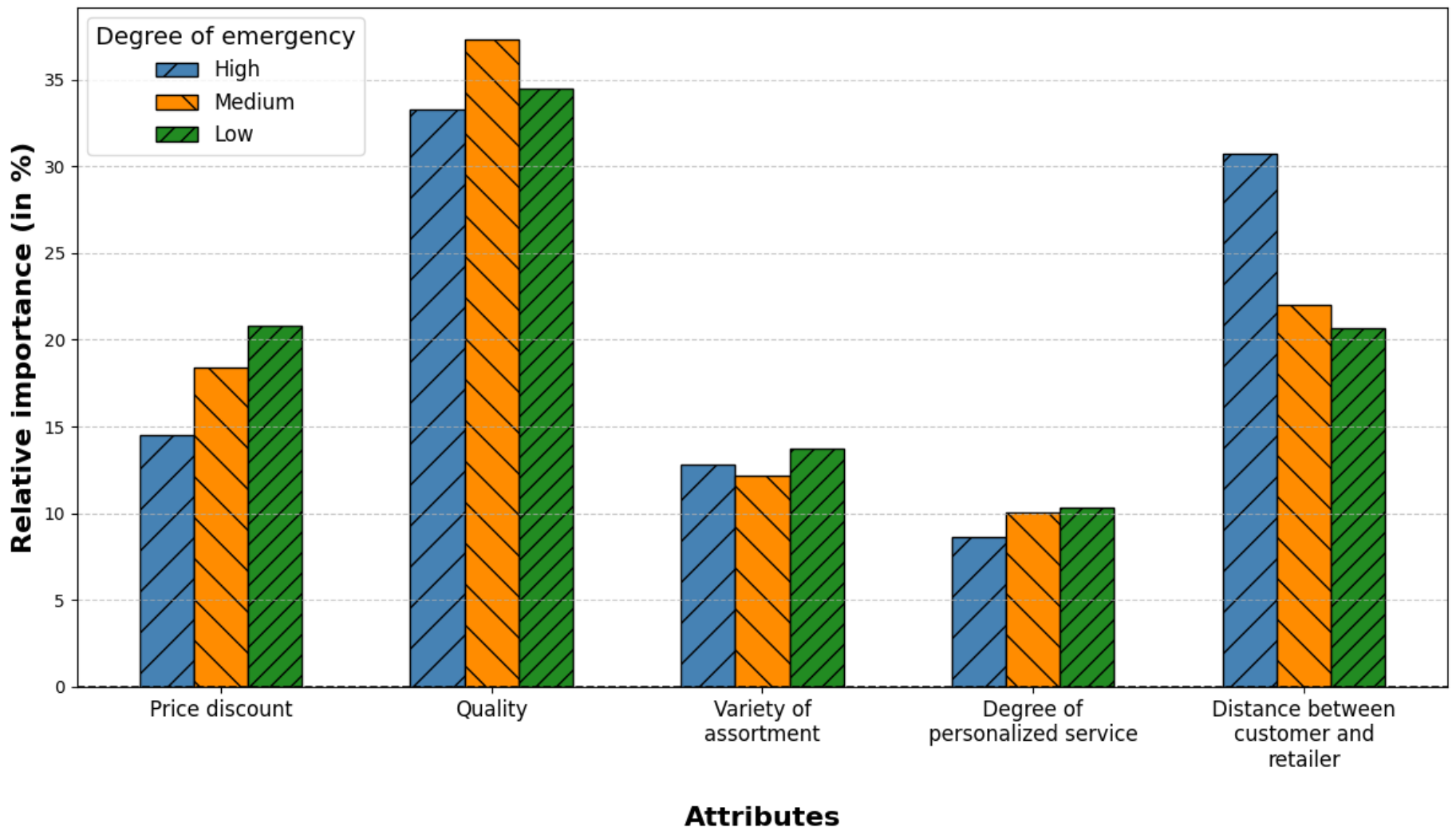

FIGURE 9 Relative importance (in %) for all the attributes

### 3.2. ABM Simulation Results

#### 3.2.1. Base-Run Results

To expand the model's explanatory scope, a standard variable configuration is established from the case study over a 312-week time horizon (Table 5) and serves as the baseline for variable variations in scenario analysis. Figure 10 provides a graphical representation of the results from the base case variable configuration. The average customer order size is Rs. 1500 obtained from the field study, the degree of mobility (n) is 0.5, and the degree of emergency ($\beta$) is uniformly distributed between 0 and 1. Unorganized pharmaceutical retailers' parameters are gross margin of 25%, total cost of Rs. 12,098 per month, and weekly minimum profits to survive of Rs. 10000. The base run results are summarized in Table 6.

TABLE 5 Base case variable configuration

| Variables | Unorganized | Organized | E-Pharmacy |
|---|---|---|---|
| Price discount (P) | 10-20% off [10-20] | More than 20% off (20-100] | More than 20% off (20-100] |
| Quality or value (ν) | High [0.7-1] | High [0.7-1] | High [0.7-1] |
| Variety of assortment (A) | Small [0-0.4) | Large [0.7-1] | Large [0.7-1] |
| Degree of personalized service (S) | High [0.7-1] | Medium [0.4-0.7) | Low [0-0.4) |

*Note:* Square bracket indicates that the value is included, and round bracket indicates not included

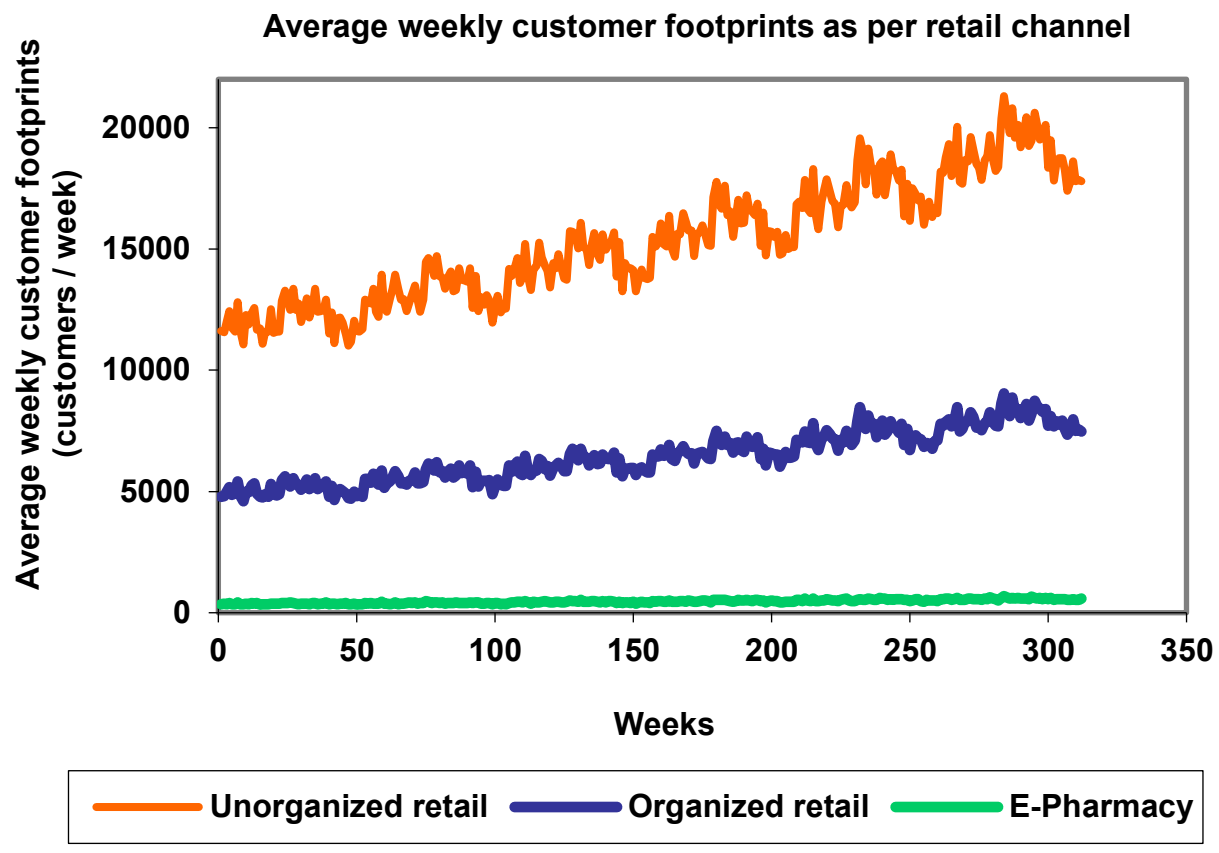

(a)

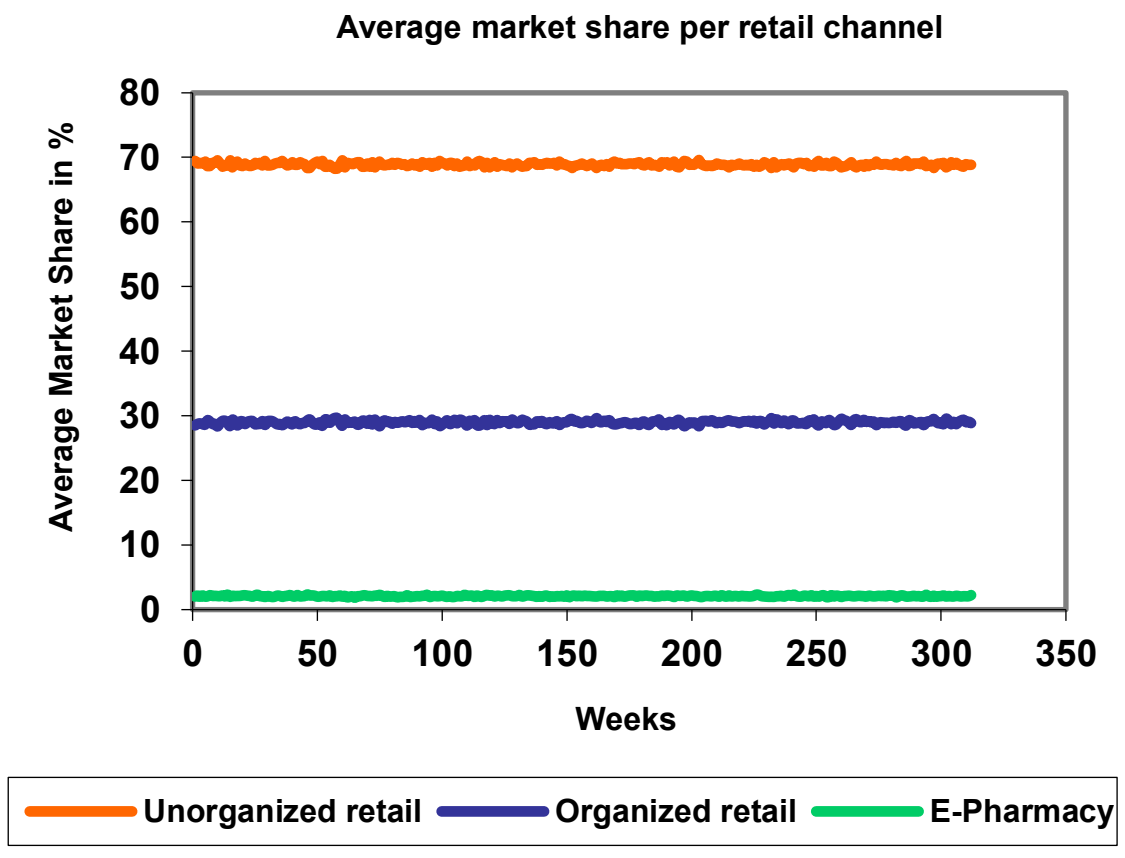

(b)

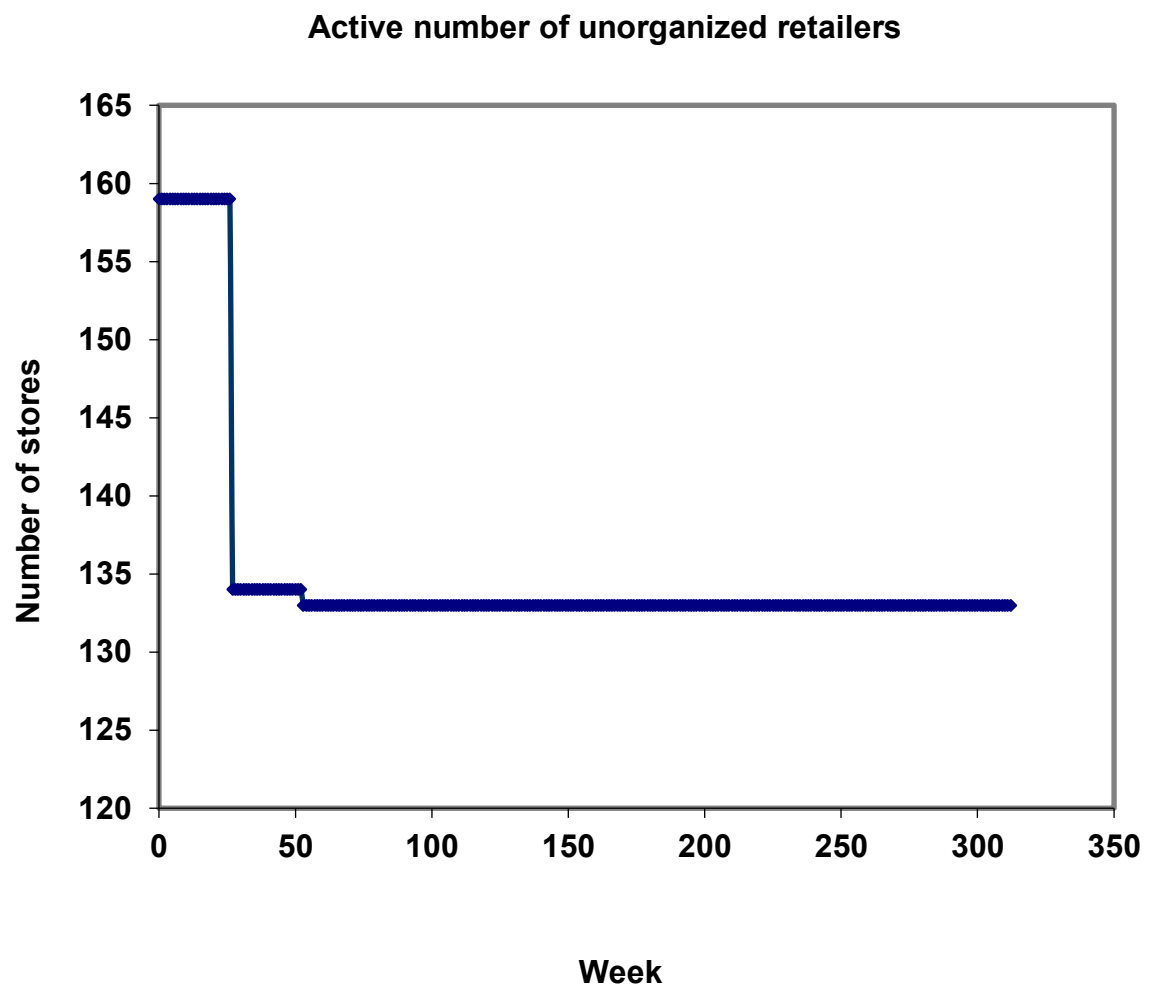

I

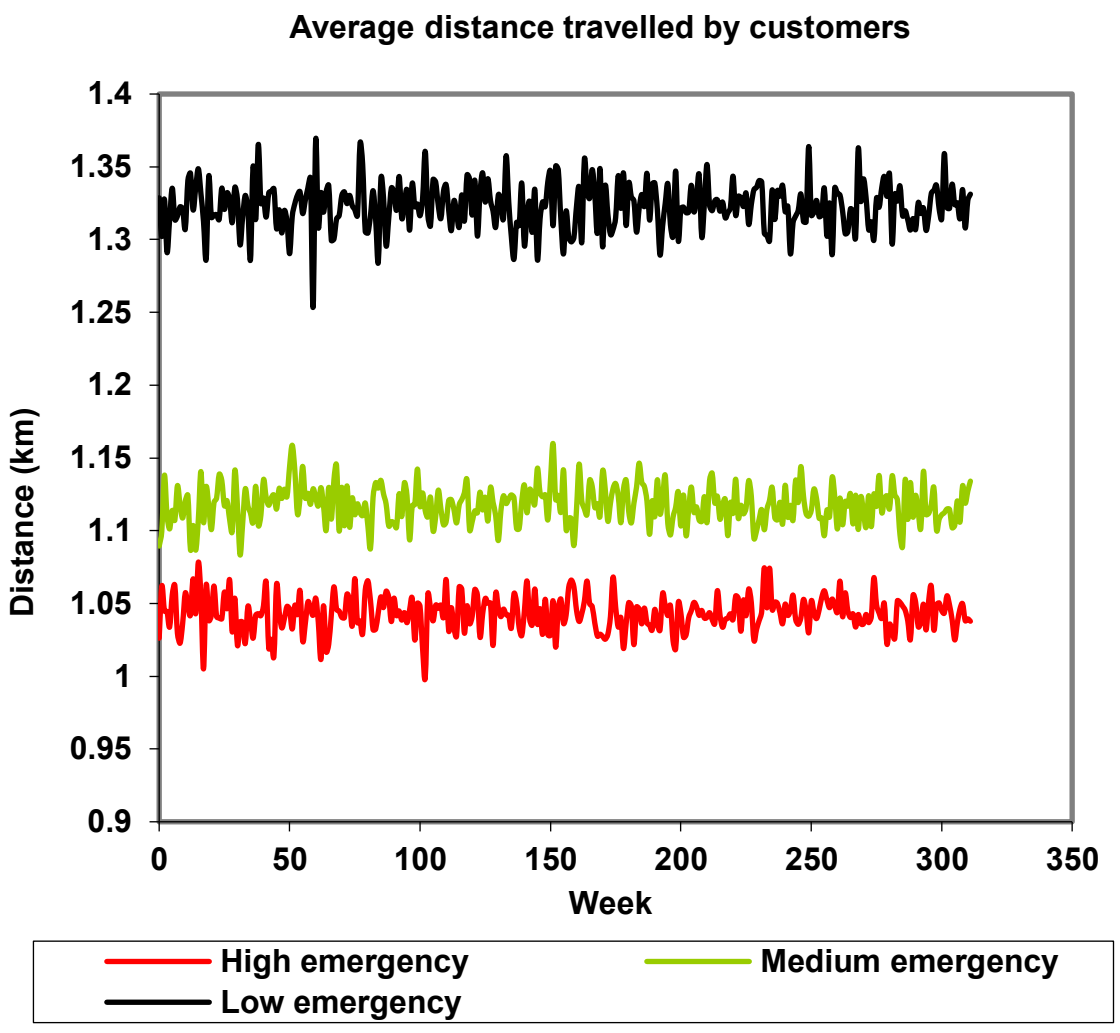

(d)

FIGURE 10 ABM base-run results: (a) average weekly customer footprints as per retail channel, (b) average market share per retail channel, (c) number of active unorganized retailer in the simulation (no. of stores surviving after shut down), (d) average distance a customer travelled.

TABLE 6 Results from base case scenario

| | Unorganized pharmaceutical retail | Organized pharmaceutical retail | E-pharmaceutical retail |
|---|---|---|---|
| Average weekly customer footprints per retail channel (customers/week) | 15331 | 6451 | 466 |
| Average Market Share per retail channel (%) | 68.91 | 28.99 | 2.10 |
| Unorganized retails shut down during simulation (no. of stores) | 26 | | |
| Average distance a high emergency customer travelled (*km*) | 1.04 | | |

### 3.2.2. Robustness of Agent-Based Model Developed: Sensitivity Analysis

Sensitivity analysis is conducted to evaluate the robustness of the ABM. Sensitivity analysis examines whether conclusions are significantly affected when assumptions are altered within a reasonable range of uncertainty (Menon and Mahanty 2015; Sterman 2000). The present work carries out the best and worst-case sensitivity analysis. All the variables are adjusted to the settings most favorable to the results one wants to test in the best-case scenario. In the worst-case situation, it is precisely the opposite. In the ABM model developed in section 2.3, the unorganized pharmaceutical retail sector is modelled in the presence of organized and e-pharmaceutical retailers in the market. The average customer footprint at the unorganized retail channel is expected to fluctuate with the changes in discount, assortment, and personalized service provided by the respective retail channels.

The base case scenario includes the values used for the base run of the ABM model. The best-case scenario for unorganized retailers might assume higher attribute values for unorganized retailers, while lesser competition from the competitors, i.e., organized and e-pharmaceutical retailers, assumes lower attribute values. The worst-case scenario for unorganized retailers is to have higher competition from organized and e-pharmaceutical retailers. In the base case, the unorganized discount is 10% to 20% while organized and e-pharmacy discounts are both more than 20%; variety of assortment of unorganized is small while the same for organized and e-pharmacy are large; and degree of personalized service is high for unorganized while it is medium for organized and low for e-pharmacy. The best-case scenario assumes price discount is more than 20% for unorganized, less than 10% for both organized and e-pharmacy; product quality is high for all the retailers; variety of assortment is large for all the three retailers; degree of personalized service is high for unorganized, low for both organized and e-pharmacy. While the worst-case scenario assumes price discount is less than 10% for unorganized, more than 20% for both organized and e-pharmacy; product quality is high for all the retailers; variety of assortment is small for unorganized but large for both organized and e-pharmacy; degree of personalized service is low for unorganized, high for both organized and e-pharmacy. Figure 11 compares the best-case and worst-case scenarios relative to the base case.

In the base case, the average customer footprint at the unorganized retail channel falls between that of the best-case and worst-case. The implications for the average customer footprint at the unorganized retail channel are very different for the best and the worst cases. In the best case, the high discount at unorganized retailers combined with a large assortment and a high degree of personalized service attracts more customers to the unorganized retailers. In contrast, the competitors, i.e., organized and e-pharmaceutical retailers, draw fewer customers with their less attractive low discount and low degree of personalized service. On the other hand, low discounts at unorganized retailers combined with a small assortment and a low degree of personalized service decrease customers' footprints to the unorganized retailers in the worst-case scenario.

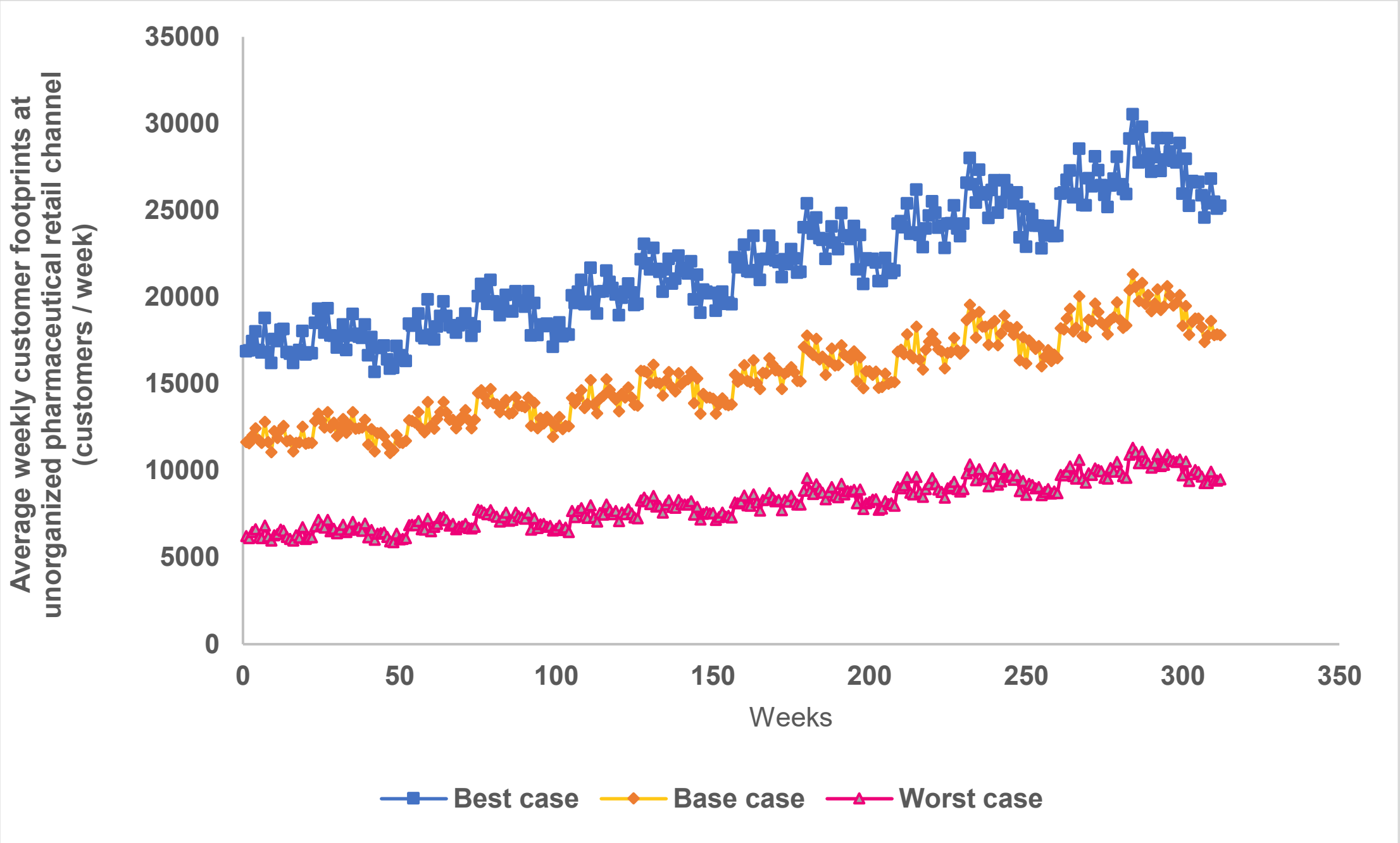

FIGURE 11 Base, best and worst-case sensitivity analysis for ABM model

### 3.2.3. ABM Experimental Runs and ANOVA Results for 'Price Discount'

As per the experimental design in Table D1 in Appendix D, the ABM simulation parameters are set for price discount, and a total of 27 experiments are conducted. Table D1 also shows the summary of experimental results from the agent-based model for price discount. The ABM experimental results for customer footprints, market share, and number of retail shutdowns are shown in Appendix E, Section E1. Further, these results from the agent-based model experimental runs are utilized to conduct ANOVA to find the most significant influencing variable (unorganized discount, organized discount, e-pharmacy discount) for the output responses of customer footprints, market share, and shutdown of the retailers.

#### 3.2.3.1. Effect of Unorganized Discount, Organized Discount, and E-Pharmacy Discount on the Average Weekly Customer Footprint and Market Share of Unorganized Pharmaceutical Retailers

Table 7 shows the ANOVA results for the average weekly customer footprint at the unorganized pharmaceutical retailer. This analysis is done at a level of 95% confidence level. The percentage contribution (%) of individual variables for the final response can be measured for all variables by the ratio of the individual sum of squares of variables to the total sum of squares.

TABLE 7 ANOVA for average weekly customer footprint at unorganized pharmaceutical retailer

| **Source** | **DF** | **Adj SS** | **Adj MS** | **F-Value** | ***p*-value** | **Contribution (%)** |
|---|---|---|---|---|---|---|
| Unorganized discount | 2 | 206219822 | 103109911 | 109.60 | 0.000 | 55.87 |
| Organized discount | 2 | 144453168 | 72226584 | 76.78 | 0.000 | 39.14 |
| E-Pharmacy discount | 2 | 3237665 | 1618832 | 1.72 | 0.239 | 0.88 |
| Unorganized discount*Organized discount | 4 | 1993580 | 498395 | 0.53 | 0.718 | 0.54 |
| Unorganized discount*E-Pharmacy discount | 4 | 1010083 | 252521 | 0.27 | 0.890 | 0.27 |
| Organized discount*E-Pharmacy discount | 4 | 4631193 | 1157798 | 1.23 | 0.371 | 1.25 |
| Error | 8 | 7525979 | 940747 | | | 2.04 |
| Total | 26 | 369071490 | | | | |
| R-Sq = 97.96% R-Sq (Adj) = 97.37% | | | | | | |

It is evident from Table 7 that unorganized discount and organized discount are statistically significant with *p*-values less than 0.05. This implies that the variables unorganized discount and organized discount have a significant influence on the average weekly customer footprint at the unorganized pharmaceutical retailer. The percentage of contribution for each factor is shown in Table 7. It is observed that the unorganized discount has the highest effect on the response. The interaction effects are not statistically significant (Figure D1).

Table 8 shows the ANOVA table for the average market share of the unorganized pharmaceutical retailers. Table 8 shows that the *p*-values for unorganized discount and organized discount are less than 0.05. This implies that the variables, unorganized discount and organized discount, have a significant influence on the average market share of the unorganized pharmaceutical retailers. It is observed that the unorganized discount has the highest effect on the response. No interaction effects are found significant (Figure D2).

TABLE 8 ANOVA for average market share of the unorganized pharmaceutical retailers

| Source | DF | Adj SS | Adj MS | F-Value | P-Value | Contribution (%) |
|---|---|---|---|---|---|---|
| Unorganized discount | 2 | 4054.71 | 2027.36 | 111.69 | 0.000 | 55.40 |
| Organized discount | 2 | 2905.00 | 1452.50 | 80.02 | 0.000 | 39.69 |
| E-Pharmacy discount | 2 | 64.79 | 32.40 | 1.78 | 0.229 | 0.88 |
| Unorganized discount*Organized discount | 4 | 38.43 | 9.61 | 0.53 | 0.718 | 0.52 |
| Unorganized discount*E-Pharmacy discount | 4 | 19.35 | 4.84 | 0.27 | 0.892 | 0.26 |
| Organized discount*E-Pharmacy discount | 4 | 90.76 | 22.69 | 1.25 | 0.364 | 1.24 |
| Error | 8 | 145.22 | 18.15 | | | 1.98 |
| Total | 26 | 7318.26 | | | | |
| R-Sq = 98.02% | | | R-Sq (Adj) = 93.55% | | | |

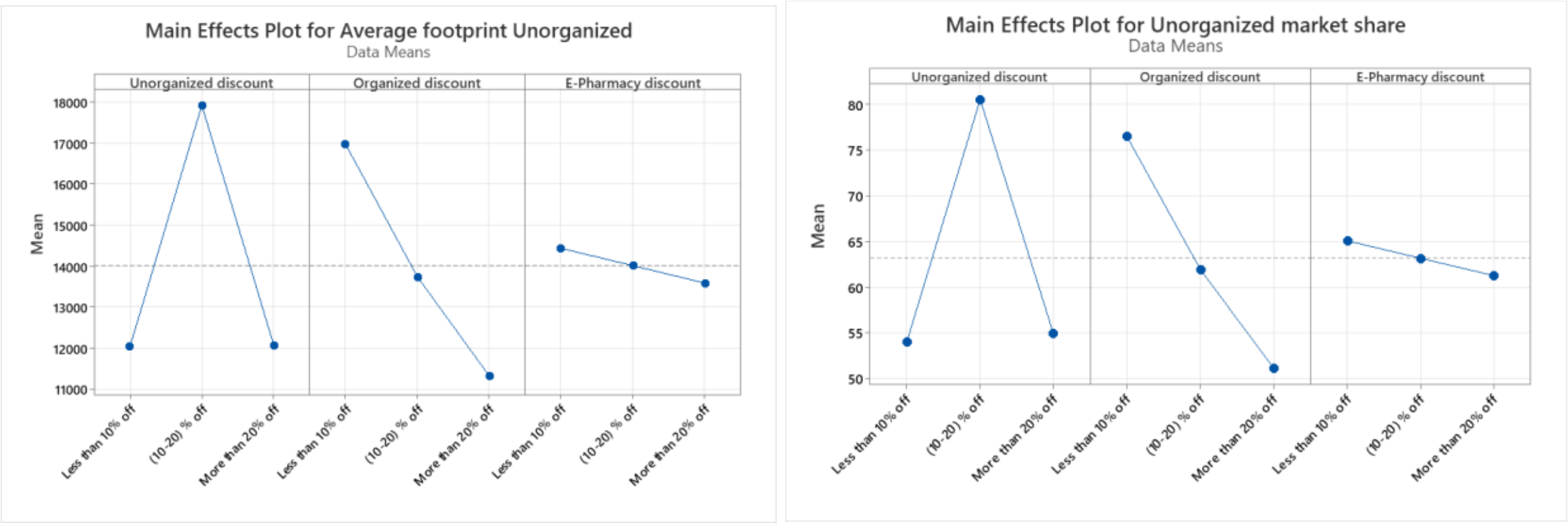

(a) (b)

FIGURE 12 Main effect plots for (a) average customer footprint, (b) average market share of the unorganized pharmaceutical retailer

The main effects plots for average customer footprint and average market share of the unorganized pharmaceutical retailer are shown in Figure 12. Further, the ANOVA results for the number of unorganized retailers shut down during simulation (Table D2), average weekly customer footprints at organized (Table D3) and e-pharmacy (Table D5) and, average market share of organized (Table D4) and e-pharmacy (Table D6) are presented in the appendix D.

### 3.2.4 ABM Experimental Runs and ANOVA Results for 'Product Quality'

As per the experimental design in Table D7 in Appendix D, the ABM simulation parameters are set for product quality, and a total of 27 experiments are conducted. Table D7 also summarizes experimental results from the agent-based model for product quality. The ABM experimental results for customer footprints, market share, and number of retail shutdowns are shown in Appendix E, Section E2. Further, these results from the agent-based model experimental runs are utilized to conduct ANOVA to find the most significant influencing variable (unorganized quality, organized quality, e-pharmacy quality) for the output responses of customer footprints, market share, and shutdown of the retailers.

#### 3.2.4.1. Effect of Unorganized Quality, Organized Quality, and E-Pharmacy Quality on Average Weekly Customer Footprint and Market Share of Unorganized Pharmaceutical Retailers

The ANOVA table for average weekly customer footprint at the unorganized pharmaceutical retailer is in Table 9. It shows that the *p*-values for unorganized, organized, and e-pharmacy quality are less than 0.05 and hence statistically significant. This implies that all three variables significantly influence the average weekly customer footprint at the unorganized pharmaceutical retailer. The percentage of contribution for each factor is shown in Table 9. It is observed that the unorganized quality has the highest effect on the response. The interaction between unorganized quality and organized quality is found to be significant (Figure D8).

TABLE 9 ANOVA table for average weekly customer footprint at unorganized pharmaceutical retailer

| **Source** | **DF** | **Adj SS** | **Adj MS** | **F-Value** | **P-Value** | **Contribution (%)** |
|---|---|---|---|---|---|---|
| Unorganized quality | 2 | 1718491901 | 859245950 | 829.22 | 0.000 | 87.71 |
| Organized quality | 2 | 132477470 | 66238735 | 63.92 | 0.000 | 6.76 |
| E-Pharmacy quality | 2 | 10659418 | 5329709 | 5.14 | 0.037 | 0.54 |
| Unorganized quality*Organized quality | 4 | 72482538 | 18120634 | 17.49 | 0.001 | 3.70 |
| Unorganized quality*E-Pharmacy quality | 4 | 4935459 | 1233865 | 1.19 | 0.385 | 0.25 |
| Organized quality*E-Pharmacy quality | 4 | 11865202 | 2966300 | 2.86 | 0.096 | 0.61 |
| Error | 8 | 8289646 | 1036206 | | | 0.42 |
| Total | 26 | 1959201633 | | | | |
| R-Sq = 99.58% R-Sq (Adj) = 98.62% | | | | | | |

Table 10 shows ANOVA results for the average market share of the unorganized pharmaceutical retailers. Table 10 shows that the *p*-values for unorganized quality, organized quality, and e-pharmacy quality are less than 0.05 and are statistically significant. This implies that all three variables have a significant influence on the average market share of the unorganized pharmaceutical retailers. It is observed that unorganized quality has the highest effect on the response. The interaction between unorganized quality and organized quality is also found to be significant (Figure D9).

TABLE 10 ANOVA for average market share of the unorganized pharmaceutical retailers

| **Source** | **DF** | **Adj SS** | **Adj MS** | **F-Value** | **P-Value** | **Contribution (%)** |
|---|---|---|---|---|---|---|
| Unorganized quality | 2 | 34715.5 | 17357.8 | 830.67 | 0.000 | 87.71 |
| Organized quality | 2 | 2675.4 | 1337.7 | 64.02 | 0.000 | 6.76 |
| E-Pharmacy quality | 2 | 215.9 | 107.9 | 5.17 | 0.036 | 0.54 |
| Unorganized quality*Organized quality | 4 | 1463.8 | 366.0 | 17.51 | 0.001 | 3.69 |
| Unorganized quality*E-Pharmacy quality | 4 | 99.9 | 25.0 | 1.20 | 0.383 | 0.25 |
| Organized quality*E-Pharmacy quality | 4 | 240.5 | 60.1 | 2.88 | 0.095 | 0.61 |
| Error | 8 | 167.2 | 20.9 | | | 0.42 |
| Total | 26 | 39578.3 | | | | |

R-Sq = 99.58% R-Sq (Adj) = 98.63%

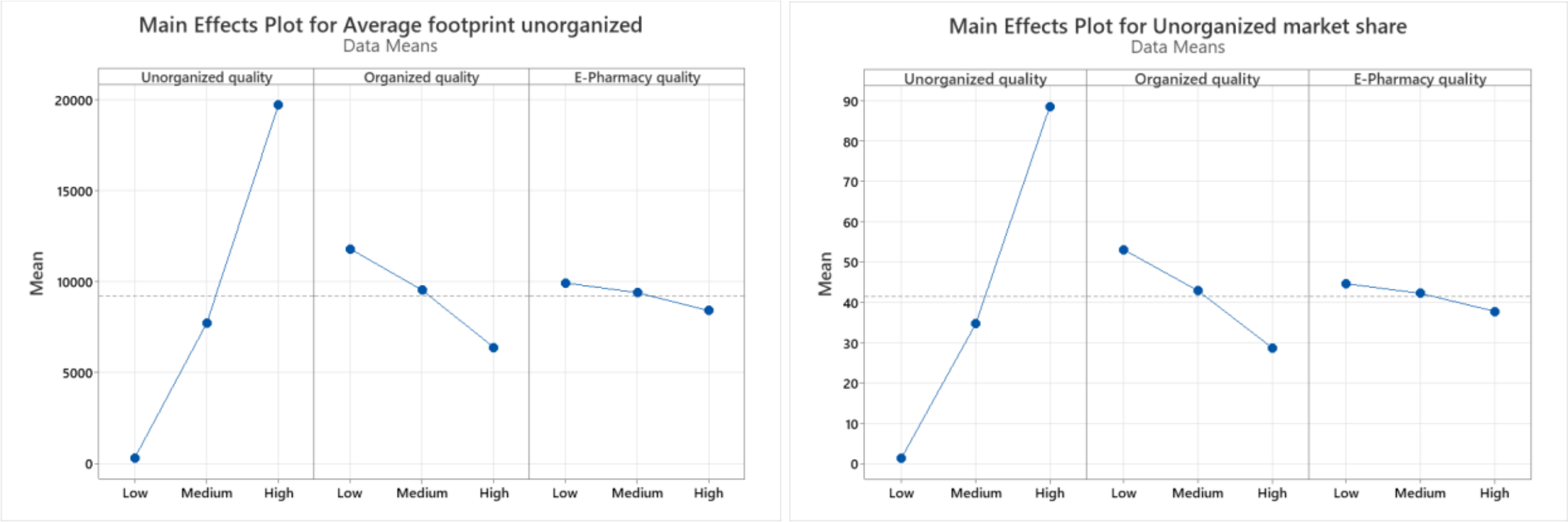

(a) (b)

FIGURE 13 Main effect and interaction effect plots for (a) average footprint at unorganized retailer, (b) average market share of the unorganized pharmaceutical retailer

The main effects plots for average customer footprint and average market share of the unorganized pharmaceutical retailer are shown in Figure 13. From the main effect plot in Figure 13(a), it is evident that with an increase in organized quality and e-pharmacy quality, there is a decrease in the average footprint at the unorganized retailer. It is also evident that with an increase in unorganized quality, there is an increase in the average footprint at the unorganized retailer. It can be concluded that the average footprint at the unorganized retailer can be maximized by keeping unorganized quality at a high level while keeping organized quality and e-pharmacy quality at a low level. From the main effect plot in Figure 13(b), it is evident that with an increase in organized quality and e-pharmacy quality, there is a decrease in the average market share of the unorganized retailers. It is also evident that with an increase in unorganized quality, there is an increase in the average market share of unorganized retailers. It can be concluded that the average market share of the unorganized retailers can be maximized by keeping unorganized quality at a high level while keeping organized quality and e-pharmacy quality at a low level. Further, the ANOVA results for the number of unorganized retailers shut down during simulation (Table D8); average weekly customer footprints at organized (Table D9) and e-pharmacy (Table D11); average market share of organized (Table D10) and e-pharmacy (Table D12); are presented in the Appendix D.

**4. Discussions**

The conjoint analysis, ABM simulation, and analysis of variance results provide a number of insights into the competitive interactions between unorganized, organized, and online e-pharmaceutical retailing channels, which are discussed below.

From conjoint analysis, it is observed that customers with high emergency prefer high-quality products from nearby pharmaceutical retailers without any influence from price discounts. In contrast, customers with low emergency look for high-quality products at better price discounts rather than the distance between the customer and the retailer. The medium emergency customers give maximum importance to product quality compared to high and low emergency customers. It is also found that high emergency customers give less importance to 'variety of assortment' than low emergency customers (Figure 9). This is a counterintuitive behavior because variety of assortment is expected to be more important to high emergency customers as it is more likely to get the exact medicine in the least possible time. This counterintuitiveness shows that the high emergency customers are willing to accept substitute products at nearby retailers with medium assortment (Figure 8c) rather than sacrificing the time and cost to travel to a distant retailer with a higher assortment.

According to the agent-based model base-run results, while the average customer footprint shows an increasing trend over time (Figure 10a), the average market share of the retail channels is found to be stable over time (Figure 10b). The reason behind this behavior is the stable customer base for each of the three retail channels, as long as the retailers do not change their attribute values. The agent-based model base-run results evidenced that initially, some of the unorganized retailers will shut down their business due to a lower profit margin over a period of time, and eventually, the market will become stable (Figure 10c). It is also found from the base-run model results (Figure 7d) that the unorganized stores that have shut down mostly fall within the catchment area of an organized retailer. These results are in line with previous studies of Kohli and Bhagwati (2012), Joseph et al. (2008), and Jerath et. al (2016). It is found from the base-run ABM simulation that the average distances travelled by the customers with different degrees of emergencies are significantly different. Moreover, the ABM base-run results show how customers' purchasing behavior is different among different degrees of emergencies (Figure 10d). High emergency customers travel the least to minimize the time needed to meet their demands. On the other hand, the low emergency customers travel significantly further to find an appropriate deal on the product. The medium emergency customers travel more than high emergency customers but less than low emergency customers.

From the ABM model experiments (Tables 7 and 8), it is found that the price discounts given by unorganized and organized retailers have a significant influence on the average weekly customer footprint and average market share of the unorganized pharmaceutical retailer. It is observed that the unorganized discount has the highest effect (i.e., 'contribution' in Tables 7 and 8) on the customer footprint and market share. The interaction effects are not statistically significant. It is also evident from the main effects plots in Figure 12 that with an increase in unorganized discount from low to medium level, there is an increase in the average footprint and average market share of the unorganized retailer. But as the unorganized discount increases further from medium to high level, a counterintuitive behavior is evidenced where the customer footprint and market share at the unorganized retailer decrease drastically.

Also, a counterintuitive behavior is evidenced from the main effects plots in Figure D3 that as the unorganized discount increases from medium to high level, the number of unorganized retailers shutting down increases drastically. These counterintuitive behaviors regarding the fall in customer footprints and retail shutdowns are attributed to the fact that as the unorganized retailers' higher price discounts lead to lower profit margins that do not tally with their operational costs, which finally results in the shutdown of their retail stores (Appendix E, Section E1.4). Moreover, it is also observed that the unorganized price discount has the highest effect (i.e., 'contribution' in Table D2) on the shutdown of unorganized retailers. Hence, these results suggest that there is a tipping point in the price discount given by unorganized retailers, beyond which significant changes in customer footprint, market share, and the shutting down of unorganized retailers are observed. This implies that an unorganized retailer's own strategy will cause more harm to themselves than from the competitors. Interestingly, from agent-based model experiments, in a competitive environment, it is evidenced that the price discounts by the e-pharmaceutical retailers have no effect on customer footprints and market share of the unorganized pharmaceutical retailers (Tables 7 and 8) and hence do not contribute to the shutting down of unorganized retailers. For unorganized discounts, the medium level shows the highest response value. It can be concluded that the average footprint and hence the market share of unorganized retailers can be maximized by keeping the unorganized discount at a medium level while the organized discount and e-pharmacy discount are at a low level.

From the experiments on the ABM model (Tables 9 and 10), it is found that the product quality offered by unorganized, organized, and e-pharmaceutical retailers has a significant influence on the average weekly customer footprint and average market share of unorganized pharmaceutical retailers. This is because customers give utmost importance to product quality (Figure 9). Any changes in product quality offered by retail channels will trigger customers to switch their preferred retailers to the ones with higher quality. It is observed that the product quality offered by unorganized retailers has the highest effect (i.e., 'contribution' in Tables 9 and 10) on the customer footprint and market share. It is also found that the interaction between unorganized quality and organized quality is statistically significant (Tables 9 and 10). The interaction plot (Figures D8 and D9 in Appendix D) shows that if organized product quality increases from medium to high, keeping unorganized product quality at a high level, a reduction in customer footprint and market share in the unorganized retailer is observed. Hence, the interaction effect nullifies the expected improvements in the customer footprints and market share of the unorganized retailers, even after a high product quality is maintained by the unorganized retailers.

The experiments with the ABM model show that the product quality offered by unorganized and organized retailers has a significant influence on the number of unorganized retailers shut down (Table D8 in Appendix D). Interestingly, it is found that the product quality offered by e-pharmaceutical retailers does not significantly affect the shutdown of unorganized retailers. It is also observed that the unorganized product quality has the highest effect (i.e., 'contribution' in Table D8) on the shutdown of unorganized retailers. From the main effect plot (Figure D10), it is evident that with an increase in organized quality, there is an increase in the number of unorganized retailers shut down. This is because customers switching to organized pharmaceutical retailers with higher quality products will decrease the customer footprints at the unorganized retailers, eventually resulting in the shutdown of these unorganized retailers due to lower profit margins. The main effect plot also shows that with an increase in unorganized quality, there is a significant decrease in the number of unorganized retailers shut down.

## 5. Managerial Implications

This study offers several practical insights that can guide decision-making for managers and policy makers across organized, unorganized, and e-pharmaceutical retail channels. First, the conjoint analysis shows that customer's degree of emergency strongly influences purchasing behavior. High emergency customers prioritize proximity and quality over discounts, whereas low emergency customers trade off distance for price savings. In this scenario, the organized retailers may benefit from implementing geographical price discounts (or location-specific price discounts), fewer discounts for convenience purchases (high emergency) at stores close to hospitals or dense urban areas, coupled with dynamic targeted discount promotions for distant, price-sensitive (low emergency) customer segments. For unorganized retailers, direct competition on price discounts may be unsustainable. Instead, keeping steady prices on emergency items in smaller lots can attract nearby high emergency consumers. In contrast, limited promotion can still attract relatively less urgent customers (medium and low emergency) within nearby neighborhoods.

Second, the results indicated that e-pharmacy price discounts have little impact on unorganized retailers' customer base or survival. This finding implies that since price promotions alone do not pull customers away from the local unorganized retailers, e-pharmacy companies may direct their promotional budget to non-price incentives like product bundling or faster delivery by introducing quick commerce or a hyperlocal delivery model. On the other hand, unorganized retailers can focus less on matching online price discounts and capitalize more on reinforcing their in-store strengths like instant access to products and the human touch of salespeople, which are still appreciated by a large chunk of their customer base.

Third, contrary to intuitive expectations, the study reveals that high emergency customers value immediate access to emergency drugs over a variety of assortments. They will promptly accept substitute products if a neighboring retailer stocks essential items. To capitalize on this behavior, unorganized and organized store managers can streamline operations by dedicating a small, visible "emergency shelf" stocked exclusively with high-turnover SKUs, reducing inventory and shelf space while ensuring critical products are always in stock. For medium emergency customers, product quality matters most. Even with limited variety, a targeted assortment of premium brands may work better if labeled clearly to accelerate customers' decision-making.

Another important implication is to use dynamic-store-network decisions based on real-time analytics. The proposed agent-based model predicts how price discount fluctuations by organized and unorganized retailers trigger unorganized store closures. To manage this risk, the independent unorganized retailers may maintain a basic spreadsheet or dashboard tracking nearby organized chain discounts. Upon identifying a surge in price discount within the neighborhood zone, unorganized retailers may immediately engage in light countermeasures like limited-time "thank-you" offers for loyal customers or temporary collaborations with e-pharmacies offering Buy Online, Pick Up in Store (BOPIS) fulfillment services. These quick, inexpensive measures may steady customer flow and postpone market exit under conditions of increased price competition.

Finally, behavioral segmentation using conjoint analysis can help improve strategic expansion planning. The results show that customers' emergency level affects both their price sensitivity and spatial preferences. Thus, the decisions to enter a new market should be guided by the local distribution of emergency profiles. Retail strategic planners may benefit from mapping local demographic indicators such as population age profiles, disease profiles, and proximity to healthcare infrastructures onto geographic information systems (GIS) to decide where to open smaller express stores versus larger full-size stores. Likewise, e-pharmacy logistics managers can utilize this GIS-based customer emergency distribution data to optimize the location of dark stores in areas of low emergency to capture discount-minded customers without cannibalizing traffic from unorganized and organized physical stores.

These implications use the agent-based model's predictive capability and the conjoint analysis's behavioral knowledge to provide actionable plans for unorganized, organized, and e-pharmaceutical retailers to calibrate price discounts, product quality, assortment, location, and service. By coordinating their decision as per customer requirements and local conditions, retailers in all formats can enhance resilience and competitiveness in a highly dynamic market.

**6. Conclusions**

The study explored the competitive dynamics among unorganized, organized and e-pharmaceutical retail channels by investigating the impact of customer preferences and various market factors on customer footprint, market share and shutdown of unorganized retailers in the Indian pharmaceutical retail sector. For that purpose, a block in the eastern state of India has been chosen as the study area. After a thorough literature review and exhaustive qualitative interviews, several factors that influence customers' channel choice behavior are identified, and subsequently, the utility functions of the customers are proposed using conjoint analysis. Further, an empirically grounded agent-based modeling of unorganized, organized and e-pharmaceutical retailers is carried out, followed by a sensitivity analysis and ANOVA. The present study has contributed the following.

1. The initial study in the Indian context that attempts to model the influence of systemic variables on the Indian pharmaceutical retail sector by considering geo-spatial attributes of customers and three types of retailers in a single framework. Hence, the study addresses the gap in existing literature which often focuses only on dyadic (two-channel) comparisons.
2. The integrated empirically grounded agent-based model framework based on field study, conjoint analysis and design of experiments proposes a new approach for understanding retail channel switching dynamics of heterogeneous consumers.
3. The present study identifies a tipping point in the price discount strategy (beyond 20% price discounts as evidenced from Figures 12, and D3 in appendix), from which the unorganized retailers begin shutting down at a fast pace, exposing the vulnerabilities of these small independent unorganized retailers to sustained price pressure. It provides new insights into the dynamics of the unorganized pharmaceutical retailer sector in the Indian context.
4. The study has modelled customer heterogeneity according to their degree of emergency. The analysis reveals the effect of the degree of emergency on the customers' choice of retail channels, price discount sensitivity, and the success of various retail strategies. This unique approach has proposed the need for a local distribution of emergency customer profiles for efficient product promotions and strategic expansion planning.

5. The study aids in understanding the levers for policy design towards improving the competition dynamics among retail channels in the pharmaceutical retail sector in India.

From a practical perspective, the findings offer clear, non-obvious guidance on issues that managers and policymakers face: when to vary pricing based on proximity and urgency; how to streamline emergency assortments; and how to anticipate vulnerable outlet closures based on competitor actions. The insights are actionable, especially in public-health-linked retail ecosystems where pharmacies are critical to urgent medicine access.

The limitations of the present study need to be addressed for future research. The results of the present study are based on specific assumptions regarding customer behavior and market factors, which may not be generalizable across all geographical regions or market conditions. However, the framework used in this study is scalable and could be applied to any specific study area of the world with modifications. The proposed framework can also be advanced to model other necessary sectors (e.g., Fast-Moving Consumer Goods (FMCG), perishable food and fresh produce, emergency home services, etc.) to generalize these dynamics in these different contexts. Future studies could examine changing regulations like constraints imposed upon discounting, and the impact of supply chain disruptions, which could provide policy insights towards an equilibrium market scenario. Additionally, a deeper exploration of customer preferences in different demographic segments, like different income groups, would offer valuable insights as their preferences for each of the attributes might be different. The agent-based model simulation shows that the average market share of the retail channels to be stable over time, which is attributed to the presence of a 'loyal customer' pool in the Indian pharmaceutical sector. But whether this is based on customer's behavioral loyalty or attitudinal loyalty or both needs further investigation, as it will aid the policy makers to devise different customer retention strategies for the long run. To make the consumer behavior data more realistic, future researchers can advance this work by introducing an adaptive learning mechanism among agents with repeated experience or word-of-mouth driven feedback loops for dynamic preference evolution over time. Moreover, the proposed empirically grounded agent-based model can be utilized to experiment with realistic and usable policies designed to further improve the performance efficiency of the pharmaceutical retail supply chains in India. Hence, this research will assist unorganized, organized and e-retail pharmaceutical sector to develop policies to reap the maximum benefits out of Indian blooming consumer market and achieve synergy and prosperity in the Indian economy.

## Online Appendix

### Appendix A: Disease Data

TABLE A1 Distribution of communicable diseases in the study area.

| **Category of morbidity** | **Diseases (ICD-11 Code)** | **Type of occurrence** | **Cases per year in India** | **Total cases per year** | **Seasonal?** | **Peak season** |
|---|---|---|---|---|---|---|
| Communicable immunizable | Tuberculosis (1B1Z) | Common | More than 1 million cases per year | 125 | no | |
| | Tetanus injection (1C14) | Very common | More than 10 million cases per year | 5824 | no | |
| | Chicken pox (1E90.0) | Rare | Fewer than 1 million cases per year | 182 | yes | Jan-march |
| | Measles (1F03) | Rare | Fewer than 1 million cases per year | 22 | yes | Jun-aug |
| Communicable, vector borne | Malaria (1F4Z) | Rare | Fewer than 1 million cases per year | 14 | Yes | Rainy |
| | Dengue (1D2Z) | Very rare | Fewer than 100 thousand cases per year | 28 | Yes | Rainy |
| | Chikungunya (1D40) | Extremely rare | Fewer than 5 thousand cases per year | 0 | Yes | Rainy |
| | Kala-azar (1F54.0) | Extremely rare | Fewer than 5 thousand cases per year | 0 | Yes | Rainy |
| | Japanese Encephalitis (1C85) | Extremely rare | Fewer than 5 thousand cases per year | 0 | Yes | Rainy |
| | Cold and flu (1E32) | Very common | More than 10 million cases per year | 58024 | Yes | Rainy |
| Communicable, water borne | Diarrheal diseases (ME05.1) | Very common | More than 10 million cases per year | 2751 | Yes | Rainy |
| | Typhoid (1A07.Z) | Very rare | Fewer than 100 thousand cases per year | 12 | Yes | Rainy |
| | Worms (1F90.Z) | Common | More than 1 million cases per year | 10805 | Yes | Rainy |
| | Jaundice (ME10.1) and Hepatitis (DB97.Z) | Rare | Fewer than 1 million cases per year | 4 | Yes | Rainy |
| Other communicable and infectious diseases | Skin infection (1B7Z) | Very common | More than 10 million cases per year | 20670 | Yes | Rainy |
| | Acute Respiratory Infection (CA4Y) | Very common | More than 10 million cases per year | 5492 | Yes | Winter and rainy |
| | Ear infection (AB0Z) | Very common | More than 10 million cases per year | 353 | No | |
| | Sore Eye & Eye complaints (MC18) | Very common | More than 10 million cases per year | 208 | No | |
| | Urinary tract infection (GC08.Z) | Very common | More than 10 million cases per year | 260 | No | |
| | Leprosy (1B20.Z) | Very rare | Fewer than 100 thousand cases per year | 5 | No | |
| Total | | | | 104779 | | |

TABLE A2 Distribution of non-communicable and other diseases in study area.

| Category of morbidity | Diseases (ICD-11 Code) | Type of occurrence | Cases per year in India | Total cases per year | Seasonal? | Peak season |
|---|---|---|---|---|---|---|
| Non-communicable diseases | Chronic obstructive pulmonary disease (CA22.Z) | Very common | More than 10 million cases per year | 5450 | yes | Rainy |
| | Gastritis (DA42.Z) | Common | More than 1 million cases per year | 9680 | no | |
| | Falls (PA6Z) /injuries (ND56.Z) /fractures (ND56.2) | Very common | More than 10 million cases per year | 230 | no | |
| | Diabetes (5A14) | Very common | More than 10 million cases per year | 22037 | No | |
| | Arthritis (FA2Z) | Very common | More than 10 million cases per year | 6400 | No | |
| | Hypertension (BA00.Z) | Very common | More than 10 million cases per year | 27414 | No | |
| | Epilepsy (8A6Z) | Common | More than 1 million cases per year | 60 | No | |
| Nutrition and metabolic disorders | Avitaminosis (5B7Y) | Very common | More than 10 million cases per year | 3030 | No | |
| | Anaemia (3A9Z) | Very common | More than 10 million cases per year | 2474 | No | |
| Other diseases | Pyrexia of unknown origin (MG26) | Very common | More than 10 million cases per year | 27733 | no | Summer |
| | Abdominal pain (MD81.4) | Very common | More than 10 million cases per year | 650 | no | |
| | Ophthalmic related (9B10.Z, 9C61.Z, 9A60.Z, 9B71.0Z, 9C20.2) | Very common | More than 10 million cases per year | 1039 | no | |
| | Toothache (DA0A.Y)/mouth pain (MD80.Y) | Very common | More than 10 million cases per year | 1104 | no | |
| | Burns and scalds (NE2Z) | Very common | More than 10 million cases per year | 20 | no | |
| | Dog bite (PA75&XE813) | Very common | More than 10 million cases per year | 253 | no | |
| | Snake bite (PA78&XE9H6) | Rare | Fewer than 1 million cases per year | 20 | yes | Rainy |
| | Rabies (1C82) and animal bite (PA75) | Rare | Fewer than 1 million cases per year | 60 | no | |
| | Obstetric complications (JB0D) | Common | More than 1 million cases per year | 1672 | no | |
| | Other diseases | | | 104951 | | |
| Total | | | | 214277 | | |

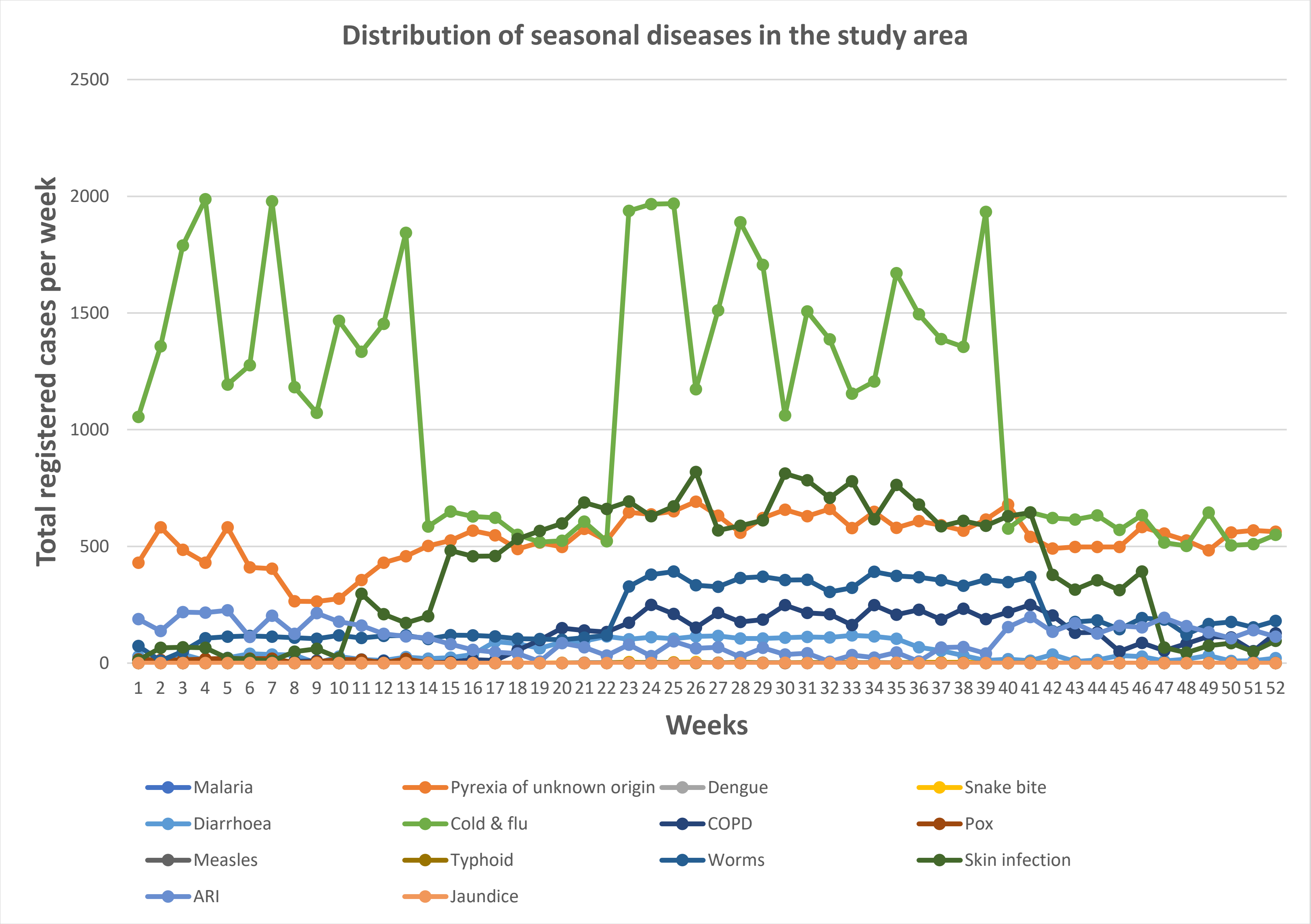

FIGURE A1 Distribution of seasonal diseases in the study area.

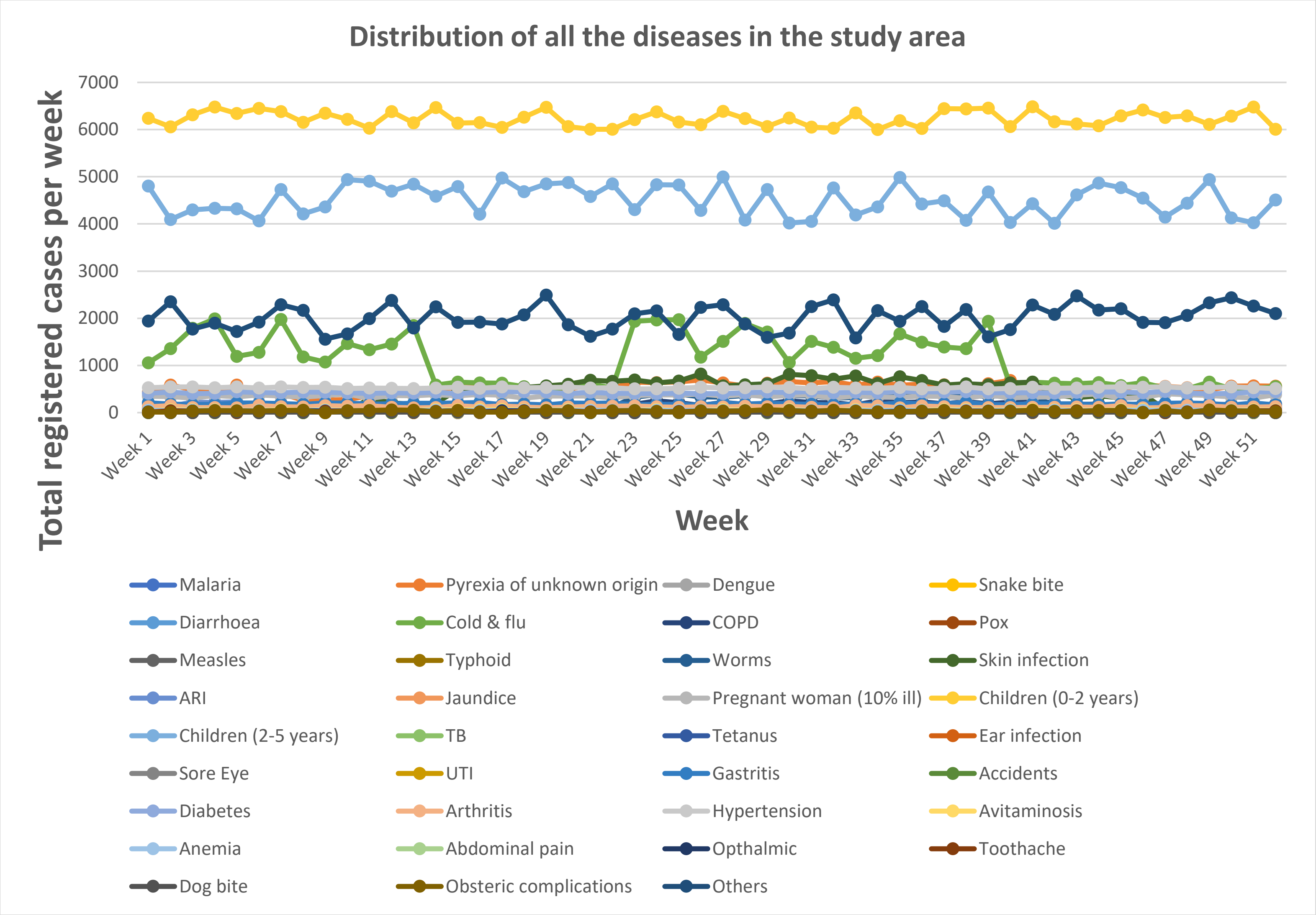

FIGURE A2 Distribution of all the diseases in the study area.

## Appendix B: Profiles of the Respondents

TABLE B1 Profiles of the respondents.

| Variable | Subgroups | Frequency | Percentage |
|---|---|---|---|
| Age | Below 18 years | 1 | 0.72 % |
| | 18–24 years | 19 | 13.77% |
| | 25–34 years | 79 | 57.25% |
| | 35–44 years | 15 | 10.87% |
| | 45–54 years | 11 | 7.97% |
| | 55–64 years | 5 | 3.62% |
| | 65 years and above | 8 | 5.80% |
| Gender | Male | 121 | 87.68% |
| | Female | 17 | 12.32% |
| Yearly household income | Less than Rs. 4 lakh | 24 | 17.39% |
| | Rs. 4 - 8 lakh | 36 | 26.09% |
| | Rs. 8 - 12 lakh | 34 | 24.64% |
| | Rs. 12 - 16 lakh | 12 | 8.70% |
| | more than Rs. 16 lakh | 27 | 19.56% |
| | Not stated | 5 | 3.62% |
| Frequency of purchase* | Daily basis | 8 | 5.80% |
| | Twice a week | 12 | 8.69% |
| | Once a week | 20 | 14.49% |
| | Twice a month | 34 | 24.64% |
| | Once a month | 53 | 38.40% |
| | Only when demand arises | 49 | 35.51% |
| Preferred retail type* | Unorganized | 97 | 70.29% |
| | Organized | 77 | 55.80% |
| | e-pharmaceutical | 25 | 18.12% |
| Reason of purchase* | Chronic diseases | 81 | 58.70% |
| | Seasonal diseases | 79 | 57.25% |
| | Over the counter (OTC) medicines | 53 | 38.40% |
| | Child related | 24 | 17.39% |
| | Any other reasons | 9 | 6.52% |
| Type of vehicle owned* | Two-wheeler | 84 | 60.87% |
| | Four-wheeler | 53 | 38.41% |
| | No personal vehicle | 37 | 26.81% |

Note: * indicates that in these questions multiple responses were allowed.

## Appendix C: Conjoint Analysis Results

TABLE C1 Partial utility and relative importance for customers with medium degree of emergency (ME).

| Attributes | Levels | Partial Utility | Std. Error | *p*-value | Relative Importance (in %) |
|---|---|---|---|---|---|
| Price discount | Less than 10% off | -0.29372 | 0.07066 | 0.000 | 18.42 |
| | [10-20] % off | 0.11950 | 0.06670 | 0.000 | |
| | More than 20% off | 0.17422 | 0.06649 | 0.000 | |
| Quality or value | Low | -0.66856 | 0.08047 | 0.000 | 37.29 |
| | Medium | -0.06568 | 0.06994 | 0.000 | |
| | High | 0.73424 | 0.06354 | 0.000 | |
| Variety of assortment | Small | -0.15353 | 0.06915 | 0.000 | 12.19 |
| | Medium | -0.04785 | 0.06819 | 0.000 | |
| | Large | 0.20138 | 0.06502 | 0.000 | |
| Degree of personalized service | Low | -0.09263 | 0.06807 | 0.000 | 10.05 |
| | Medium | 0.10396 | 0.06648 | 0.000 | |
| | High | -0.01133 | 0.06742 | 0.000 | |
| Distance between customer and retailer | Less than or equal to 2 km | 0.34517 | 0.06562 | 0.000 | 22.05 |
| | (2-10] km | 0.18748 | 0.06652 | 0.000 | |
| | More than 10 km | -0.53265 | 0.07589 | 0.000 | |

TABLE C2 Partial utility and relative importance for customers with low degree of emergency (LE).

| Attributes | Levels | Partial Utility | Std. Error | *p*-value | Relative Importance (in %) |
|---|---|---|---|---|---|
| Price discount | Less than 10% off | -0.38597 | 0.07276 | 0.000 | 20.84 |
| | [10-20] % off | -0.03937 | 0.0688 | 0.000 | |
| | More than 20% off | 0.42534 | 0.06488 | 0.000 | |
| Quality or value | Low | -0.56592 | 0.0776 | 0.000 | 34.45 |
| | Medium | -0.14698 | 0.07068 | 0.000 | |
| | High | 0.7129 | 0.06339 | 0.000 | |
| Variety of assortment | Small | -0.10238 | 0.06845 | 0.000 | 13.73 |
| | Medium | -0.03235 | 0.0678 | 0.000 | |
| | Large | 0.13473 | 0.06578 | 0.000 | |
| Degree of personalized service | Low | -0.13485 | 0.06884 | 0.000 | 10.33 |
| | Medium | 0.09805 | 0.06646 | 0.000 | |
| | High | 0.0368 | 0.06696 | 0.000 | |
| Distance between customer and retailer | Less than or equal to 2 km | 0.31971 | 0.0656 | 0.000 | 20.65 |
| | (2-10] km | 0.14914 | 0.06667 | 0.000 | |
| | More than 10 km | -0.46885 | 0.07447 | 0.000 | |

## Appendix D: ANOVA Results for Price Discount and Quality

### Section D1. ANOVA Results for Price Discount

Analysis of variance (ANOVA) is used to find the most significant influencing parameter for the output response. As per the experimental design in Table D1, the simulation model parameters are set and total 27 experiments are conducted. Table D1 shows the summary of experimental results for price discount. The simulations results are shown in appendix E.

TABLE D1 Summary of simulation results for discount.

| Exp no. | Input parameters | | | Response variables | | | | | | |
|---|---|---|---|---|---|---|---|---|---|---|
| | Unorganized discount | Organized discount | E-Pharmacy discount | Average weekly customer footprints per retail channel (customers / week) | | | Average market share per retail channel (%) | | | Unorganized retails shut down during simulation (no. of stores) |
| | | | | Unorganized | Organized | E-Pharmacy | Unorganized | Organized | E-Pharmacy | |
| 1 | Less than 10% off | Less than 10% off | Less than 10% off | 15609 | 6381 | 252 | 70.18 | 28.69 | 1.13 | 16 |
| 2 | Less than 10% off | Less than 10% off | (10-20) % off | 15247 | 5022 | 1991 | 68.5 | 22.56 | 8.94 | 17 |
| 3 | Less than 10% off | Less than 10% off | More than 20% off | 14290 | 4563 | 3399 | 64.22 | 20.51 | 15.27 | 16 |
| 4 | Less than 10% off | (10-20) % off | Less than 10% off | 11637 | 10607 | 0 | 52.31 | 47.69 | 0 | 23 |
| 5 | Less than 10% off | (10-20) % off | (10-20) % off | 11631 | 10100 | 507 | 52.30 | 45.42 | 2.28 | 24 |
| 6 | Less than 10% off | (10-20) % off | More than 20% off | 11304 | 9127 | 1807 | 50.83 | 41.04 | 8.13 | 24 |
| 7 | Less than 10% off | More than 20% off | Less than 10% off | 9513 | 12727 | 0 | 42.78 | 57.22 | 0 | 28 |
| 8 | Less than 10% off | More than 20% off | (10-20) % off | 9519 | 12405 | 328 | 42.78 | 55.74 | 1.48 | 27 |
| 9 | Less than 10% off | More than 20% off | More than 20% off | 9529 | 12095 | 632 | 42.82 | 54.34 | 2.84 | 26 |
| 10 | (10-20) % off | Less than 10% off | Less than 10% off | 21601 | 636 | 12 | 97.08 | 2.86 | 0.06 | 9 |
| 11 | (10-20) % off | Less than 10% off | (10-20) % off | 21072 | 635 | 543 | 94.70 | 2.86 | 2.44 | 7 |
| 12 | (10-20) % off | Less than 10% off | More than 20% off | 20385 | 625 | 1239 | 91.62 | 2.81 | 5.57 | 7 |
| 13 | (10-20) % off | (10-20) % off | Less than 10% off | 17459 | 4791 | 0 | 78.47 | 21.53 | 0 | 22 |
| 14 | (10-20) % off | (10-20) % off | (10-20) % off | 17430 | 4548 | 267 | 78.35 | 20.44 | 1.21 | 20 |
| 15 | (10-20) % off | (10-20) % off | More than 20% off | 17128 | 4211 | 917 | 76.95 | 18.93 | 4.12 | 20 |
| 16 | (10-20) % off | More than 20% off | Less than 10% off | 15404 | 6837 | 0 | 69.26 | 30.74 | 0 | 23 |
| 17 | (10-20) % off | More than 20% off | (10-20) % off | 15400 | 6703 | 162 | 69.16 | 30.10 | 0.74 | 24 |
| 18 | (10-20) % off | More than 20% off | More than 20% off | 15332 | 6457 | 464 | 68.90 | 29.02 | 2.08 | 24 |
| 19 | More than 20% off | Less than 10% off | Less than 10% off | 16960 | 5170 | 114 | 76.71 | 22.79 | 0.5 | 149 |
| 20 | More than 20% off | Less than 10% off | (10-20) % off | 12995 | 5993 | 3250 | 59.23 | 26.43 | 14.34 | 154 |
| 21 | More than 20% off | Less than 10% off | More than 20% off | 14509 | 4046 | 3694 | 65.85 | 17.84 | 16.31 | 154 |
| 22 | More than 20% off | (10-20) % off | Less than 10% off | 11301 | 10941 | 0 | 51.55 | 48.45 | 0 | 151 |
| 23 | More than 20% off | (10-20) % off | (10-20) % off | 13782 | 8231 | 242 | 62.44 | 36.48 | 1.08 | 146 |
| 24 | More than 20% off | (10-20) % off | More than 20% off | 11915 | 9267 | 1075 | 54.21 | 41.02 | 4.77 | 147 |
| 25 | More than 20% off | More than 20% off | Less than 10% off | 10368 | 11872 | 0 | 47.31 | 52.69 | 0 | 153 |
| 26 | More than 20% off | More than 20% off | (10-20) % off | 8962 | 12948 | 332 | 41.13 | 57.39 | 1.48 | 149 |
| 27 | More than 20% off | More than 20% off | More than 20% off | 7826 | 13781 | 632 | 36.10 | 61.09 | 2.81 | 152 |

**D1.1. Effect of Unorganized Discount, Organized Discount and E-Pharmacy Discount on the Average Weekly Customer Footprint at Unorganized Pharmaceutical Retailers**

For details kindly refer to the section 3.2.3.1., Table 7, Figure 12 (a). The interaction plots are shown in Figure D1. The interaction effects are not statistically significant.

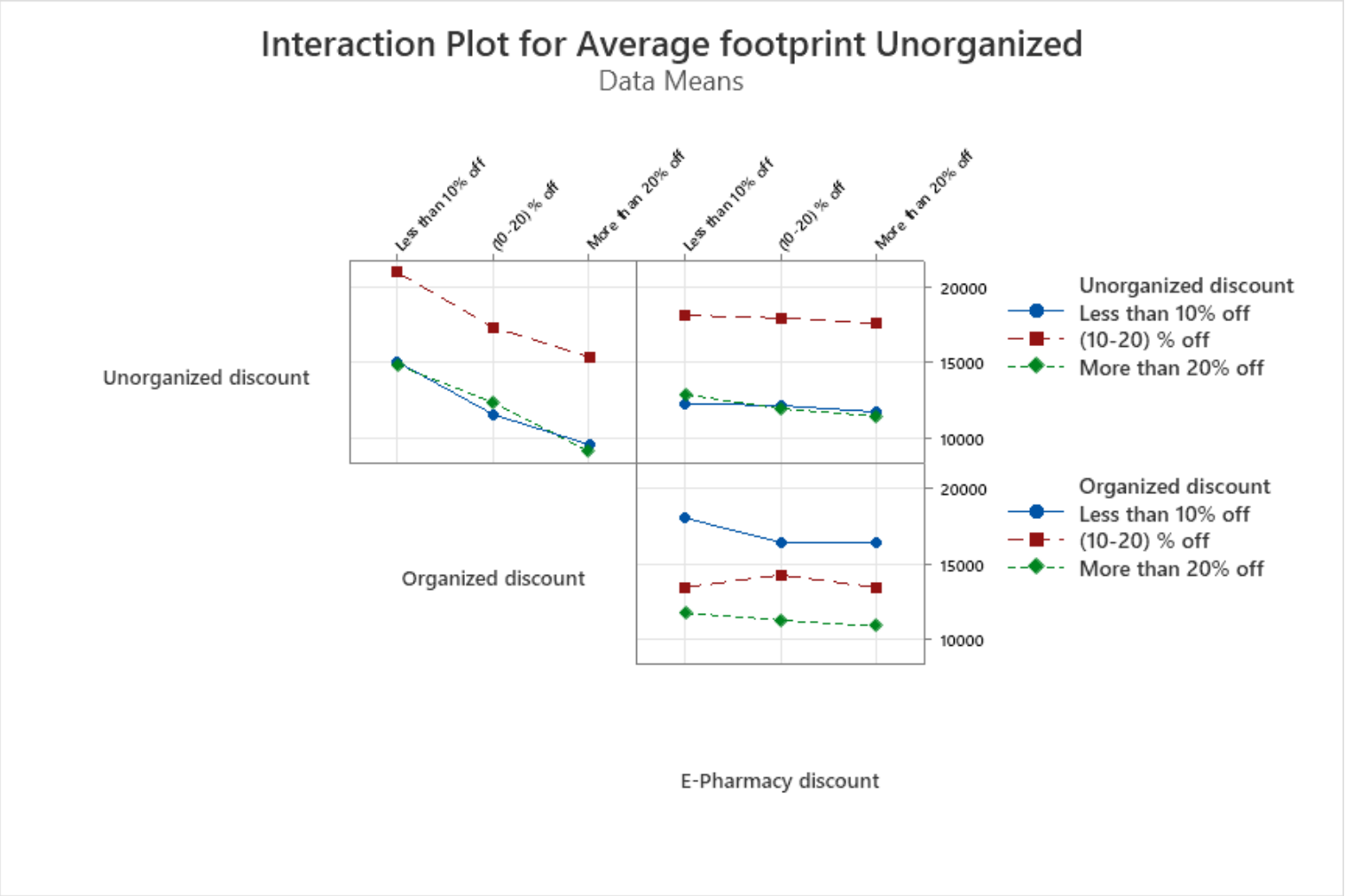

FIGURE D1 Interaction effect plots for average footprint at unorganized retailers.

**D1.2. Effect of Unorganized Discount, Organized Discount and E-Pharmacy Discount on the Average Market Share of the Unorganized Pharmaceutical Retailers**

For details kindly refer to the section 3.2.3.1., Table 8, Figure 12 (b). The interaction plots are shown in Figure D2. The interaction effects are not statistically significant.

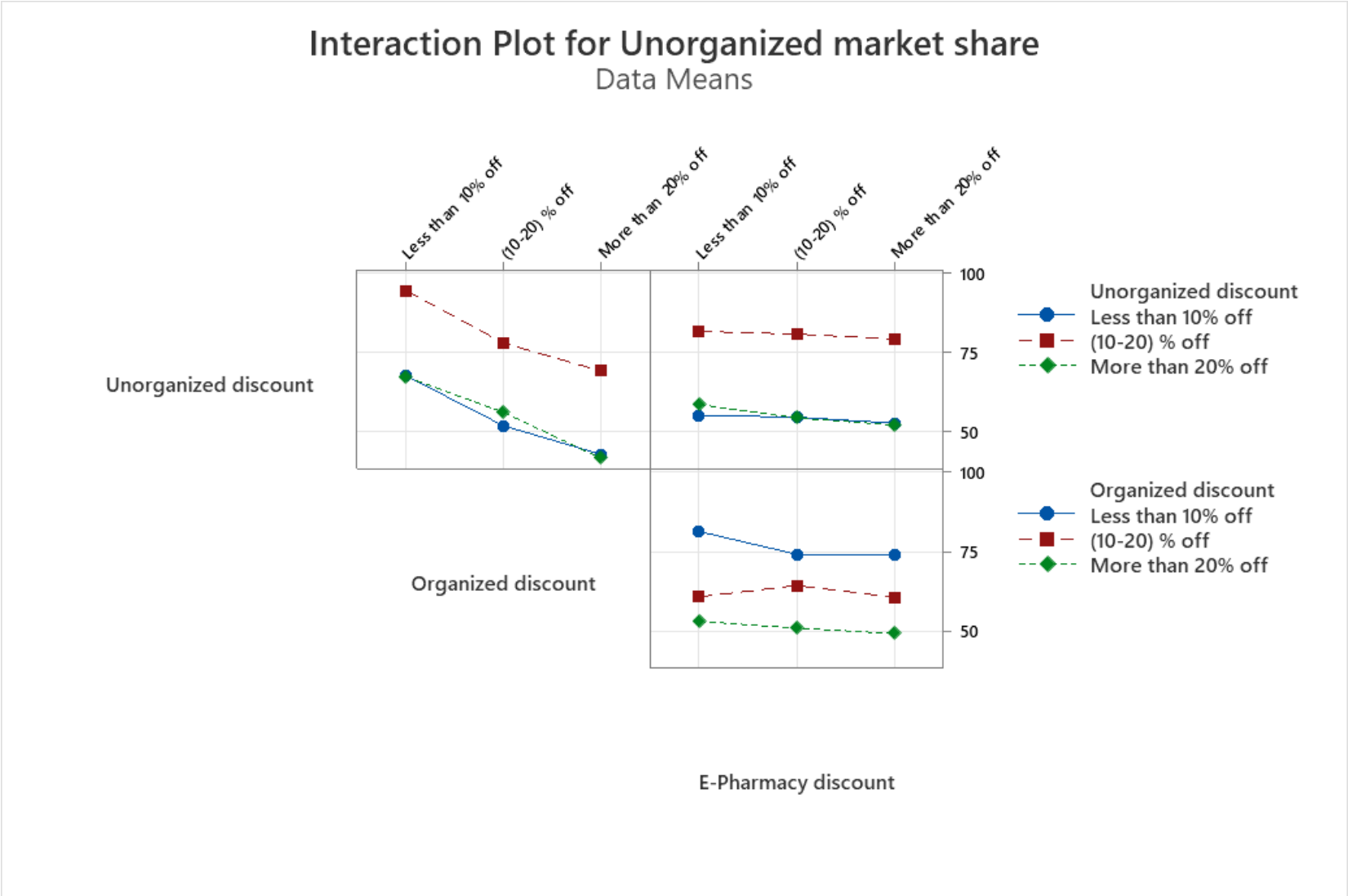

FIGURE D2 Interaction effect plots for average market share of unorganized retailers.

### D1.3. Effect of Unorganized Discount, Organized Discount and E-Pharmacy Discount on the Number of Unorganized Retailers Shut Down During Simulation

Table D2 shows the ANOVA table for the number of unorganized retailers shut down during simulation. Table D2 shows that the *p*-values for unorganized discount and organized discount are less than 0.05. This implies that the variables, unorganized discount and organized discount have a significant influence on the number of unorganized retailers shut down during simulation. It is also observed that the unorganized discount has the highest effect on the response. The interaction between unorganized discount and organized discount is found to be significant.

TABLE D2 ANOVA for number of unorganized retailers shut down during simulation.

| Source | DF | Adj SS | Adj MS | F-Value | *P*-Value | Contribution (%) |
|---|---|---|---|---|---|---|
| Unorganized discount | 2 | 102642 | 51321.1 | 12800.66 | 0.000 | 99.33 |
| Organized discount | 2 | 336 | 168.0 | 41.91 | 0.000 | 0.32 |
| E-Pharmacy discount | 2 | 2 | 1.0 | 0.26 | 0.778 | 0.001 |
| Unorganized discount*Organized discount | 4 | 307 | 76.9 | 19.17 | 0.000 | 0.29 |
| Unorganized discount*E-Pharmacy discount | 4 | 4 | 1.0 | 0.26 | 0.896 | 0.008 |
| Organized discount*E-Pharmacy discount | 4 | 10 | 2.6 | 0.65 | 0.645 | 0.009 |
| Error | 8 | 32 | 4.0 | | | 0.03 |
| Total | 26 | 103335 | | | | |

R-Sq = 99.97%    R-Sq (Adj) = 99.90%

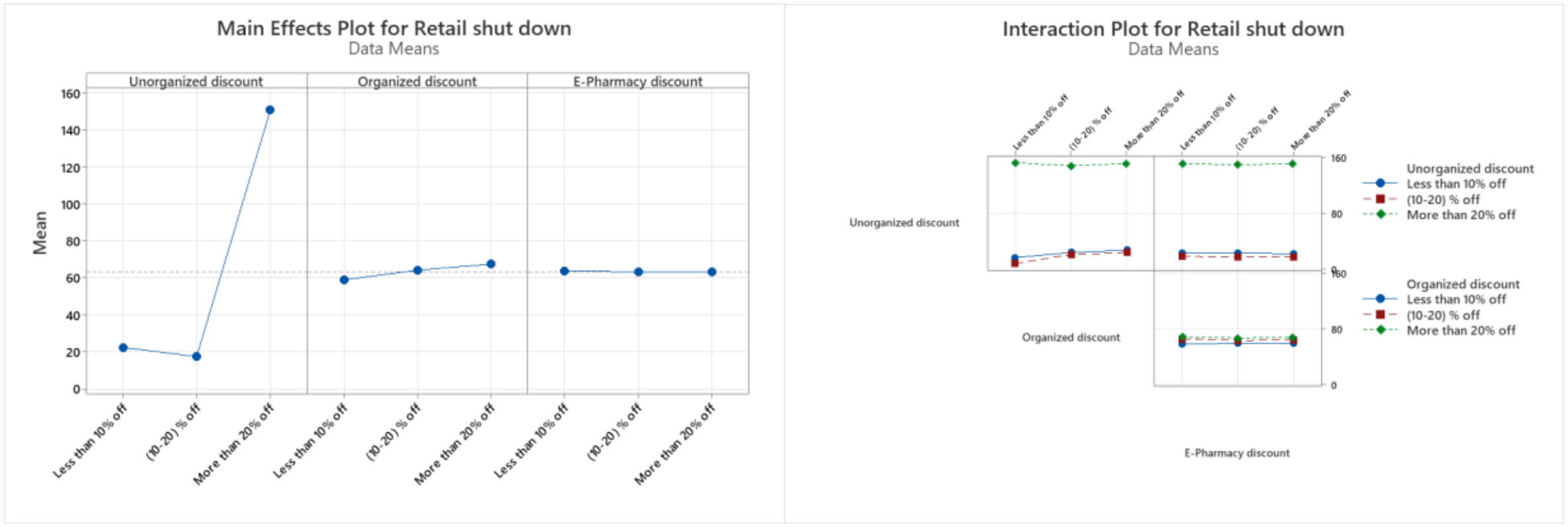

FIGURE D3 Main effect and interaction effect plots for the number of unorganized retailers shut down during simulation.

The main effects and interaction plots are shown in Figure D3. From the main effect plot it is evident that with an increase in organized discount and e-pharmacy discount, there is slight increase in the number of unorganized retailers shut down. It is also evident that with an increase in unorganized discount from low to medium level, there is decrease in the number of unorganized retailers shut down.

But as the unorganized discount increase further from medium to high level, a counter intuitive behavior is seen, the number of unorganized retailers shut down increases drastically. For unorganized discount the medium level is showing the lowest response value. It can be concluded that the number of unorganized retailers shut down can be minimized by keeping unorganized discount at medium level while organized discount at a low level.

### D1.4. Effect of Unorganized Discount, Organized Discount and E-Pharmacy Discount on the Average Weekly Customer Footprint at Organized Pharmaceutical Retailers

Table D3 shows the ANOVA table for average weekly customer footprint at organized pharmaceutical retailer. Table D3 shows that the *p*-values for unorganized discount and organized discount are less than 0.05. This implies that the parameters, unorganized discount and organized discount have a significant influence on the average weekly customer footprint at organized pharmaceutical retailer. Percentage of contribution for each factor is shown in Table D3. It is observed that the organized discount has the highest effect on the response. The interaction effects are not significant.

TABLE D3 ANOVA for average weekly customer footprint at organized pharmaceutical retailers.

| Source | DF | Adj SS | Adj MS | F-Value | *P*-Value | Contribution (%) |
|---|---|---|---|---|---|---|
| Unorganized discount | 2 | 165023851 | 82511925 | 128.41 | 0.000 | 41.11 |
| Organized discount | 2 | 222810297 | 111405149 | 173.38 | 0.000 | 55.51 |
| E-Pharmacy discount | 2 | 1879659 | 939830 | 1.46 | 0.287 | 0.47 |
| Unorganized discount*Organized discount | 4 | 2691761 | 672940 | 1.05 | 0.441 | 0.67 |
| Unorganized discount*E-Pharmacy discount | 4 | 1026819 | 256705 | 0.40 | 0.804 | 0.26 |
| Organized discount*E-Pharmacy discount | 4 | 2798208 | 699552 | 1.09 | 0.424 | 0.69 |
| Error | 8 | 5140419 | 642552 | | | 1.28 |
| Total | 26 | 401371015 | | | | |
| R-Sq = 98.72% | | | R-Sq (Adj) = 95.84% | | | |

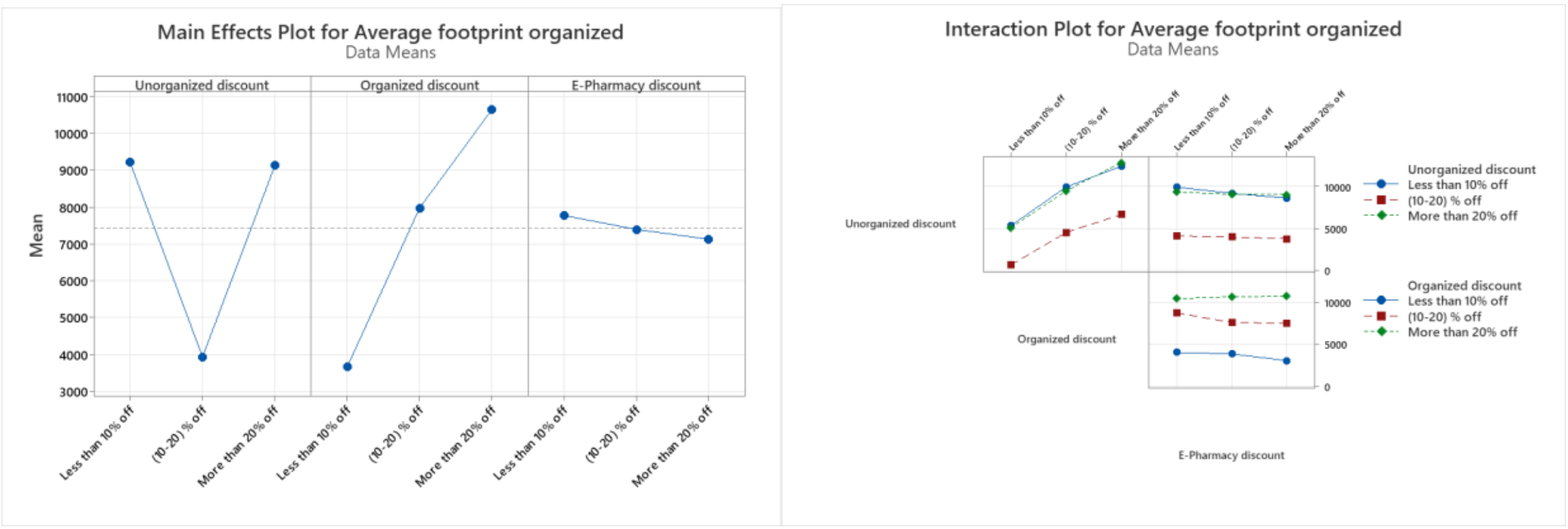

FIGURE D4 Main effect and interaction effect plots for average footprint at organized retailers.

The main effects and interaction plots are shown in Figure D4. From the main effect plot it is evident that with an increase in organized discount, there is an increase in the average footprint at organized retailer. But with an increase in e-pharmacy discount, there is a decrease in the response. It is also evident that with an increase in unorganized discount from low to medium level, there is decrease in the average footprint at organized retailer. But as the unorganized discount increase further from medium to high level, a counter intuitive behavior is seen, the average footprint at organized retailer increases. For unorganized discount the low and high level are showing the highest response value. It can be concluded that the average footprint at organized retailer can be maximized by keeping unorganized discount at either low or high level while organized discount at high level and e-pharmacy discount at a low level.

### D1.5. Effect of Unorganized Discount, Organized Discount and E-Pharmacy Discount on the Average Market Share of the Organized Pharmaceutical Retailers

Table D4 shows the ANOVA table for average market share of the organized pharmaceutical retailers. Table D4 shows that the *p*-values for unorganized discount and organized discount are less than 0.05. This implies that the parameters, unorganized discount and organized discount have a significant influence on the average market share of the organized pharmaceutical retailers. Percentage of contribution for each factor is shown in Table D4. It is observed that the organized discount has the highest effect on the response. The interaction effects are not significant.

TABLE D4 ANOVA for average market share of the organized pharmaceutical retailers.

| **Source** | **DF** | **Adj SS** | **Adj MS** | **F-Value** | ***P*-Value** | **Contribution (%)** |
|---|---|---|---|---|---|---|
| Unorganized discount | 2 | 3252.71 | 1626.36 | 129.45 | 0.000 | 40.73 |
| Organized discount | 2 | 4467.55 | 2233.77 | 177.80 | 0.000 | 55.95 |
| E-Pharmacy discount | 2 | 38.09 | 19.05 | 1.52 | 0.277 | 0.48 |
| Unorganized discount*Organized discount | 4 | 51.00 | 12.75 | 1.01 | 0.454 | 0.64 |
| Unorganized discount*E-Pharmacy discount | 4 | 20.88 | 5.22 | 0.42 | 0.793 | 0.26 |
| Organized discount*E-Pharmacy discount | 4 | 54.33 | 13.58 | 1.08 | 0.427 | 0.68 |
| Error | 8 | 100.51 | 12.56 | | | 1.26 |
| Total | 26 | 7985.08 | | | | |
| R-Sq = 98.74% | | | R-Sq (Adj) = 95.91% | | | |

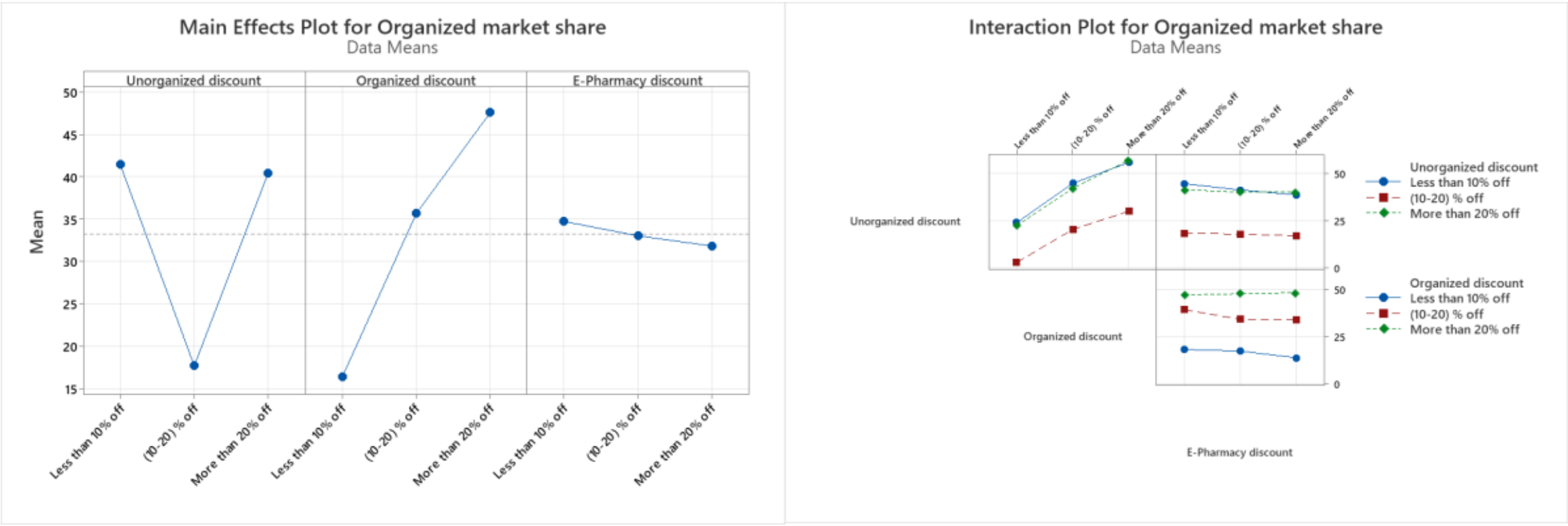

FIGURE D5 Main effect and interaction effect plots for average market share of organized retailers.

The main effects and interaction plots are shown in Figure D5. From the main effect plot it is evident that with an increase in organized discount, there is an increase in the average market share of organized retailers. But with an increase in e-pharmacy discount, there is a decrease in the response. It is also evident that with an increase in unorganized discount from low to medium level, there is decrease in the average market share of organized retailers. But as the unorganized discount increase further from medium to high level, a counter intuitive behavior is seen, the average market share of organized retailers increases. For unorganized discount the low level is showing the highest response value. It can be concluded that the average market share of organized retailers can be maximized by keeping unorganized discount at either low level while organized discount at high level and e-pharmacy discount at a low level.

### D1.6. Effect of Unorganized Discount, Organized Discount and E-Pharmacy Discount on the Average Weekly Customer Footprint at E-Pharmaceutical Retailers

Table D5 shows the ANOVA table for average weekly customer footprint at e-pharmaceutical retailer. Table D5 shows that the *p*-values for unorganized discount, organized discount and e-pharmacy discount are less than 0.05. This implies that the parameters, unorganized discount, organized discount and e-pharmacy discount have a significant influence on the average weekly customer footprint at e-pharmaceutical retailer. Percentage of contribution for each factor is shown in Table D5. It is observed that the e-pharmacy discount has the highest effect on the response. Two interaction effects are found to be significant. One is between unorganized discount and organized discount. The other is the interaction between organized discount and e-pharmacy discount.

TABLE D5 ANOVA for average weekly customer footprint at e-pharmaceutical retailers.

| **Source** | **DF** | **Adj SS** | **Adj MS** | **F-Value** | ***P*-Value** | **Contribution (%)** |
|---|---|---|---|---|---|---|
| Unorganized discount | 2 | 2269870 | 1134935 | 6.35 | 0.022 | 7.36 |
| Organized discount | 2 | 8943422 | 4471711 | 25.02 | 0.000 | 28.99 |
| E-Pharmacy discount | 2 | 10115299 | 5057649 | 28.30 | 0.000 | 32.79 |
| Unorganized discount*Organized discount | 4 | 2956382 | 739095 | 4.13 | 0.042 | 9.59 |
| Unorganized discount*E-Pharmacy discount | 4 | 1167321 | 291830 | 1.63 | 0.257 | 3.78 |
| Organized discount*E-Pharmacy discount | 4 | 3959757 | 989939 | 5.54 | 0.020 | 12.84 |
| Error | 8 | 1429956 | 178745 | | | 4.64 |
| Total | 26 | 30842009 | | | | |

R-Sq = 95.36% R-Sq (Adj) = 84.93%

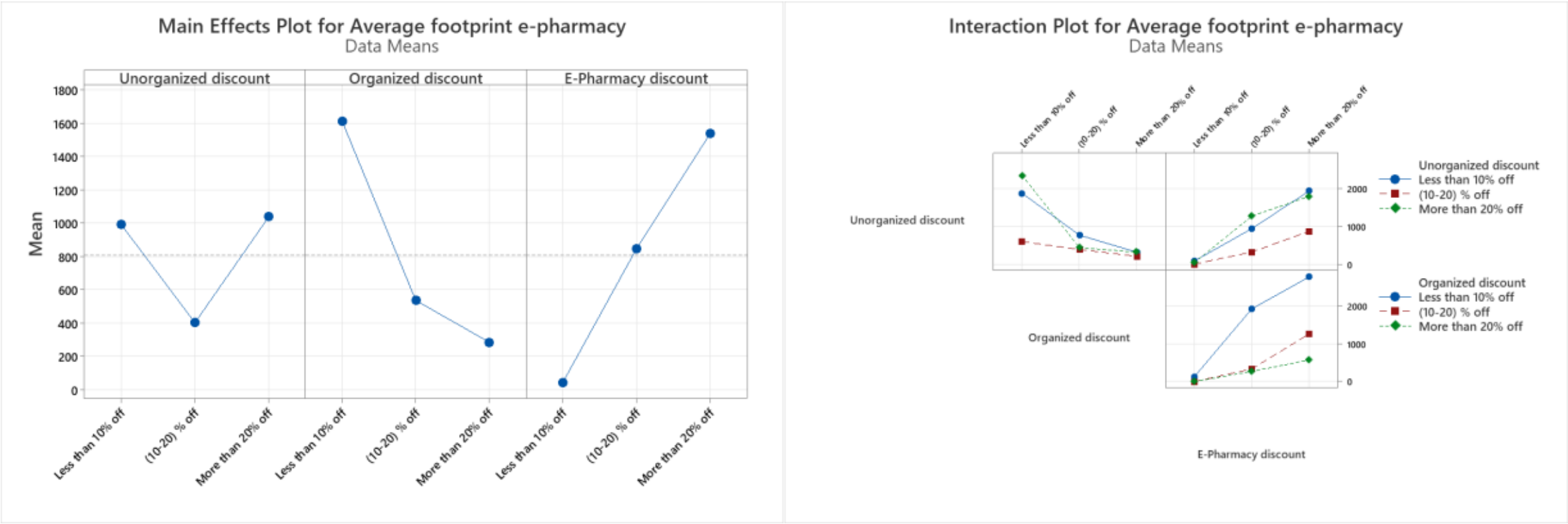

FIGURE D6 Main effect and interaction effect plots for average footprint at e-pharmaceutical retailers.

The main effects and interaction plots are shown in Figure D6. From the main effect plot it is evident that with an increase in e-pharmacy discount, there is an increase in the average footprint at e-pharmaceutical retailer. But with an increase in organized discount, there is a decrease in the response. It is also evident that with an increase in unorganized discount from low to medium level, there is decrease in the average footprint at e-pharmaceutical retailer. But as the unorganized discount increase further from medium to high level, a counter intuitive behavior is seen, the average footprint at e-pharmaceutical retailer increases. For unorganized discount the high level is showing the highest response value. It can be concluded that the average footprint at e-pharmaceutical retailer can be maximized by keeping unorganized discount at high level while organized discount at low level and e-pharmacy discount at a high level.

### D1.7. Effect of Unorganized Discount, Organized Discount and E-Pharmacy Discount on the Average Market Share of the E-Pharmaceutical Retailers

Table D6 shows the ANOVA table for the average market share of the e-pharmaceutical retailers. Table D6 shows that the *p*-values for unorganized discount, organized discount and e-pharmacy discount are less than 0.05. This implies that the parameters, unorganized discount, organized discount and e-pharmacy discount have a significant influence on the average market share of the e-pharmaceutical retailers. Percentage of contribution for each factor is shown in Table D6. It is observed that the e-pharmacy discount has the highest effect on the response. Two interaction effects are found to be significant. One is between unorganized discount and organized discount. The other is the interaction between organized discount and e-pharmacy discount.

TABLE D6 ANOVA for average market share of the e-pharmaceutical retailers.

| **Source** | **DF** | **Adj SS** | **Adj MS** | **F-Value** | ***P*-Value** | **Contribution (%)** |
|---|---|---|---|---|---|---|
| Unorganized discount | 2 | 44.40 | 22.200 | 6.38 | 0.022 | 7.28 |
| Organized discount | 2 | 176.76 | 88.379 | 25.40 | 0.000 | 29.00 |
| E-Pharmacy discount | 2 | 201.76 | 100.880 | 28.99 | 0.000 | 33.10 |
| Unorganized discount*Organized discount | 4 | 57.37 | 14.344 | 4.12 | 0.042 | 9.41 |
| Unorganized discount*E-Pharmacy discount | 4 | 23.01 | 5.753 | 1.65 | 0.252 | 3.77 |
| Organized discount*E-Pharmacy discount | 4 | 78.32 | 19.581 | 5.63 | 0.019 | 12.85 |
| Error | 8 | 27.84 | 3.479 | | | 4.57 |
| Total | 26 | 609.47 | | | | |
| R-Sq = 95.43% | | | R-Sq (Adj) = 85.16% | | | |

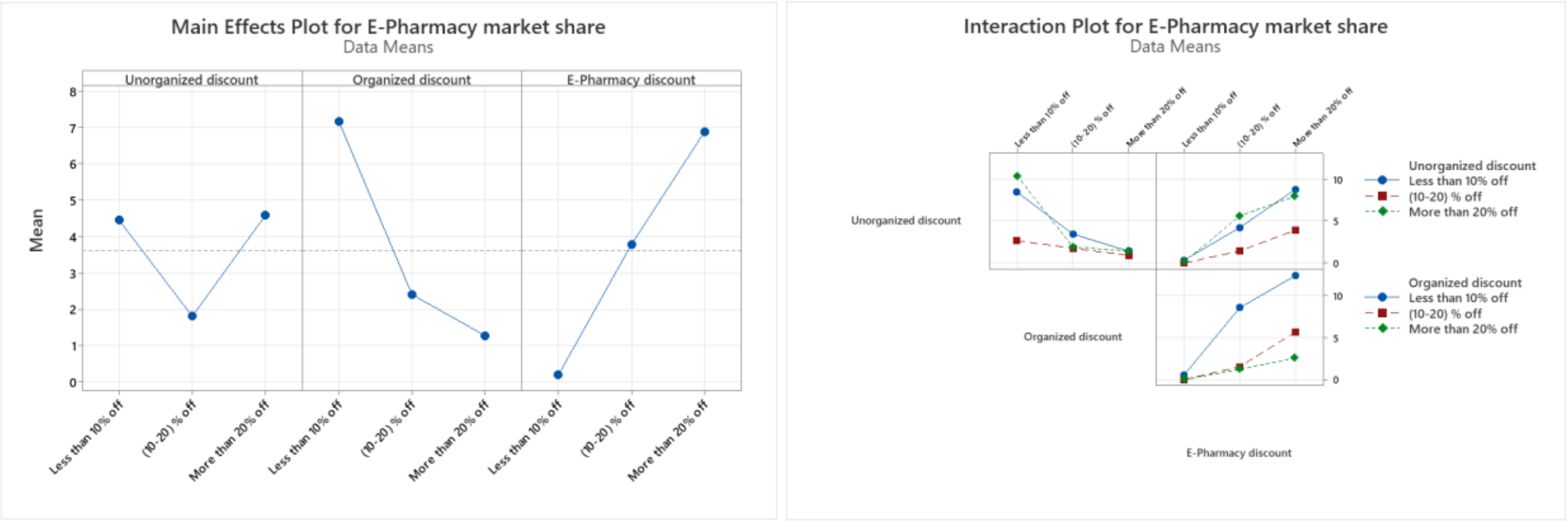

FIGURE D7 Main effect and interaction effect plots for average market share of e-pharmaceutical retailers.

The main effects and interaction plots are shown in Figure D7. From the main effect plot it is evident that with an increase in e-pharmacy discount, there is an increase in the average market share of the e-pharmaceutical retailers. But with an increase in organized discount, there is a decrease in the response.

It is also evident that with an increase in unorganized discount from low to medium level, there is decrease in the average market share of the e-pharmaceutical retailers. But as the unorganized discount increase further from medium to high level, a counter intuitive behavior is seen, the average market share of the e-pharmaceutical retailers increases. For unorganized discount the high level is showing the highest response value. It can be concluded that the average market share of the e-pharmaceutical retailers can be maximized by keeping unorganized discount at high level while organized discount at low level and e-pharmacy discount at a high level.

**Section D2. ANOVA Results for Quality**

In this section, the ANOVA for quality of products offered by the retailers is carried out. As per the experimental design in Table D7, the simulation model parameters are set and total 27 experiments are conducted. Table D7 shows the summary of experimental results for quality. This analysis is done at a level of 95% confidence level. The percentage contribution (%) of individual parameters for the final response can be measured for all variables by the ratio of the individual sum of squares of parameters to the total sum of squares. The simulations results are shown in appendix E.

TABLE D7 Summary of simulation results for quality.

| Exp no. | Input parameters | | | Response variables | | | | | | |
|---|---|---|---|---|---|---|---|---|---|---|
| | Unorganized quality | Organized quality | E-Pharmacy quality | Average weekly customer footprints per retail channel (customers / week) | | | Average market share per retail channel (%) | | | Unorganized retails shut down during simulation (no. of stores) |
| | | | | Unorganized | Organized | E-Pharmacy | Unorganized | Organized | E-Pharmacy | |
| 1 | Low | Low | Low | 2823 | 18212 | 1213 | 12.69 | 81.86 | 5.45 | 103 |
| 2 | Low | Low | Medium | 68 | 6871 | 15313 | 0.30 | 30.88 | 68.82 | 157 |
| 3 | Low | Low | High | 2 | 1440 | 20800 | 0.01 | 6.48 | 93.51 | 159 |
| 4 | Low | Medium | Low | 0 | 22248 | 0 | 0 | 100 | 0 | 159 |
| 5 | Low | Medium | Medium | 4 | 22214 | 29 | 0.02 | 99.84 | 0.14 | 158 |
| 6 | Low | Medium | High | 2 | 7871 | 14370 | 0.01 | 35.39 | 64.60 | 159 |
| 7 | Low | High | Low | 0 | 22238 | 0 | 0 | 100 | 0 | 159 |
| 8 | Low | High | Medium | 0 | 22252 | 0 | 0 | 100 | 0 | 159 |
| 9 | Low | High | High | 4 | 21385 | 854 | 0.02 | 96.14 | 3.84 | 159 |
| 10 | Medium | Low | Low | 15644 | 6592 | 0 | 70.35 | 29.65 | 0 | 9 |
| 11 | Medium | Low | Medium | 13630 | 2864 | 5760 | 61.24 | 12.88 | 25.88 | 9 |
| 12 | Medium | Low | High | 8421 | 863 | 12966 | 37.84 | 3.88 | 58.28 | 8 |
| 13 | Medium | Medium | Low | 7097 | 15148 | 0 | 31.91 | 68.09 | 0 | 40 |
| 14 | Medium | Medium | Medium | 7113 | 15114 | 29 | 31.96 | 67.90 | 0.14 | 38 |
| 15 | Medium | Medium | High | 6071 | 6585 | 9602 | 27.27 | 29.59 | 43.14 | 41 |
| 16 | Medium | High | Low | 3785 | 18464 | 0 | 17.02 | 82.98 | 0 | 70 |
| 17 | Medium | High | Medium | 3785 | 18461 | 0 | 17.02 | 82.98 | 0 | 73 |
| 18 | Medium | High | High | 3762 | 17798 | 679 | 16.92 | 80.03 | 3.05 | 72 |
| 19 | High | Low | Low | 22243 | 0 | 0 | 100 | 0 | 0 | 5 |
| 20 | High | Low | Medium | 22249 | 0 | 0 | 100 | 0 | 0 | 5 |
| 21 | High | Low | High | 21011 | 0 | 1242 | 94.42 | 0 | 5.58 | 5 |
| 22 | High | Medium | Low | 22251 | 1 | 0 | 99.99 | 0.01 | 0 | 4 |
| 23 | High | Medium | Medium | 22251 | 1 | 0 | 99.99 | 0.01 | 0 | 6 |
| 24 | High | Medium | High | 20999 | 1 | 1244 | 94.40 | 0.01 | 5.59 | 6 |
| 25 | High | High | Low | 15406 | 6839 | 0 | 69.25 | 30.75 | 0 | 21 |
| 26 | High | High | Medium | 15402 | 6841 | 0 | 69.24 | 30.76 | 0 | 25 |
| 27 | High | High | High | 15334 | 6455 | 465 | 68.91 | 29.00 | 2.09 | 24 |

**D2.1. Effect of Unorganized Quality, Organized Quality and E-Pharmacy Quality on the Average Weekly Customer Footprint at Unorganized Pharmaceutical Retailers**

For details refer to section 3.2.4.1., Table 9 and Figure 13 (a). The interaction plots are shown in Figure D8. The interaction between unorganized quality and organized quality is found significant.

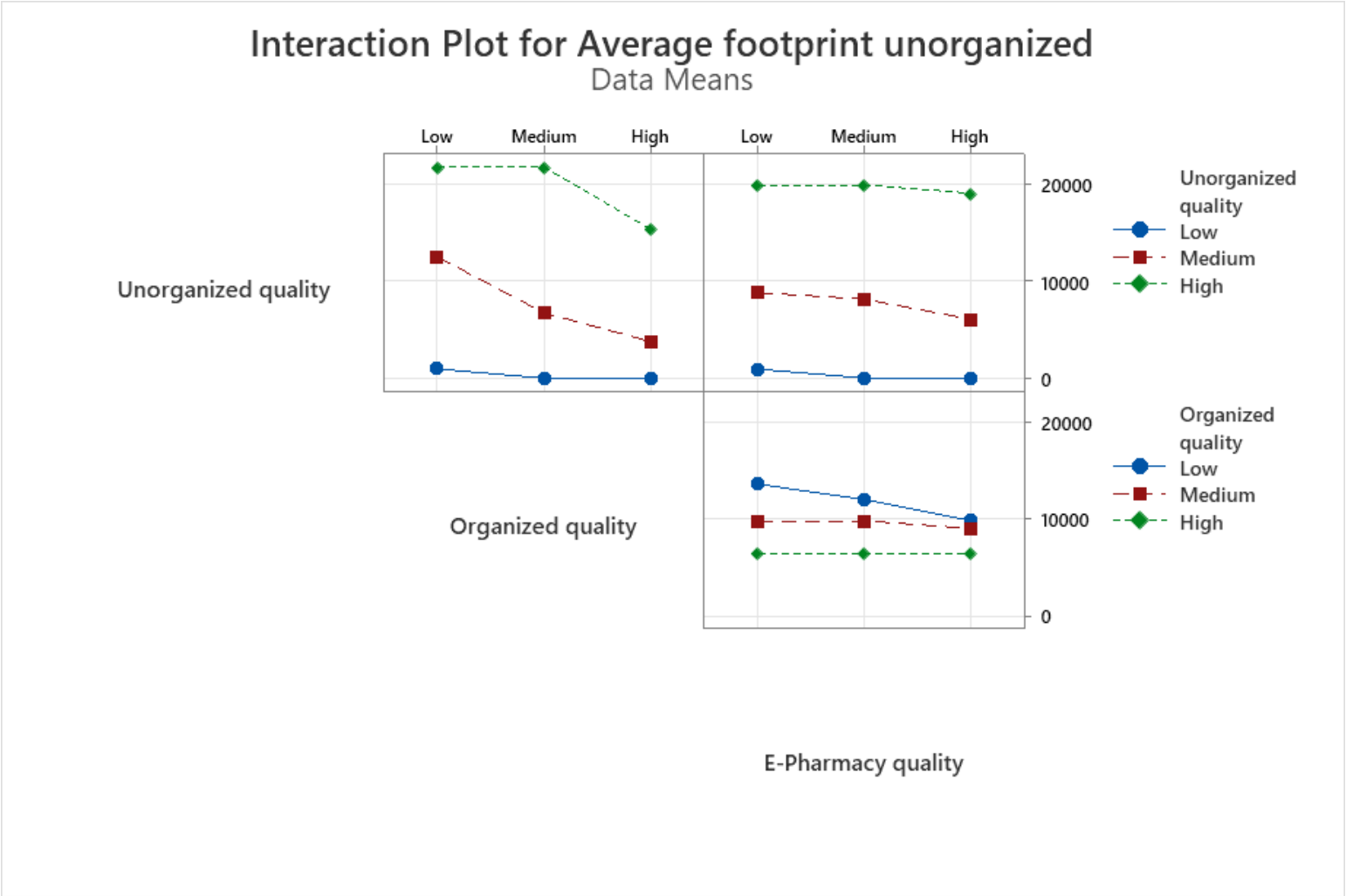

FIGURE D8 Interaction effect plots for average footprint at unorganized retailers.

**D2.2. Effect of Unorganized Quality, Organized Quality and E-Pharmacy Quality on the Average Market Share of the Unorganized Pharmaceutical Retailers**

For details refer to the section 3.2.4.1., Table 10 and Figure 13 (b). The interaction plots are shown in Figure D9. The interaction between unorganized quality and organized quality is found significant.

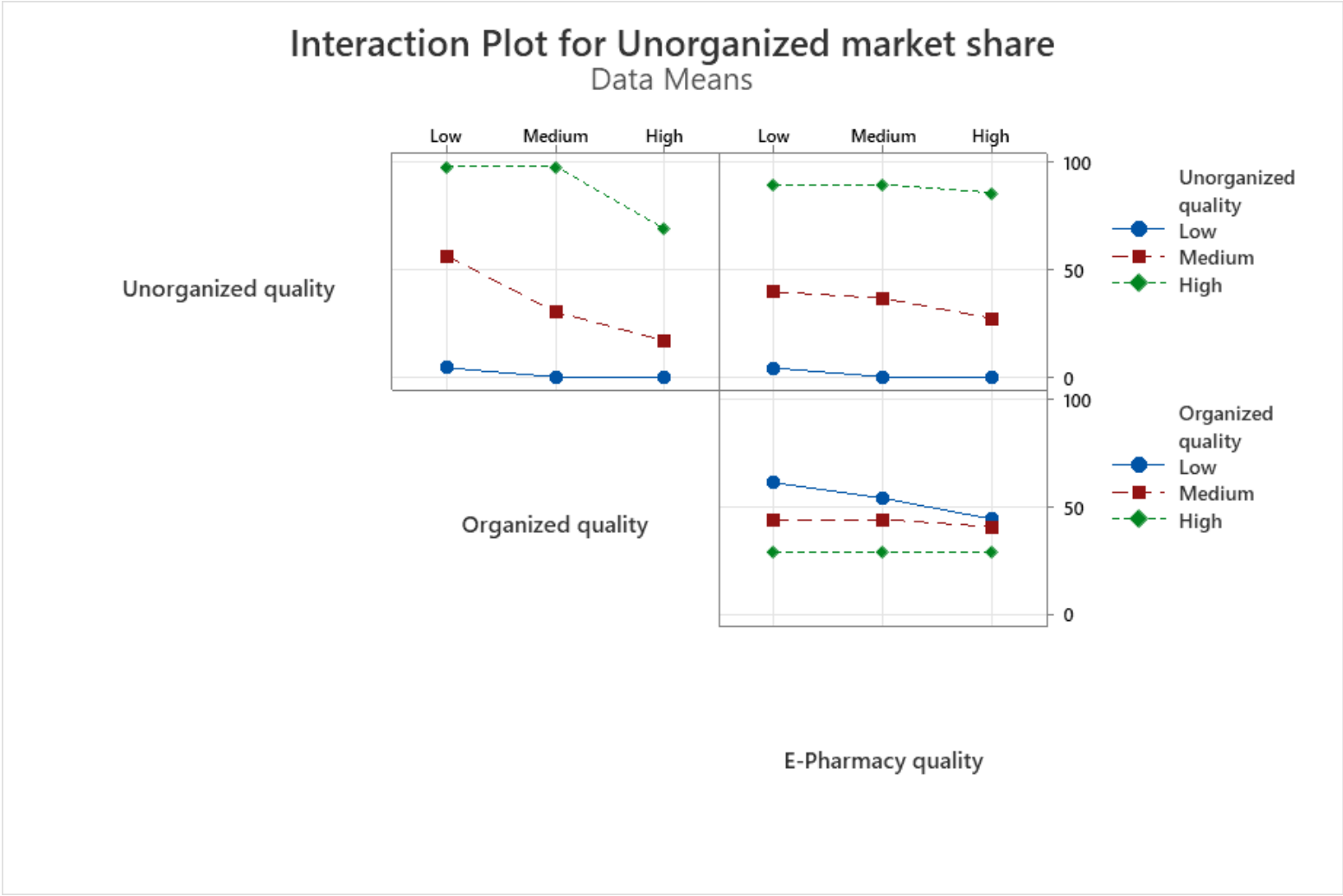

FIGURE D9 Interaction effect plots for average market share of unorganized retailers.

### D2.3. Effect of Unorganized Quality, Organized Quality and E-Pharmacy Quality on the Number of Unorganized Retailers Shut Down During Simulation

Table D8 shows the ANOVA table for the number of unorganized retailers shut down during simulation. Table D8 shows that the *p*-values for unorganized quality and organized quality are less than 0.05. This implies that the variables, unorganized quality and organized quality have a significant influence on the number of unorganized retailers shut down. It is also observed that the unorganized quality has the highest effect on the response. The interaction between unorganized quality and organized quality is found to be significant.

TABLE D8 ANOVA for number of unorganized retailers shut down during simulation.

| **Source** | **DF** | **Adj SS** | **Adj MS** | **F-Value** | ***P*-Value** | **Contribution (%)** |
|---|---|---|---|---|---|---|
| Unorganized quality | 2 | 100247 | 50123.4 | 408.99 | 0.000 | 91.43 |
| Organized quality | 2 | 5067 | 2533.4 | 20.67 | 0.001 | 4.62 |
| E-Pharmacy quality | 2 | 281 | 140.3 | 1.15 | 0.365 | 0.26 |
| Unorganized quality*Organized quality | 4 | 2282 | 570.6 | 4.66 | 0.031 | 2.08 |
| Unorganized quality*E-Pharmacy quality | 4 | 388 | 97.1 | 0.79 | 0.562 | 0.35 |
| Organized quality*E-Pharmacy quality | 4 | 391 | 97.8 | 0.80 | 0.559 | 0.36 |
| Error | 8 | 980 | 122.6 | | | 0.89 |
| Total | 26 | 109637 | | | | |

R-Sq = 99.11%  R-Sq (Adj) = 97.09%

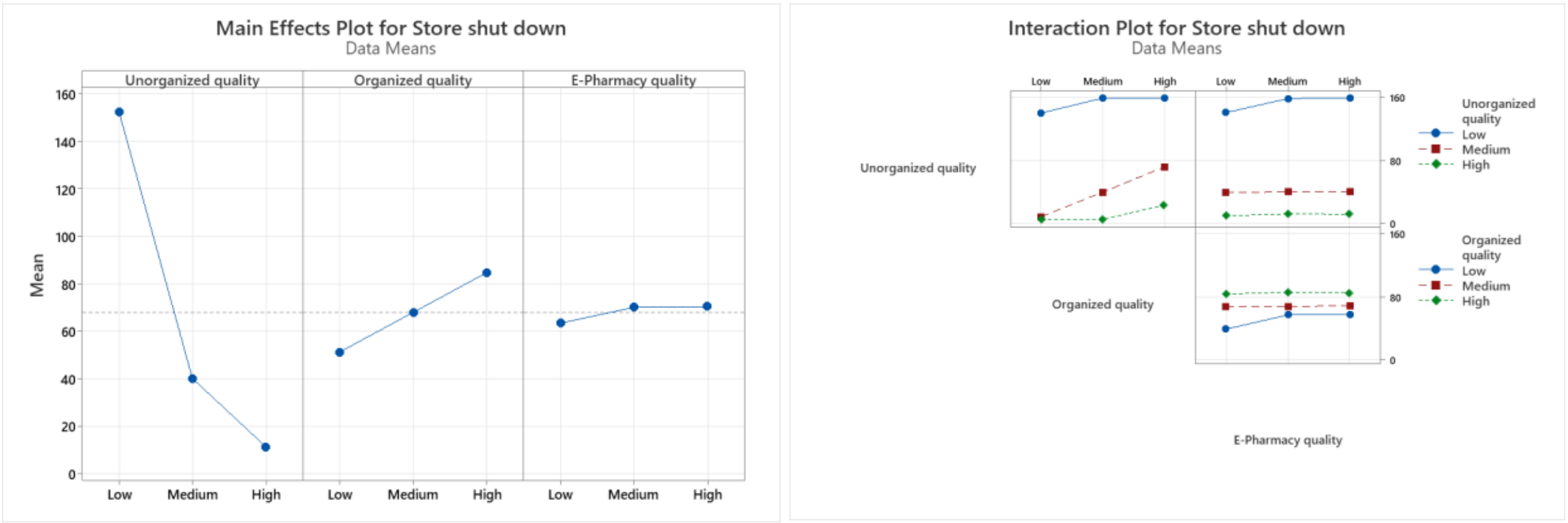

FIGURE D10 Main effect and interaction effect plots for the number of unorganized retailers shut down.

The main effects and interaction plots are shown in Figure D10. From the main effect plot it is evident that with an increase in organized quality and e-pharmacy quality, there is an increase in the number of unorganized retailers shut down. It is also evident that with an increase in unorganized quality, there is significant decrease in the number of unorganized retailers shut down. It can be concluded that the number of unorganized retailers shut down can be minimized by keeping unorganized quality at high level while organized quality and e-pharmacy quality at a low level.

### D2.4. Effect of Unorganized Quality, Organized Quality and E-Pharmacy Quality on the Average Weekly Customer Footprint at Organized Pharmaceutical Retailers

Table D9 shows the ANOVA table for average weekly customer footprint at organized pharmaceutical retailer. Table D9 shows that the *p*-values for unorganized quality, organized quality and e-pharmacy quality are less than 0.05. This implies that all the three parameters have a significant influence on the average weekly customer footprint at organized pharmaceutical retailer. Percentage of contribution for each factor is shown in Table D9. It is observed that the unorganized quality has the highest effect on the response. The interaction effects are found not significant.

TABLE D9 ANOVA for average weekly customer footprint at organized pharmaceutical retailers.

| **Source** | **DF** | **Adj SS** | **Adj MS** | **F-Value** | ***P*-Value** | **Contribution (%)** |
|---|---|---|---|---|---|---|
| Unorganized quality | 2 | 890447356 | 445223678 | 59.89 | 0.000 | 46.13 |
| Organized quality | 2 | 599641580 | 299820790 | 40.33 | 0.000 | 31.07 |
| E-Pharmacy quality | 2 | 129937708 | 64968854 | 8.74 | 0.010 | 6.73 |
| Unorganized quality*Organized quality | 4 | 89673842 | 22418461 | 3.02 | 0.086 | 4.65 |
| Unorganized quality*E-Pharmacy quality | 4 | 85923170 | 21480792 | 2.89 | 0.094 | 4.45 |
| Organized quality*E-Pharmacy quality | 4 | 75092695 | 18773174 | 2.53 | 0.123 | 3.89 |
| Error | 8 | 59475572 | 7434447 | | | 3.08 |
| Total | 26 | 1930191924 | | | | |

R-Sq = 96.92%  R-Sq (Adj) = 89.99%

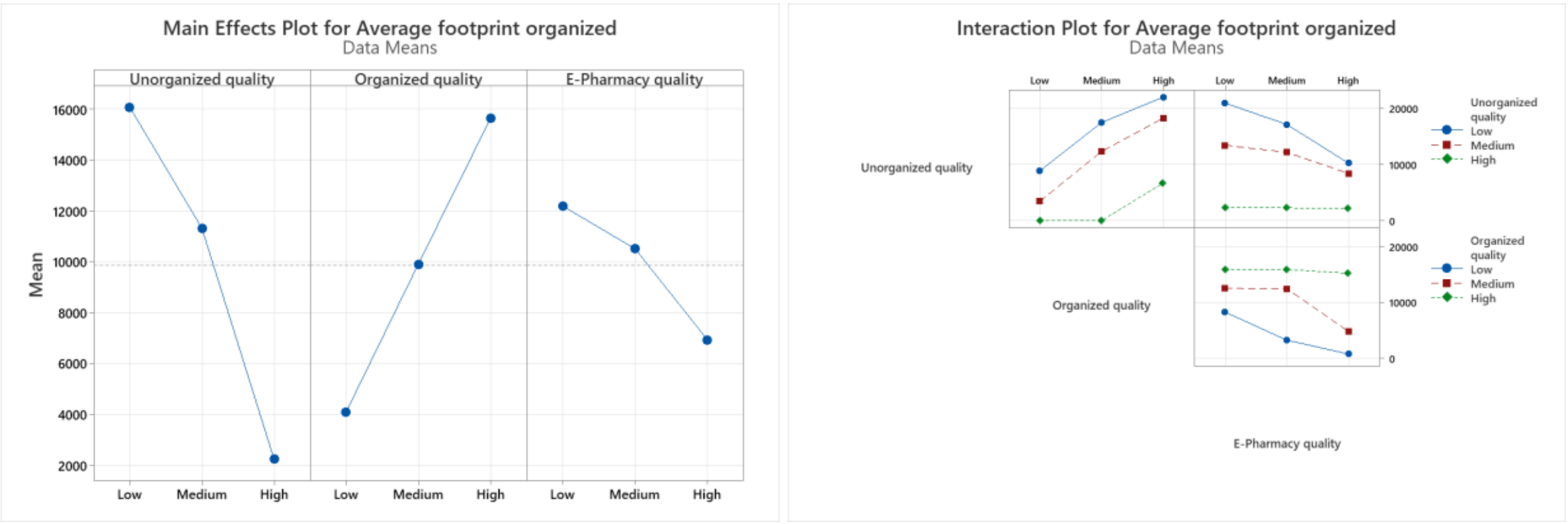

FIGURE D11 Main effect and interaction effect plots for average footprint at organized retailers.

The main effects and interaction plots are shown in Figure D11. From the main effect plot it is evident that with an increase in unorganized quality and e-pharmacy quality, there is a decrease in the average footprint at organized retailer. It is also evident that with an increase in organized quality, there is increase in the average footprint at organized retailer. It can be concluded that the average footprint at organized retailer can be maximized by keeping organized quality at high level while unorganized quality and e-pharmacy quality at a low level.

### D2.5. Effect of Unorganized Quality, Organized Quality and E-Pharmacy Quality on the Average Market Share of the Organized Pharmaceutical Retailers

Table D10 shows the ANOVA table for the average market share of the organized pharmaceutical retailers. Table D10 shows that the *p*-values for unorganized quality, organized quality and e-pharmacy quality are less than 0.05. This implies that all the three parameters have a significant influence on the average market share of the organized pharmaceutical retailers. It is observed that the unorganized quality has the highest effect on the response. The interaction effects are found to be not significant.

TABLE D10 ANOVA for average market share of the organized pharmaceutical retailers.

| Source | DF | Adj SS | Adj MS | F-Value | *P*-Value | Contribution (%) |
|---|---|---|---|---|---|---|
| Unorganized quality | 2 | 17991 | 8995.7 | 59.91 | 0.000 | 46.13 |
| Organized quality | 2 | 12117 | 6058.4 | 40.35 | 0.000 | 31.07 |
| E-Pharmacy quality | 2 | 2625 | 1312.5 | 8.74 | 0.010 | 6.73 |
| Unorganized quality*Organized quality | 4 | 1811 | 452.7 | 3.01 | 0.086 | 4.64 |
| Unorganized quality*E-Pharmacy quality | 4 | 1735 | 433.8 | 2.89 | 0.094 | 4.45 |
| Organized quality*E-Pharmacy quality | 4 | 1517 | 379.2 | 2.53 | 0.123 | 3.89 |
| Error | 8 | 1201 | 150.2 | | | 3.08 |
| Total | 26 | 38997 | | | | |
| R-Sq = 96.92% | | | R-Sq (Adj) = 89.99% | | | |

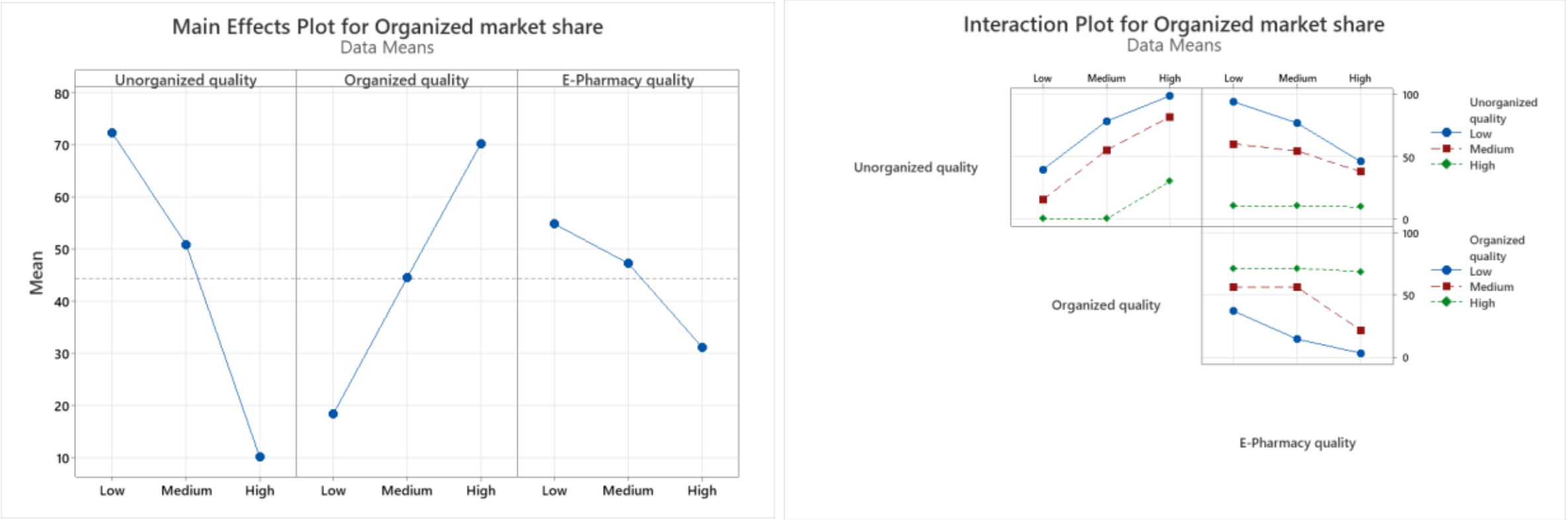

FIGURE D12 Main effect and interaction effect plots for average market share of organized retailers.

The main effects and interaction plots are shown in Figure D12. From the main effect plot it is evident that with an increase in unorganized quality and e-pharmacy quality, there is a decrease in the average market share of the organized retailers. It is also evident that with an increase in organized quality, there is an increase in the average market share of organized retailers. It can be concluded that the average market share of the organized retailers can be maximized by keeping organized quality at high level while unorganized quality and e-pharmacy quality at a low level.

### D2.6. Effect of Unorganized Quality, Organized Quality and E-Pharmacy Quality on the Average Weekly Customer Footprint at E-Pharmaceutical Retailers

Table D11 shows the ANOVA table for average weekly customer footprint at e-pharmaceutical retailer. Table D11 shows that the *p*-values for unorganized quality, organized quality and e-pharmacy quality are less than 0.05. This implies that all the three parameters have a significant influence on the average weekly customer footprint at e-pharmaceutical retailer. Percentage of contribution for each factor is shown in Table D11. It is observed that the e-pharmacy quality has the highest effect on the response. The interaction effects are found to be not significant.

TABLE D11 ANOVA for average weekly customer footprint at e-pharmaceutical retailers.

| **Source** | **DF** | **Adj SS** | **Adj MS** | **F-Value** | ***P*-Value** | **Contribution (%)** |
|---|---|---|---|---|---|---|
| Unorganized quality | 2 | 136949573 | 68474786 | 8.14 | 0.012 | 15.08 |
| Organized quality | 2 | 171285192 | 85642596 | 10.19 | 0.006 | 18.86 |
| E-Pharmacy quality | 2 | 215085003 | 107542502 | 12.79 | 0.003 | 23.68 |
| Unorganized quality*Organized quality | 4 | 109724058 | 27431014 | 3.26 | 0.073 | 12.08 |
| Unorganized quality*E-Pharmacy quality | 4 | 88838728 | 22209682 | 2.64 | 0.113 | 9.78 |
| Organized quality*E-Pharmacy quality | 4 | 119077757 | 29769439 | 3.54 | 0.060 | 13.11 |
| Error | 8 | 67260475 | 8407559 | | | 7.40 |
| Total | 26 | 908220786 | | | | |

R-Sq = 92.59% R-Sq (Adj) = 75.93%

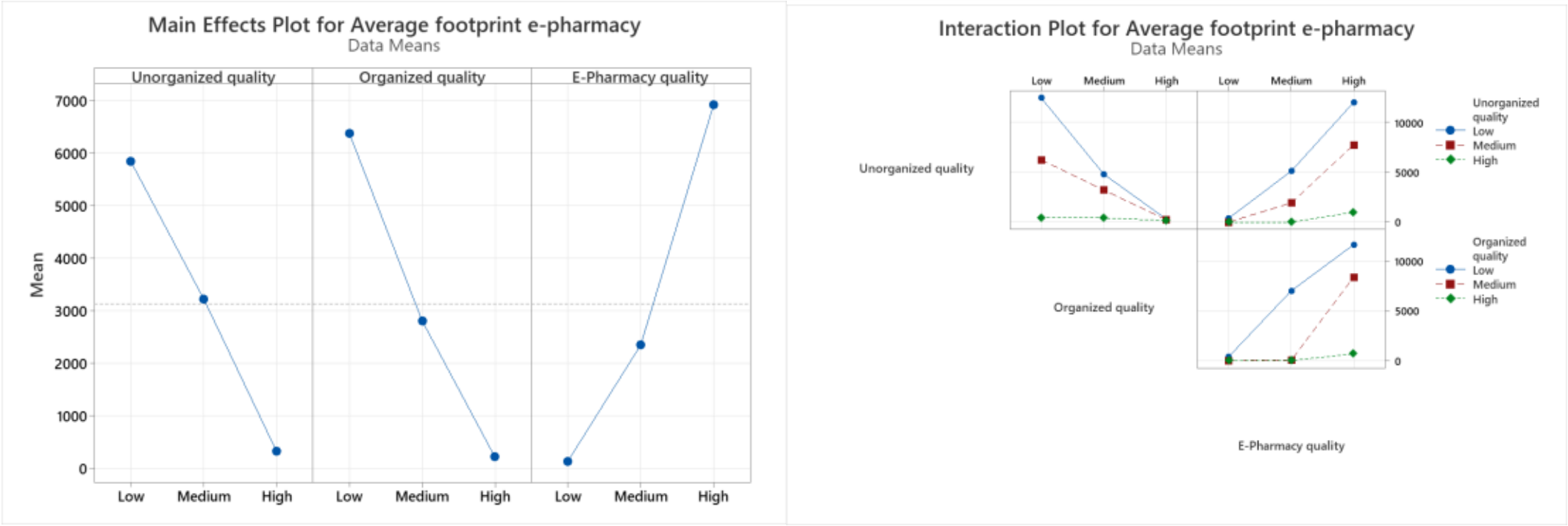

FIGURE D13 Main effect and interaction effect plots for average footprint at e-pharmaceutical retailers.

The main effects and interaction plots are shown in Figure D13. From the main effect plot it is evident that with an increase in unorganized quality and organized quality, there is a decrease in the average footprint at e-pharmaceutical retailer. It is also evident that with an increase in e-pharmacy quality, there is an increase in the average footprint at e-pharmaceutical retailer. It can be concluded that the average footprint at e-pharmaceutical retailer can be maximized by keeping e-pharmacy quality at high level while unorganized quality and organized quality at a low level.

### D2.7. Effect of Unorganized Quality, Organized Quality and E-Pharmacy Quality on the Average Market Share of the E-Pharmaceutical Retailers

Table D12 shows the ANOVA table for the average market share of the e-pharmaceutical retailers. Table D12 shows that the *p*-values for unorganized quality, organized quality and e-pharmacy quality are less than 0.05. This implies that all the three parameters have a significant influence on the average market share of the e-pharmaceutical retailers. It is observed that the e-pharmacy quality has the highest effect on the response. The interaction effects are found to be not significant.

TABLE D12 ANOVA for average market share of the e-pharmaceutical retailers.

| Source | DF | Adj SS | Adj MS | F-Value | *P*-Value | Contribution (%) |
|---|---|---|---|---|---|---|
| Unorganized quality | 2 | 2768 | 1383.8 | 8.15 | 0.012 | 15.08 |
| Organized quality | 2 | 3460 | 1730.2 | 10.19 | 0.006 | 18.85 |
| E-Pharmacy quality | 2 | 4346 | 2172.8 | 12.79 | 0.003 | 23.68 |
| Unorganized quality*Organized quality | 4 | 2217 | 554.2 | 3.26 | 0.073 | 12.08 |
| Unorganized quality*E-Pharmacy quality | 4 | 1795 | 448.9 | 2.64 | 0.113 | 9.78 |
| Organized quality*E-Pharmacy quality | 4 | 2405 | 601.3 | 3.54 | 0.060 | 13.10 |
| Error | 8 | 1359 | 169.9 | | | 7.40 |
| Total | 26 | 18350 | | | | |
| R-Sq = 92.59% | | | R-Sq (Adj) = 75.93% | | | |

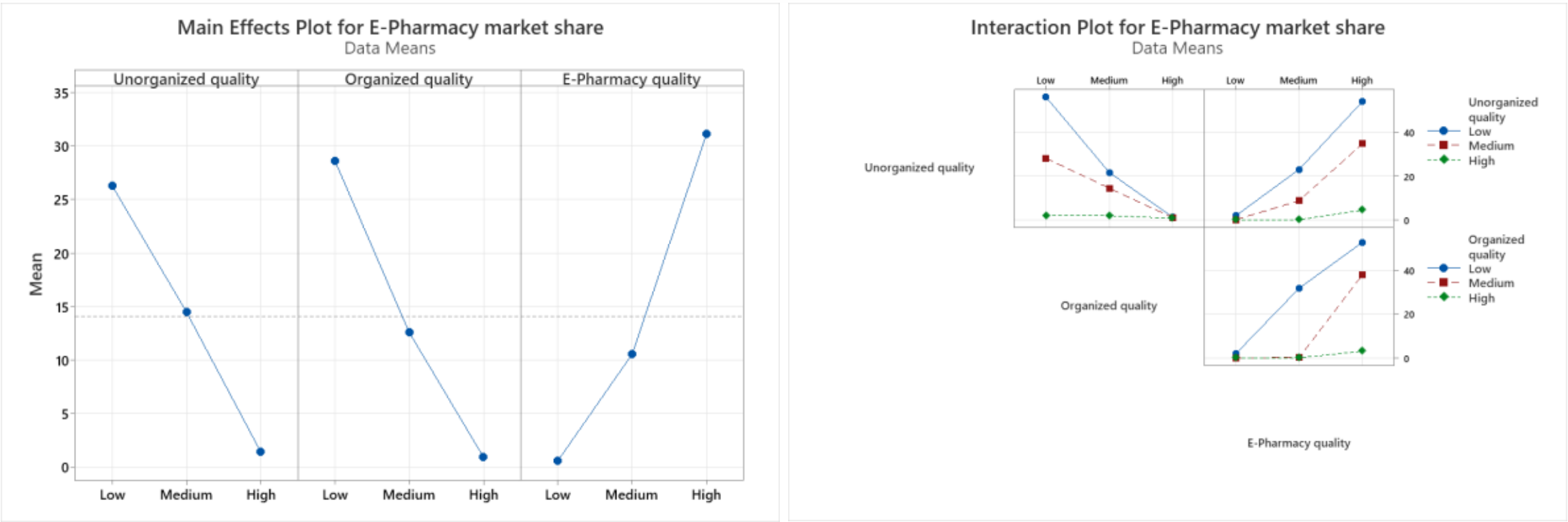

FIGURE D14 Main effect and interaction effect plots for average market share of e-pharmaceutical retailers.

The main effects and interaction plots are shown in Figure D14. From the main effect plot it is evident that with an increase in unorganized quality and organized quality, there is a decrease in the average market share of the e-pharmaceutical retailers. It is also evident that with an increase in e-pharmacy quality, there is an increase in the average market share of e-pharmaceutical retailers. It can be concluded that the average market share of the e-pharmaceutical retailers can be maximized by keeping e-pharmacy quality at high level while unorganized quality and organized quality at a low level.

## Appendix E: Simulation Results for Price Discount and Quality

### Section E1. Simulation Results for Price Discount

As per the experimental design in Table D1, the simulation model parameters are set and total 27 experiments are conducted. The simulation results for price discount are shown in this section.

### E1.1. Pictorial Representation of ABM Experiments Utilising ANOVA for Price Discount

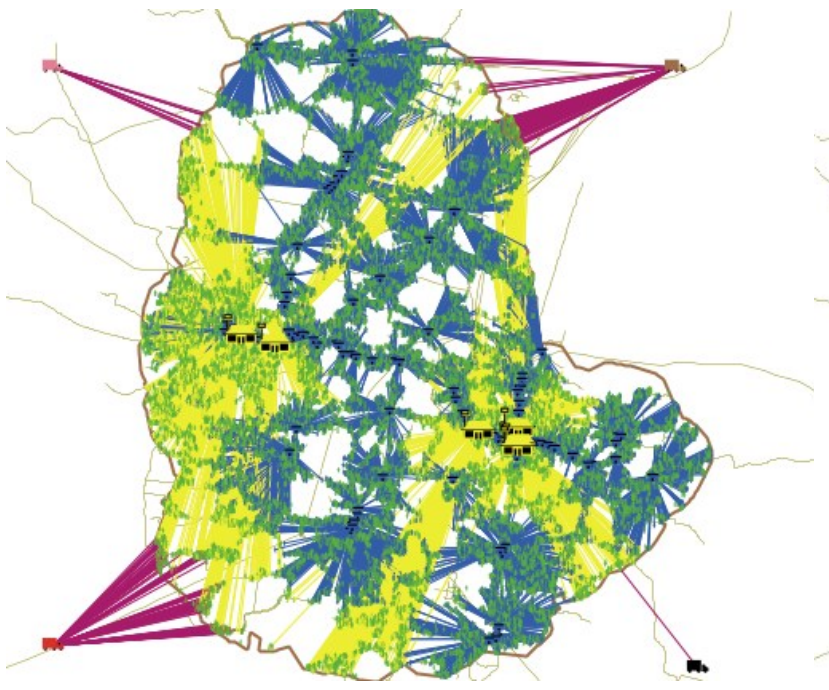

Experiment 1

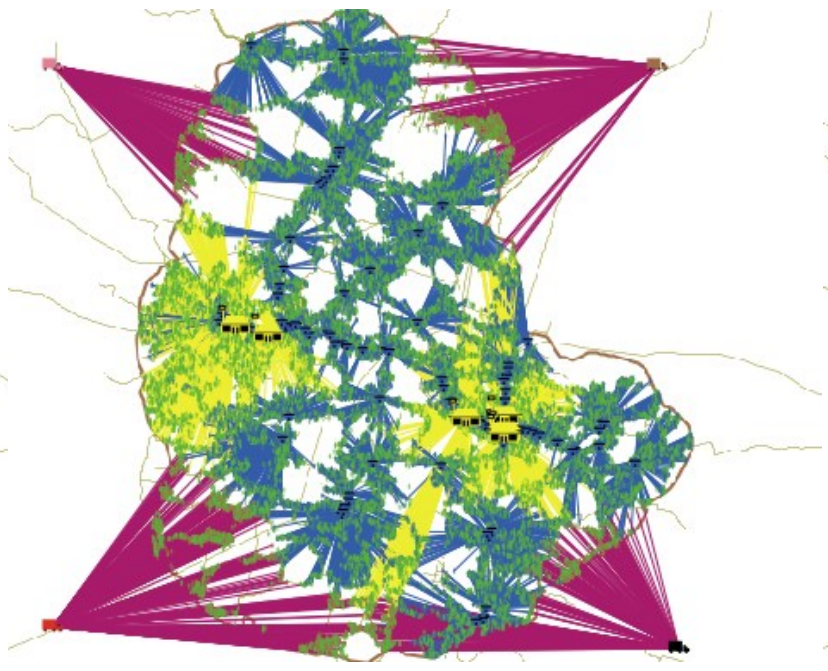

Experiment 2

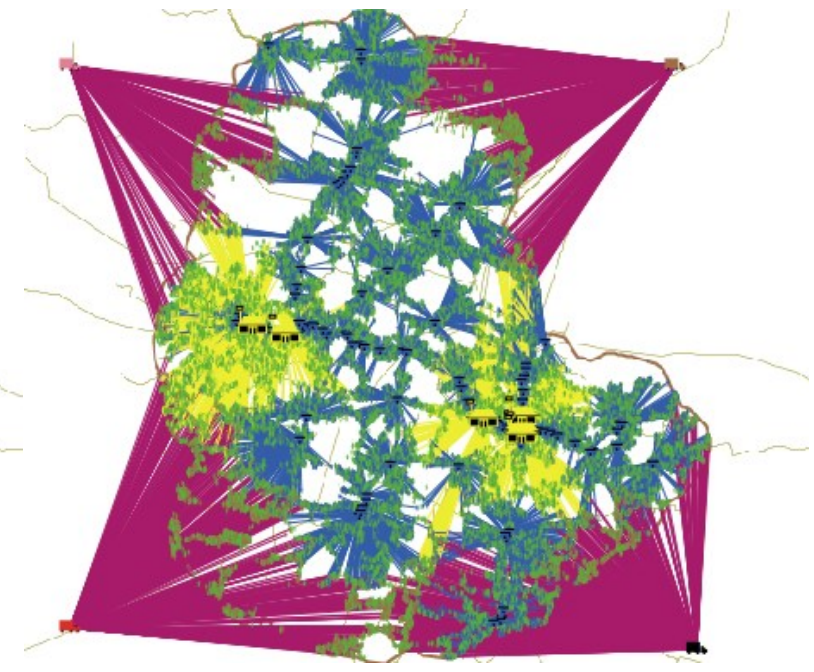

Experiment 3

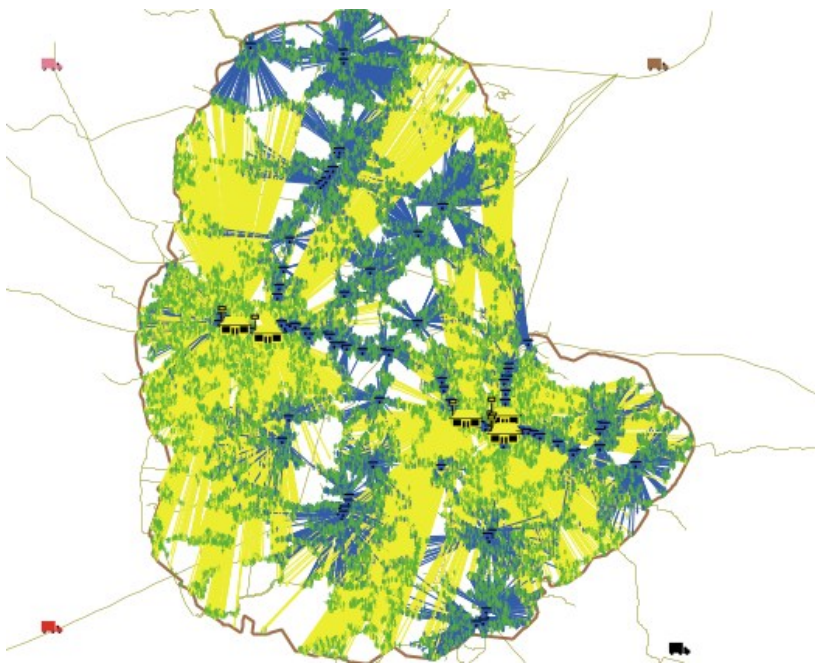

Experiment 4

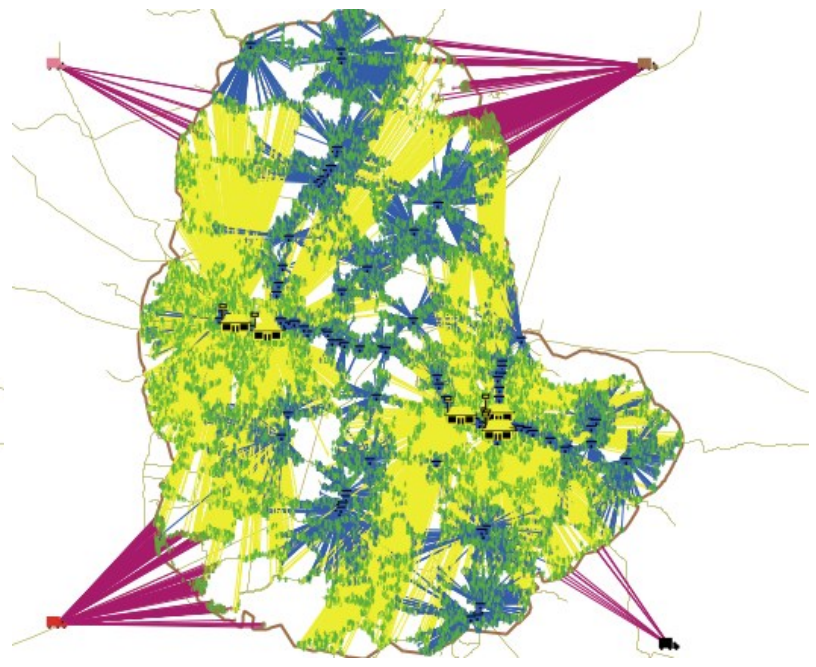

Experiment 5

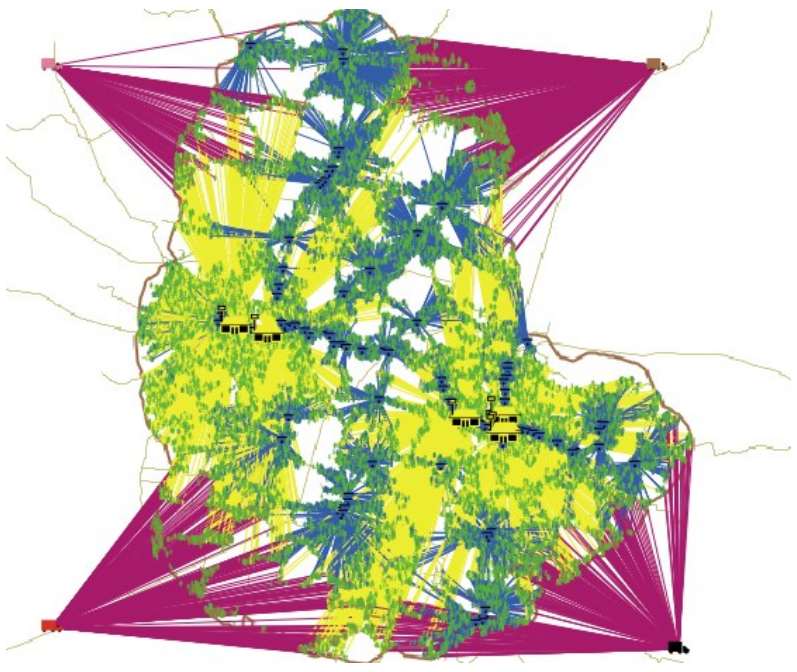

Experiment 6

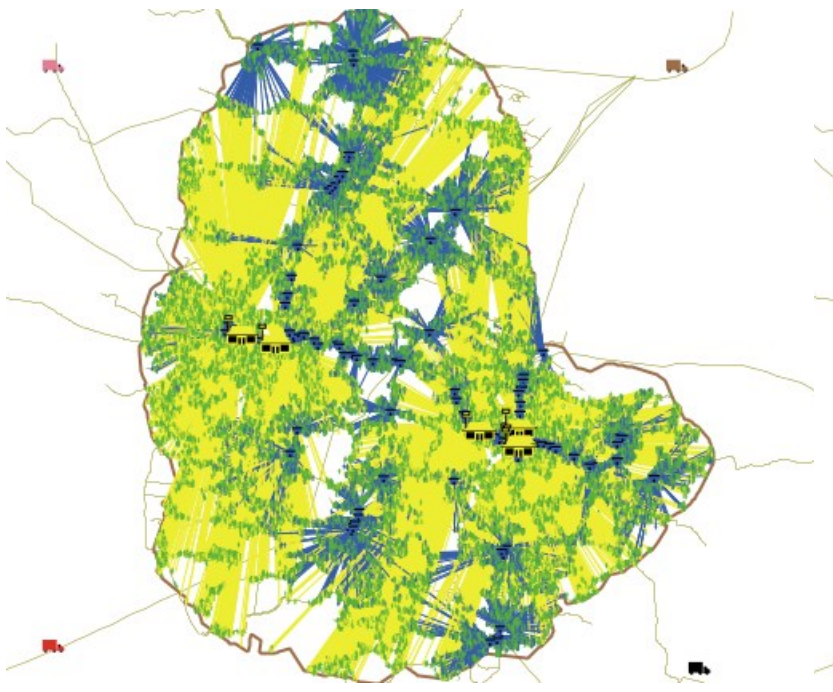

Experiment 7

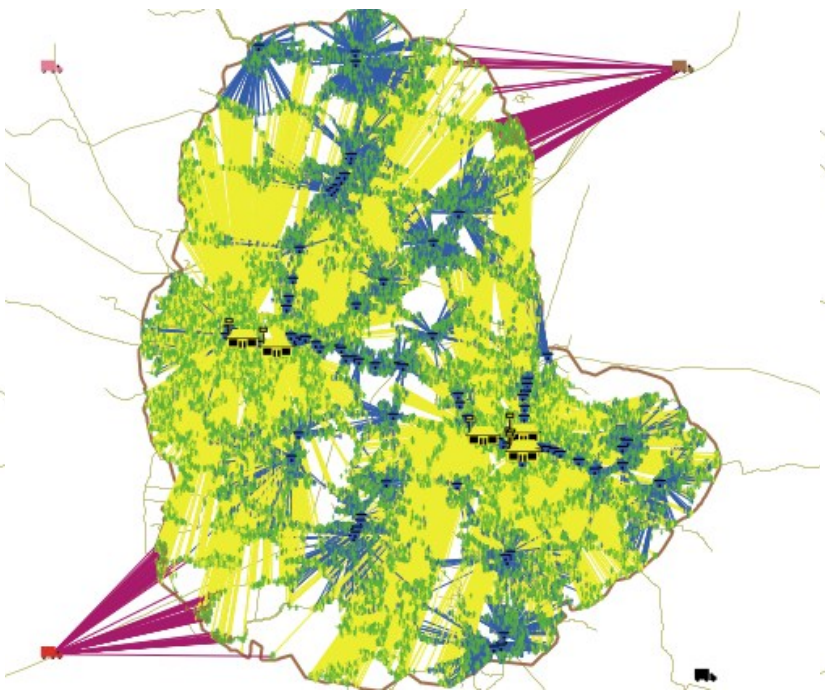

Experiment 8

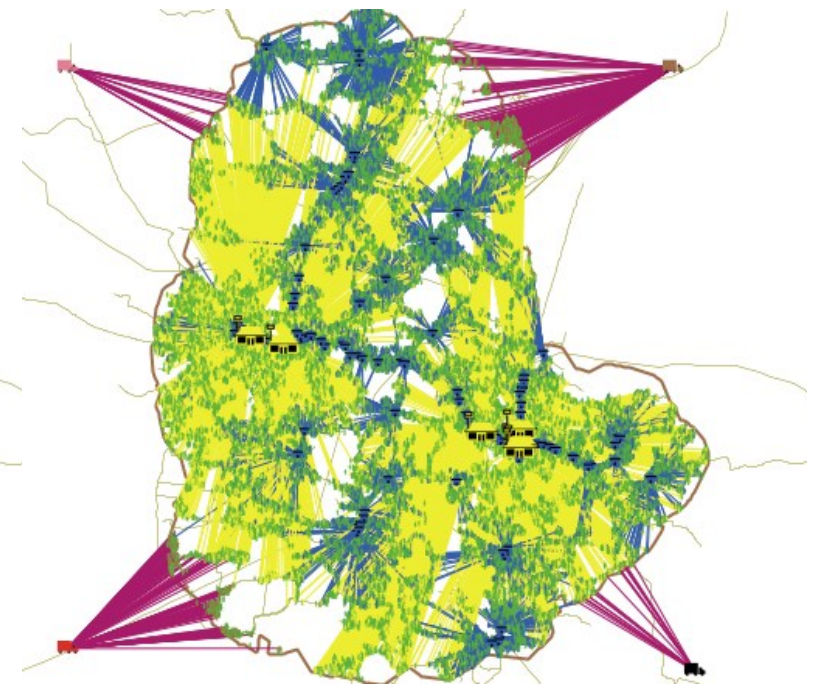

Experiment 9

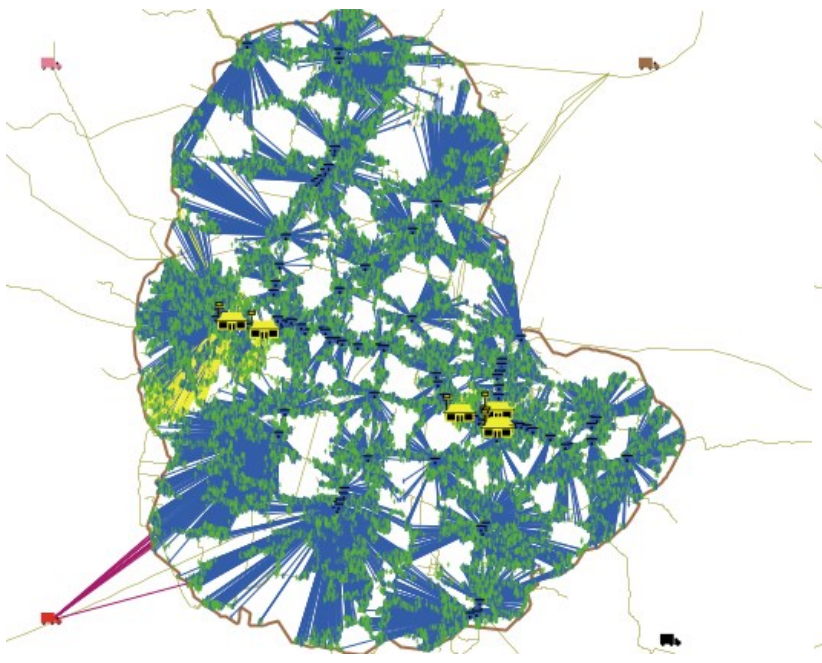

Experiment 10

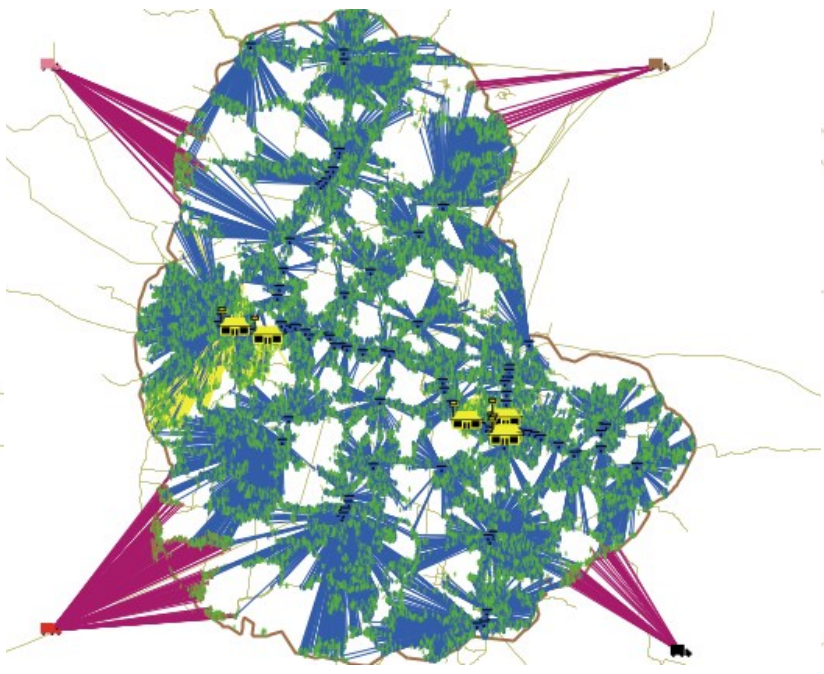

Experiment 11

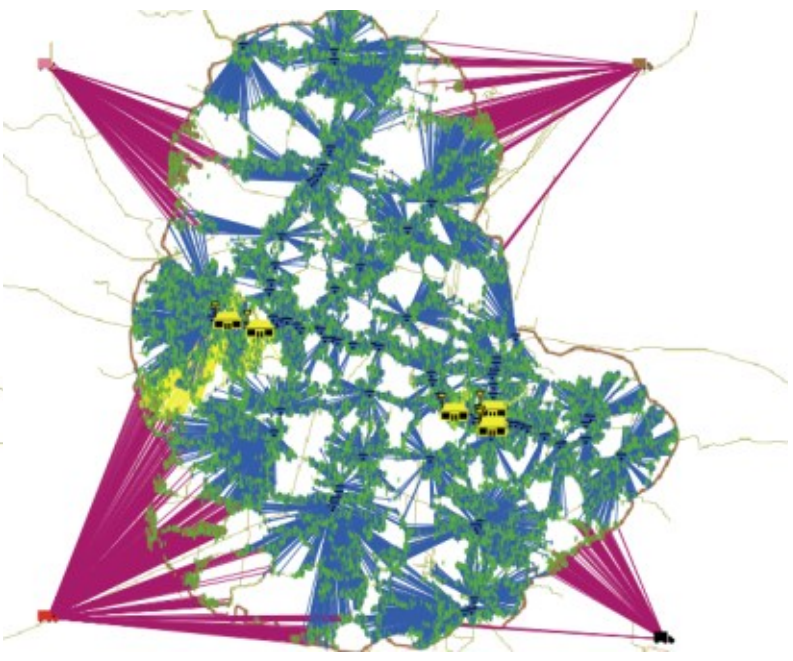

Experiment 12

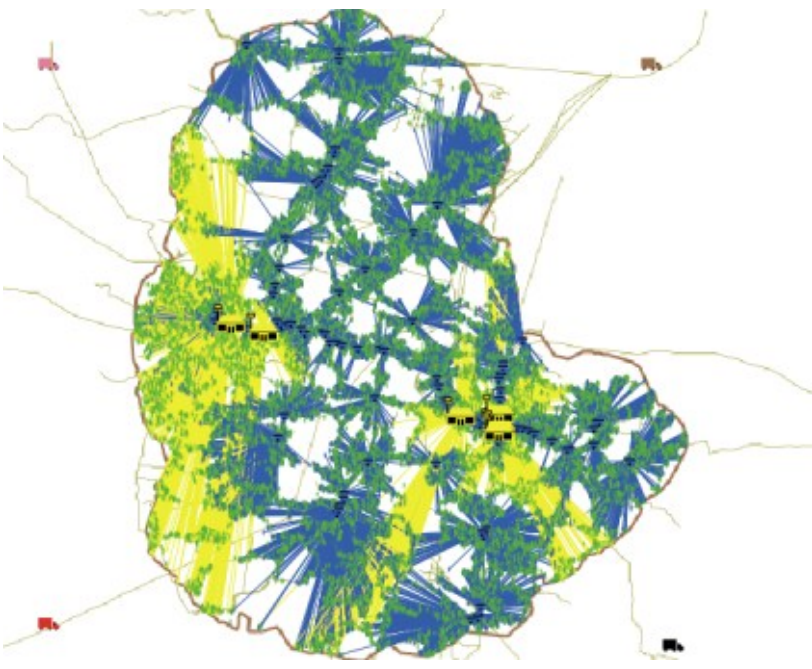

Experiment 13

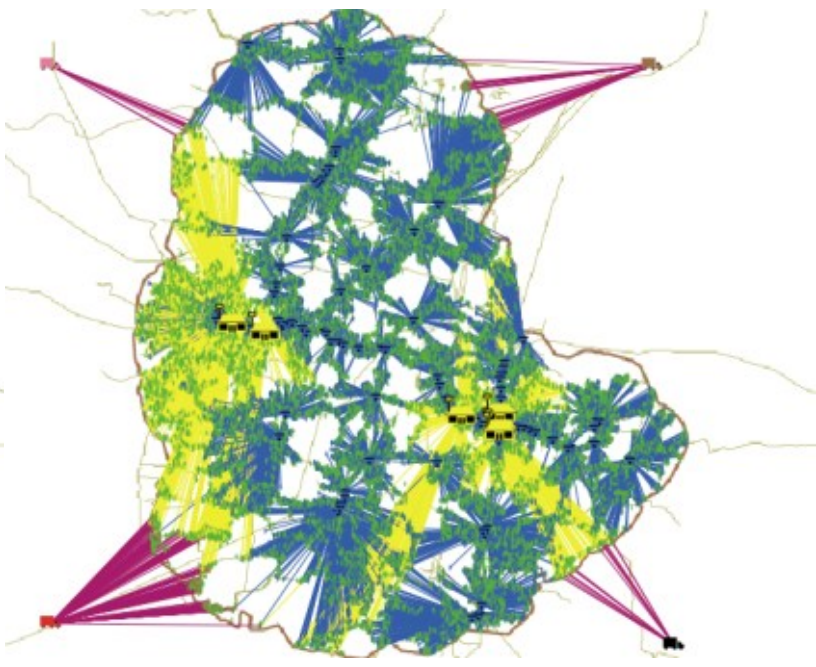

Experiment 14

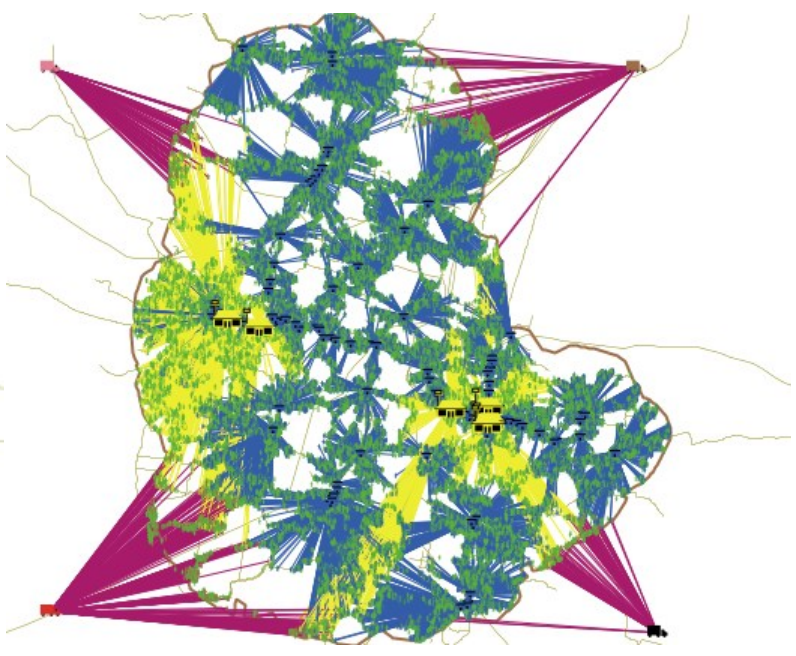

Experiment 15

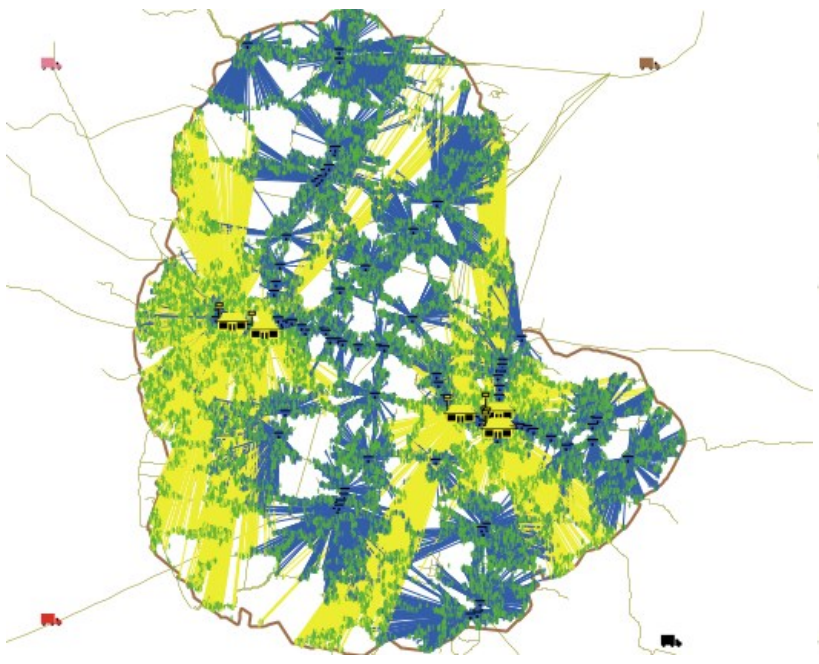

Experiment 16

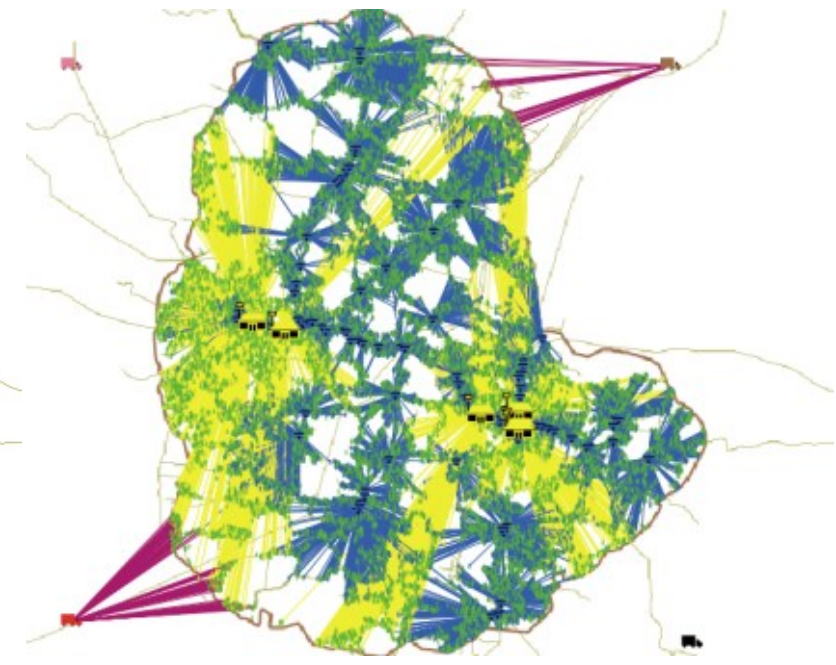

Experiment 17

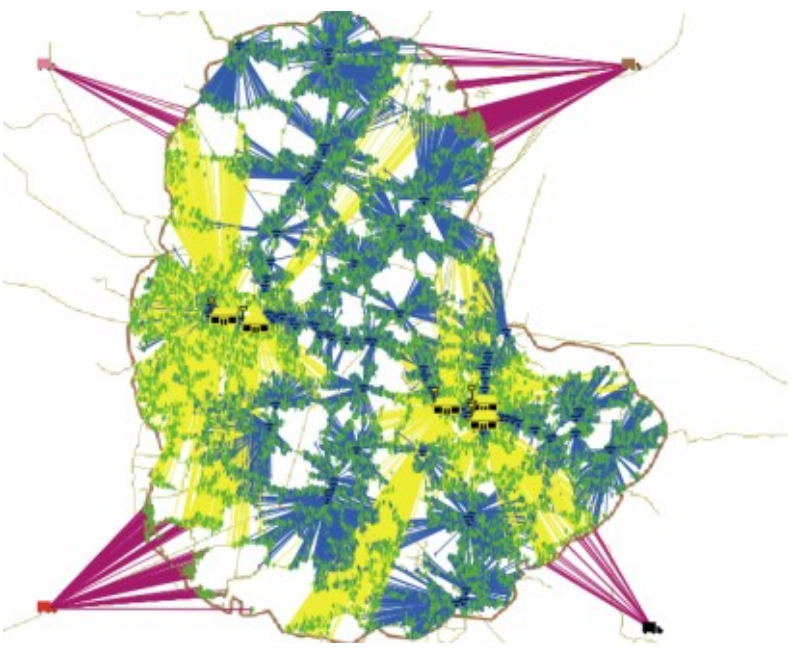

Experiment 18

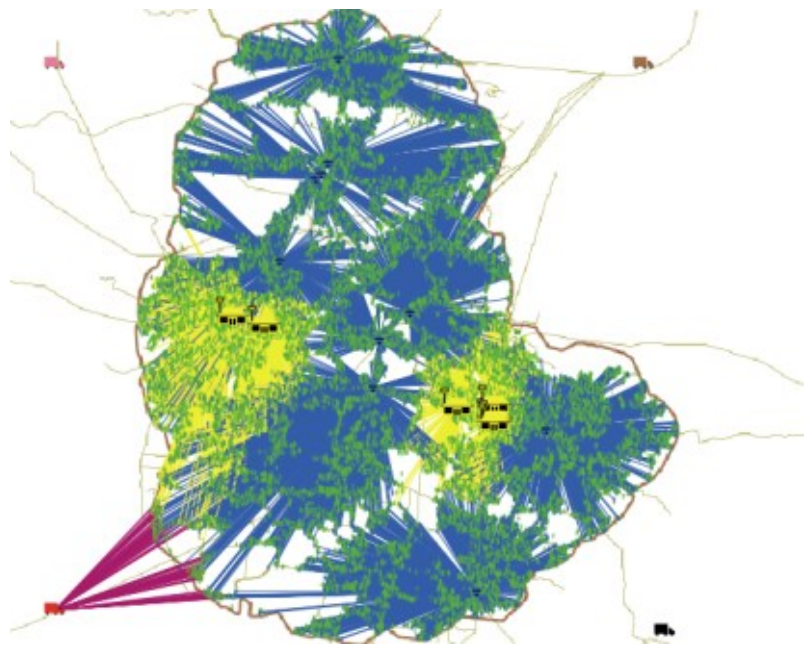

Experiment 19

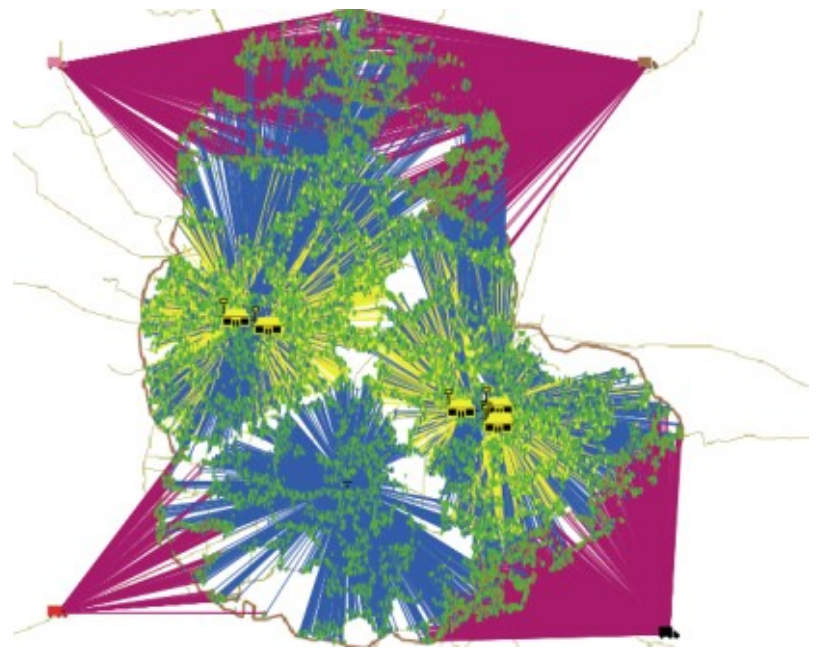

Experiment 20

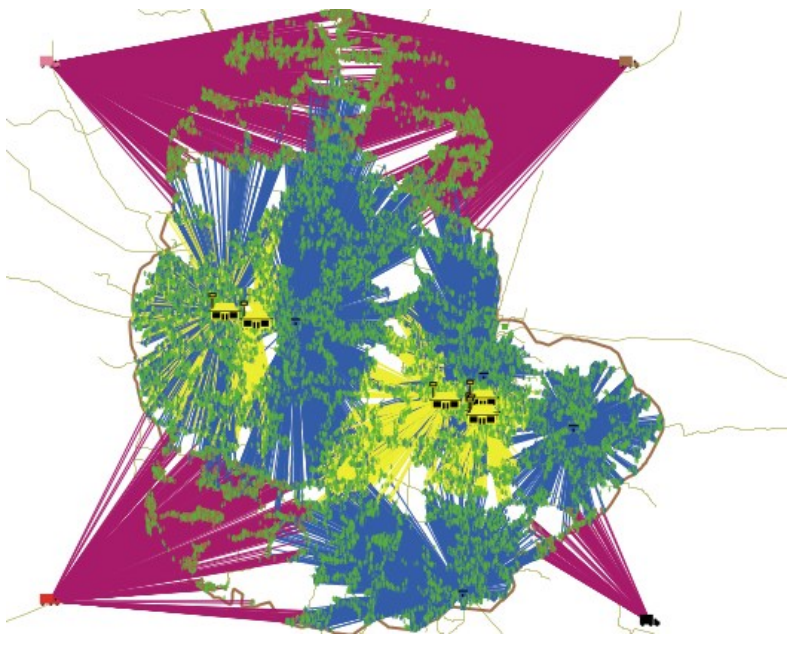

Experiment 21

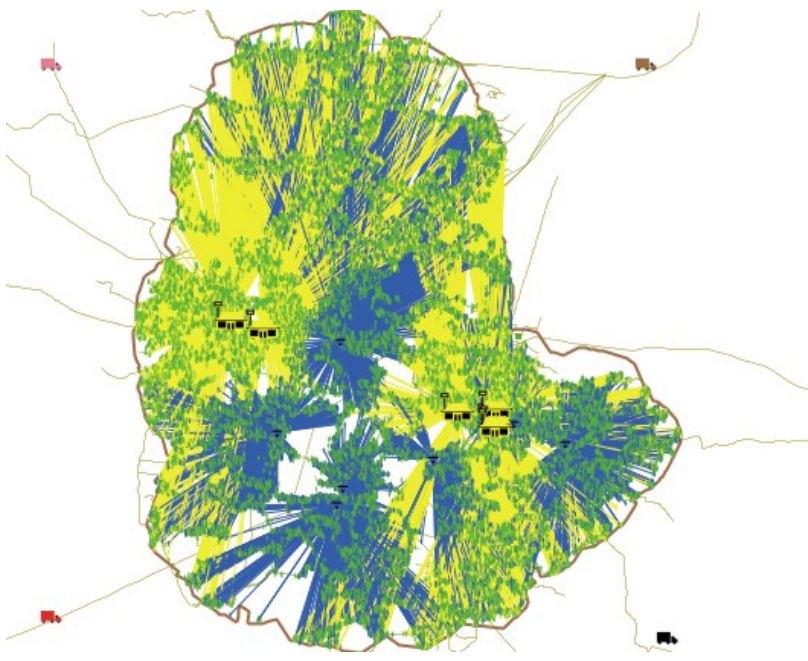

Experiment 22

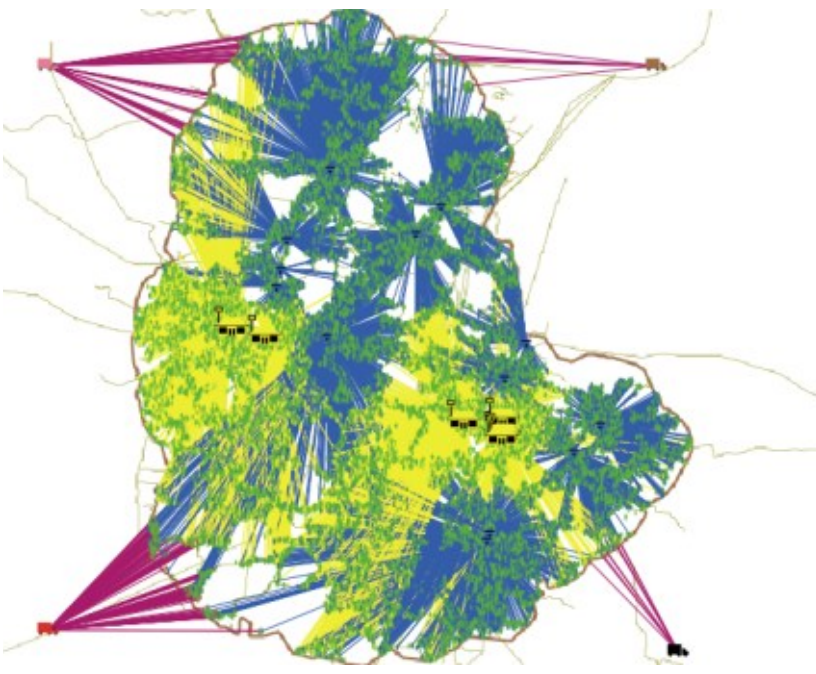

Experiment 23

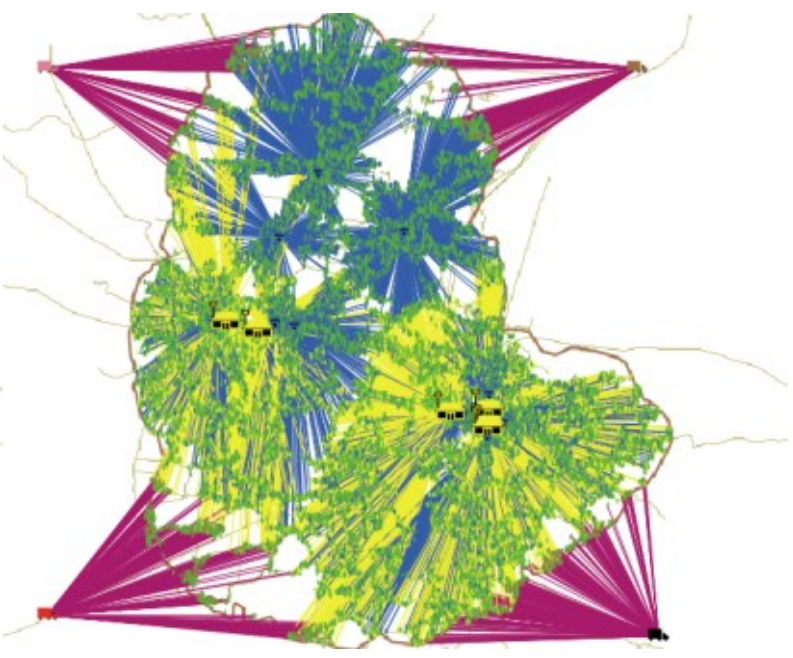

Experiment 24

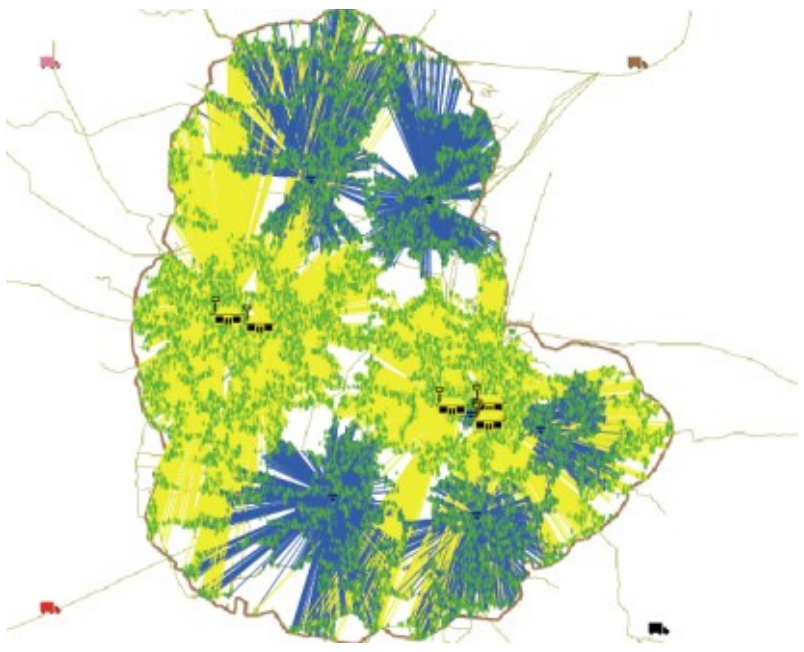

Experiment 25

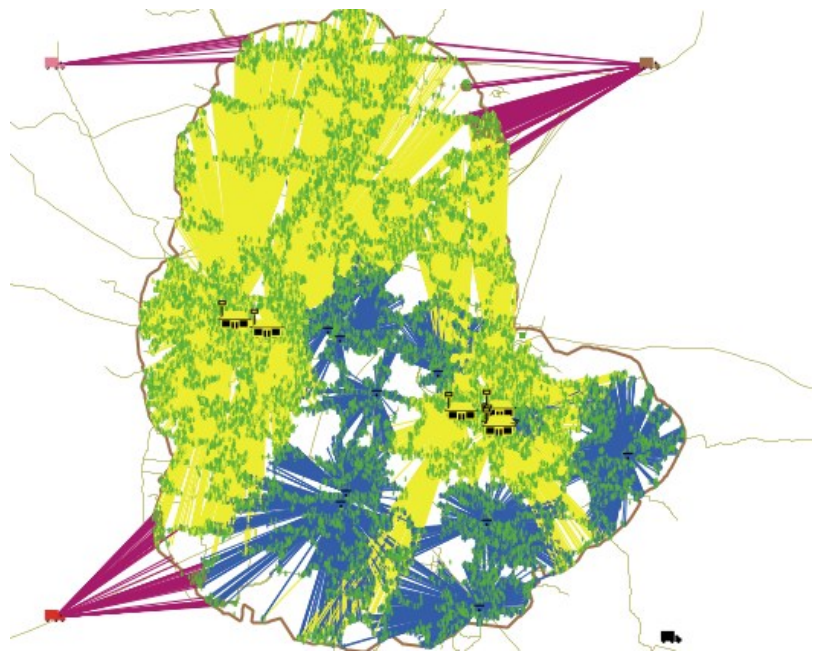

Experiment 26

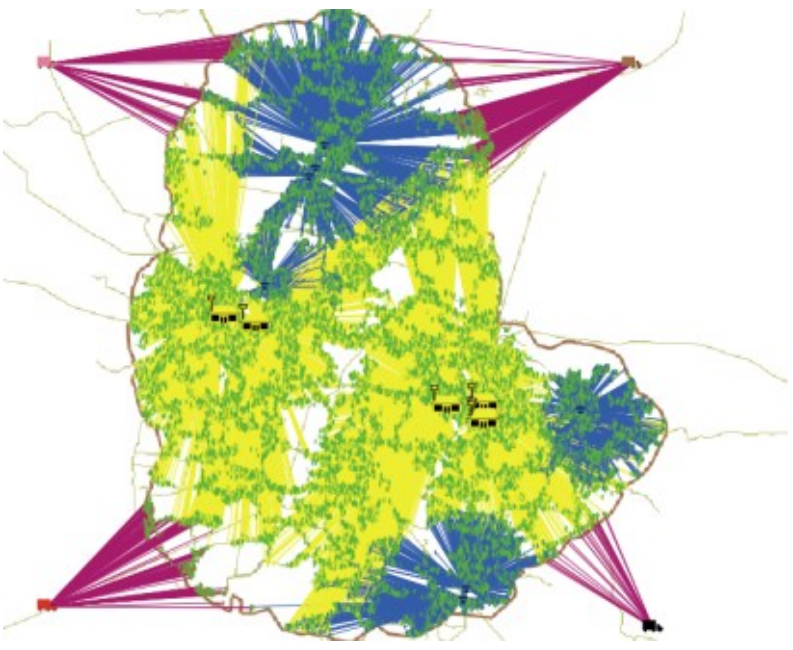

Experiment 27

FIGURE E1 Simulation results as per experimental design for price discount.

### E1.2. Simulation Results of Weekly Customer Footprint as per 27 Experimental Runs Utilising ANOVA for Price Discount

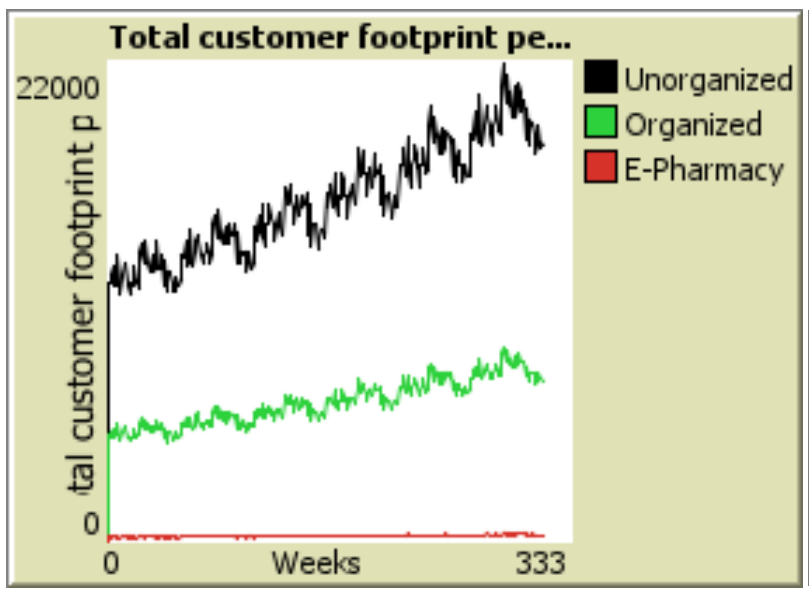

Experiment 1

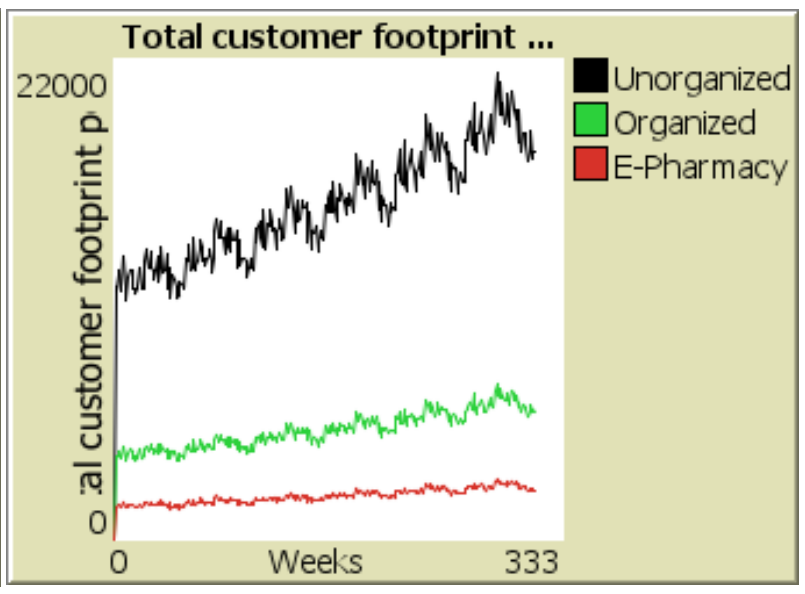

Experiment 2

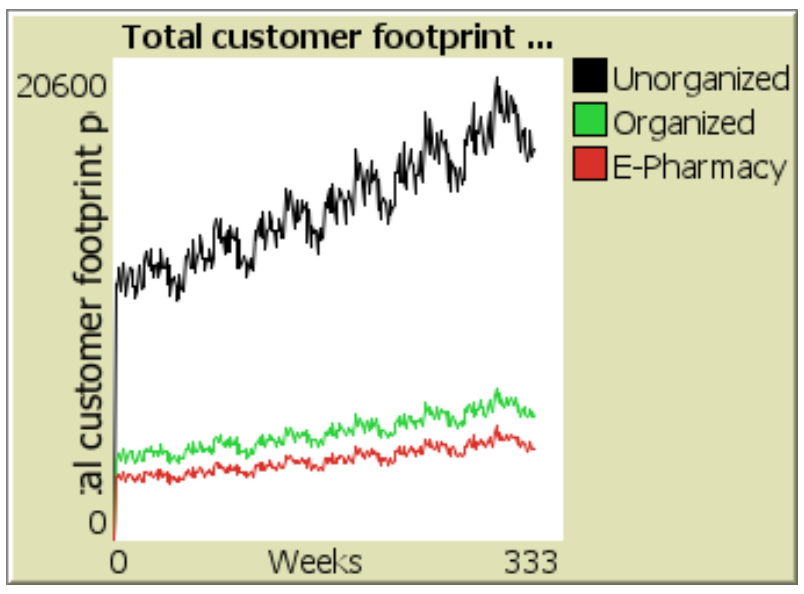

Experiment 3

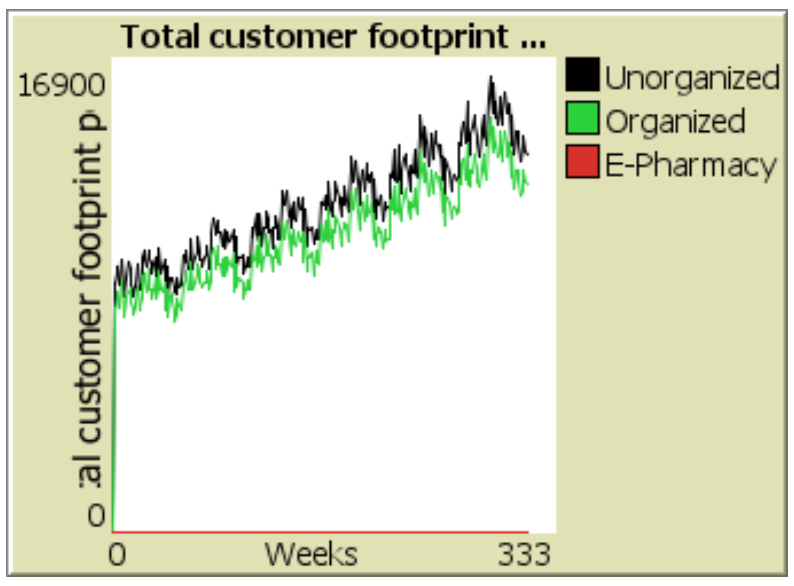

Experiment 4

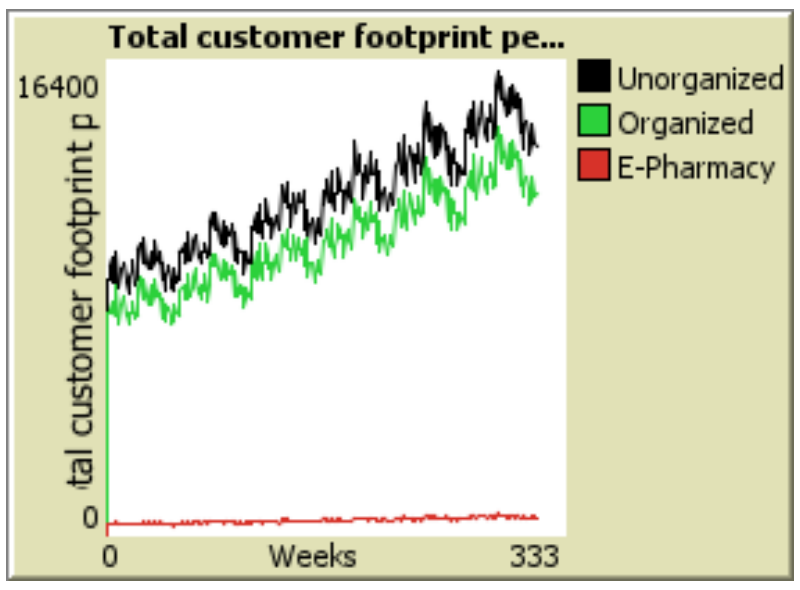

Experiment 5

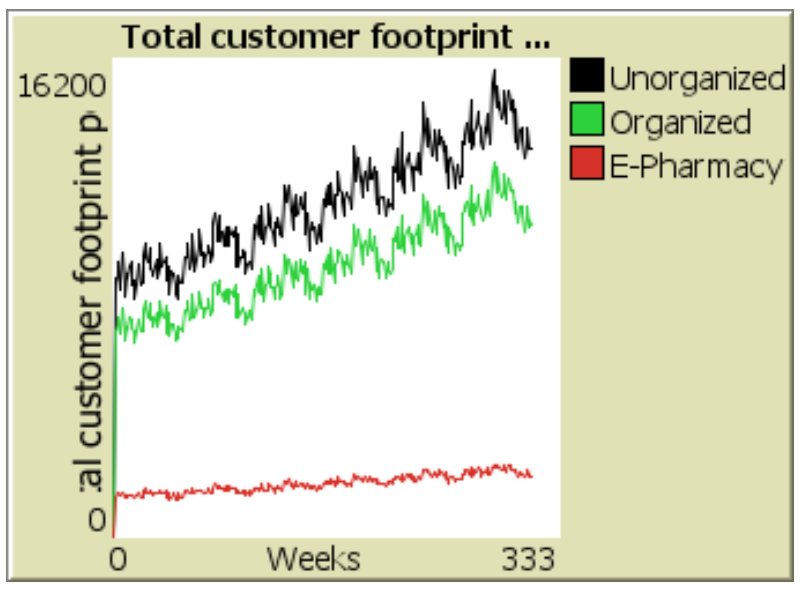

Experiment 6

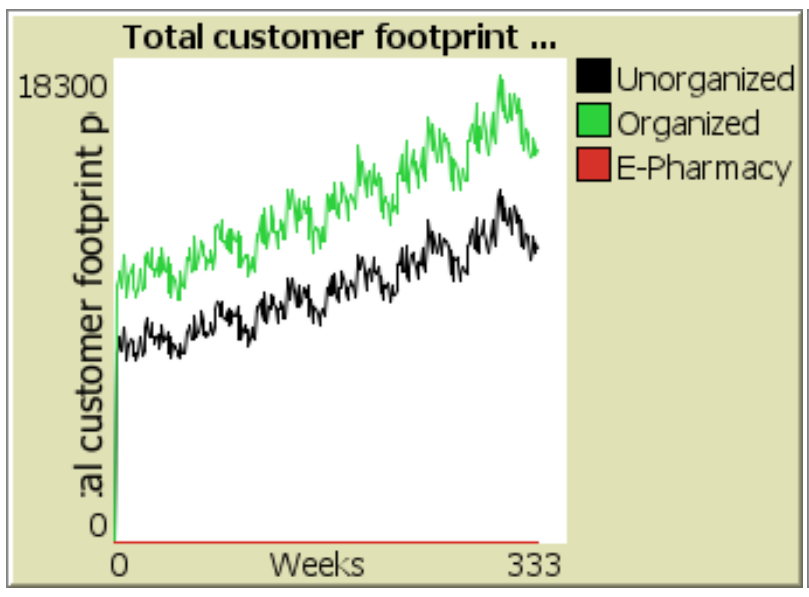

Experiment 7

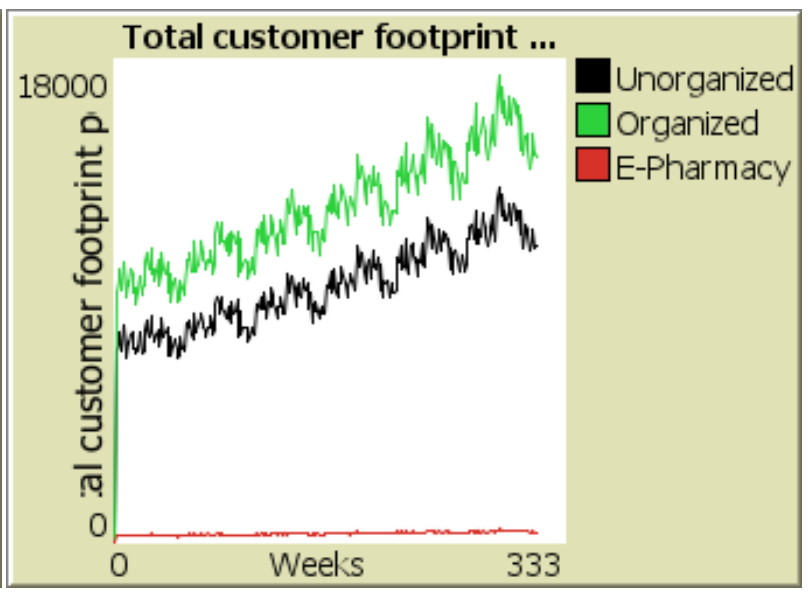

Experiment 8

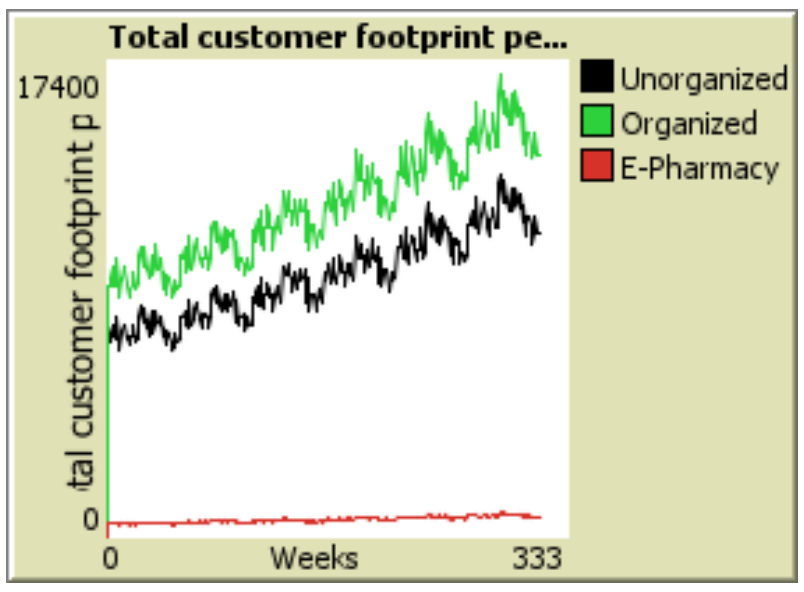

Experiment 9

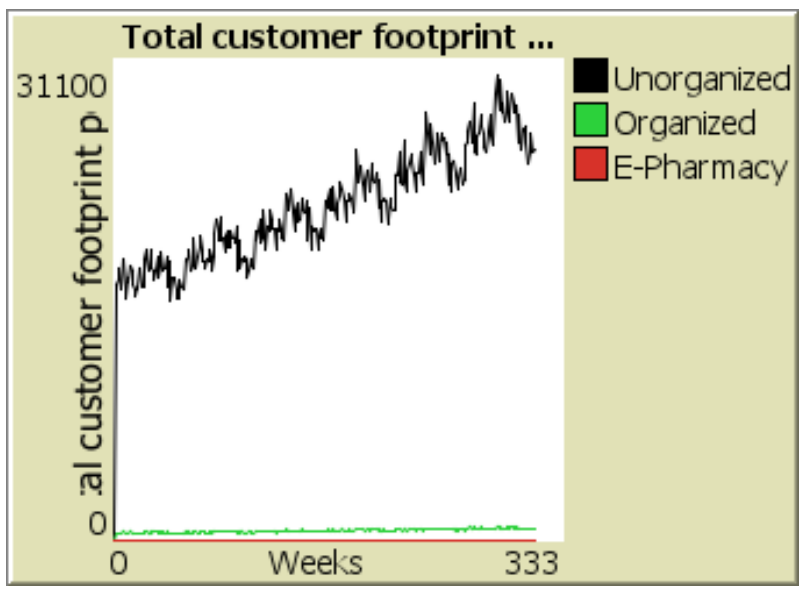

Experiment 10

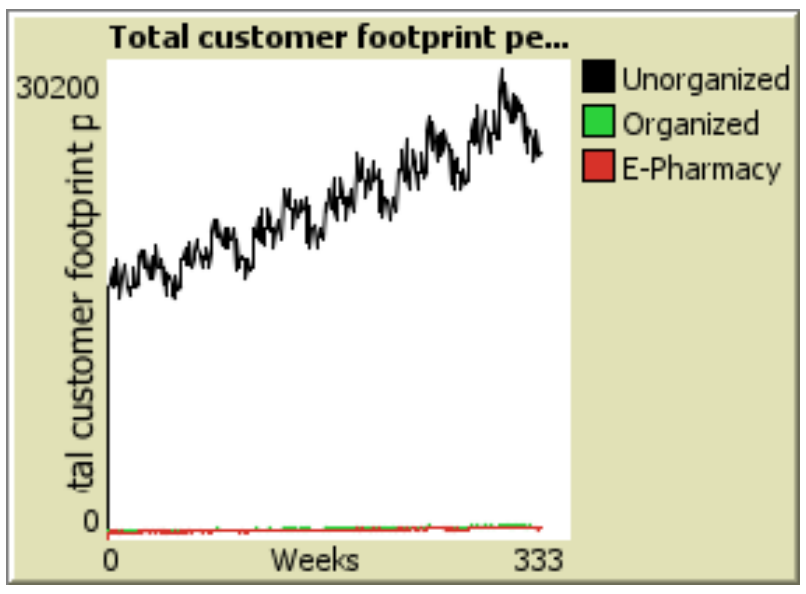

Experiment 11

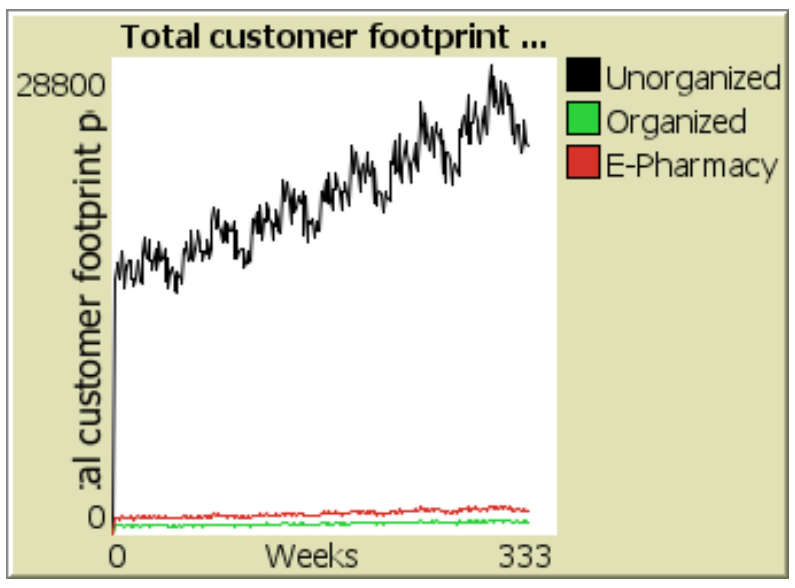

Experiment 12

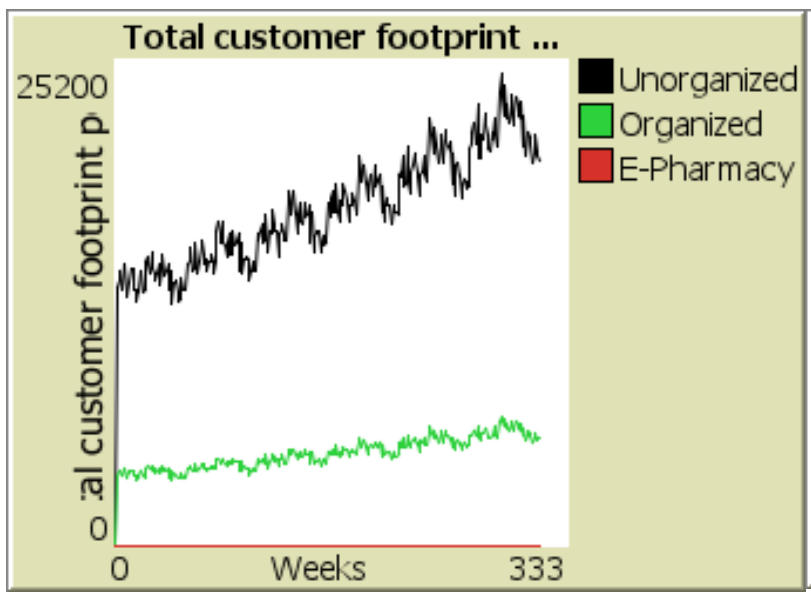

Experiment 13

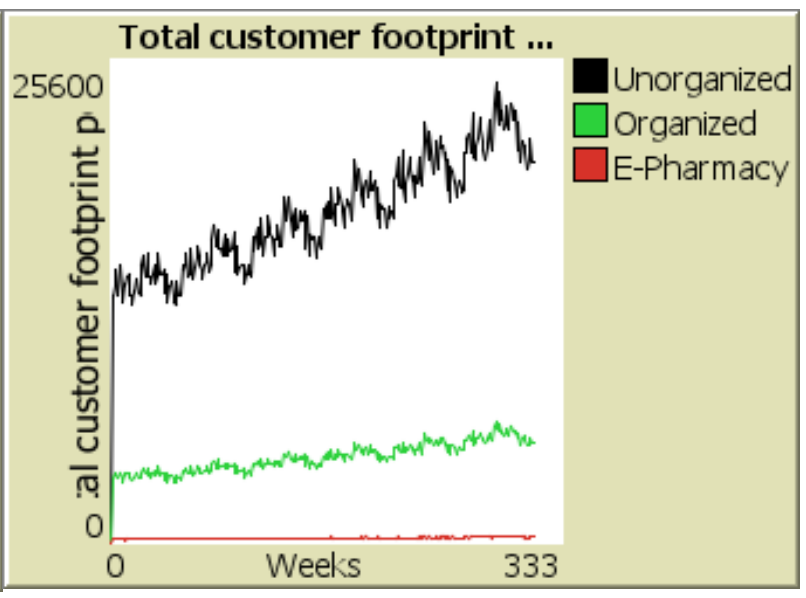

Experiment 14

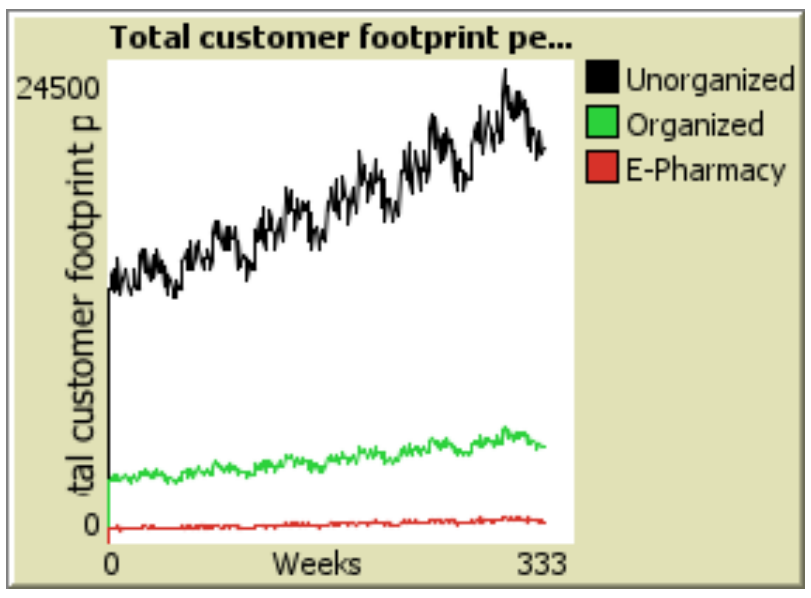

Experiment 15

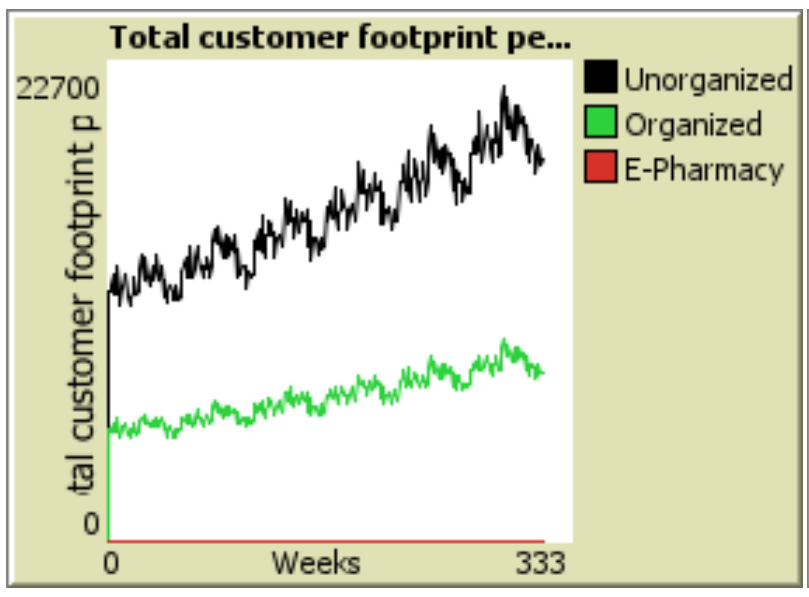

Experiment 16

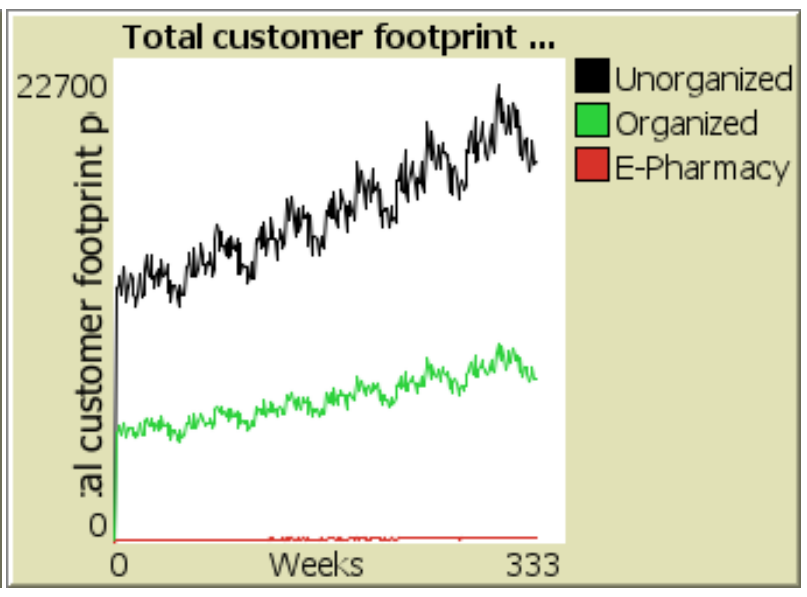

Experiment 17

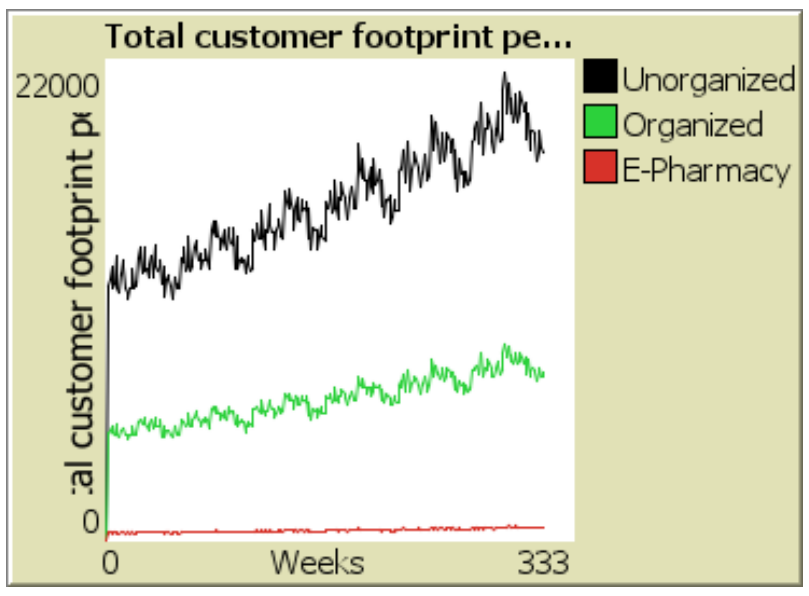

Experiment 18

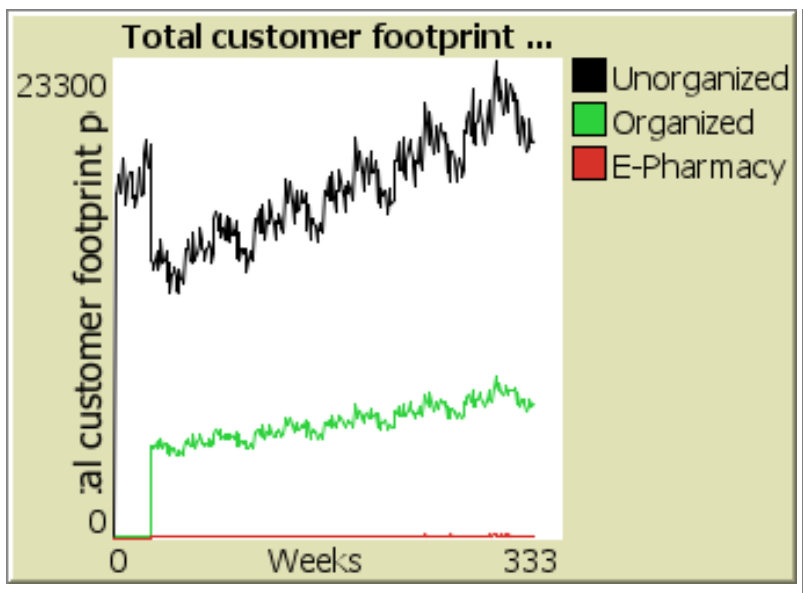

Experiment 19

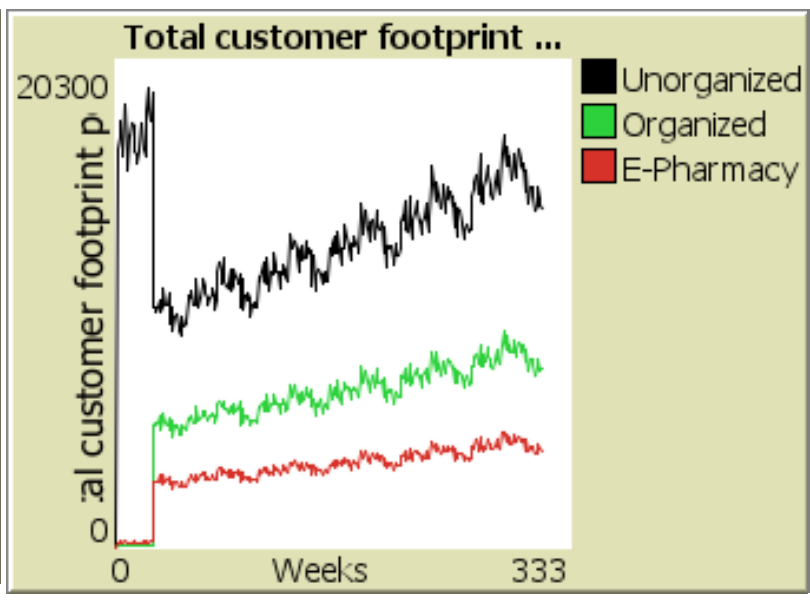

Experiment 20

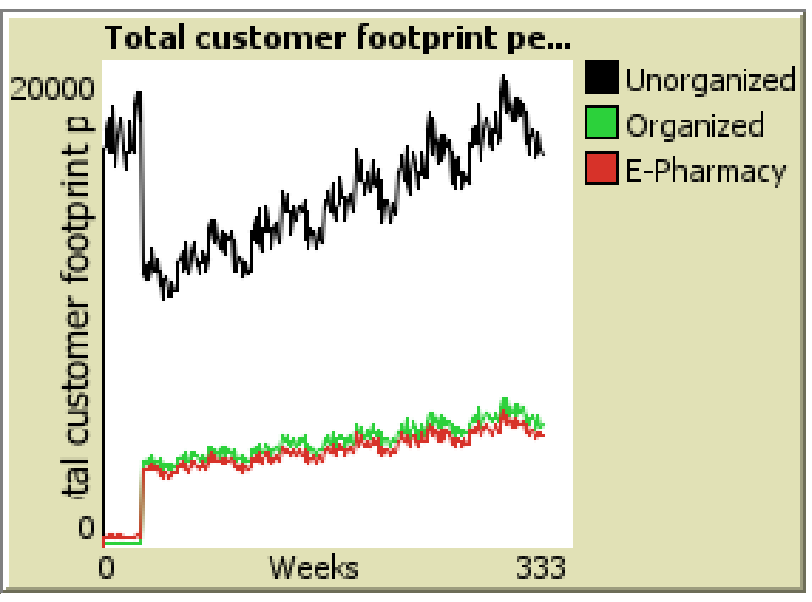

Experiment 21

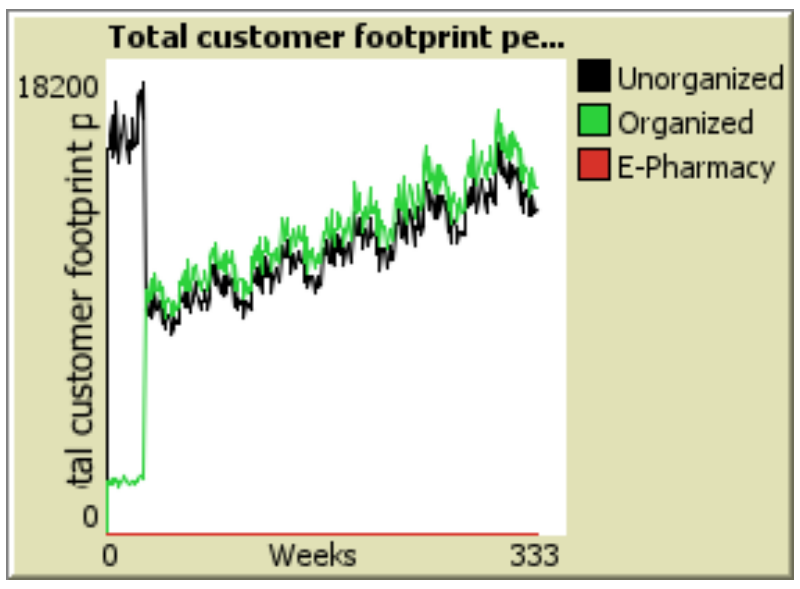

Experiment 22

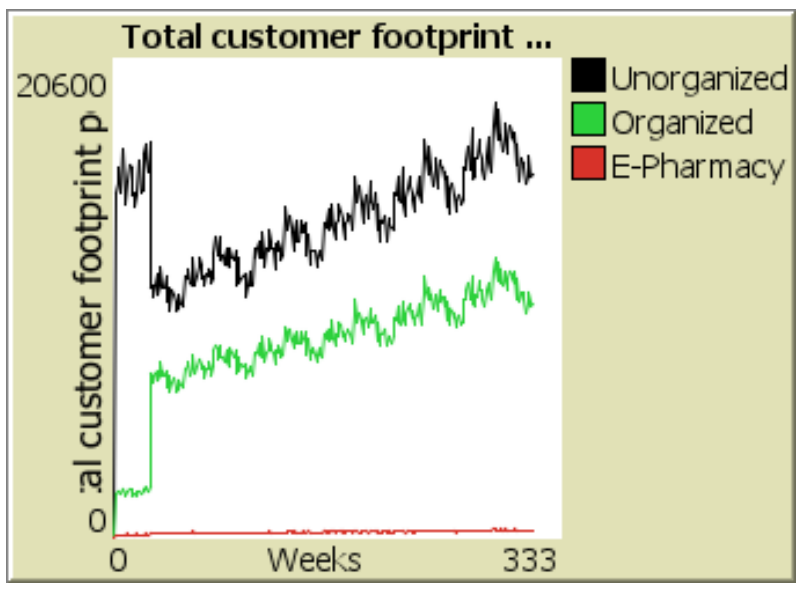

Experiment 23

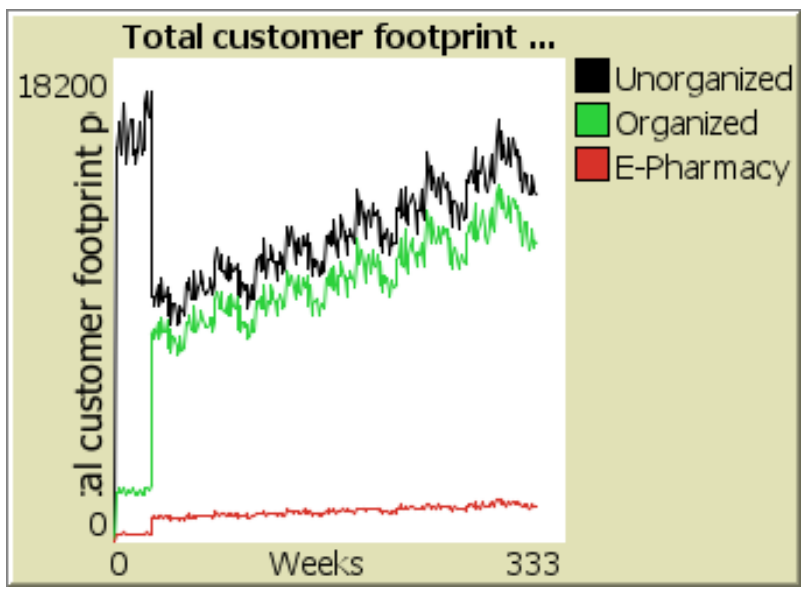

Experiment 24

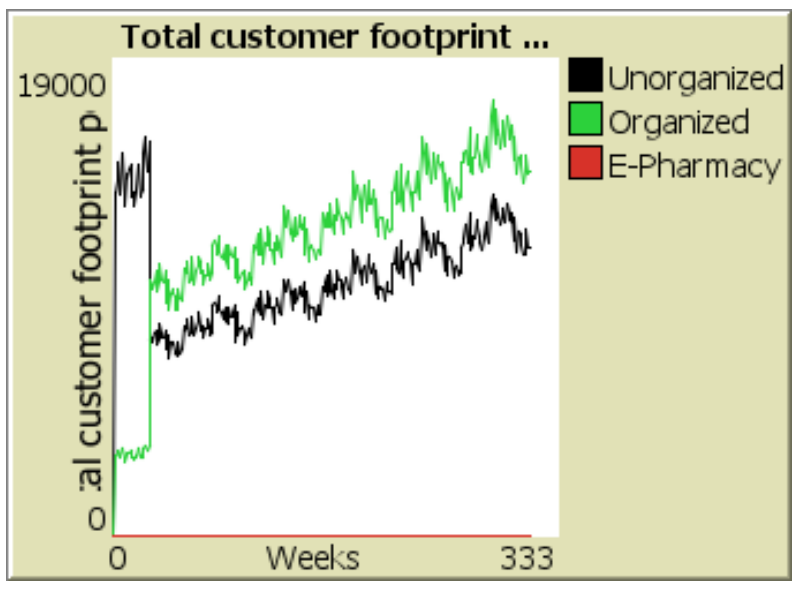

Experiment 25

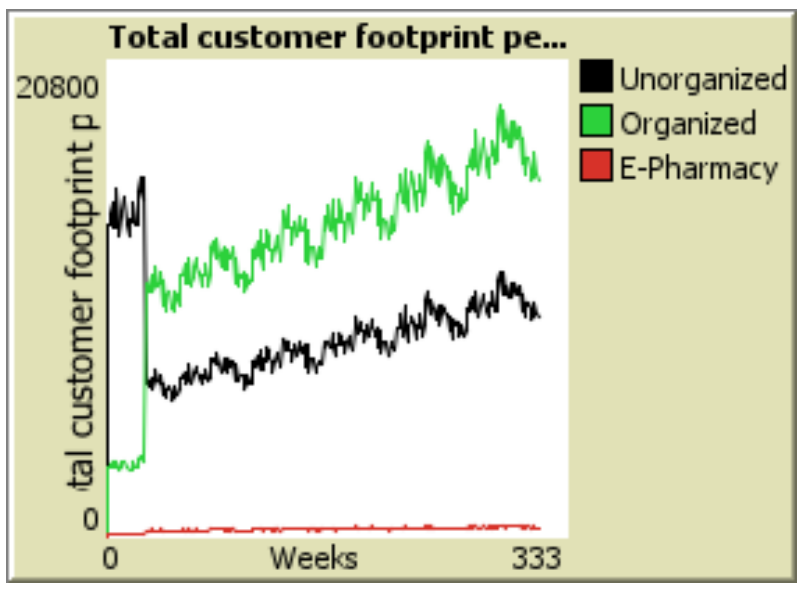

Experiment 26

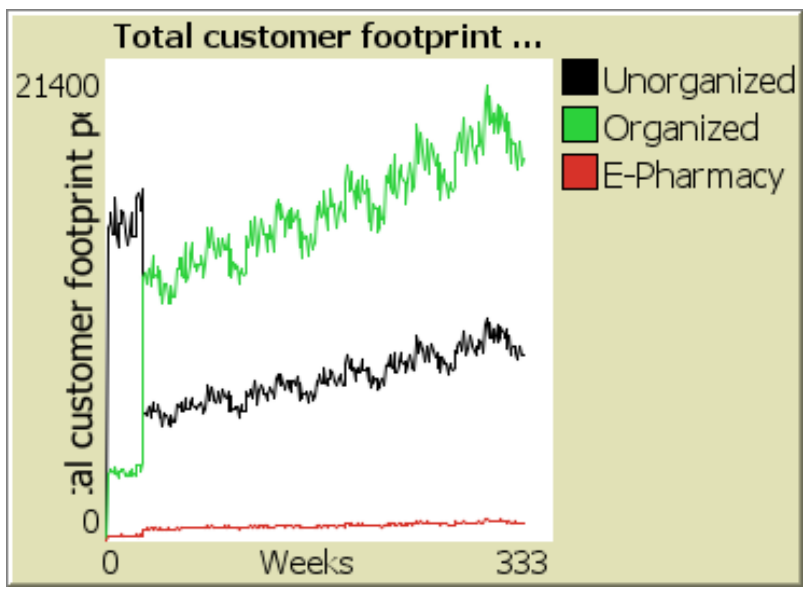

Experiment 27

FIGURE E2 Simulation results of weekly customer footprint as per experimental design for price discount.

## E1.3. Simulation Results of Market Share as per 27 Experimental Runs Utilising ANOVA for Price Discount

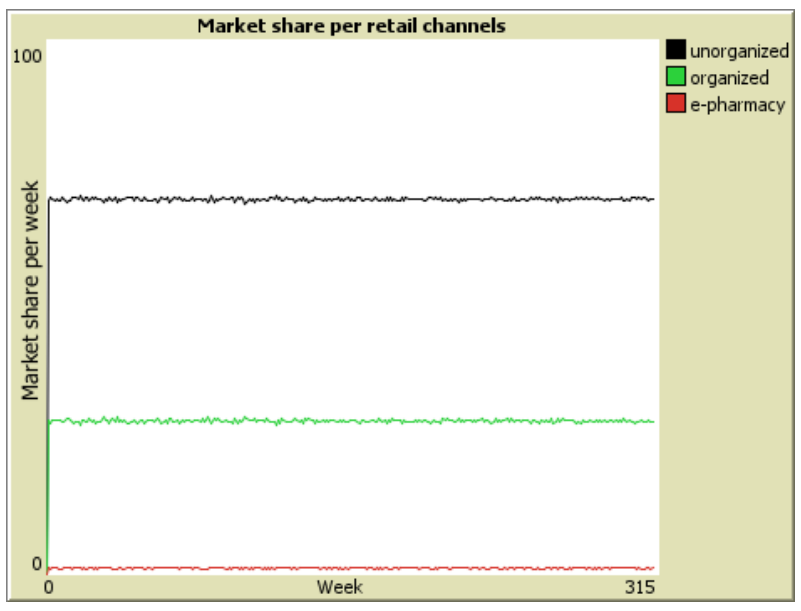

Experiment 1

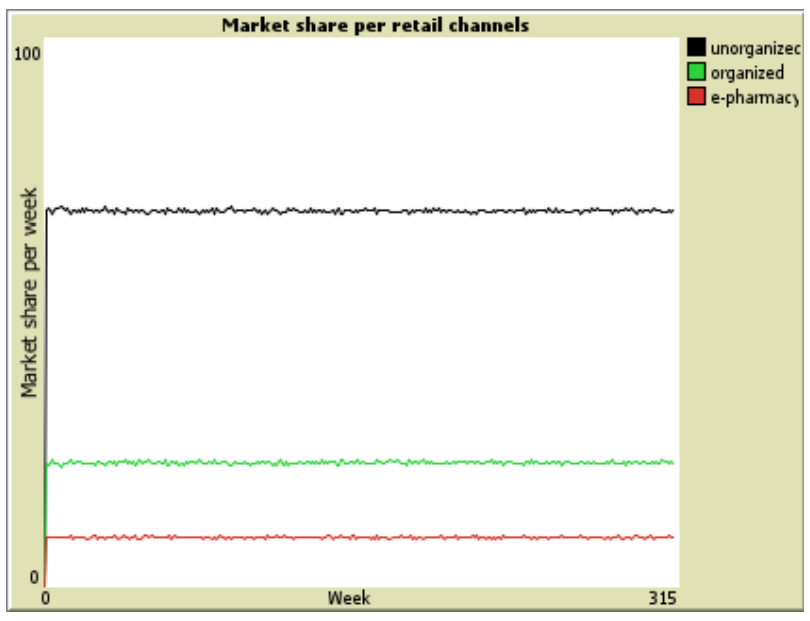

Experiment 2

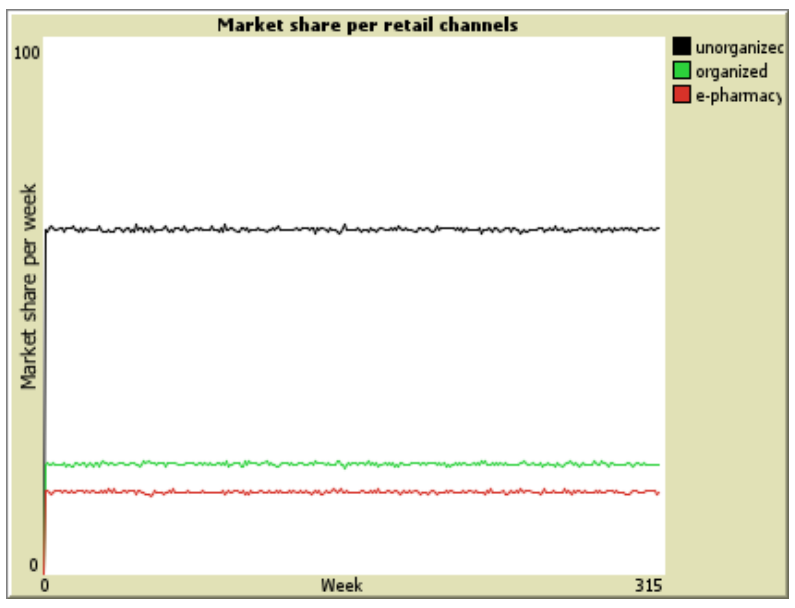

Experiment 3

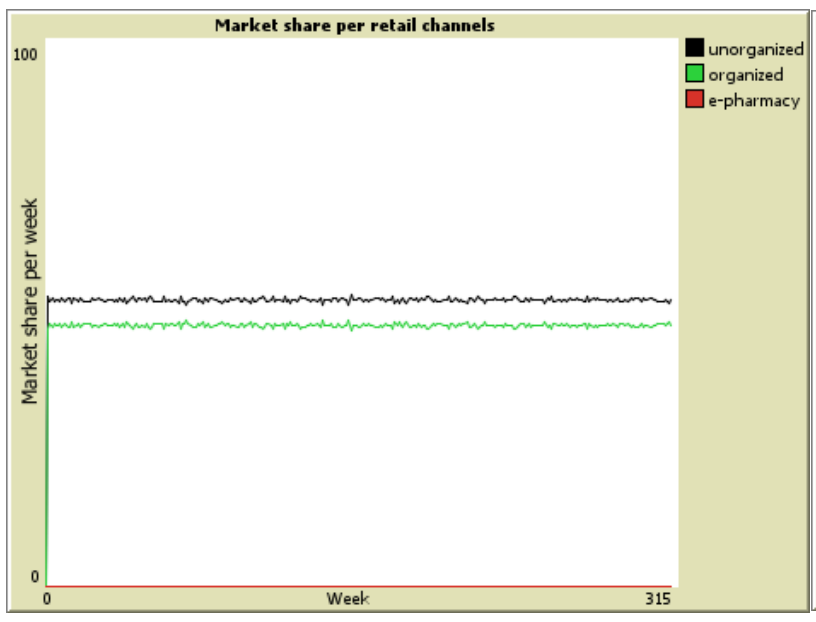

Experiment 4

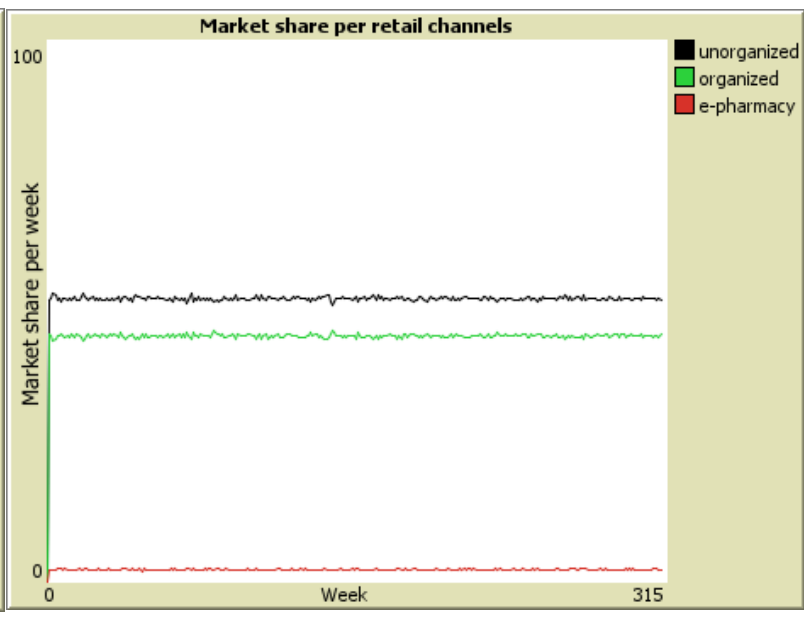

Experiment 5

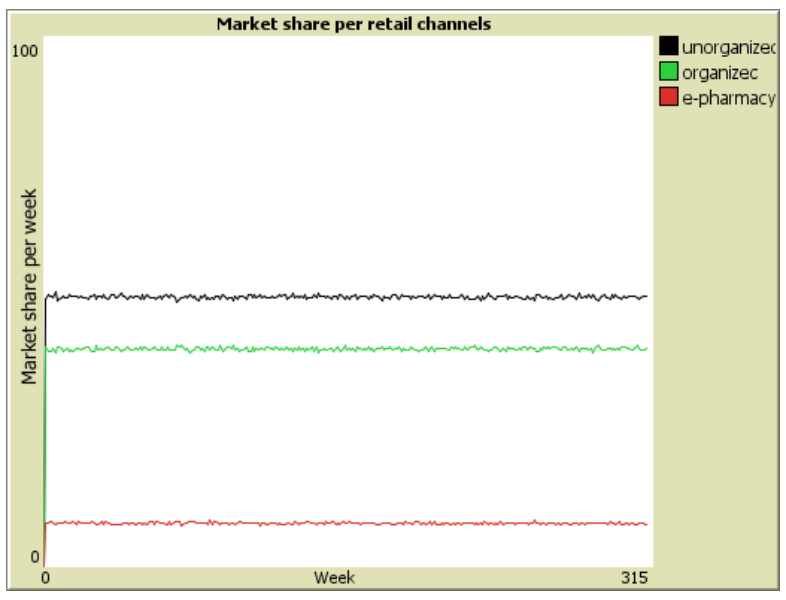

Experiment 6

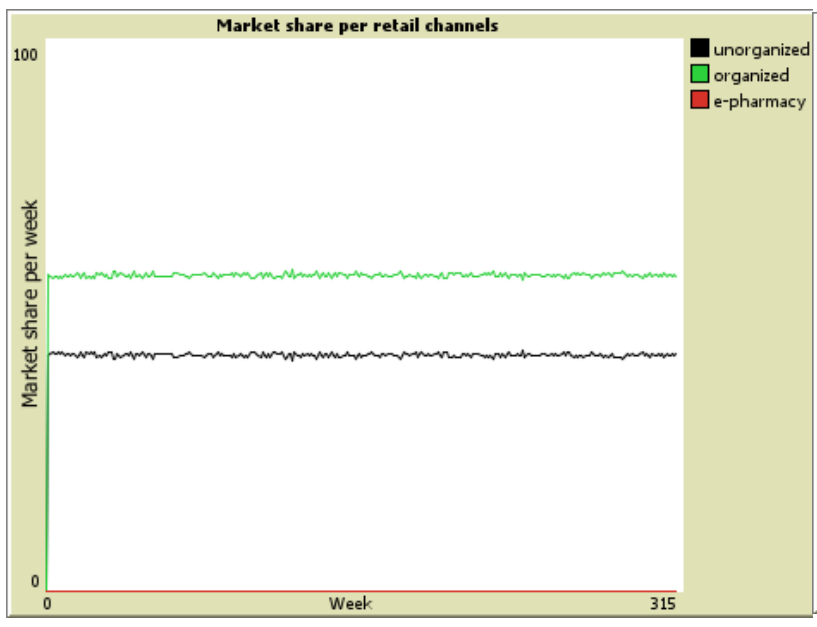

Experiment 7

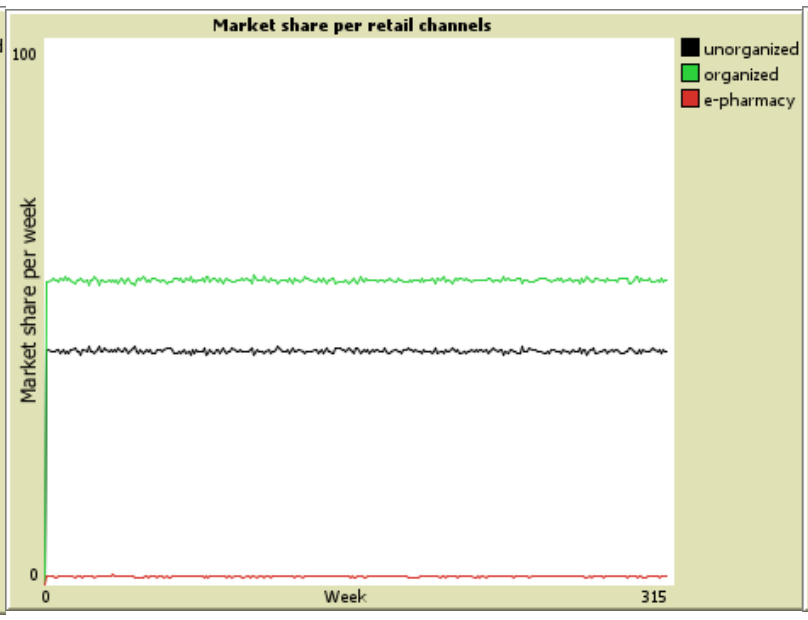

Experiment 8

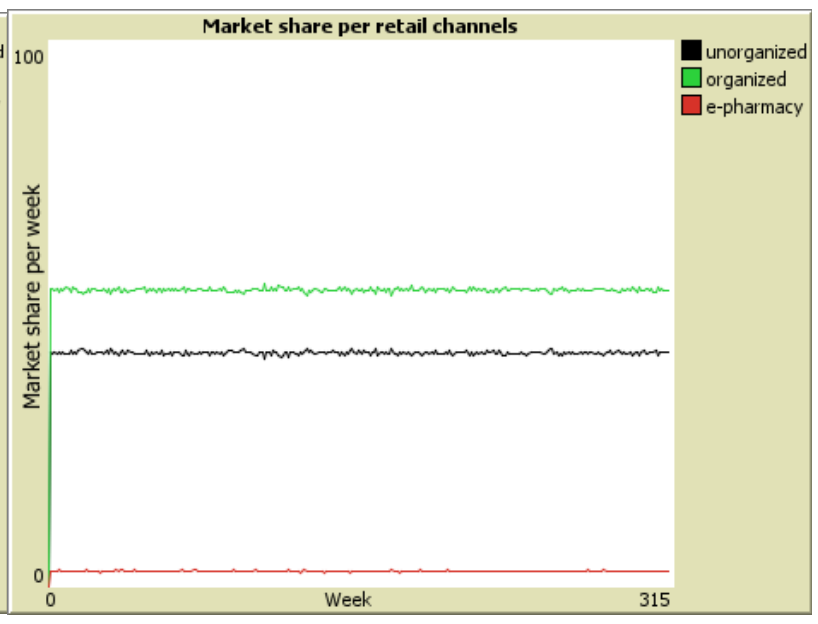

Experiment 9

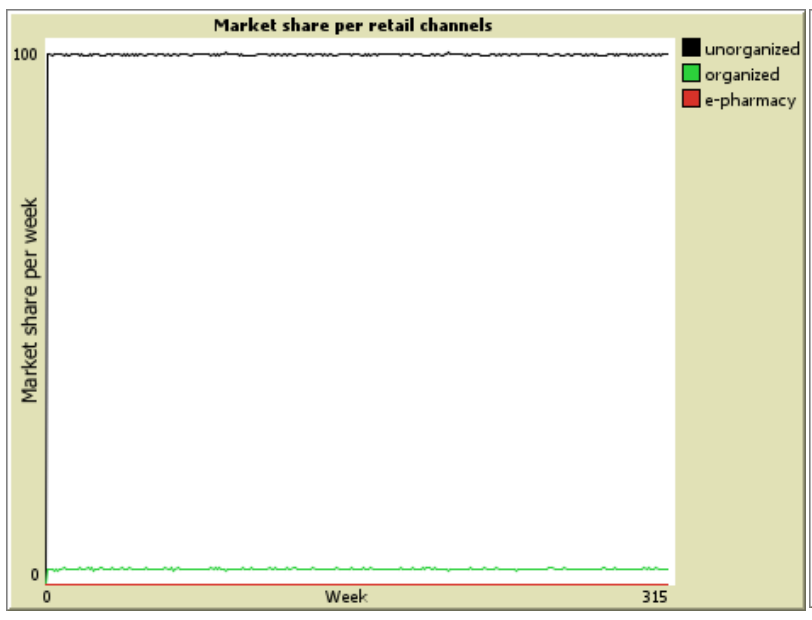

Experiment 10

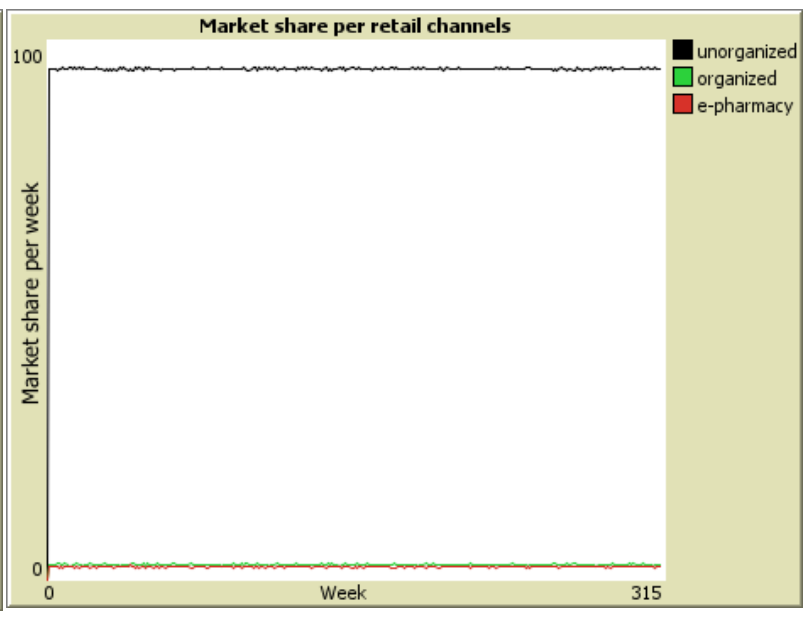

Experiment 11

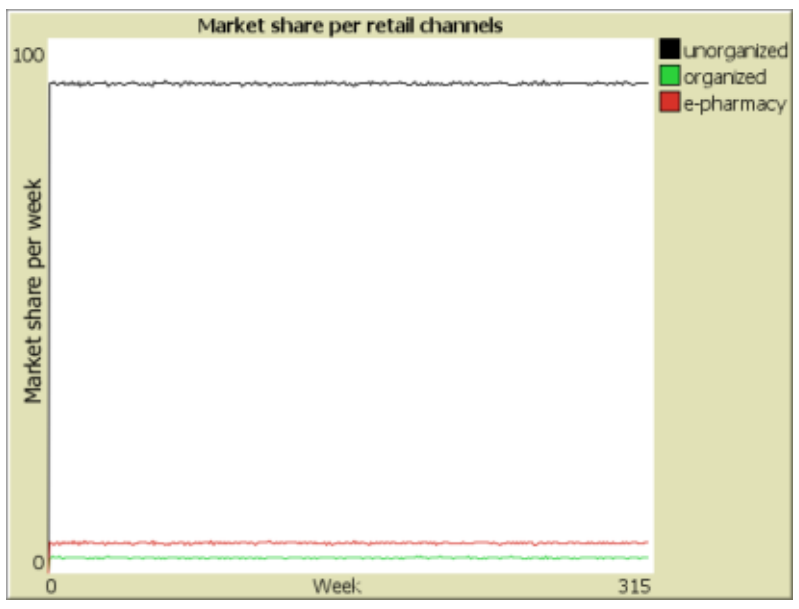

Experiment 12

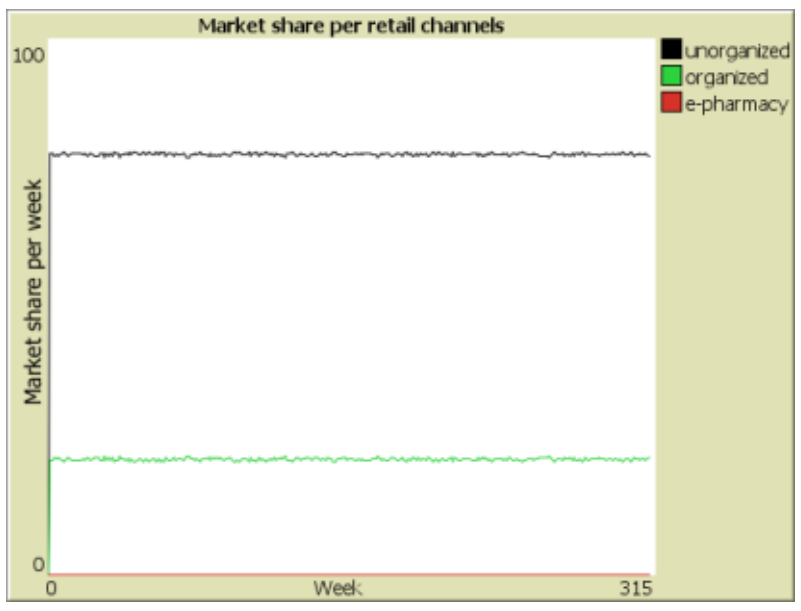

Experiment 13

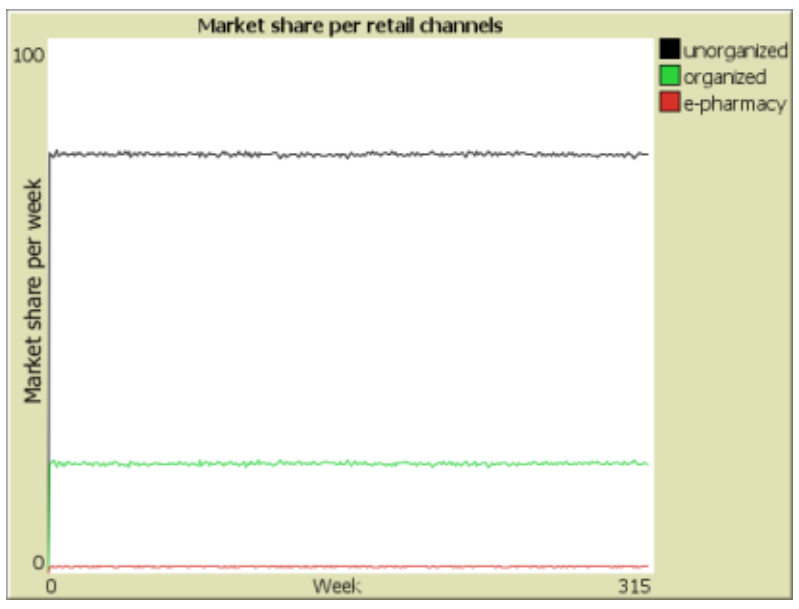

Experiment 14

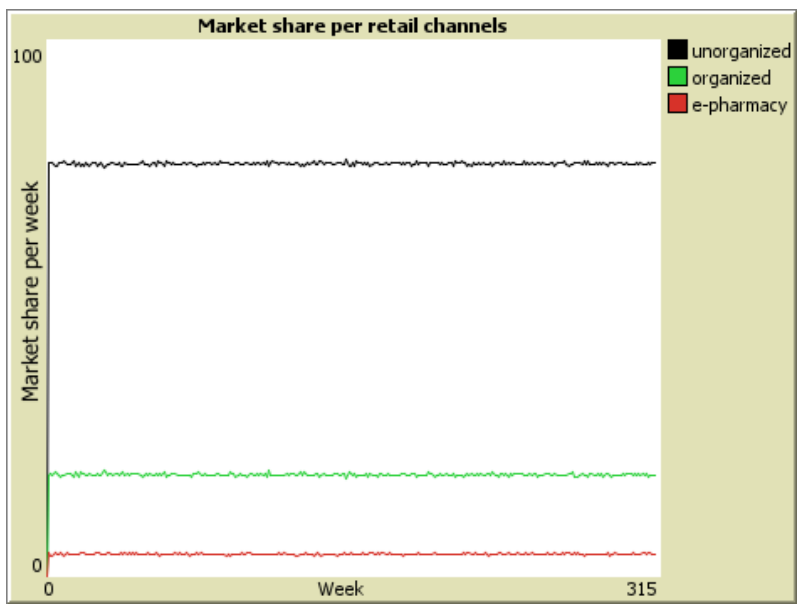

Experiment 15

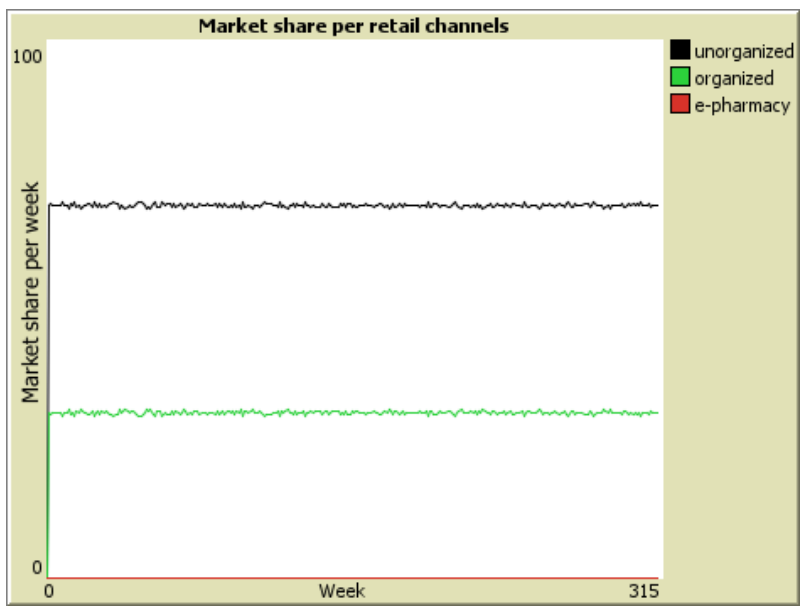

Experiment 16

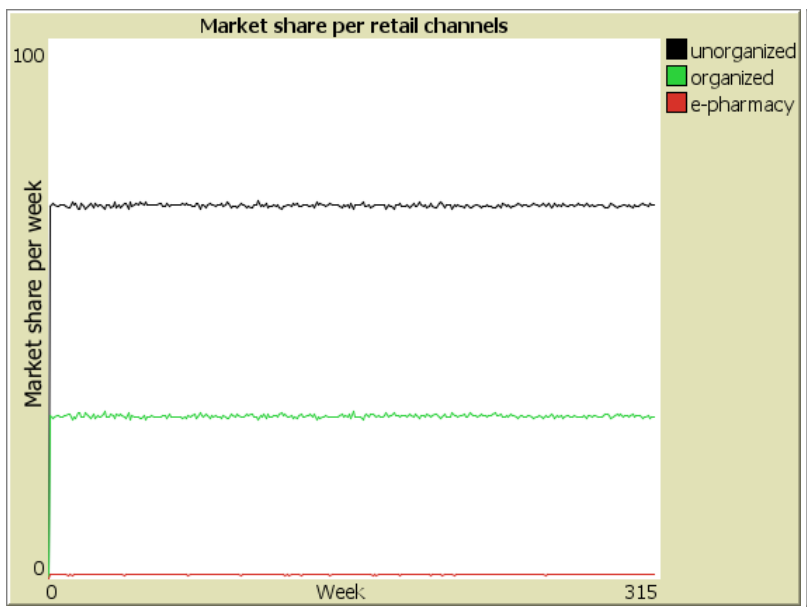

Experiment 17

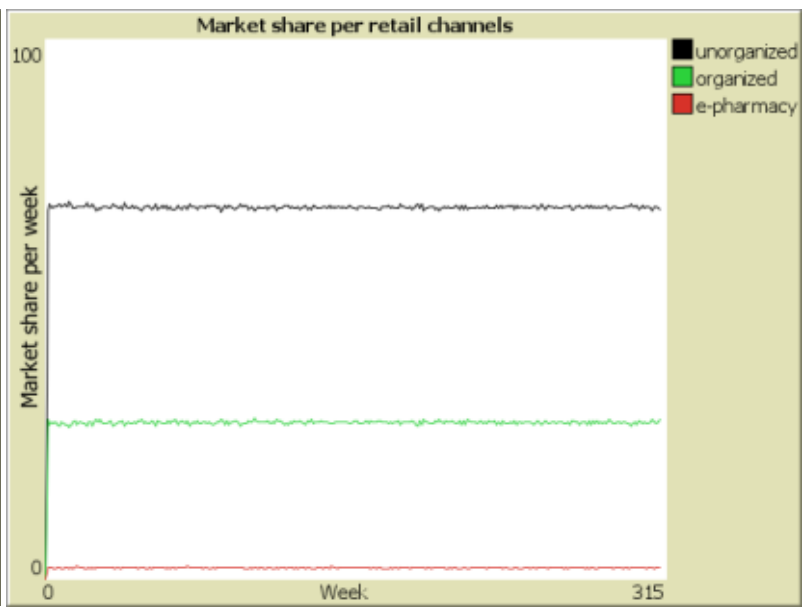

Experiment 18

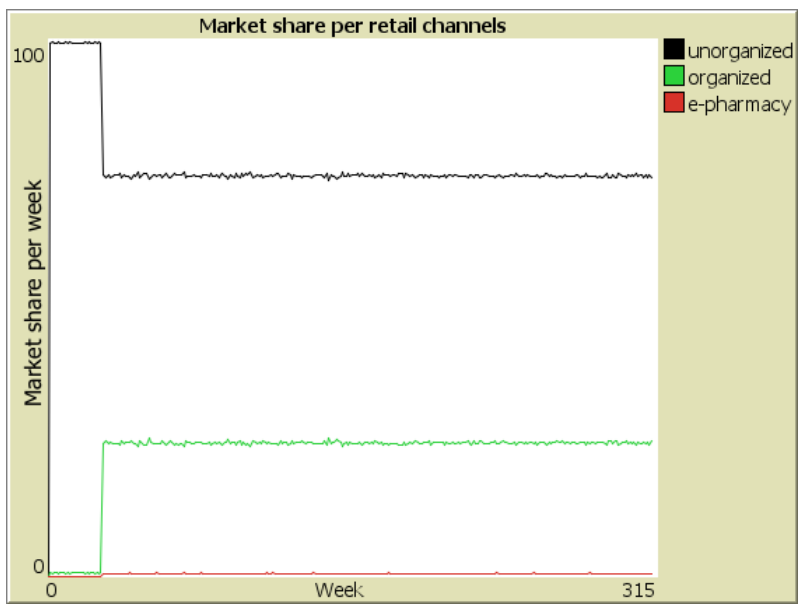

Experiment 19

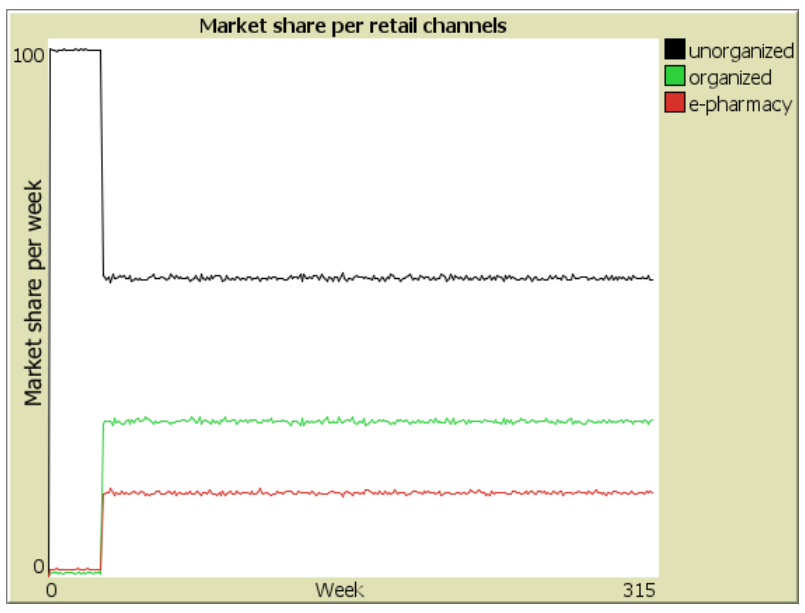

Experiment 20

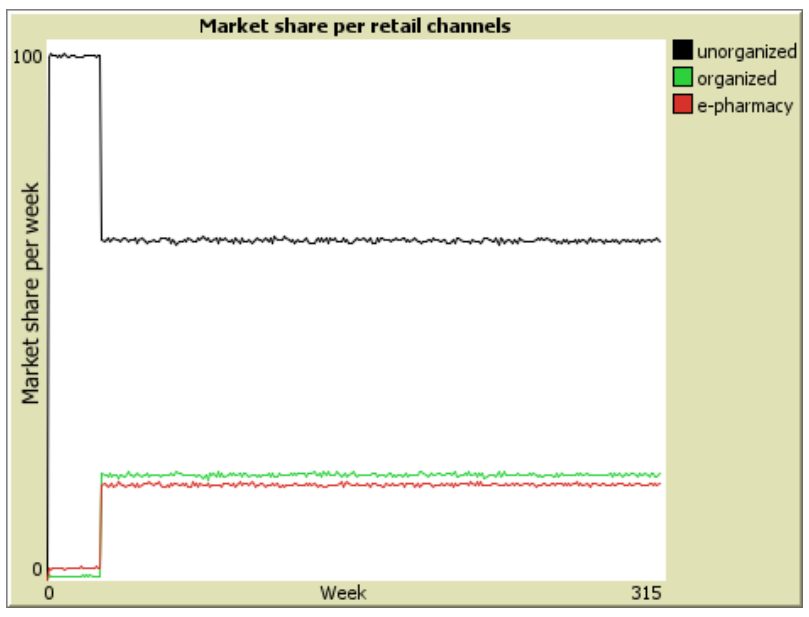

Experiment 21

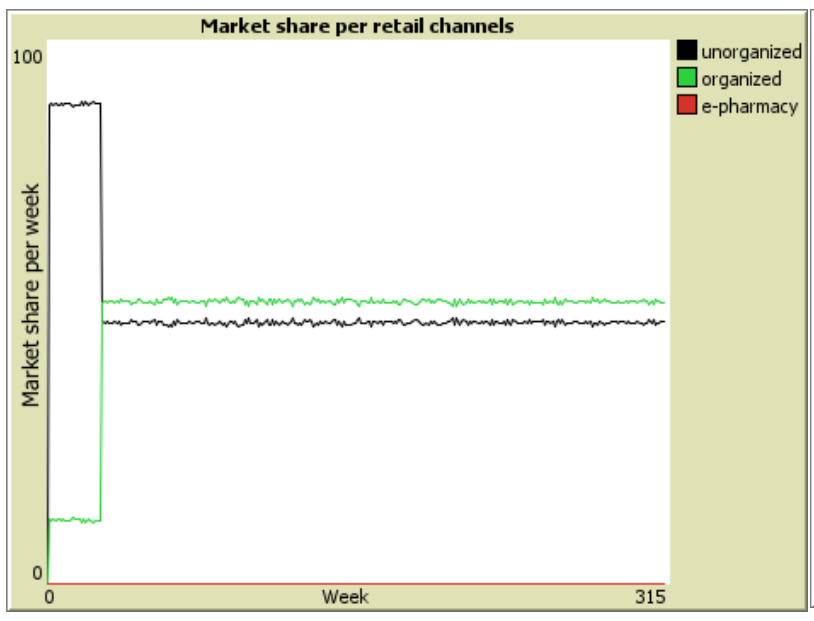

Experiment 22

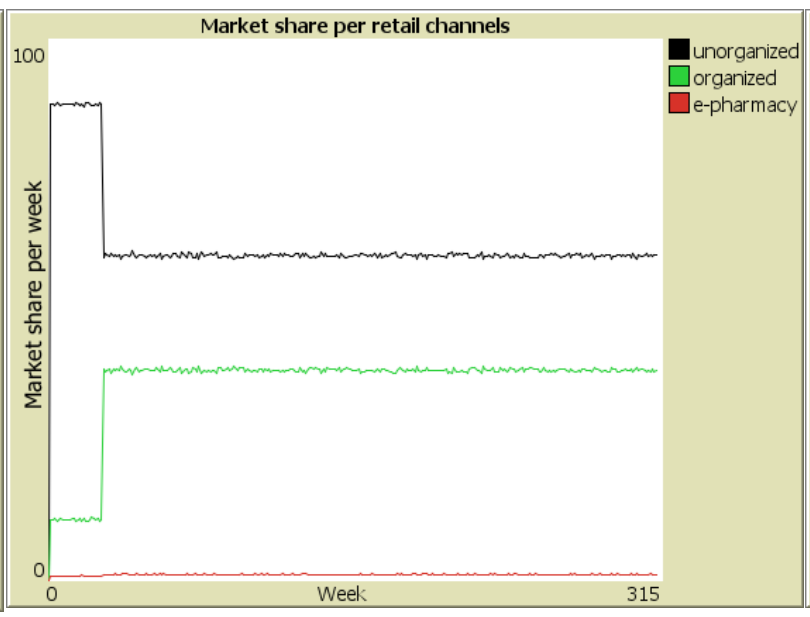

Experiment 23

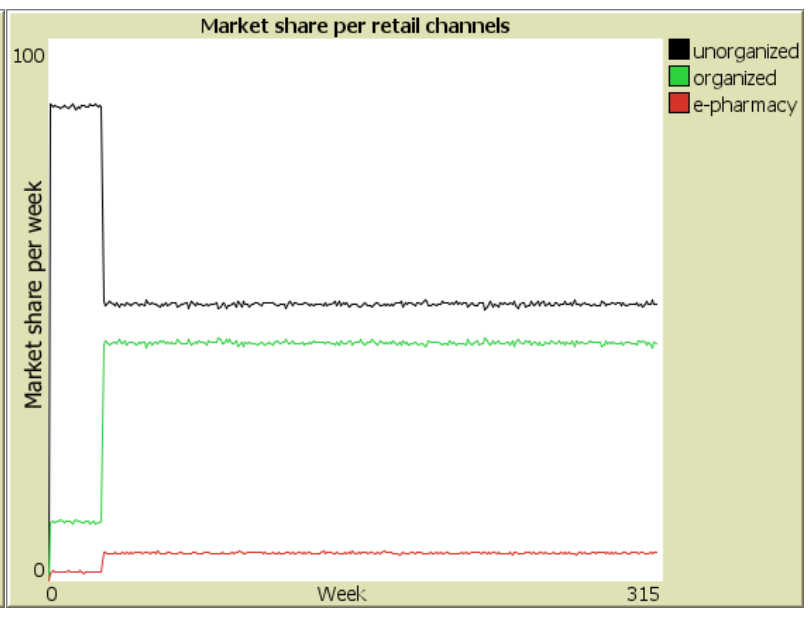

Experiment 24

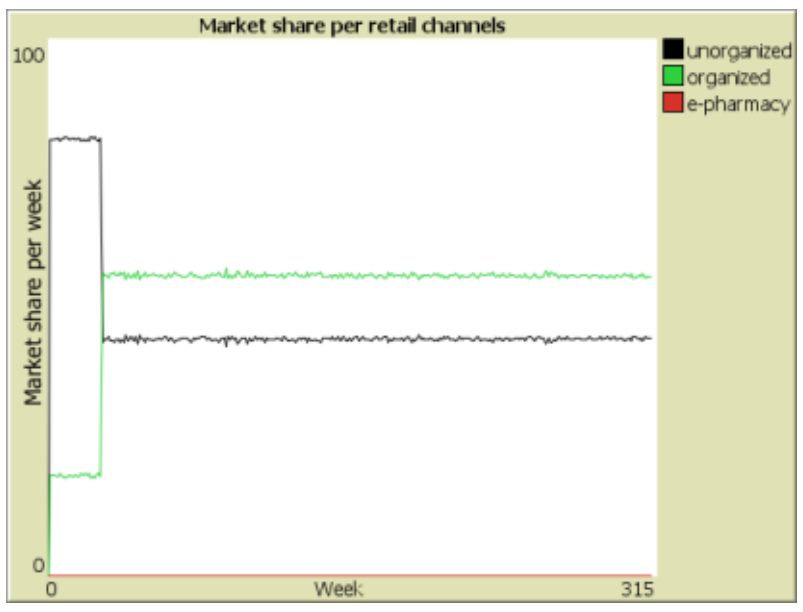

Experiment 25

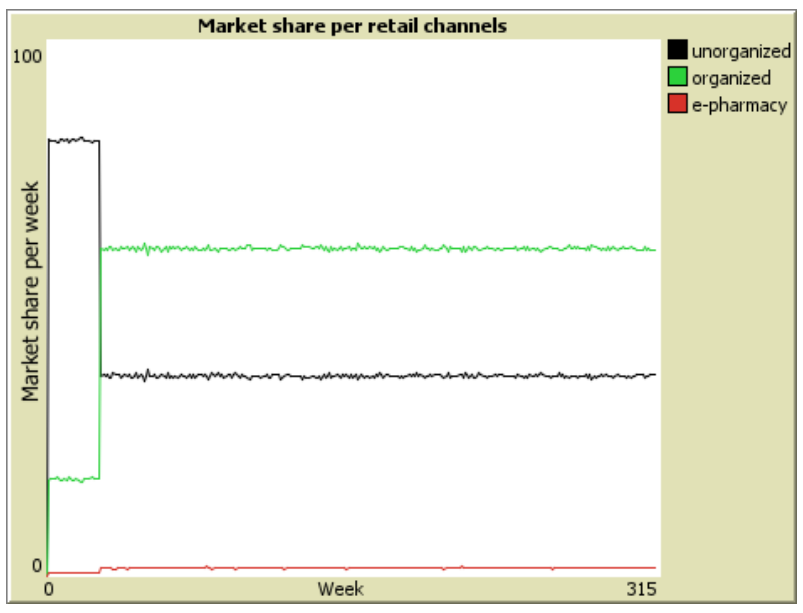

Experiment 26

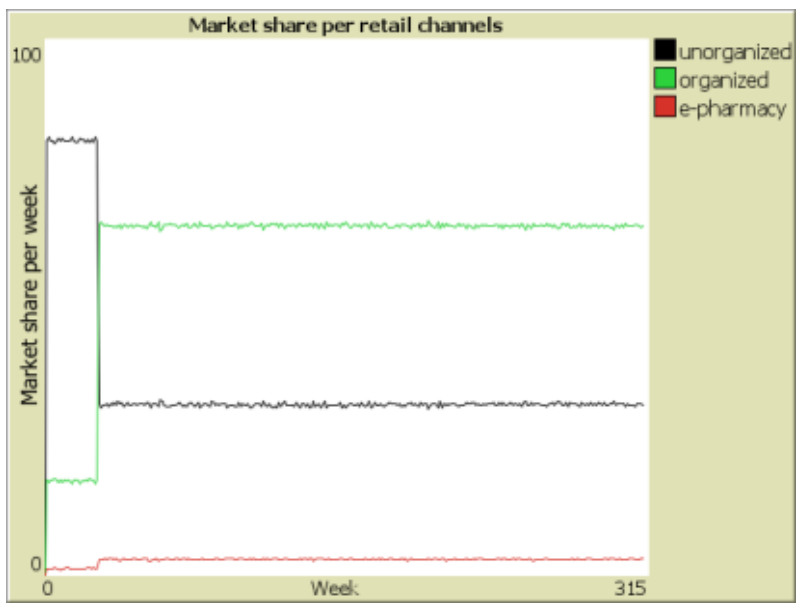

Experiment 27

FIGURE E3 Simulation results of market share as per experimental design for price discount.

## E1.4. Simulation Results of Unorganized Retails Present in the Market (No. of Stores Surviving After Shut Down) as per 27 Experimental Runs Utilising ANOVA for Price Discount

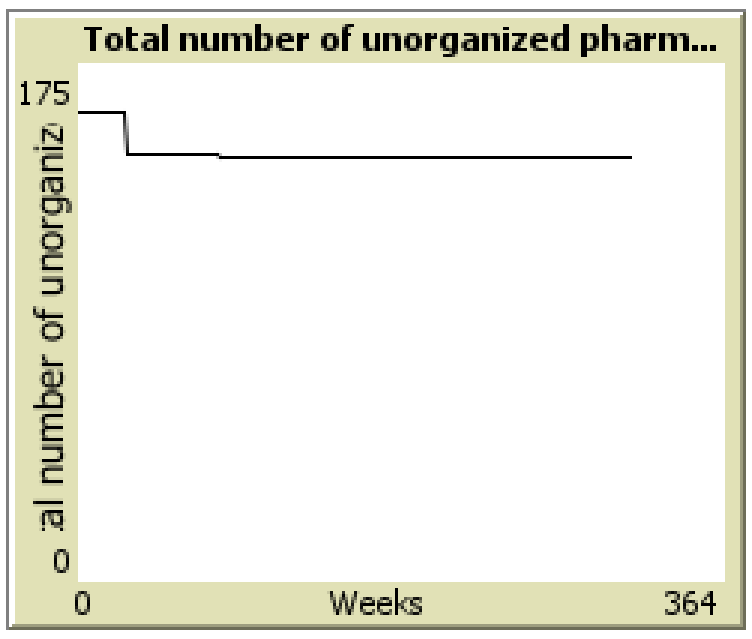

Experiment 1

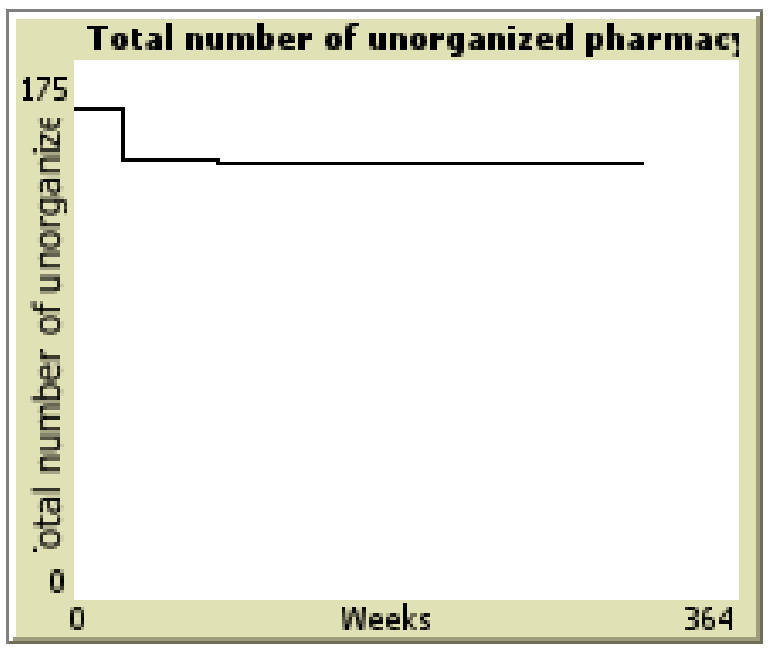

Experiment 2

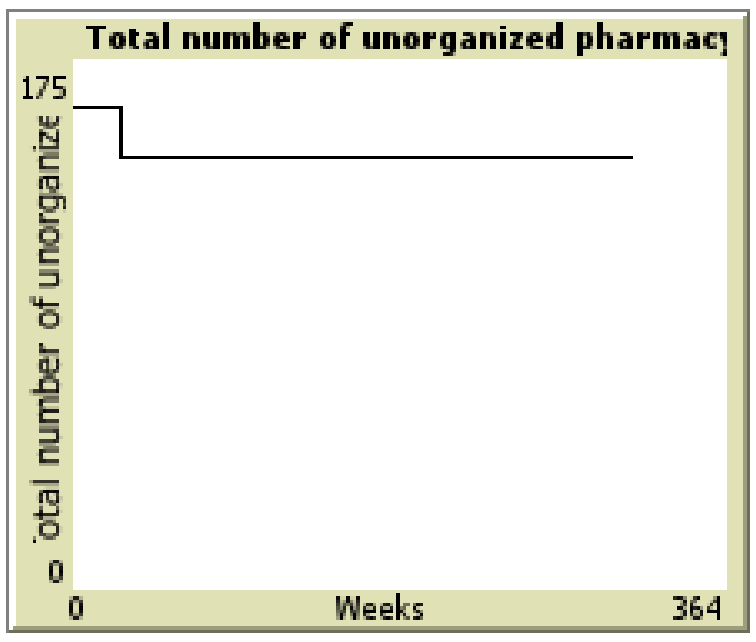

Experiment 3

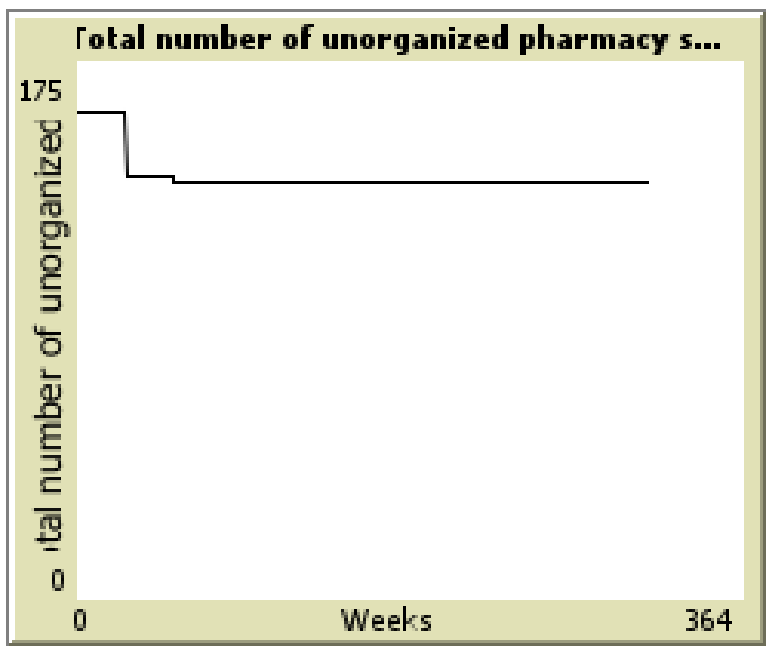

Experiment 4

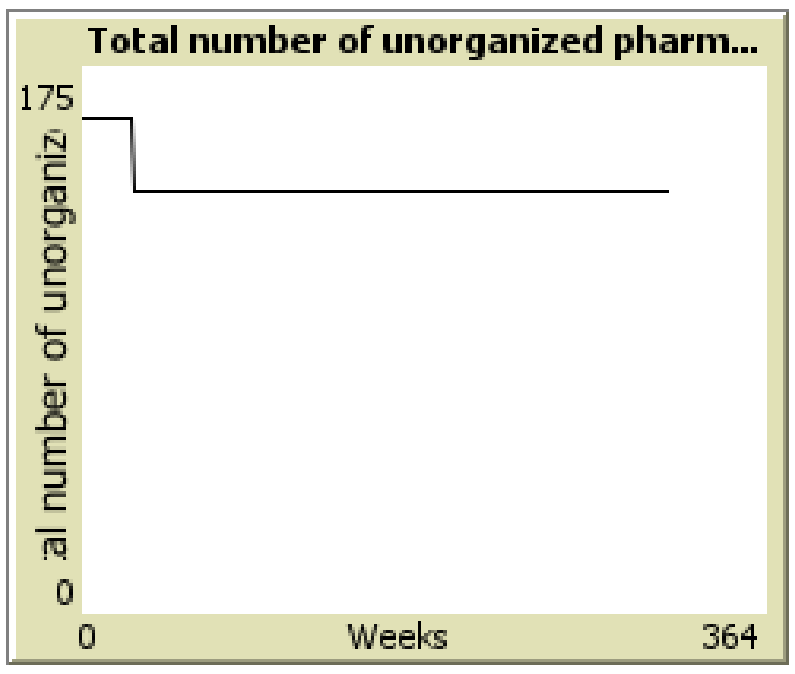

Experiment 5

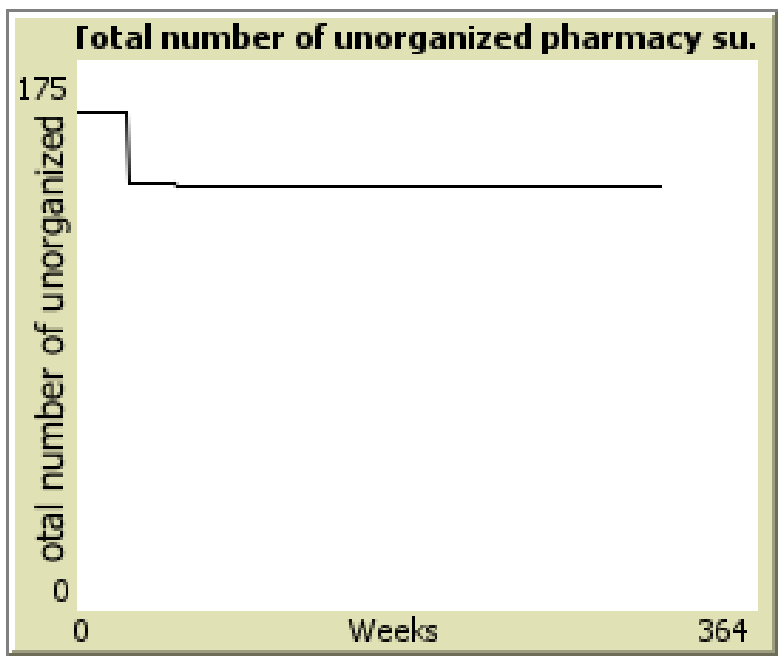

Experiment 6

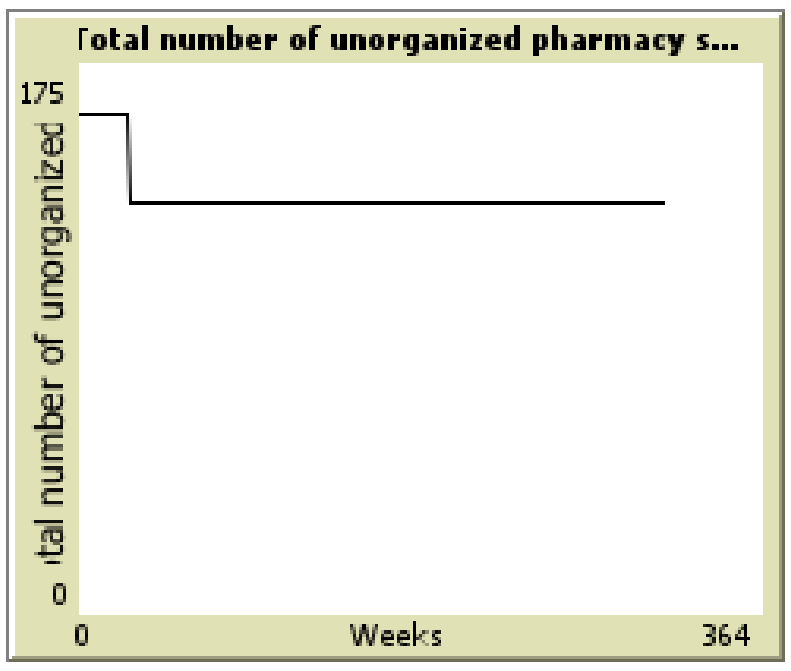

Experiment 7

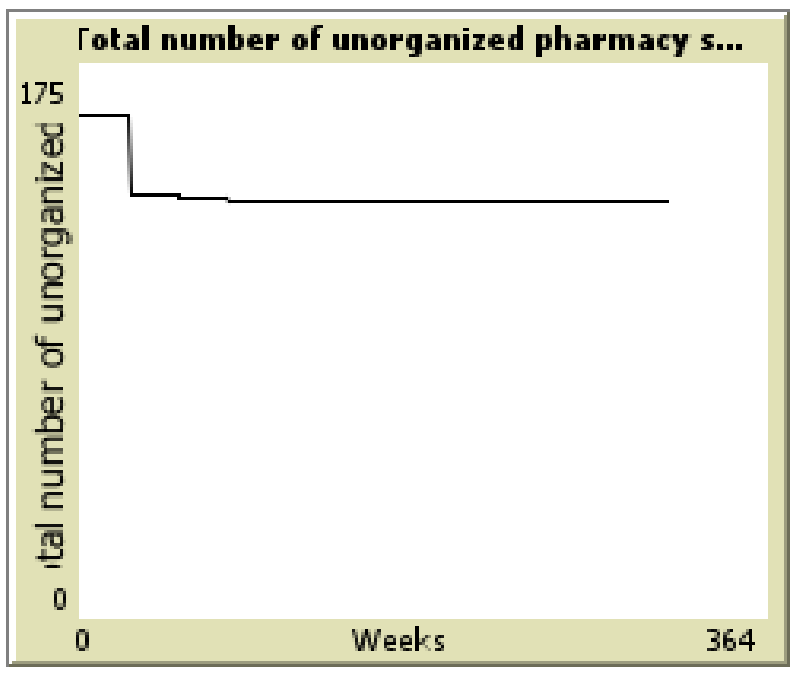

Experiment 8

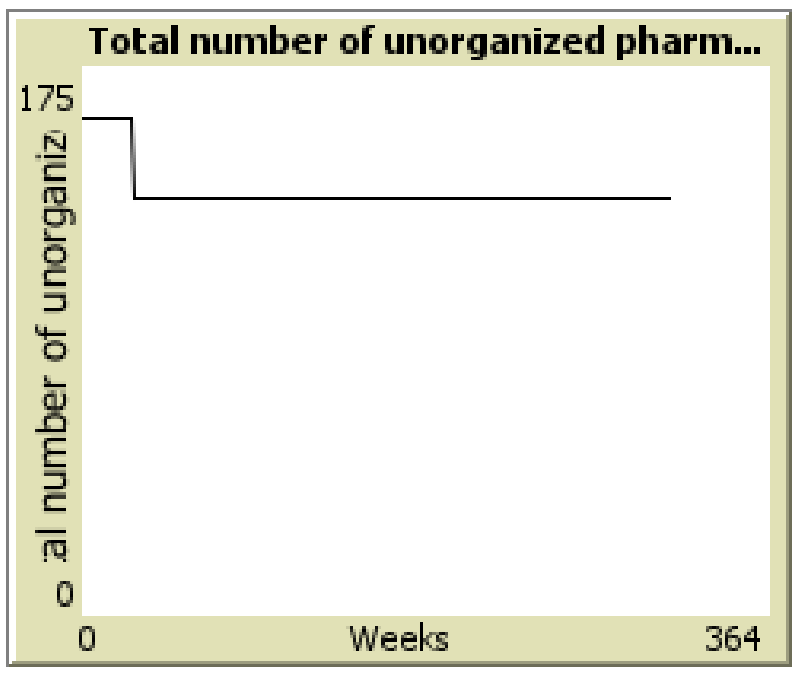

Experiment 9

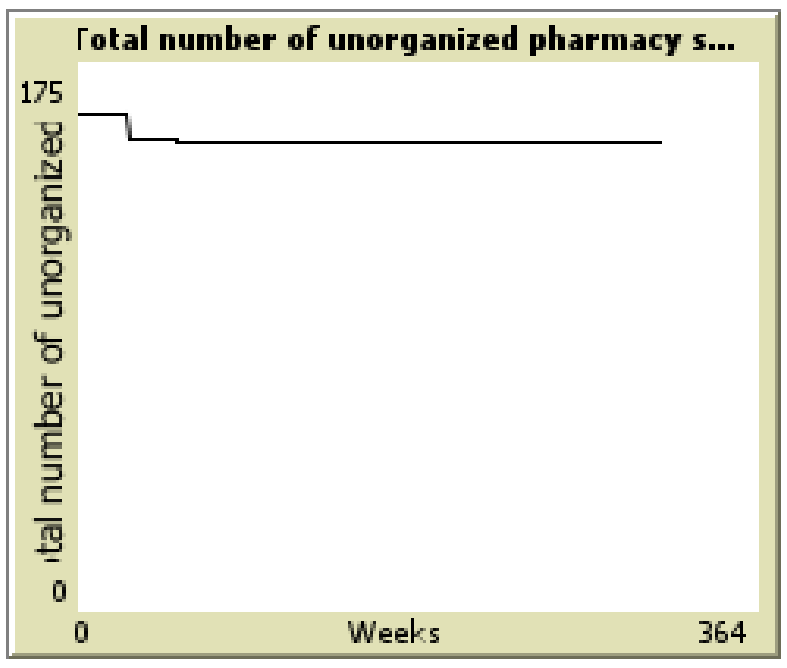

Experiment 10

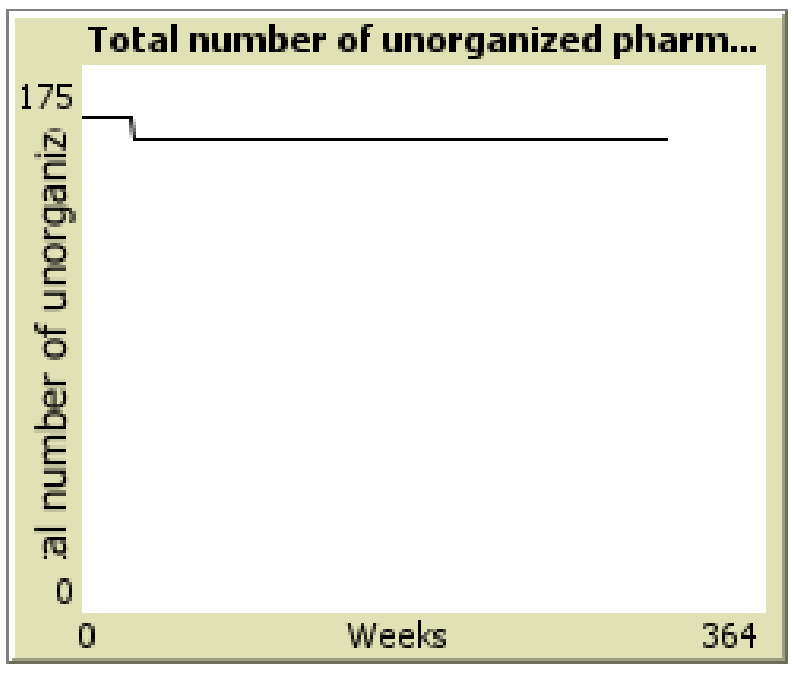

Experiment 11

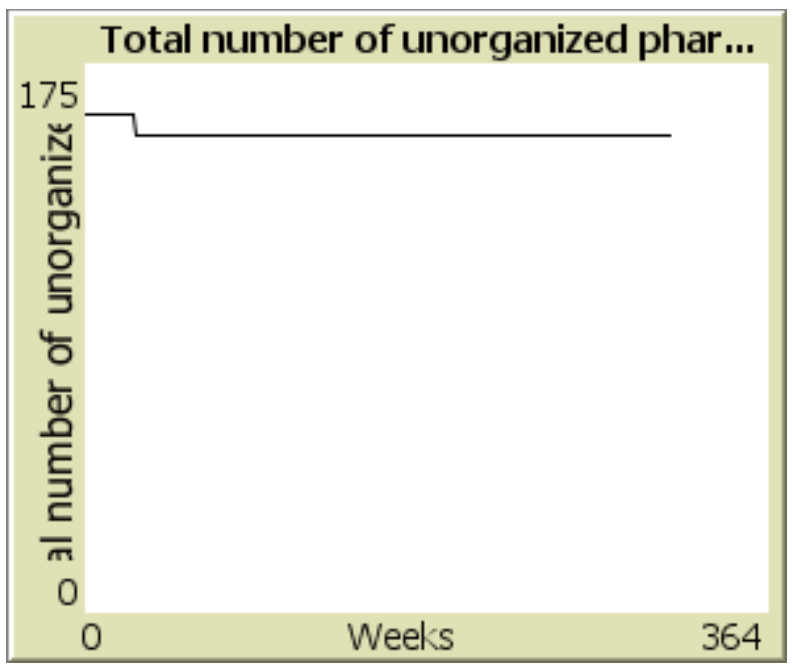

Experiment 12

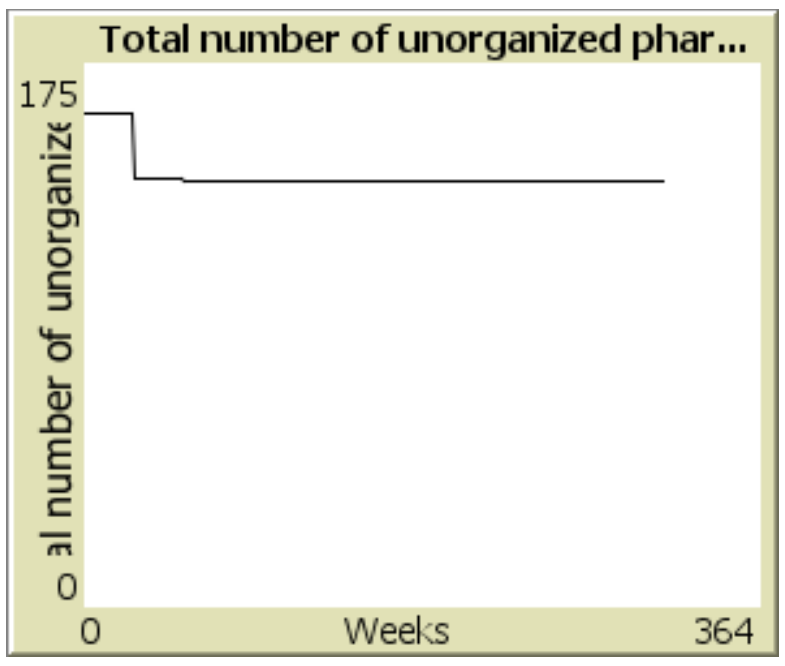

Experiment 13

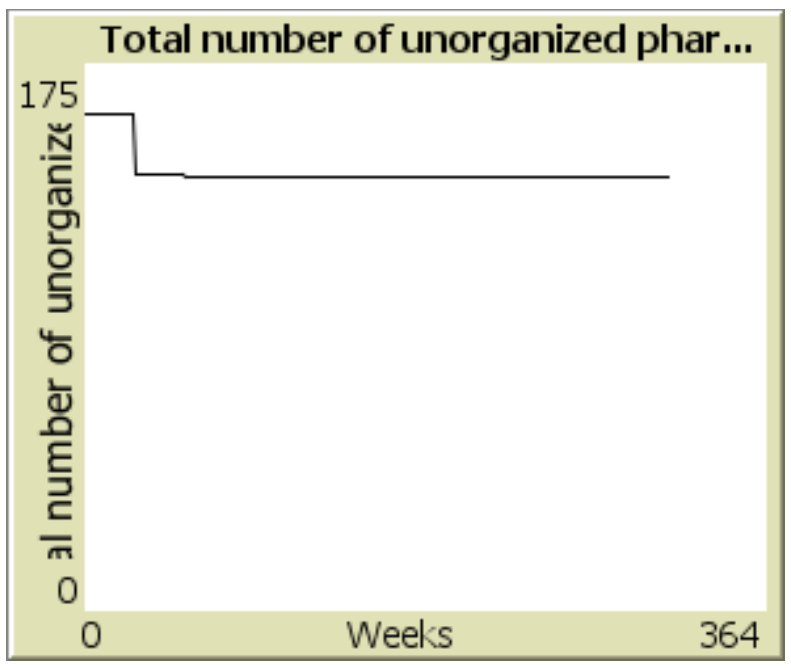

Experiment 14

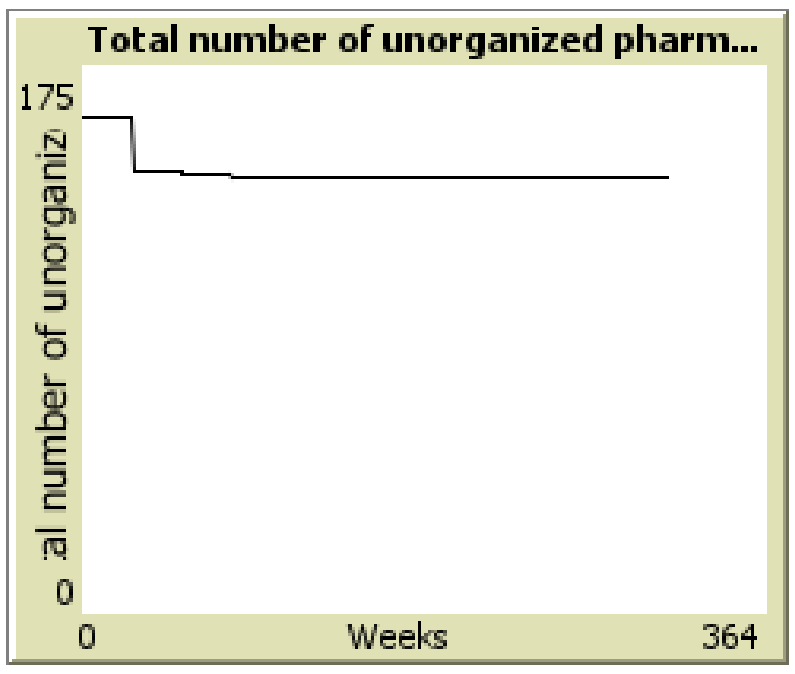

Experiment 15

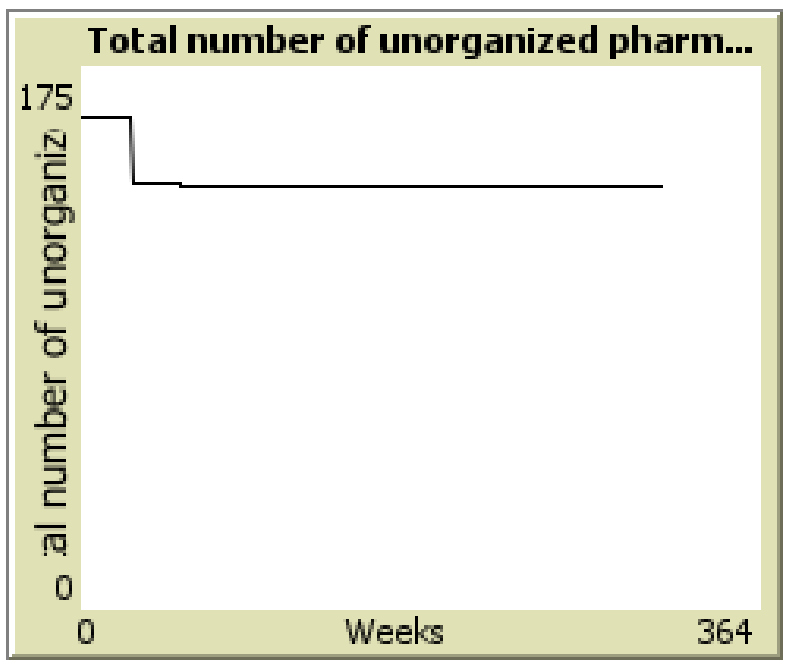

Experiment 16

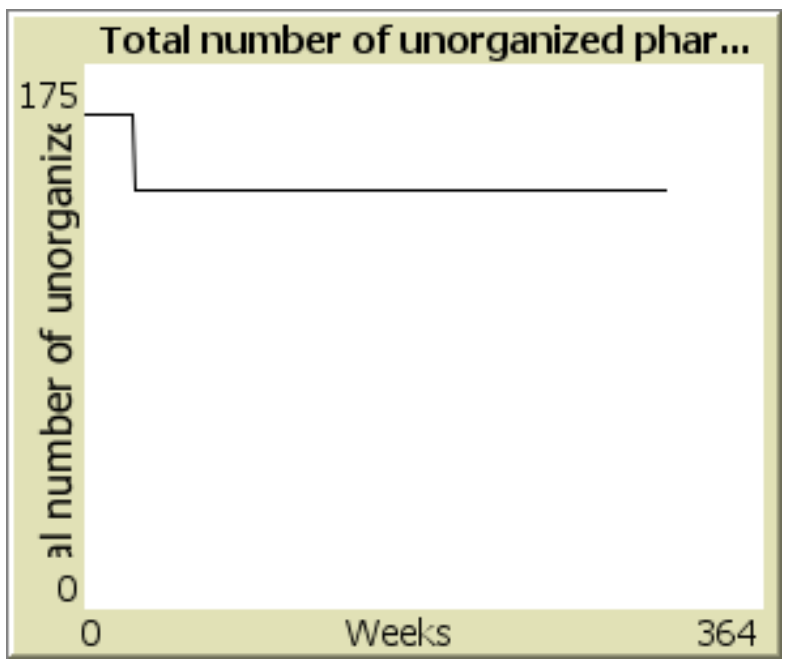

Experiment 17

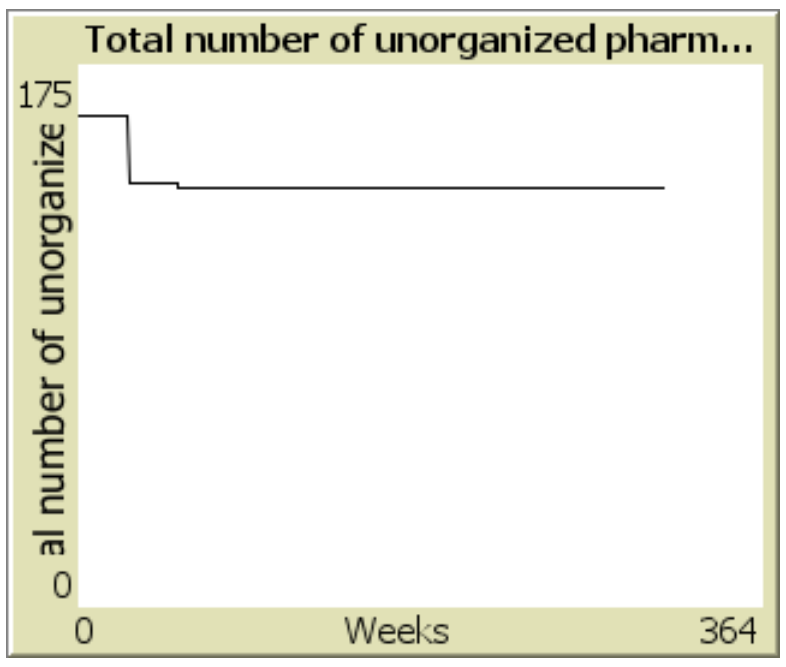

Experiment 18

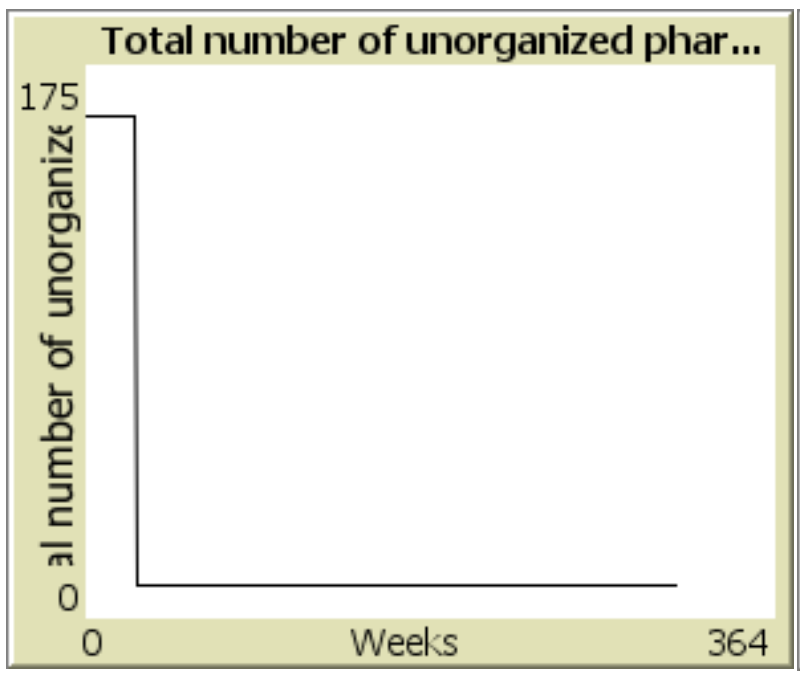

Experiment 19

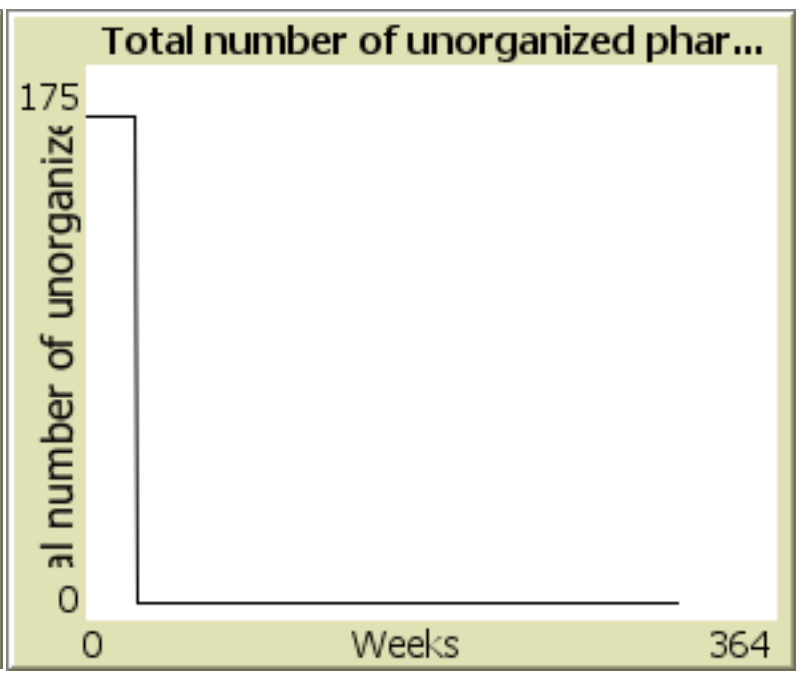

Experiment 20

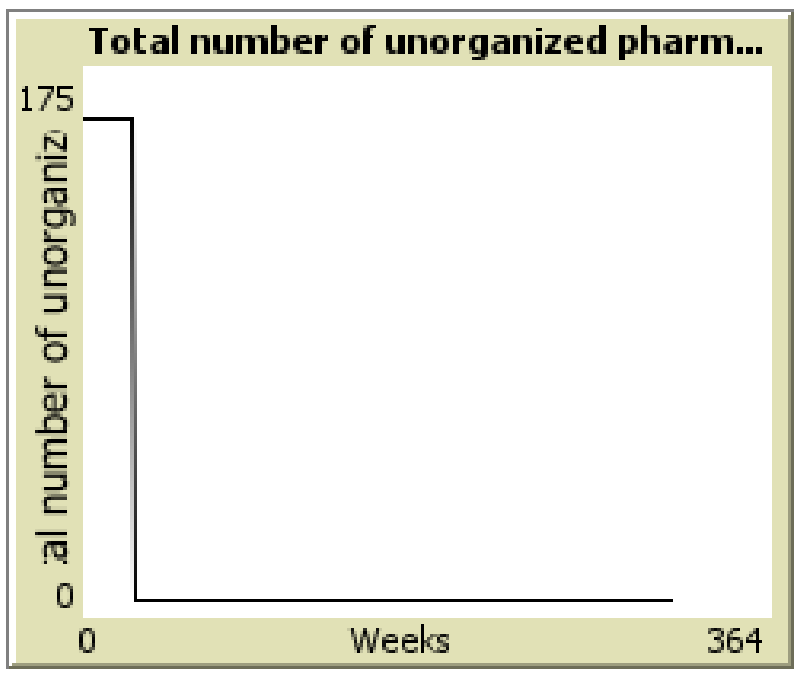

Experiment 21

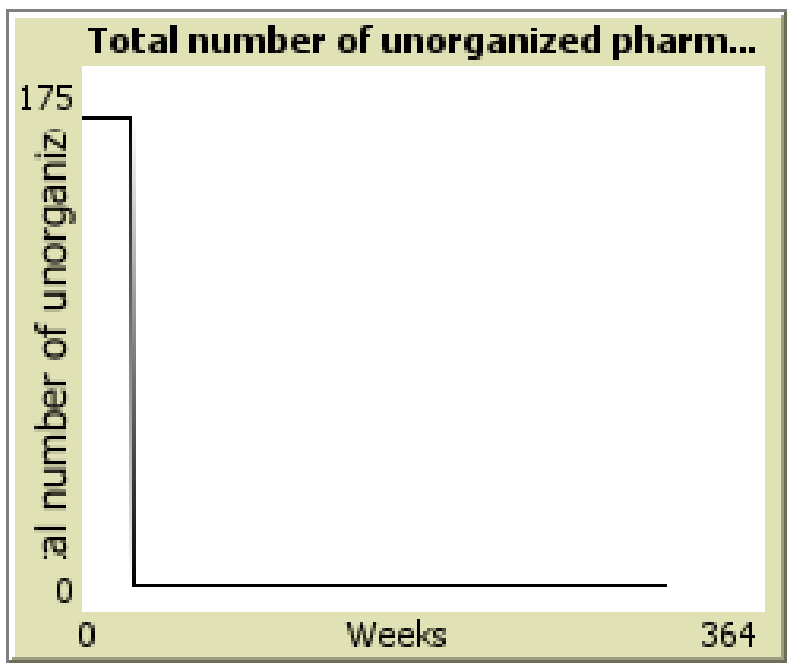

Experiment 22

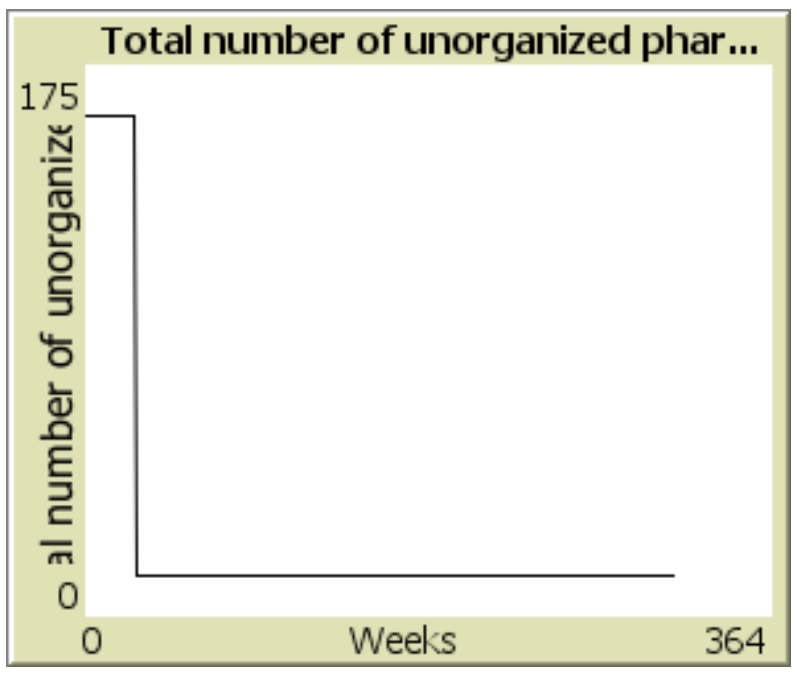

Experiment 23

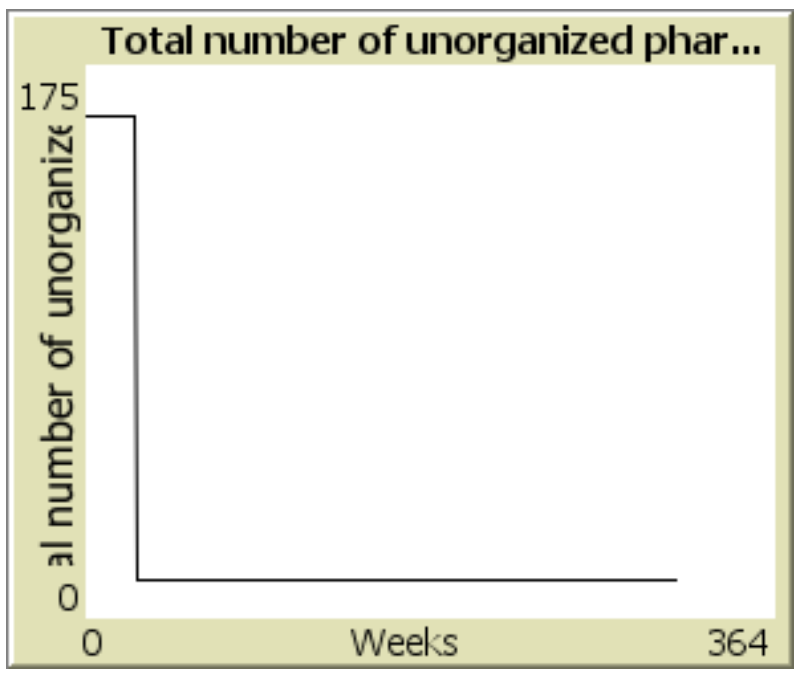

Experiment 24

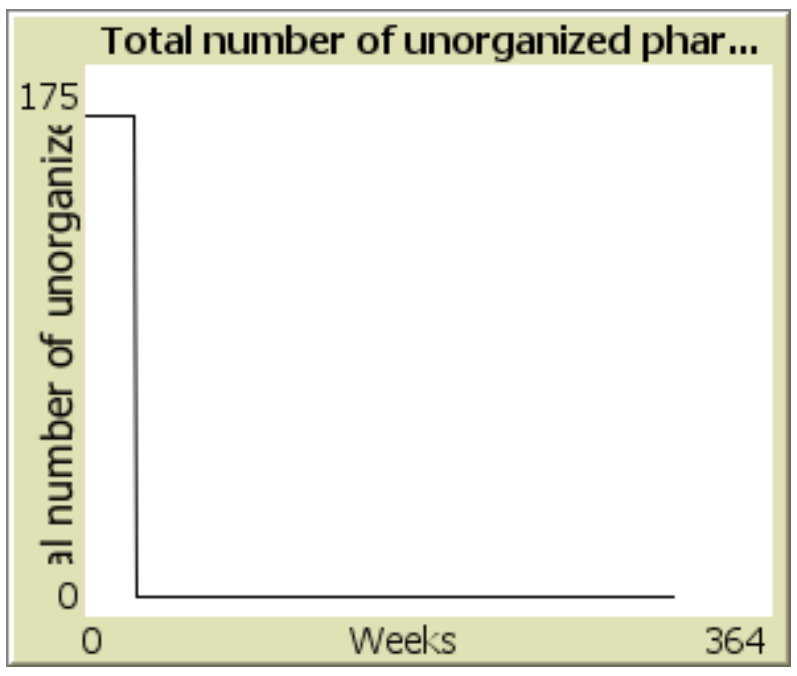

Experiment 25

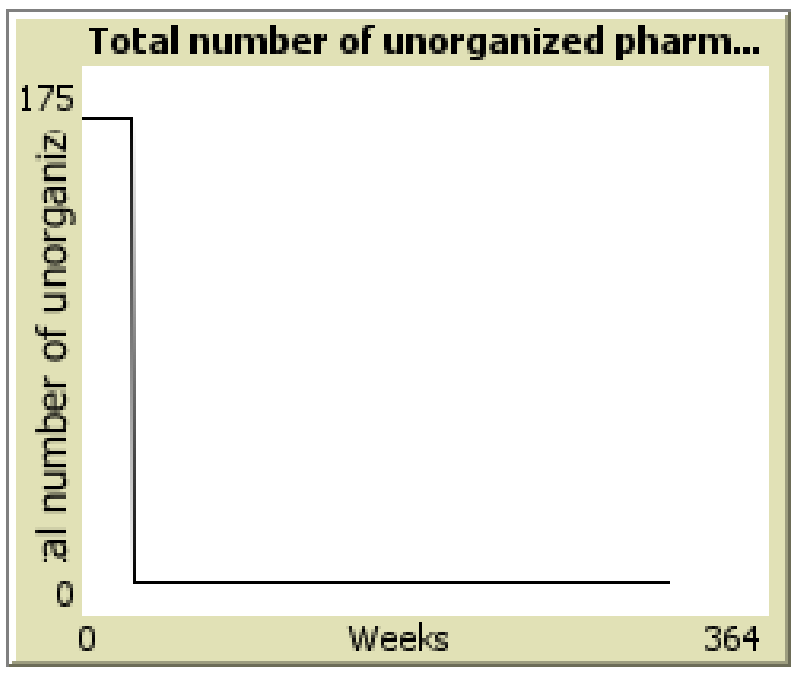

Experiment 26

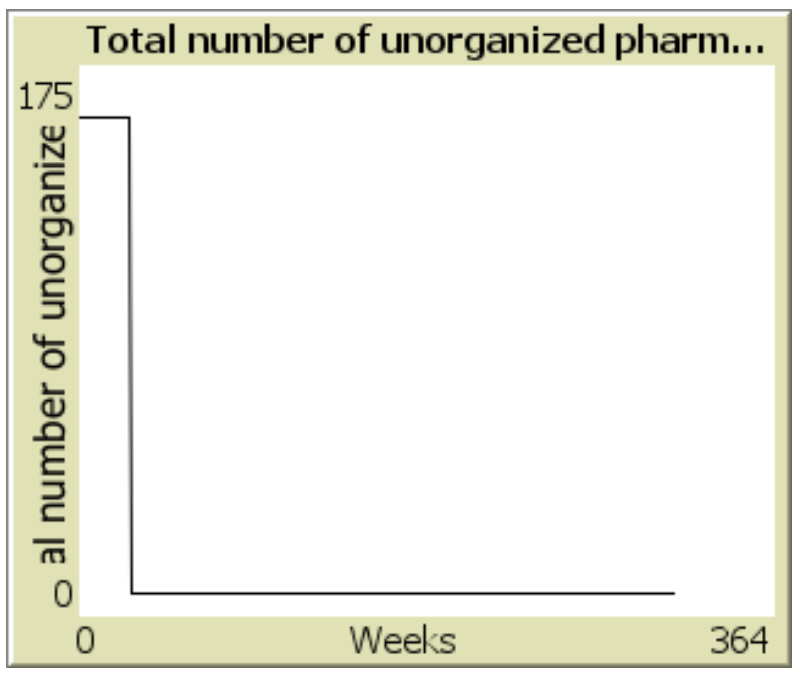

Experiment 27

FIGURE E4 Simulation results of unorganized retails present in the market (no. of stores) as per experimental design for price discount.

### E1.5. Simulation Results of Distance (km) a Customer Travelled as per 27 Experimental Runs Utilising ANOVA for Price Discount

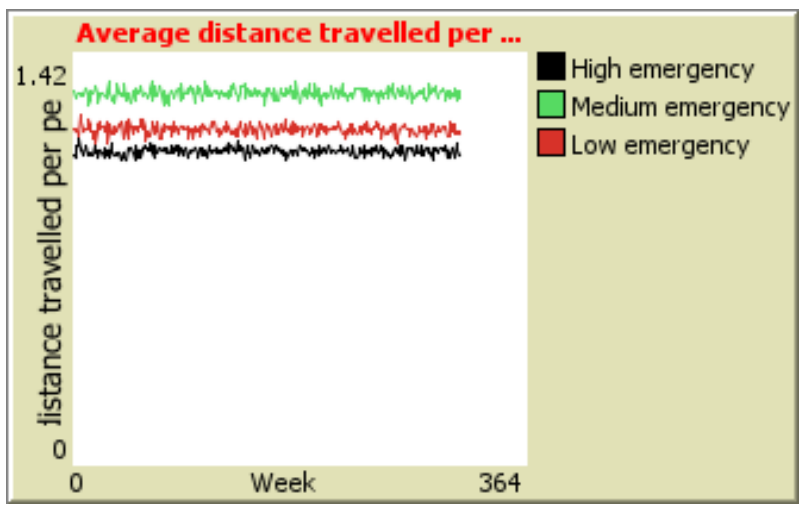

Experiment 1

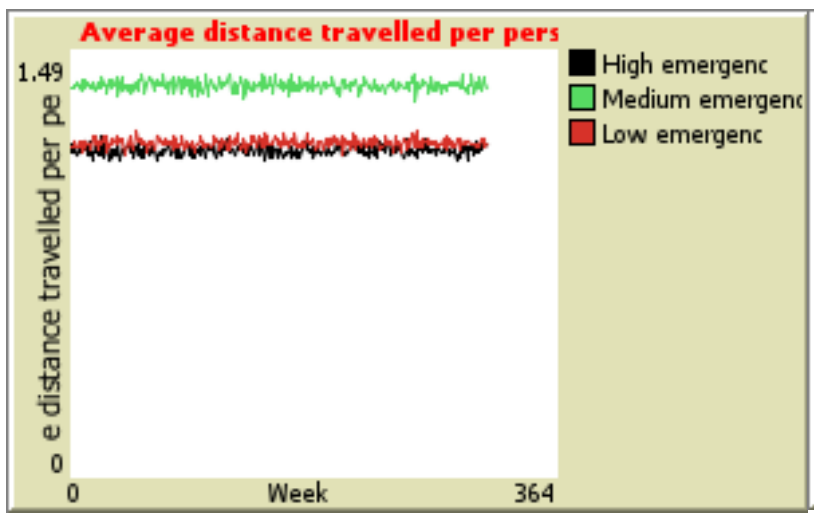

Experiment 2

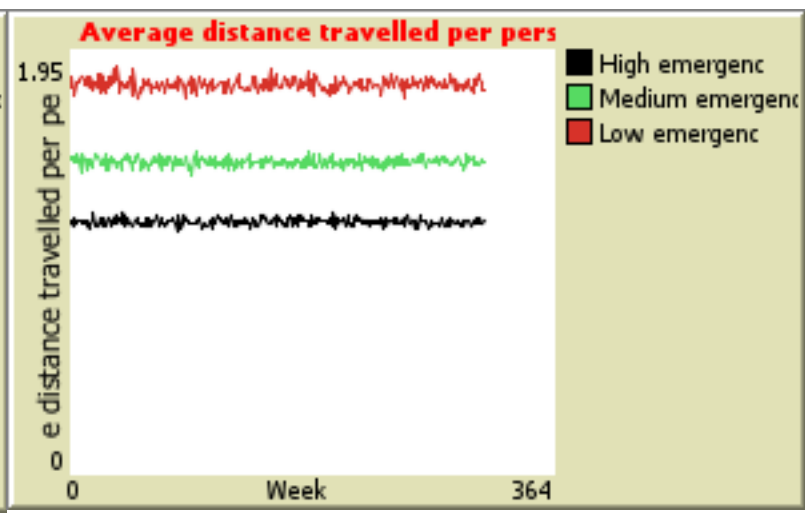

Experiment 3

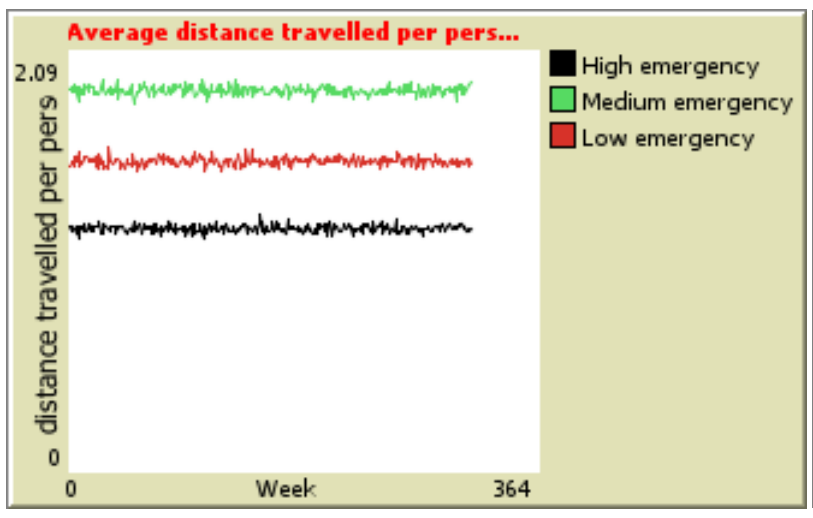

Experiment 4

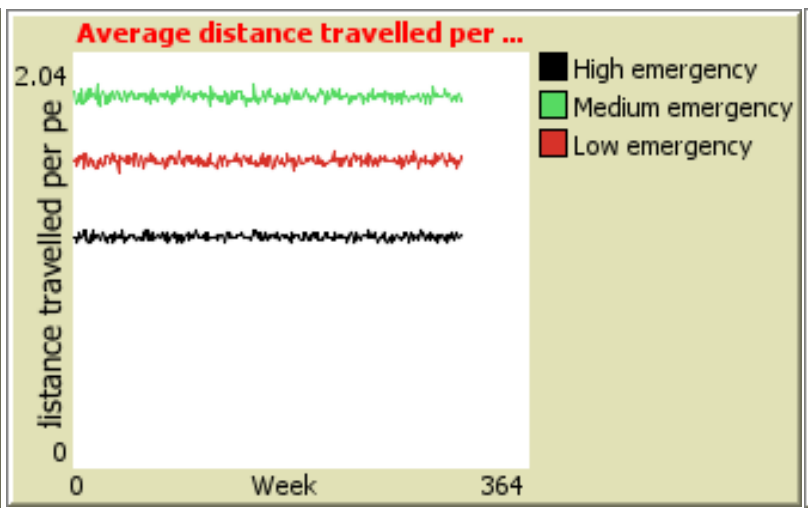

Experiment 5

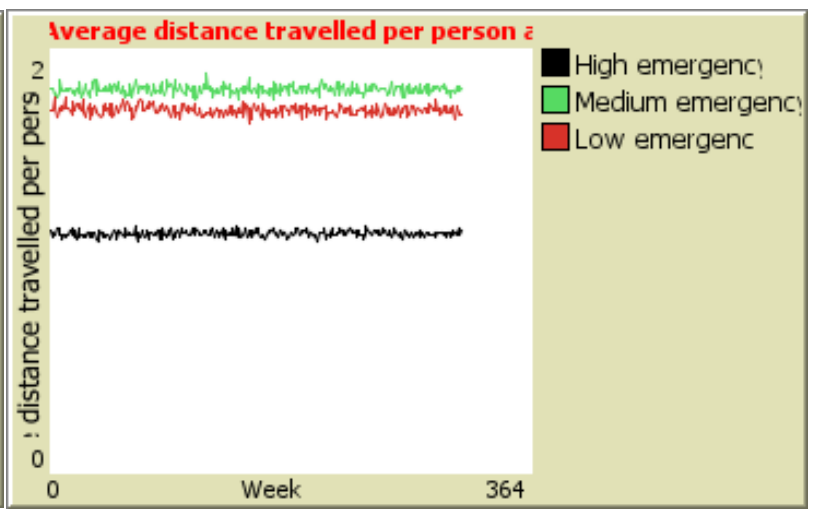

Experiment 6

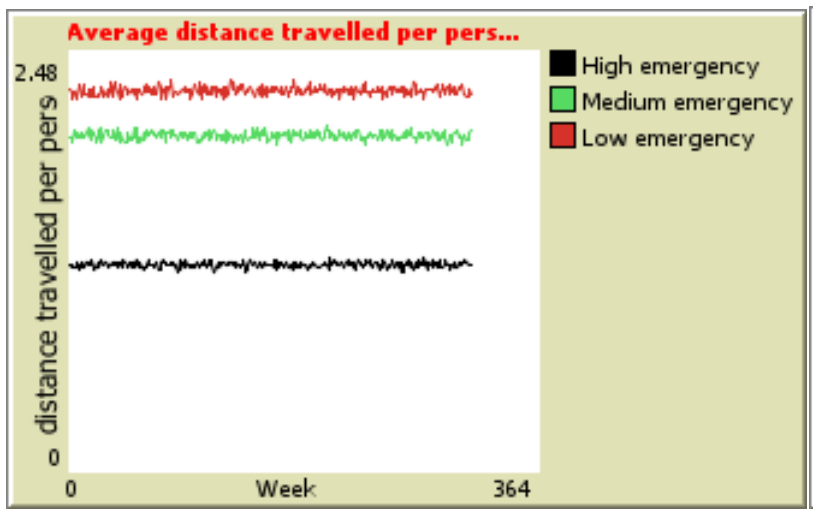

Experiment 7

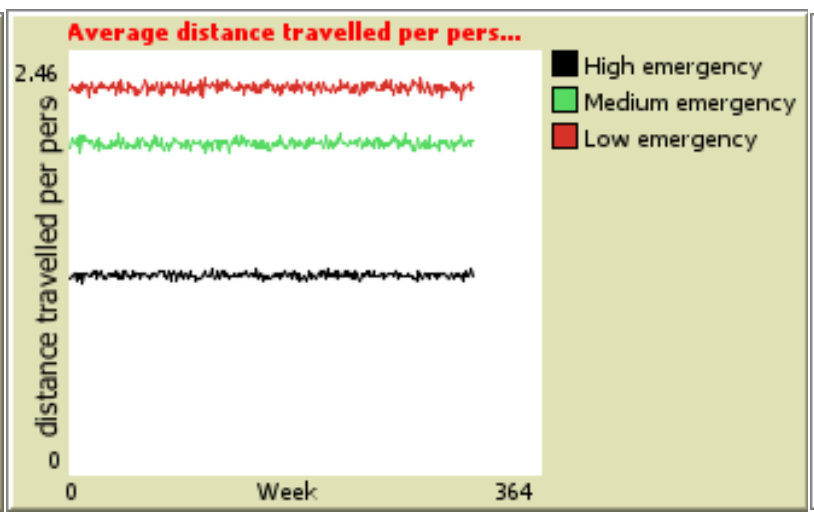

Experiment 8

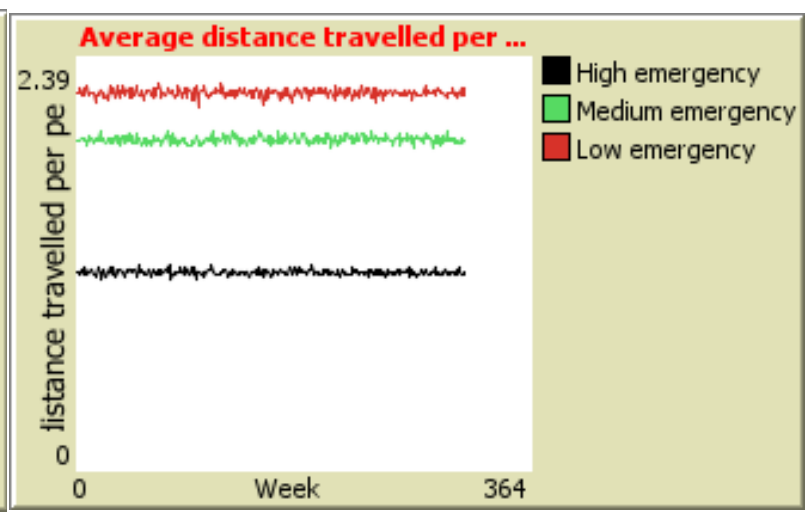

Experiment 9

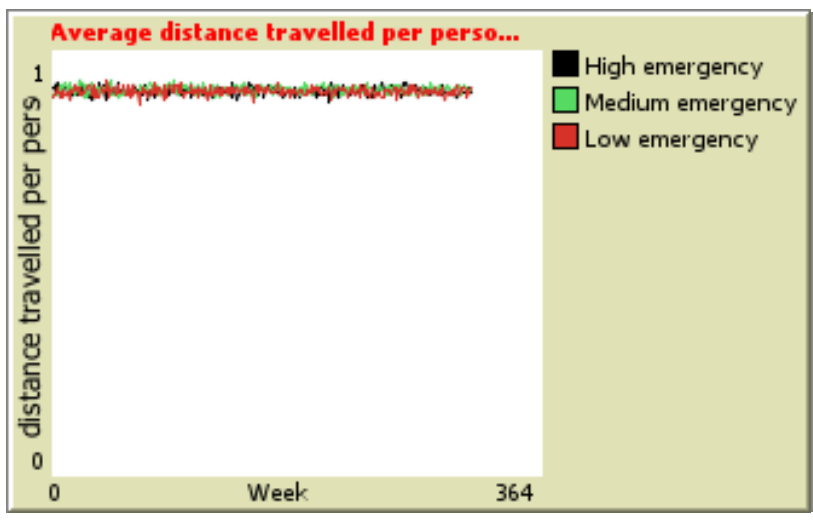

Experiment 10

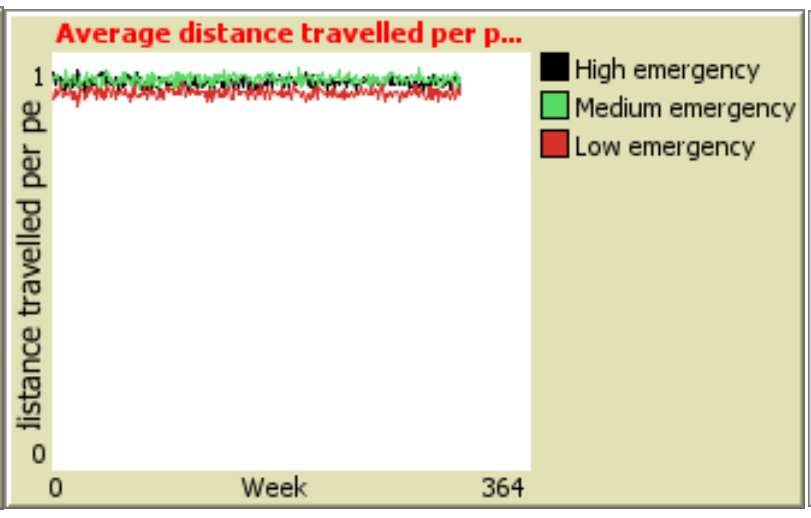

Experiment 11

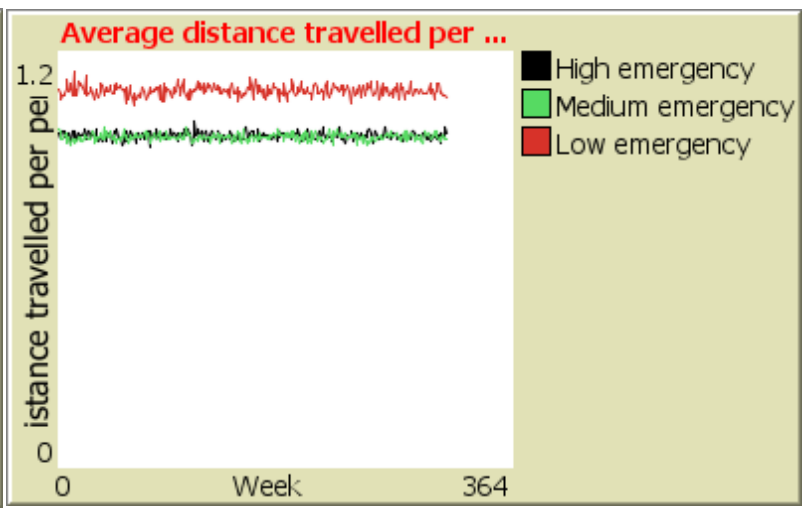

Experiment 12

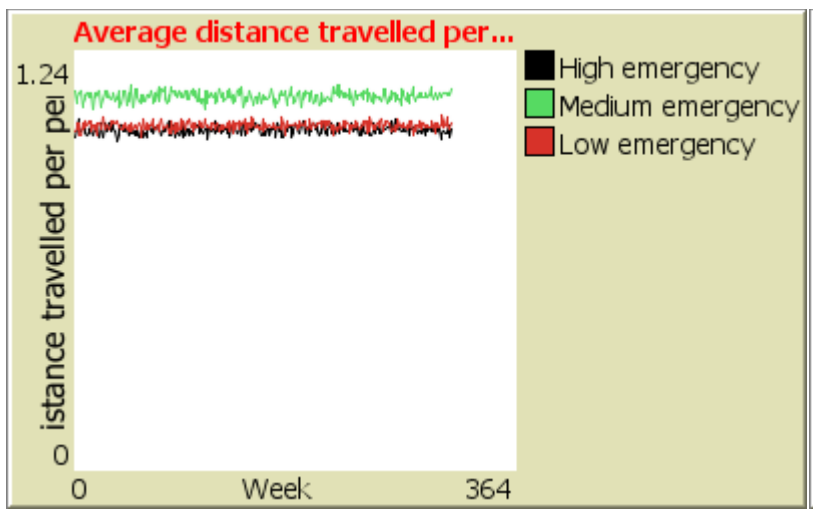

Experiment 13

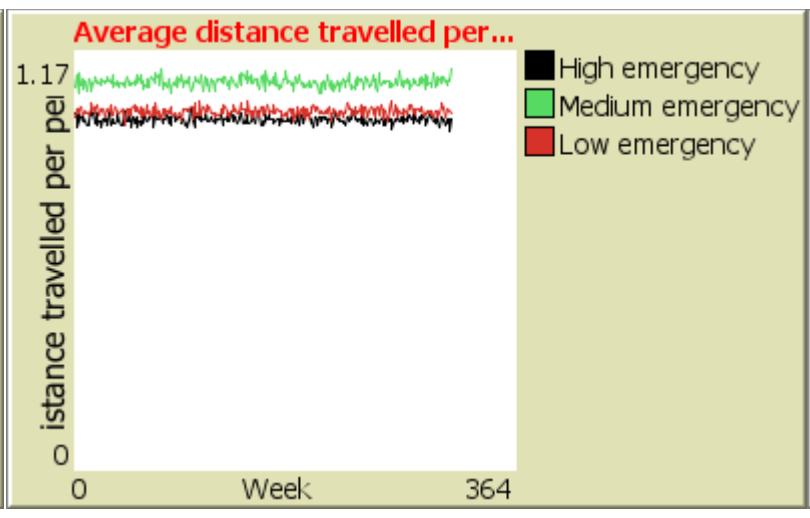

Experiment 14

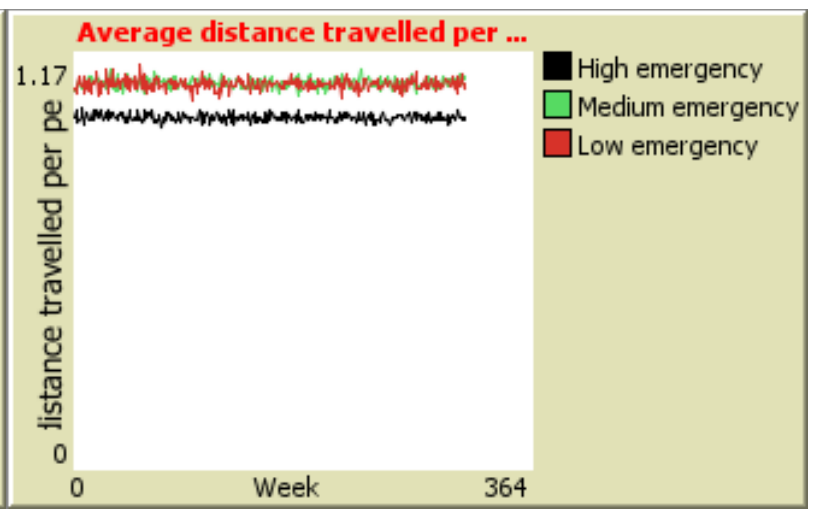

Experiment 15

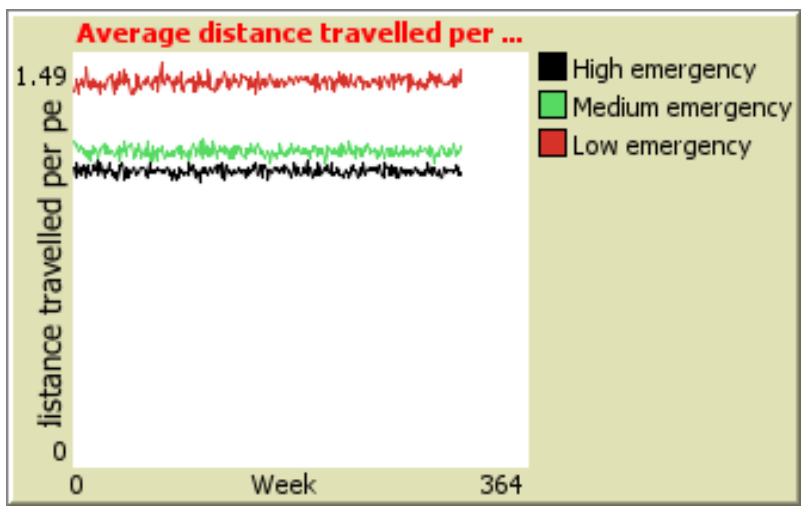

Experiment 16

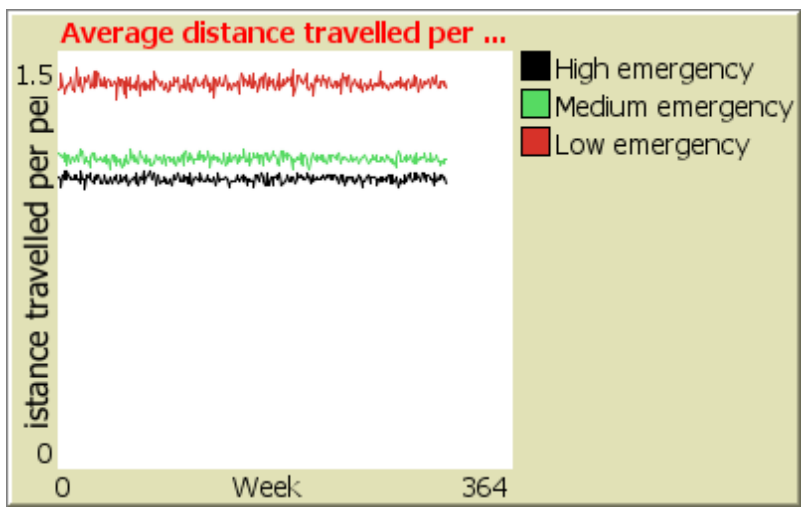

Experiment 17

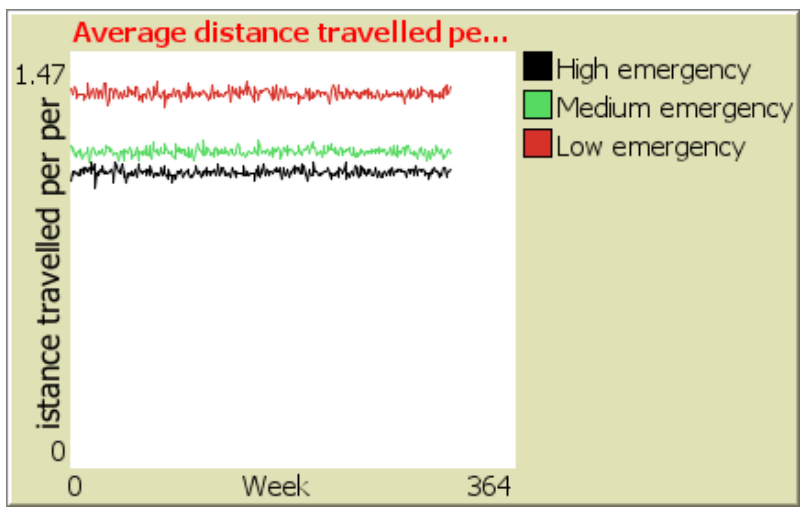

Experiment 18

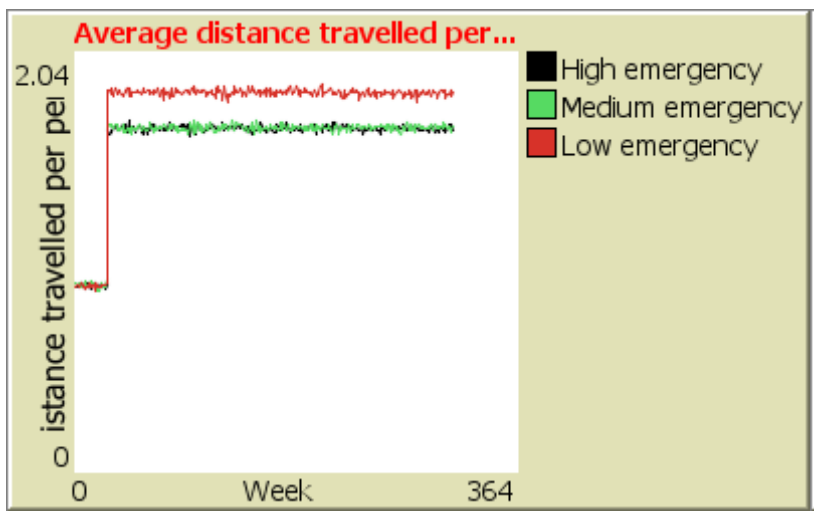

Experiment 19

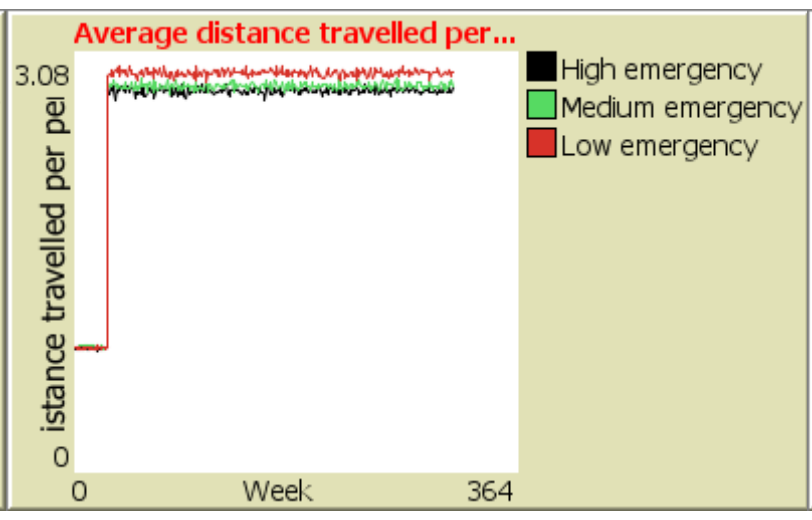

Experiment 20

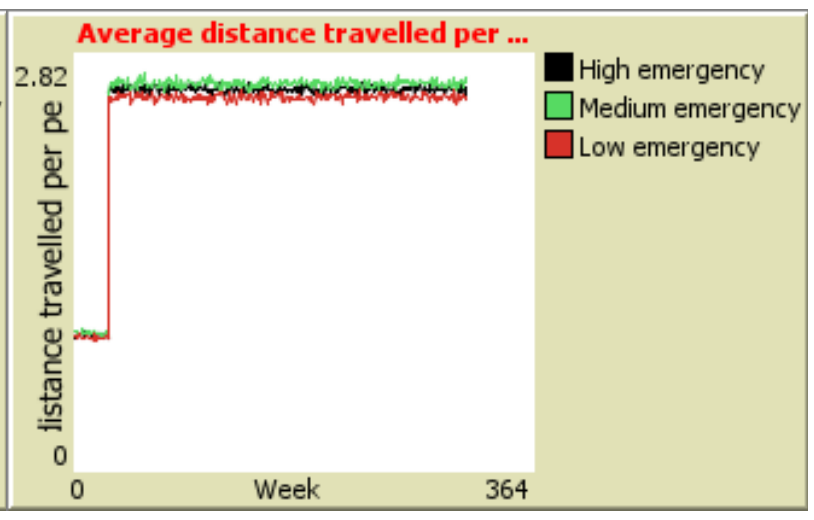

Experiment 21

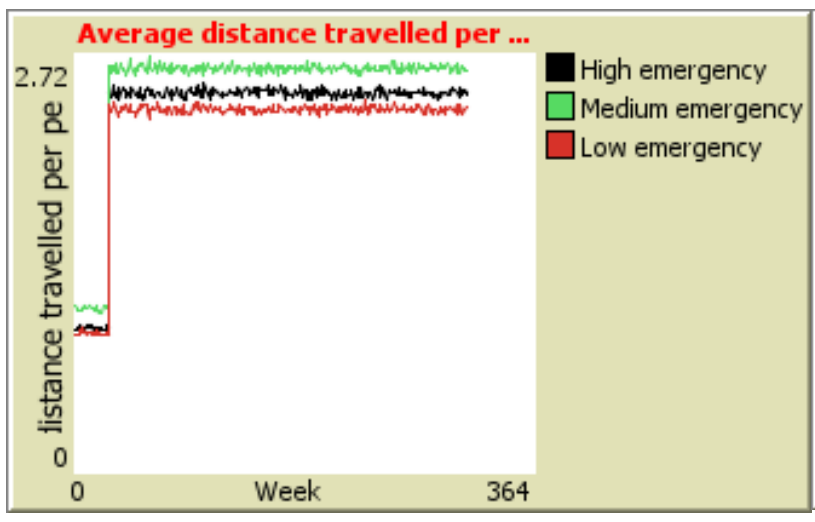

Experiment 22

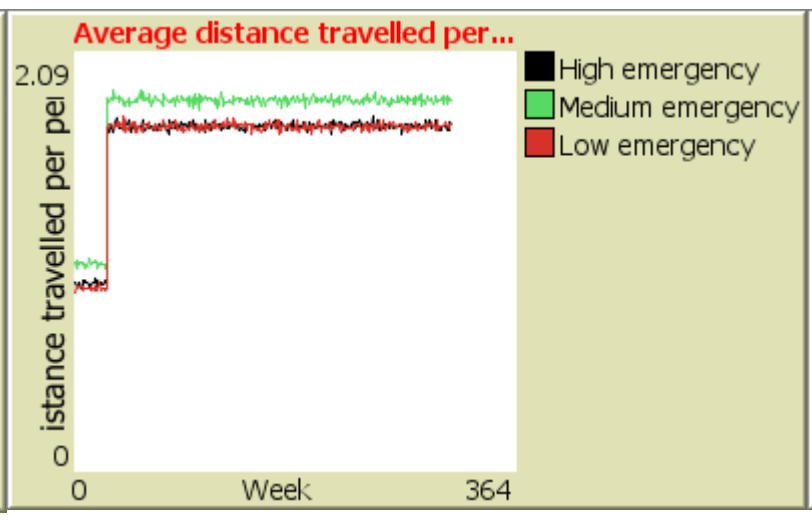

Experiment 23

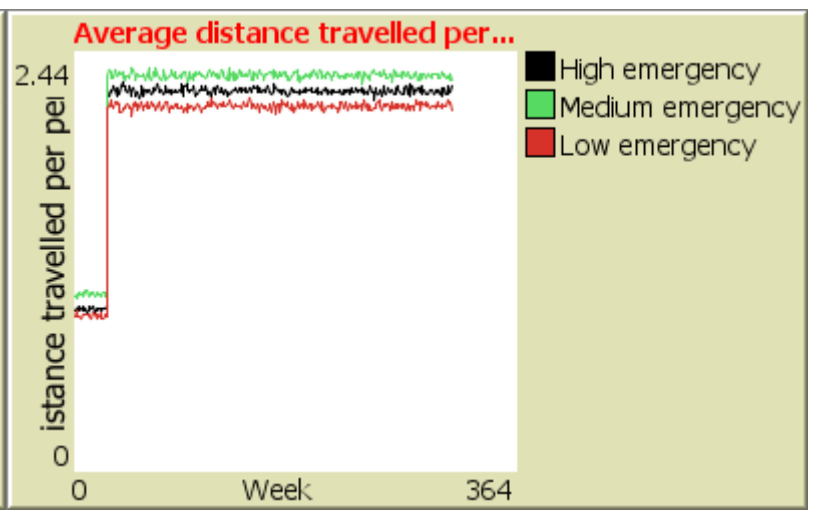

Experiment 24

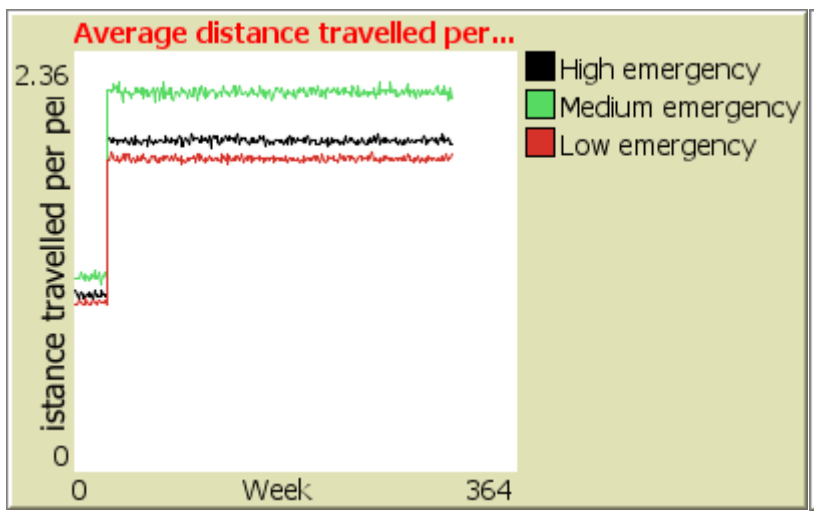

Experiment 25

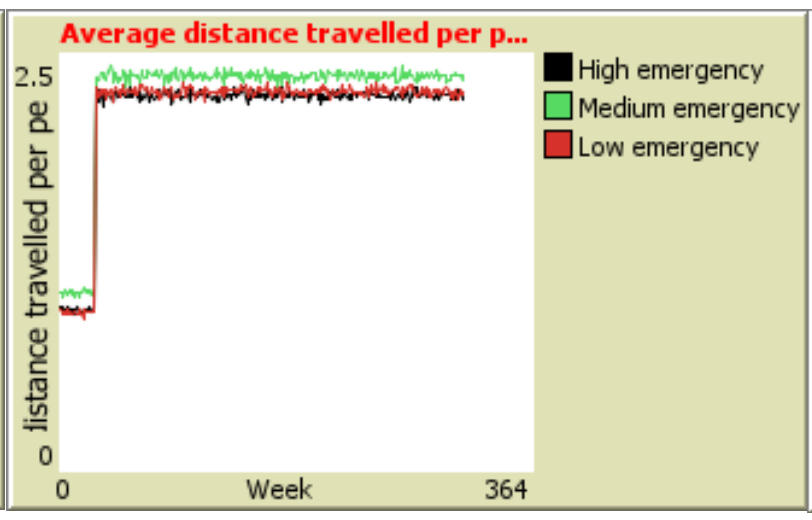

Experiment 26

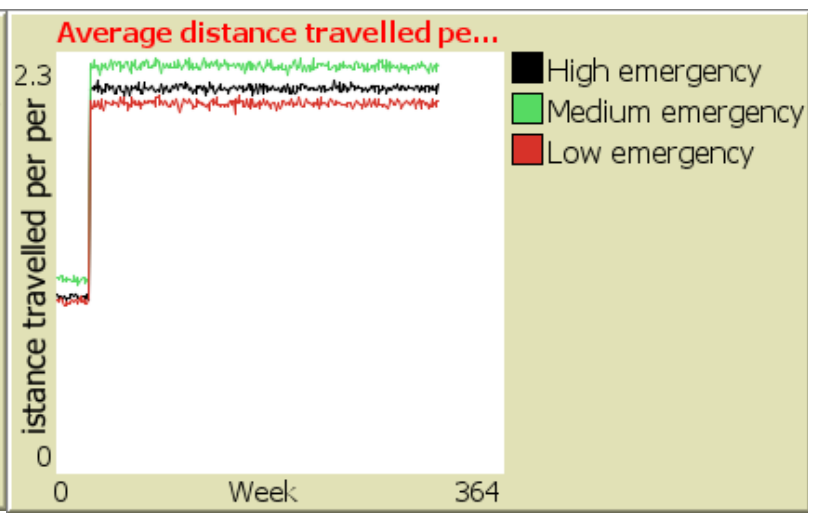

Experiment 27

FIGURE E5 Simulation results of distance (km) a customer travelled as per experimental design for price discount.

## Section E2. Simulation Results for Quality of Product

As per the experimental design in Table D10, the simulation model parameters are set and total 27 experiments are conducted. The simulation results for quality of the products are shown in this section.

### E2.1. Pictorial Representation of ABM Experiments Utilising ANOVA for Quality of Product

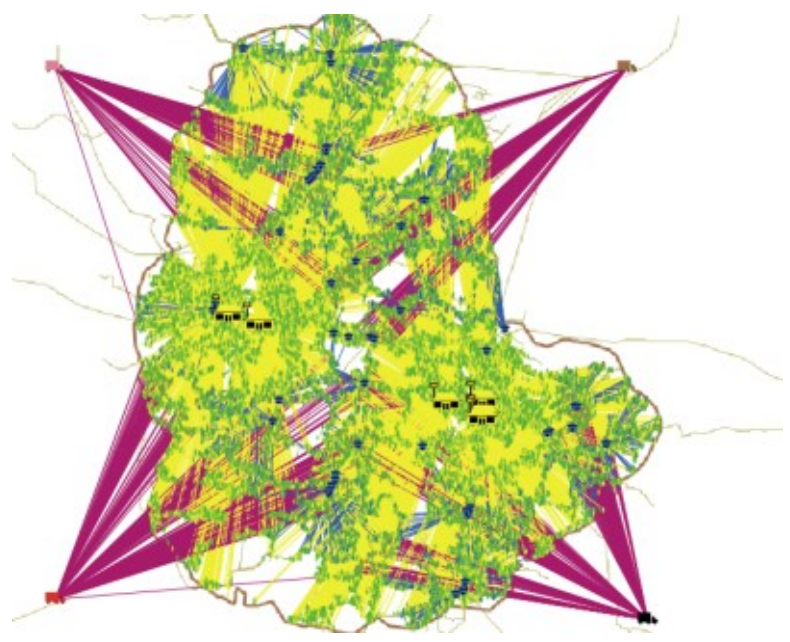

Experiment 1

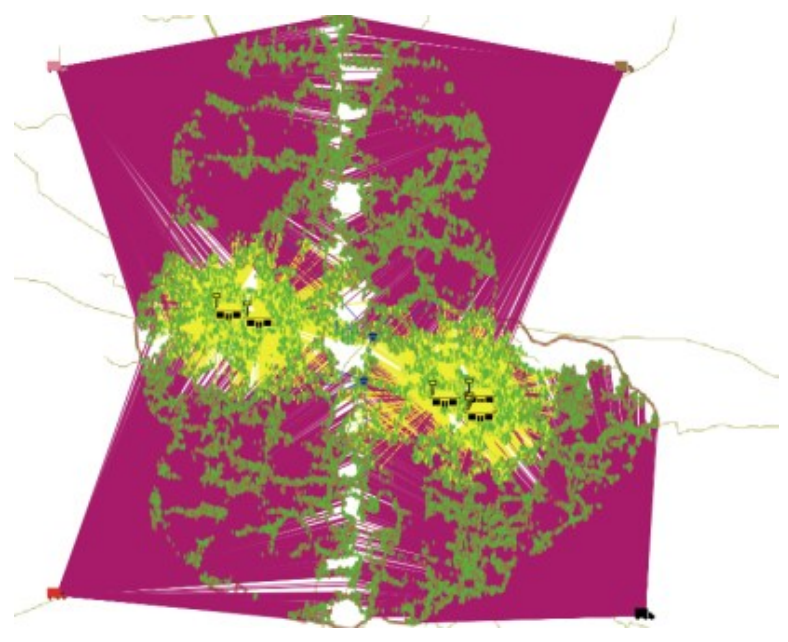

Experiment 2

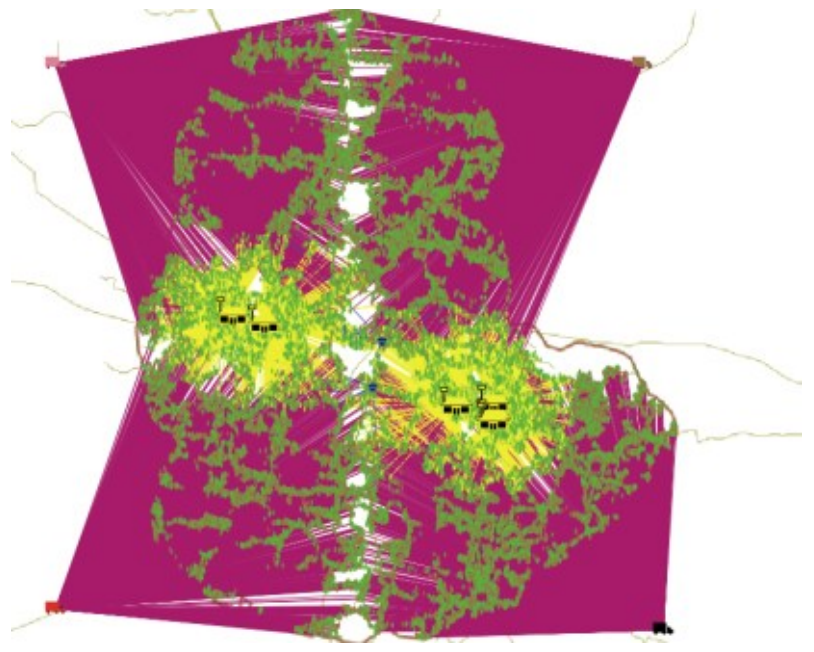

Experiment 3

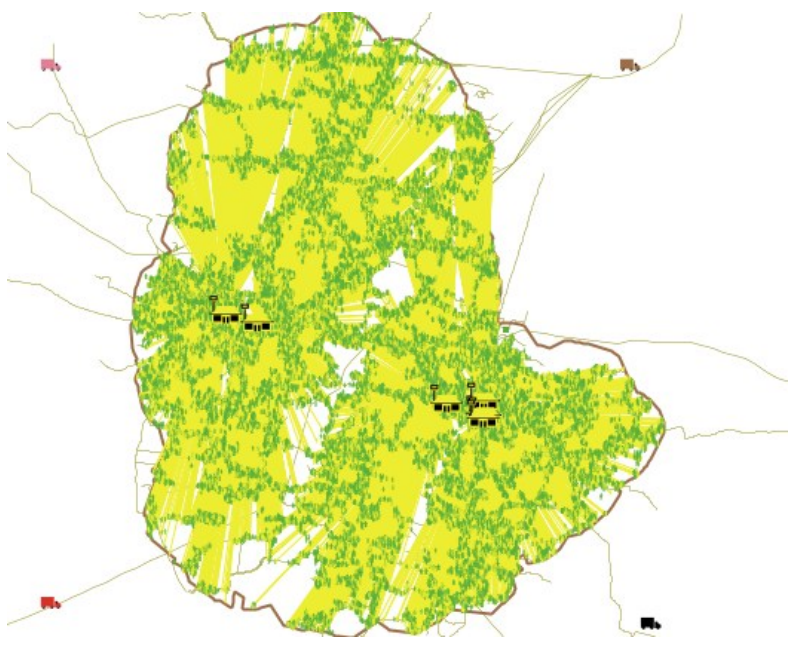

Experiment 4

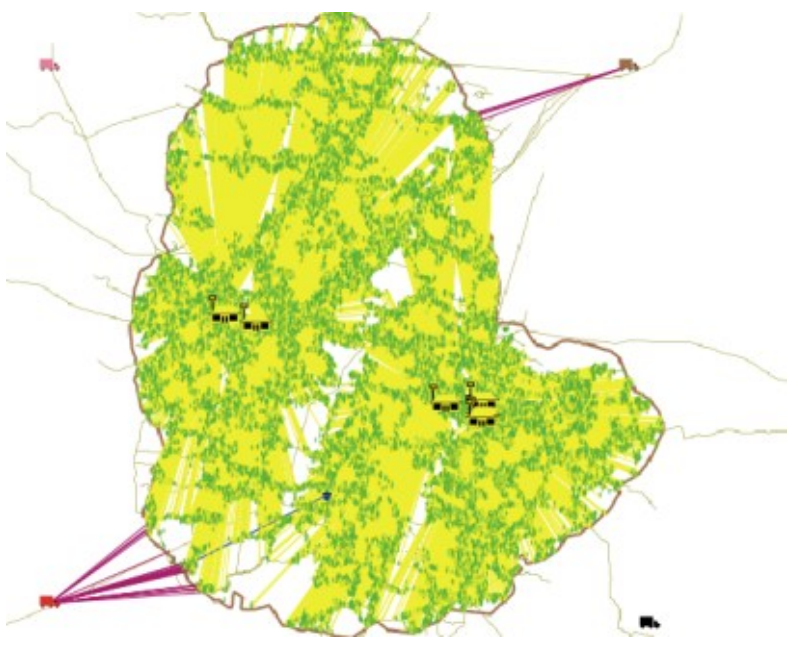

Experiment 5

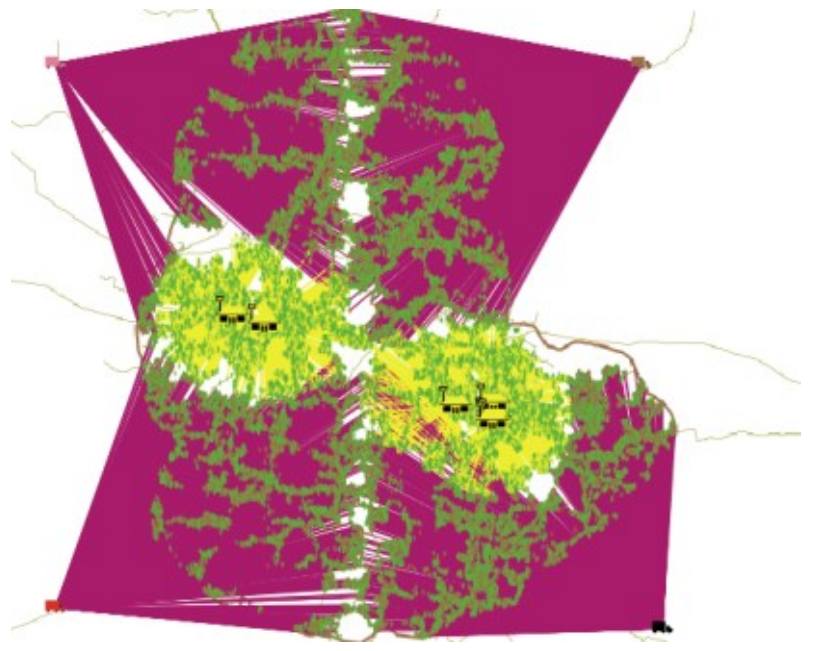

Experiment 6

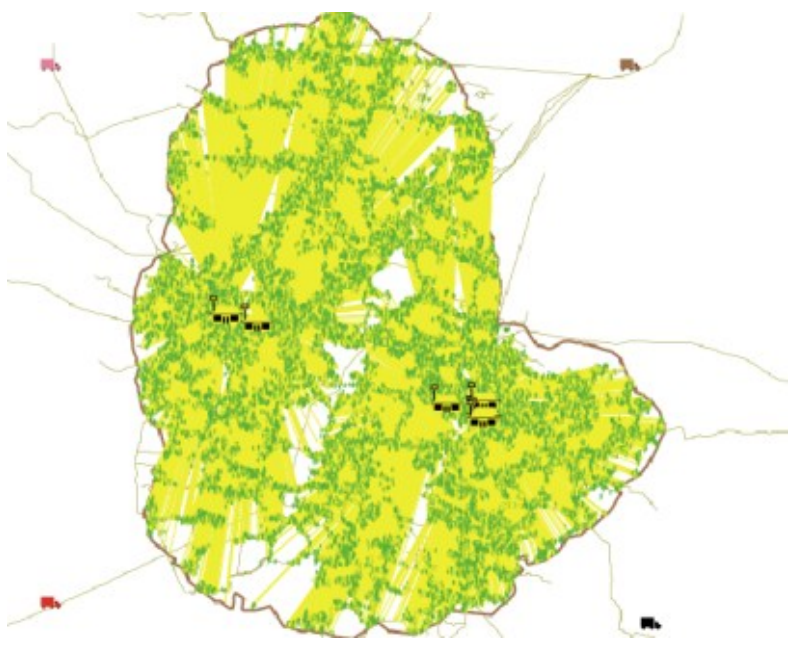

Experiment 7

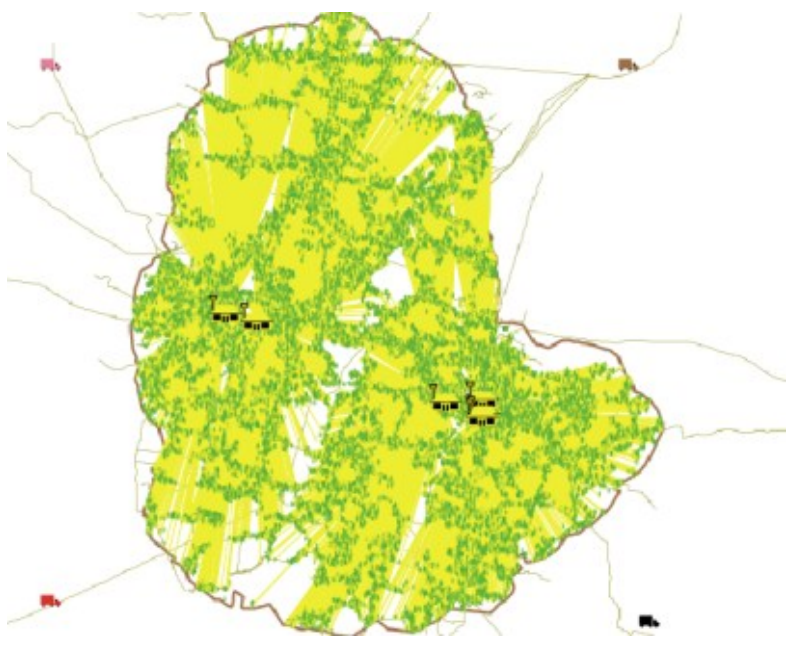

Experiment 8

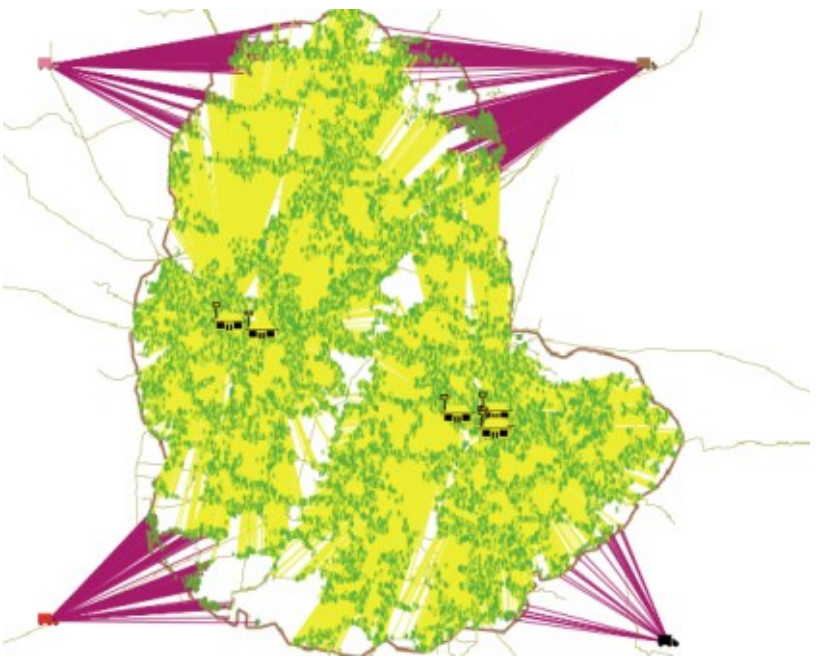

Experiment 9

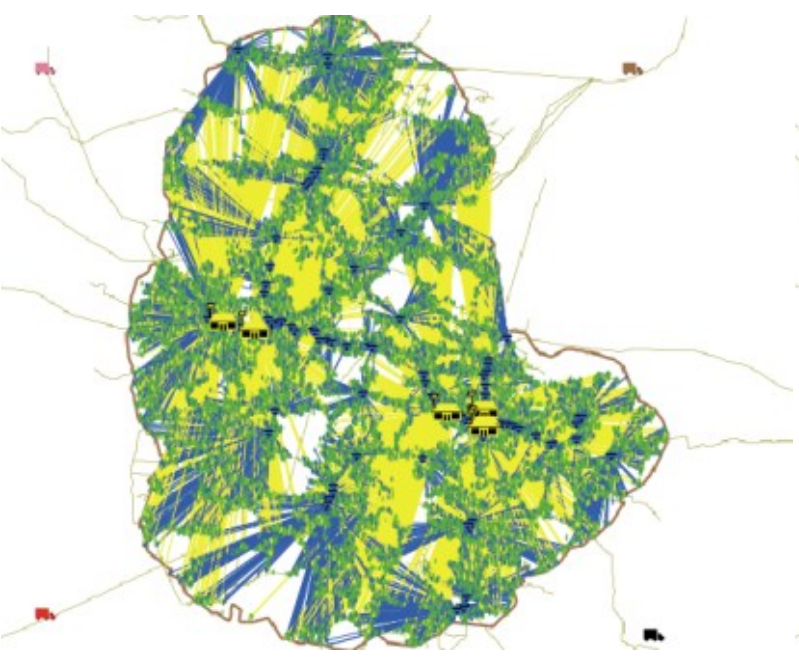

Experiment 10

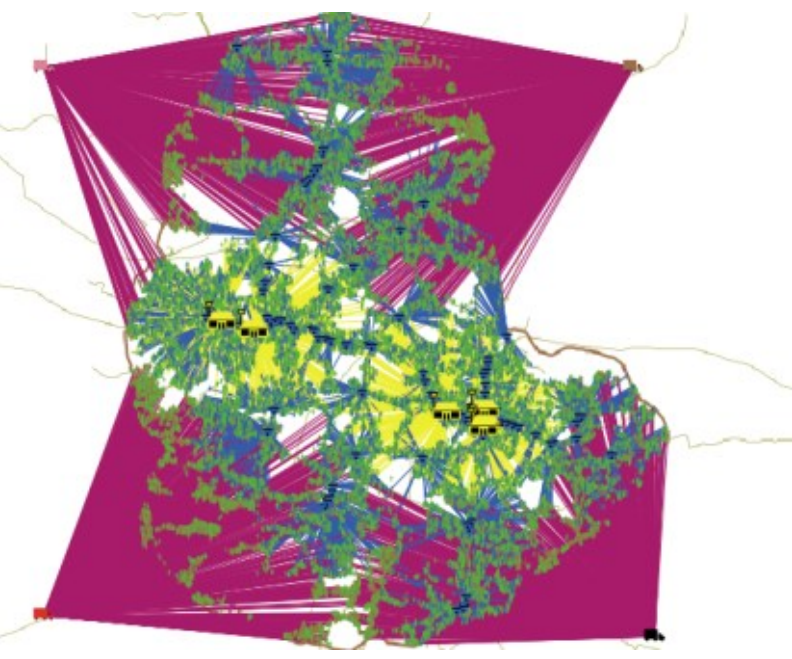

Experiment 11

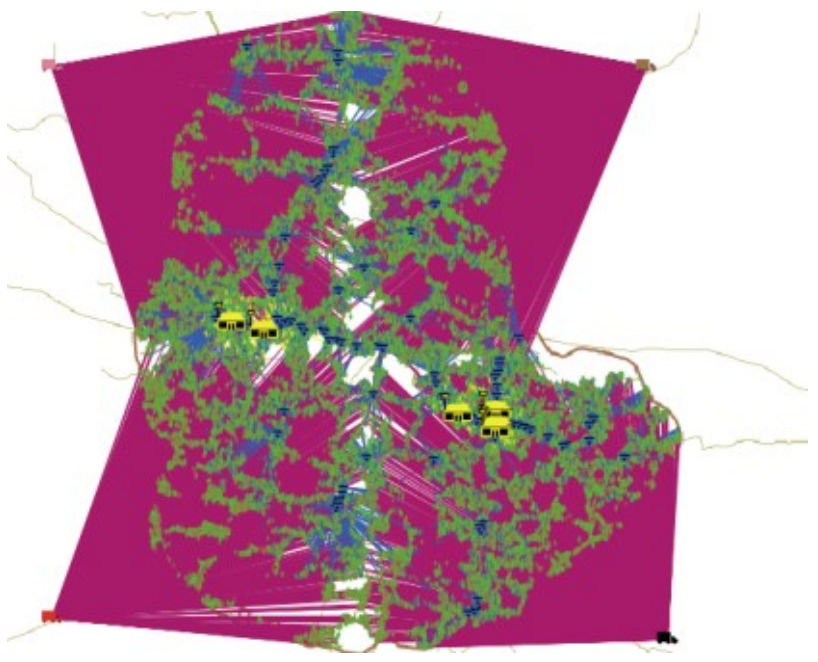

Experiment 12

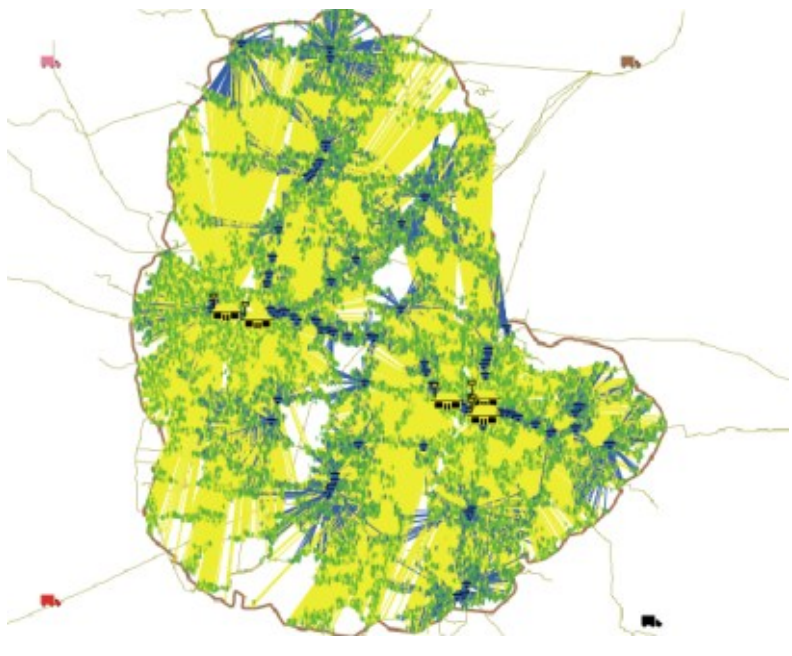

Experiment 13

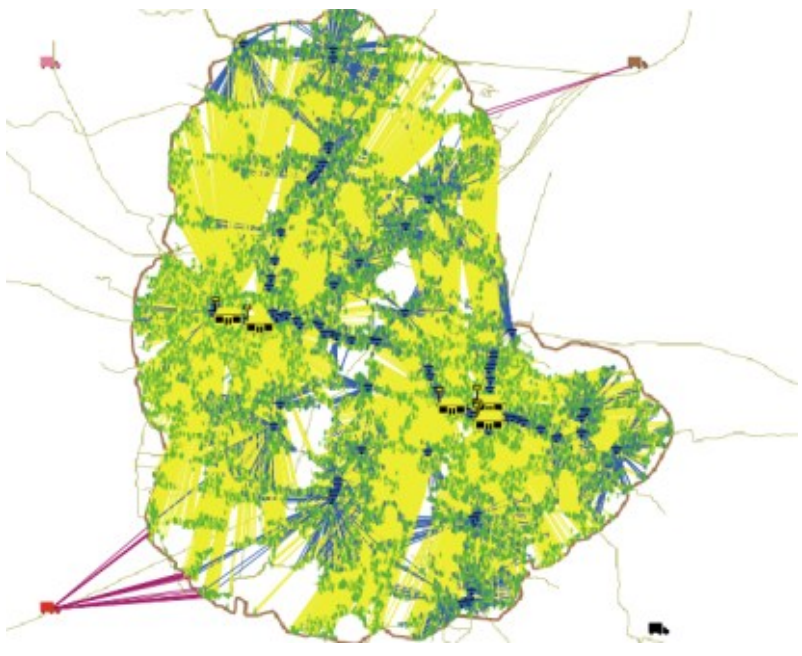

Experiment 14

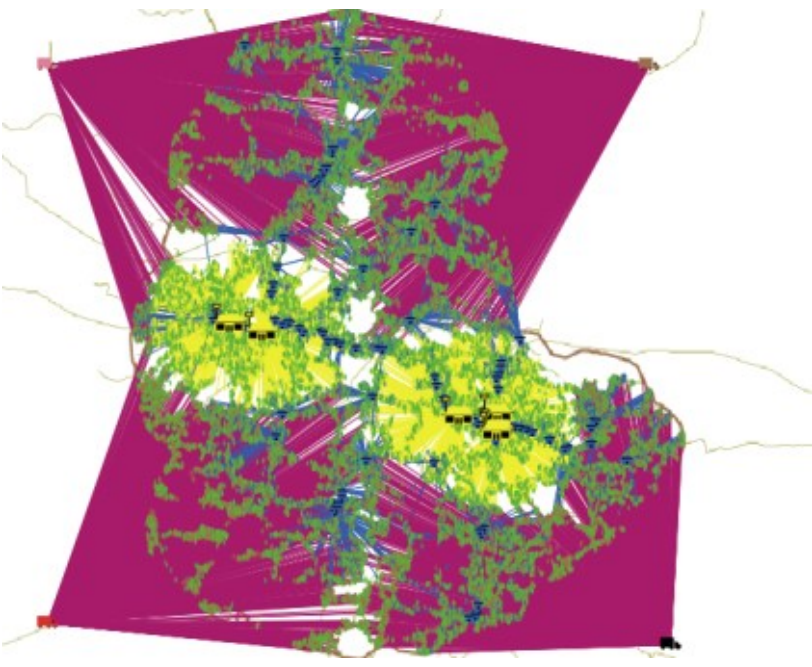

Experiment 15

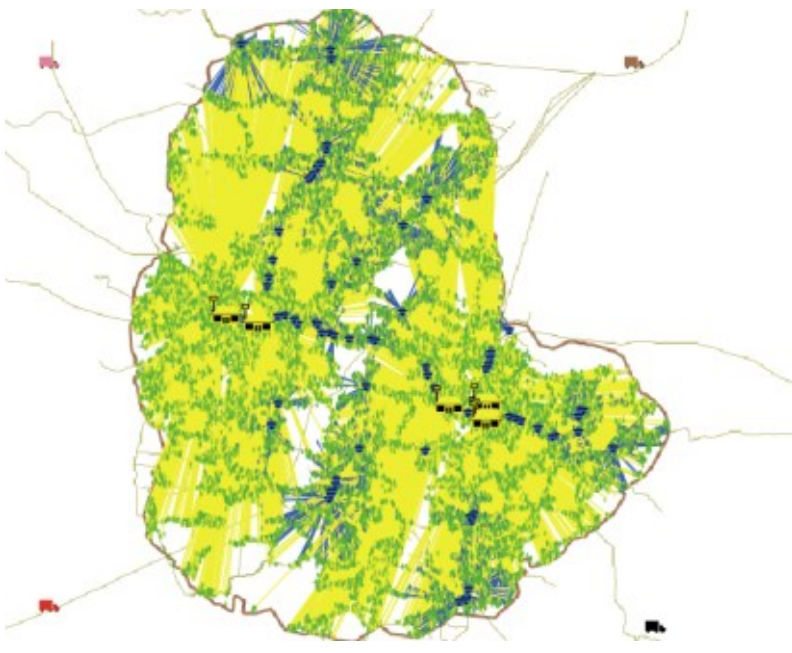

Experiment 16

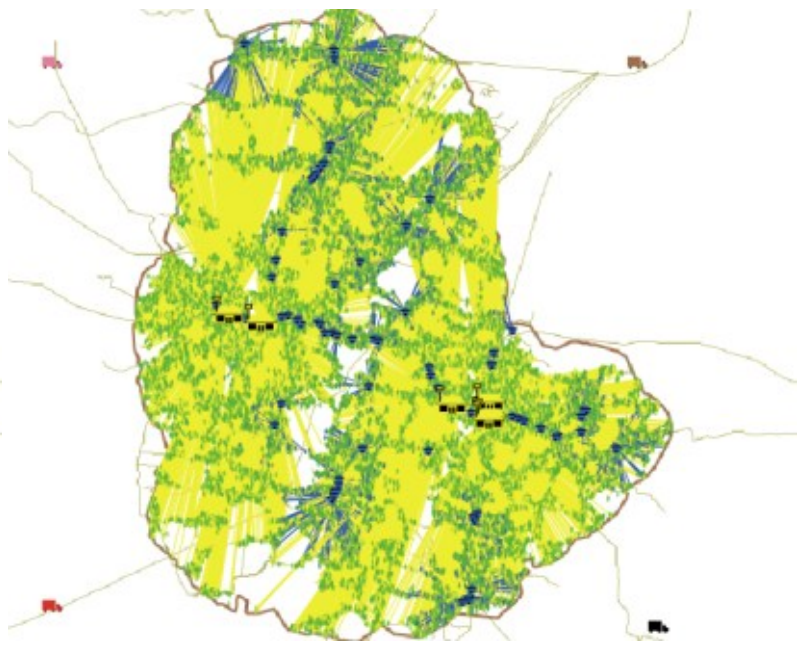

Experiment 17

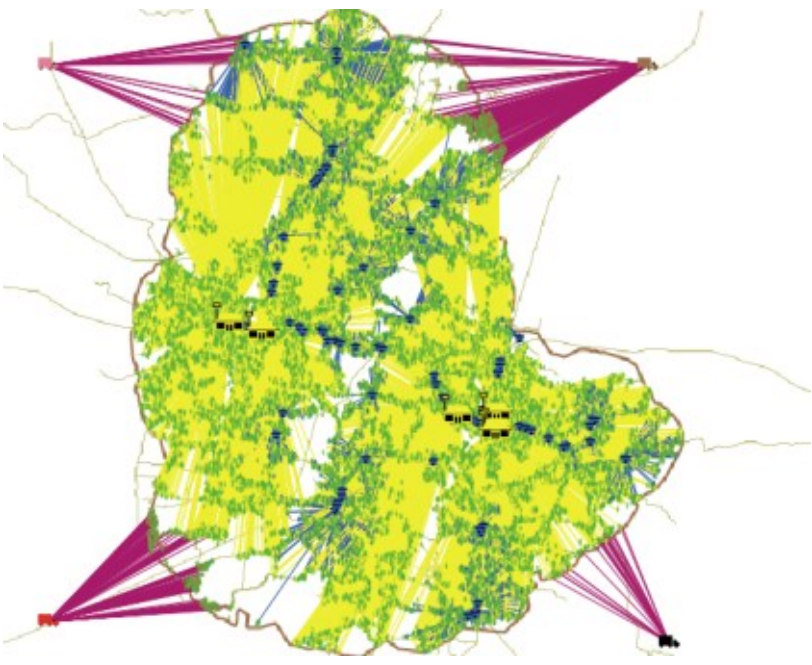

Experiment 18

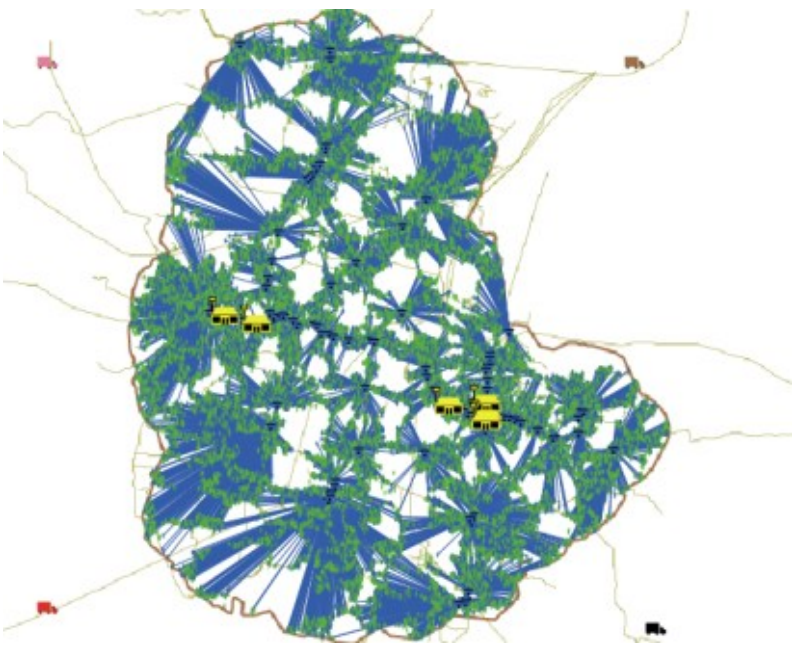

Experiment 19

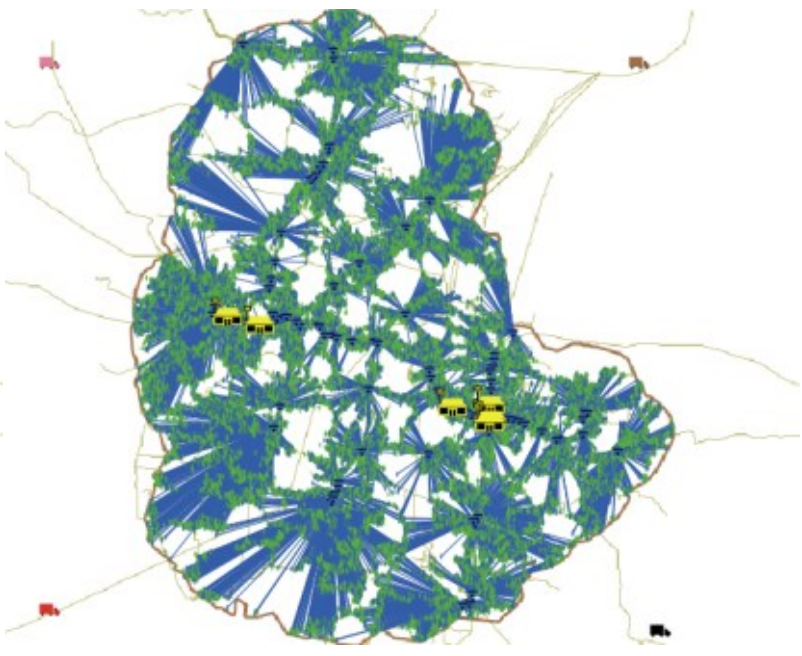

Experiment 20

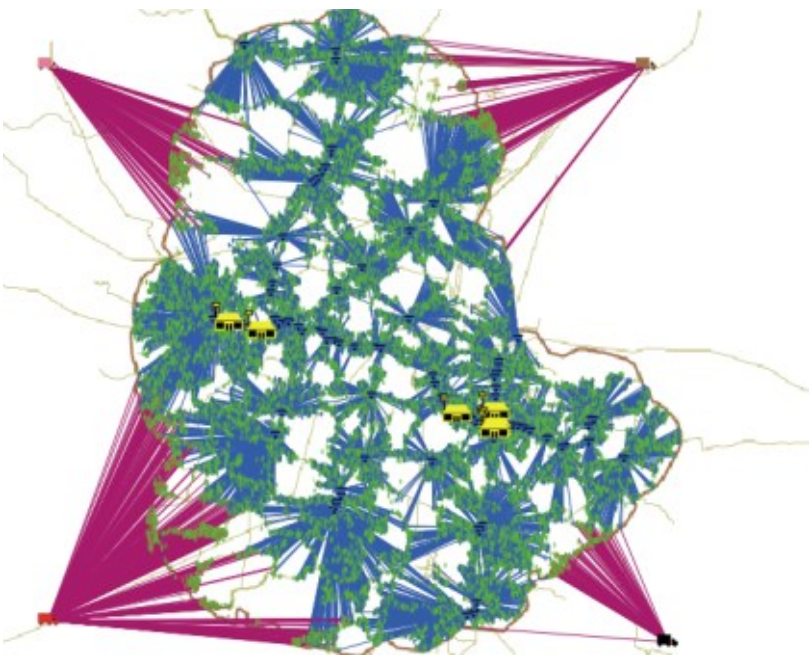

Experiment 21

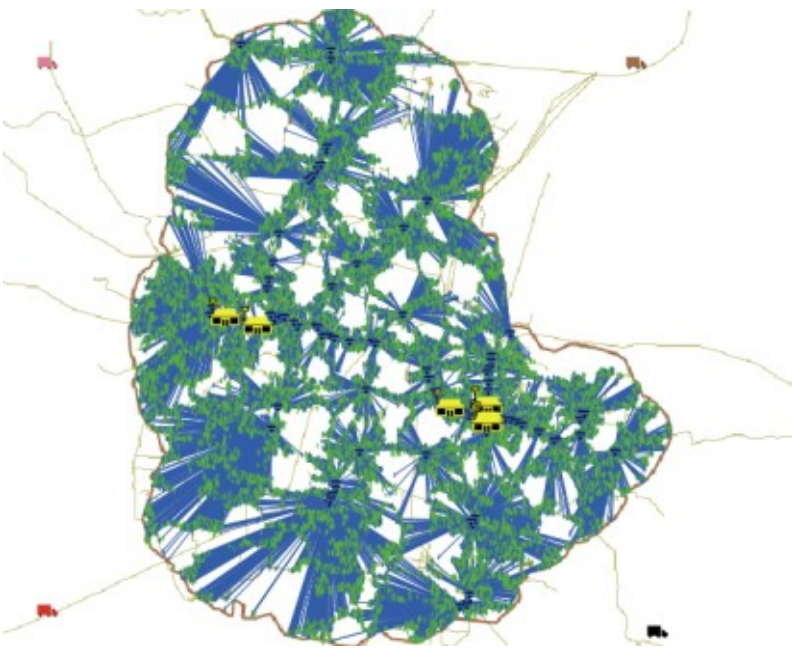

Experiment 22

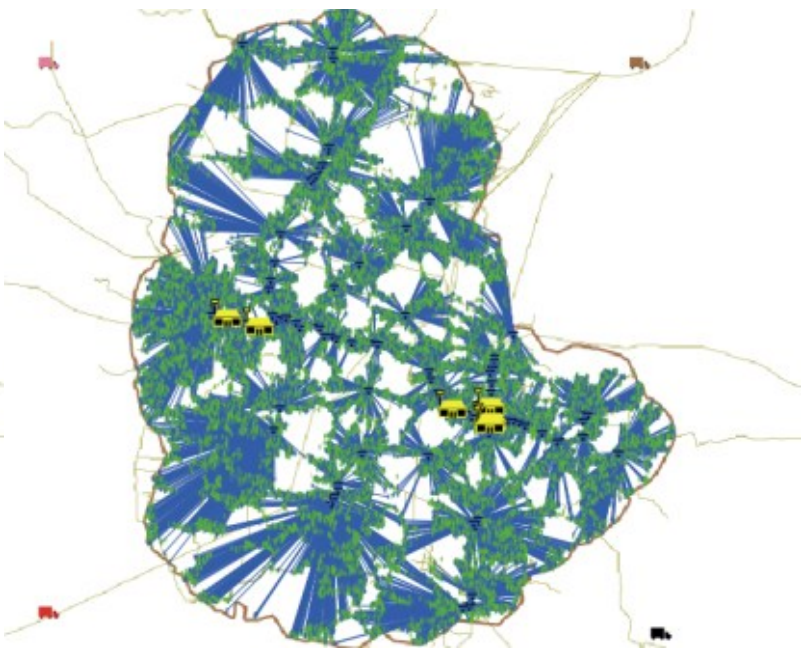

Experiment 23

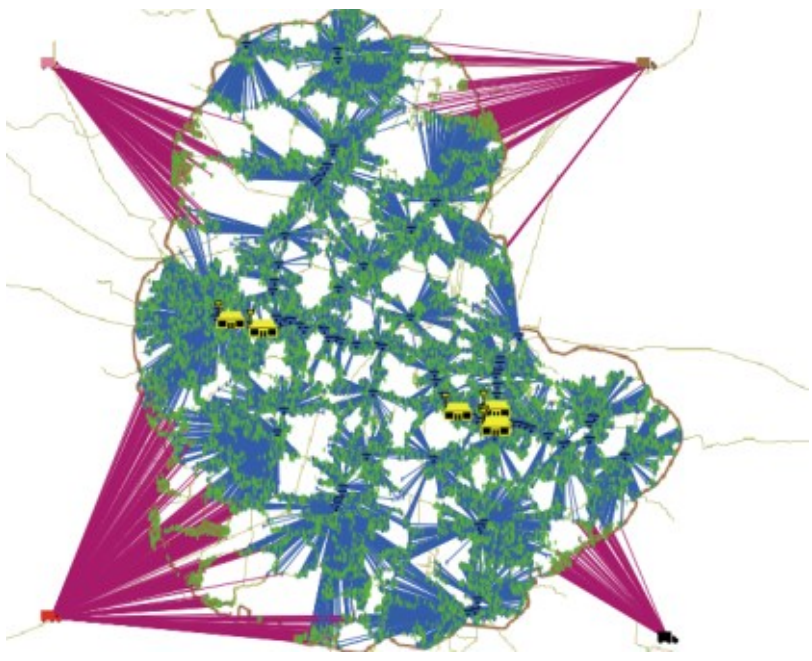

Experiment 24

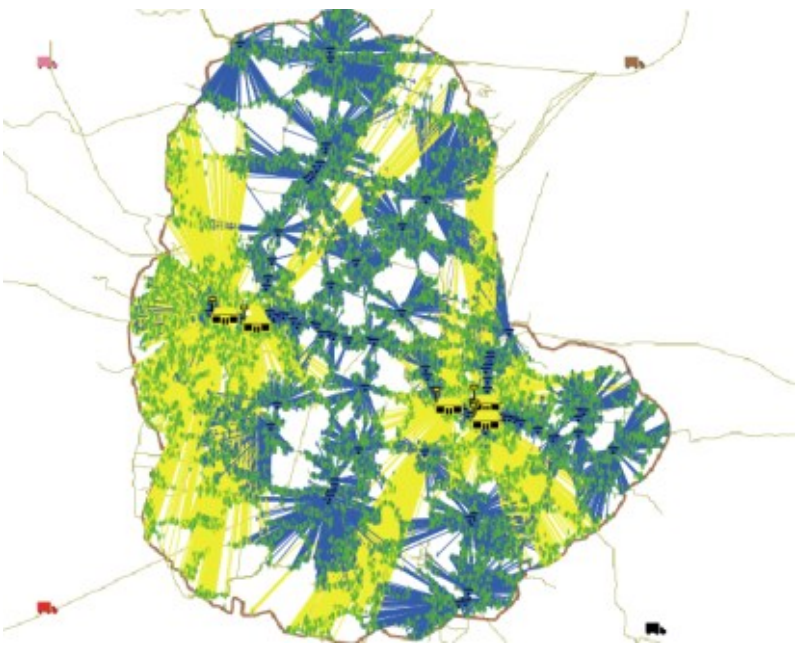

Experiment 25

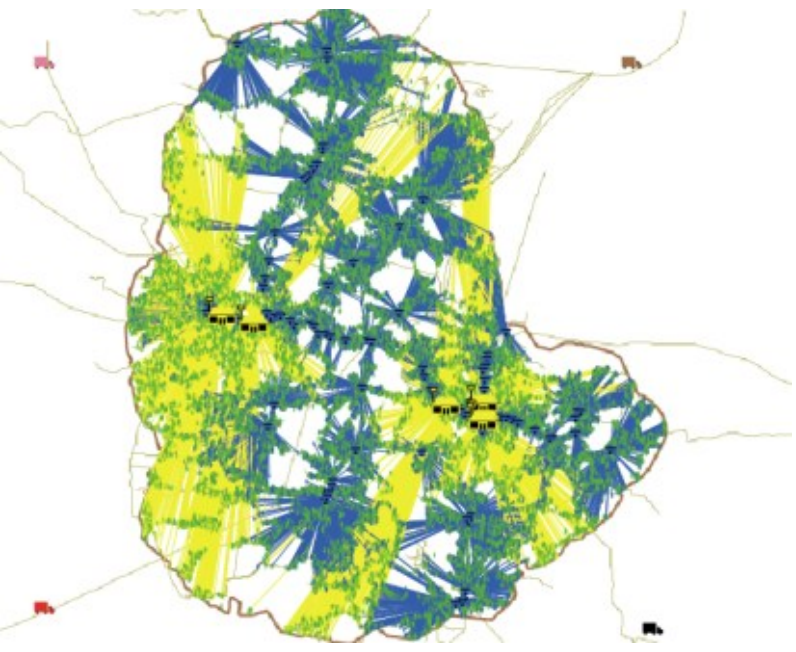

Experiment 26

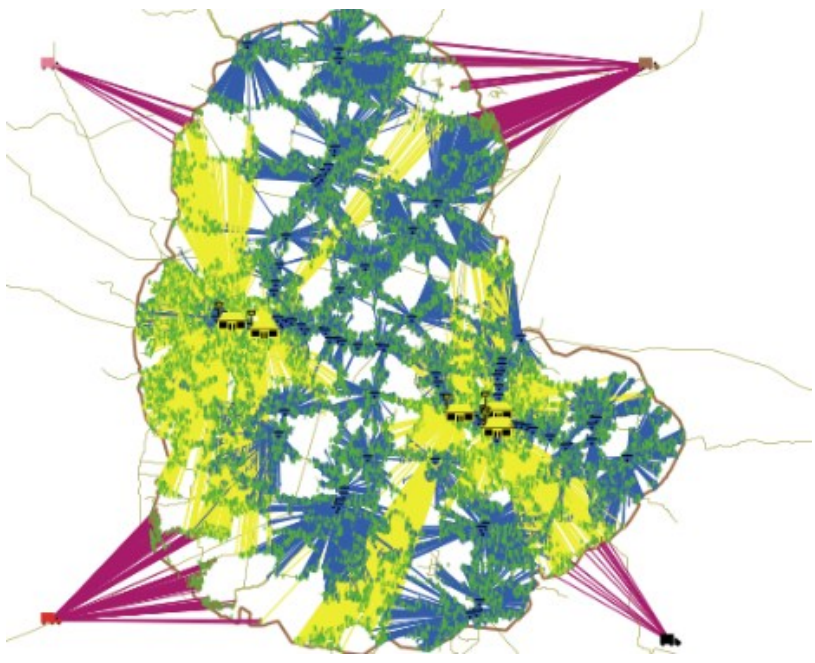

Experiment 27

FIGURE E6 Simulation results as per experimental design for product quality.

### E2.2. Simulation Results of Weekly Customer Footprint as per 27 Experimental Runs Utilising ANOVA for Quality of Product

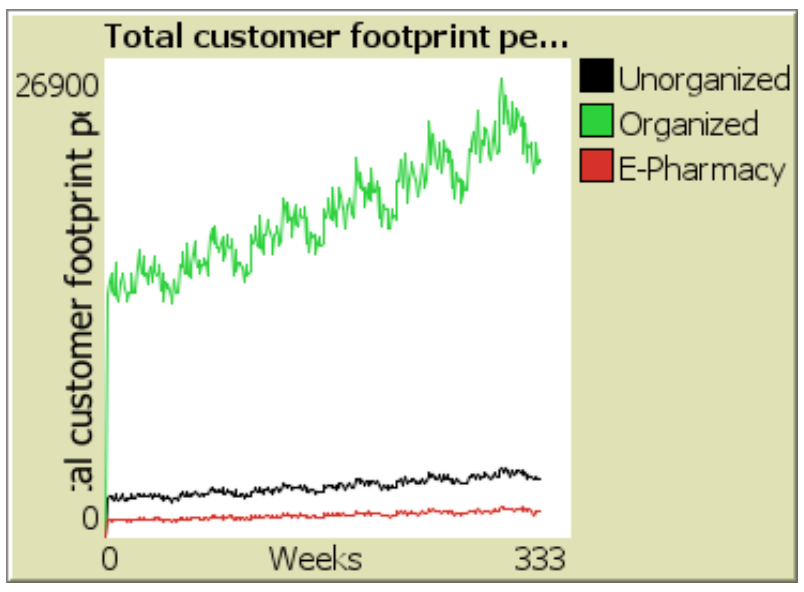

Experiment 1

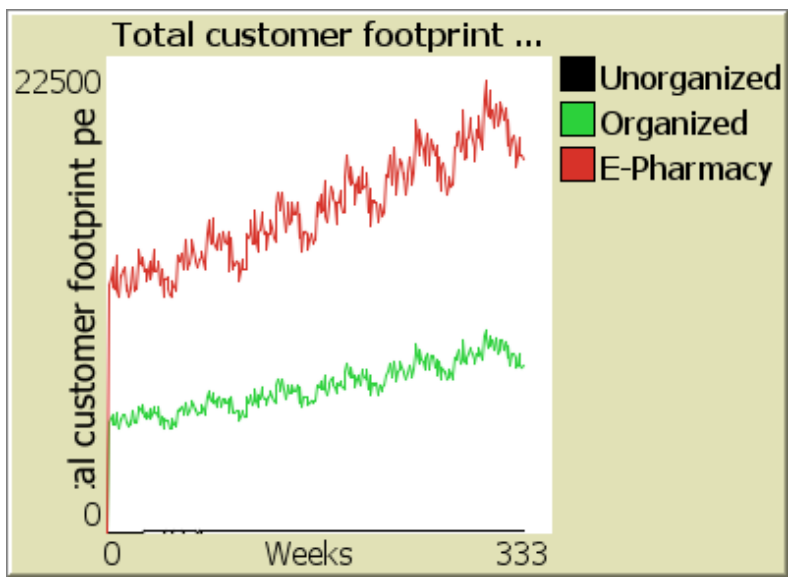

Experiment 2

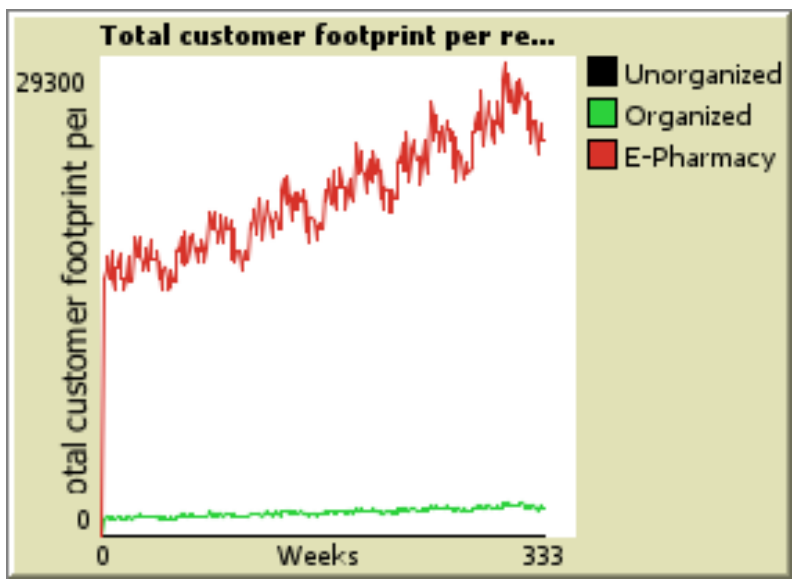

Experiment 3

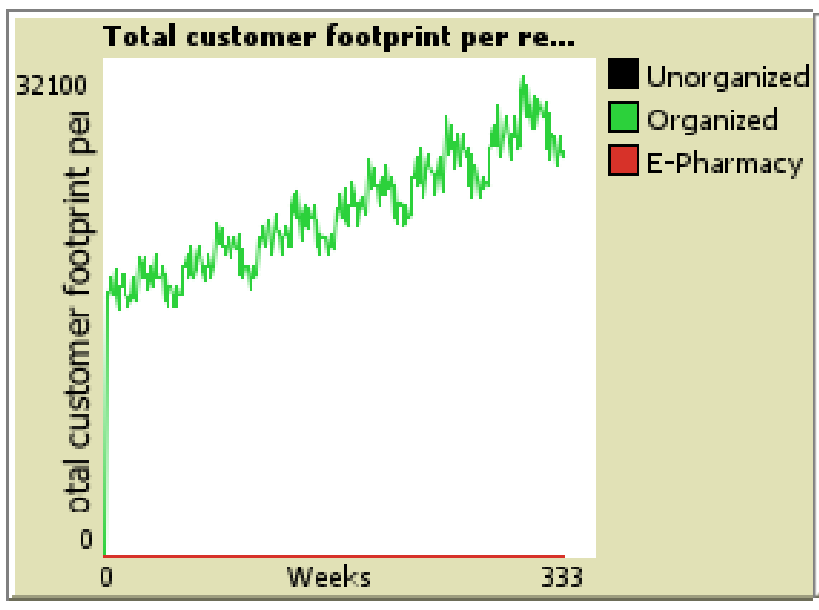

Experiment 4

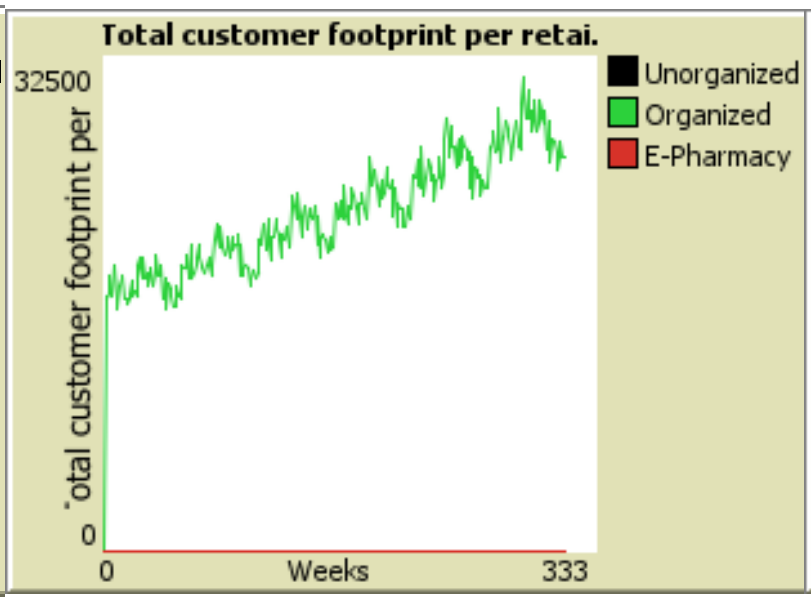

Experiment 5

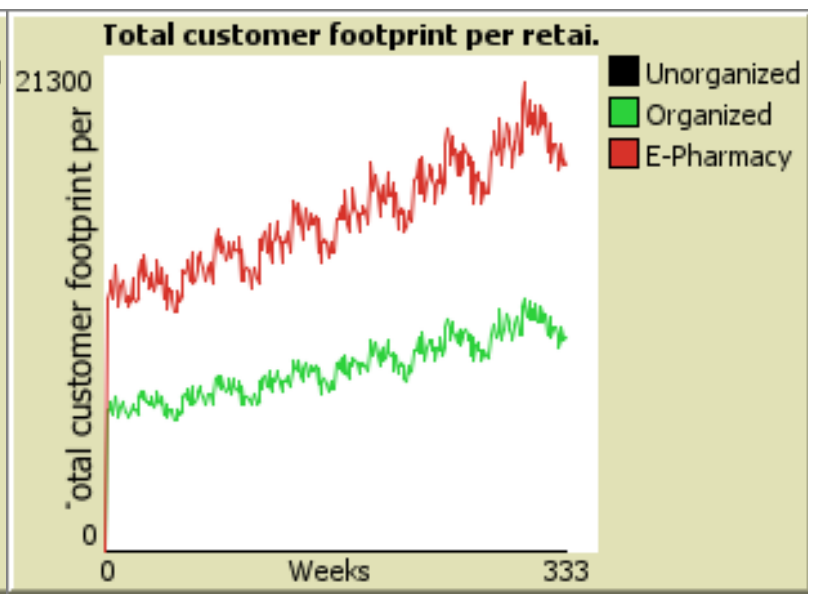

Experiment 6

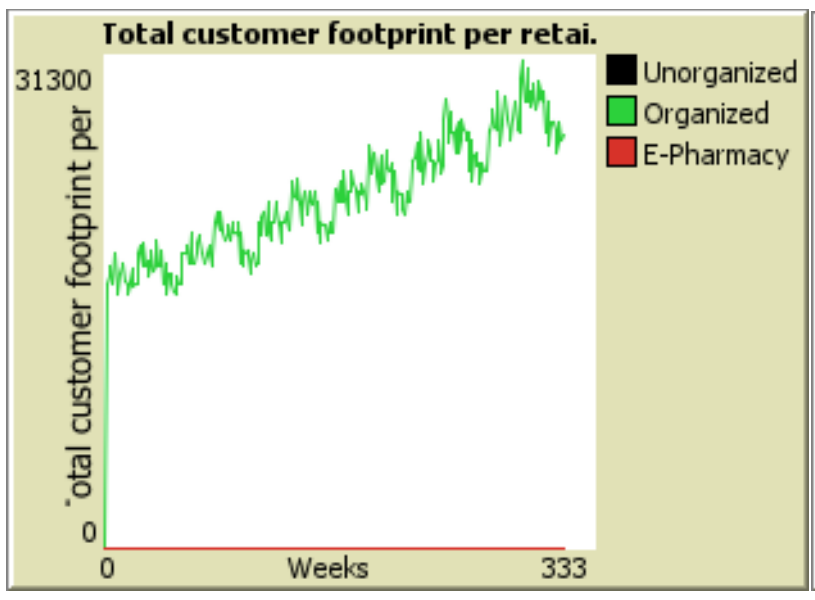

Experiment 7

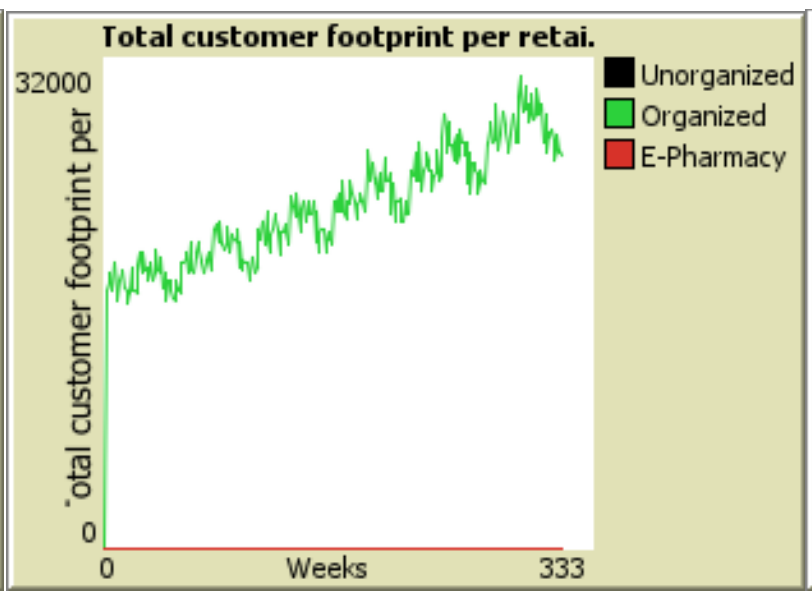

Experiment 8

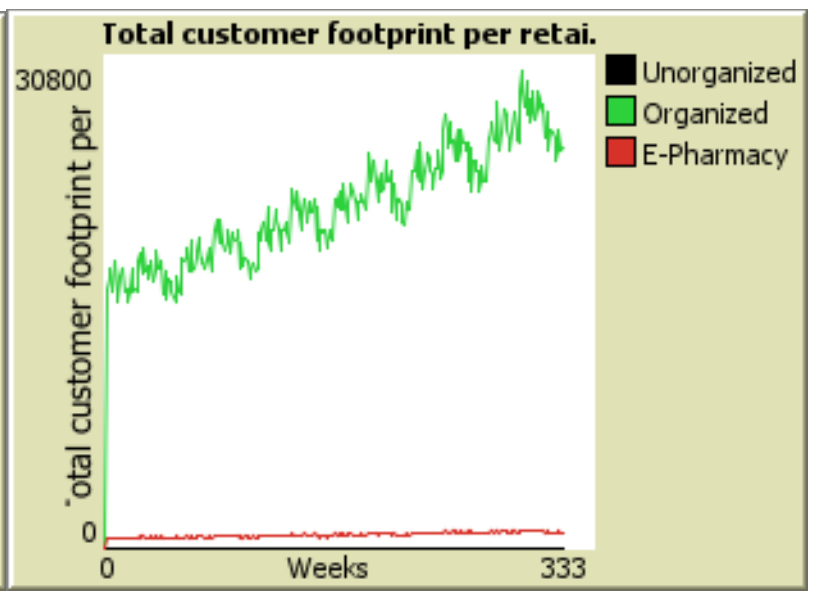

Experiment 9

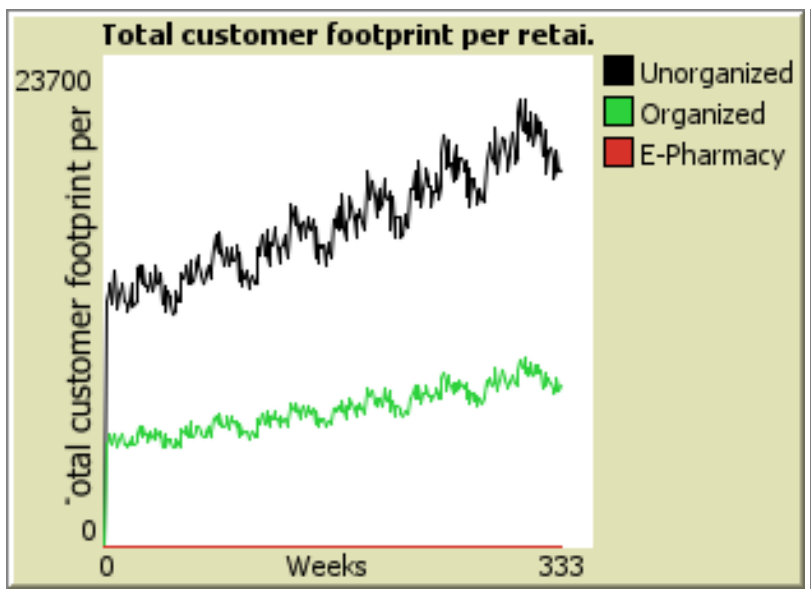

Experiment 10

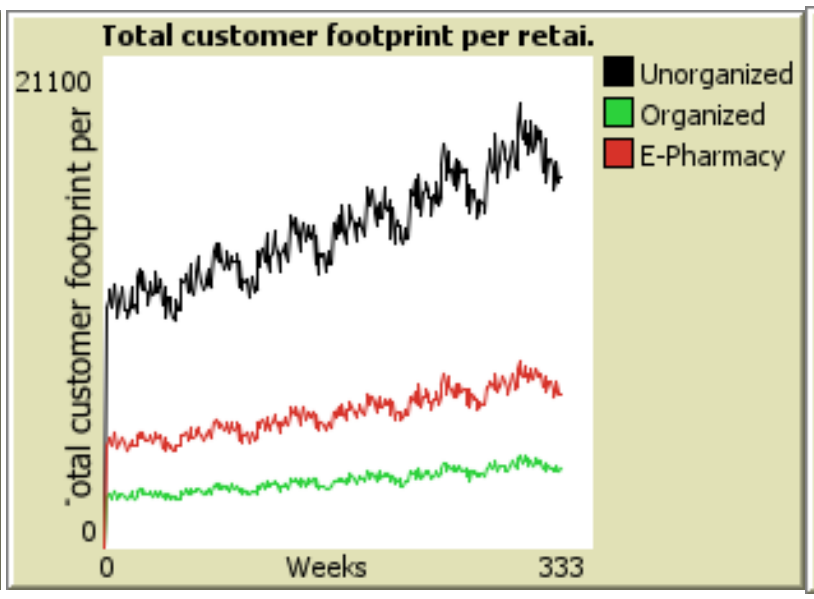

Experiment 11

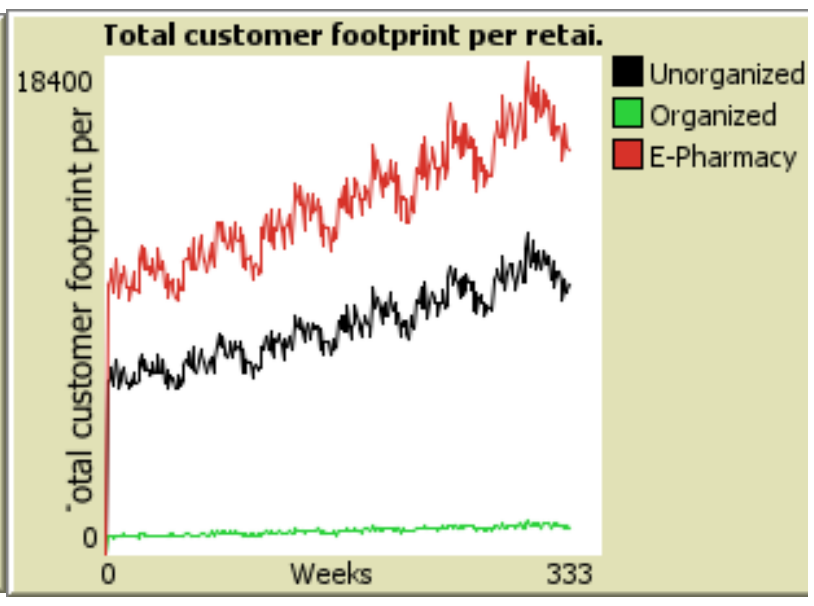

Experiment 12

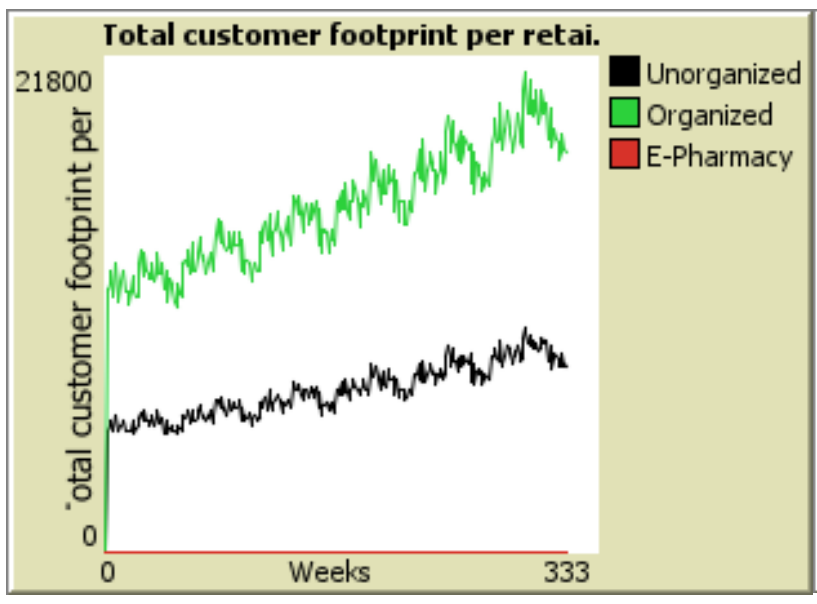

Experiment 13

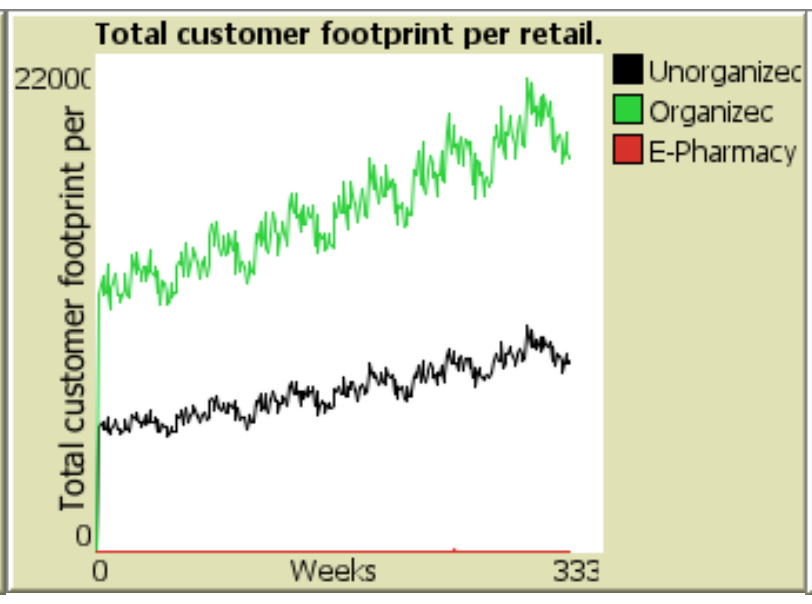

Experiment 14

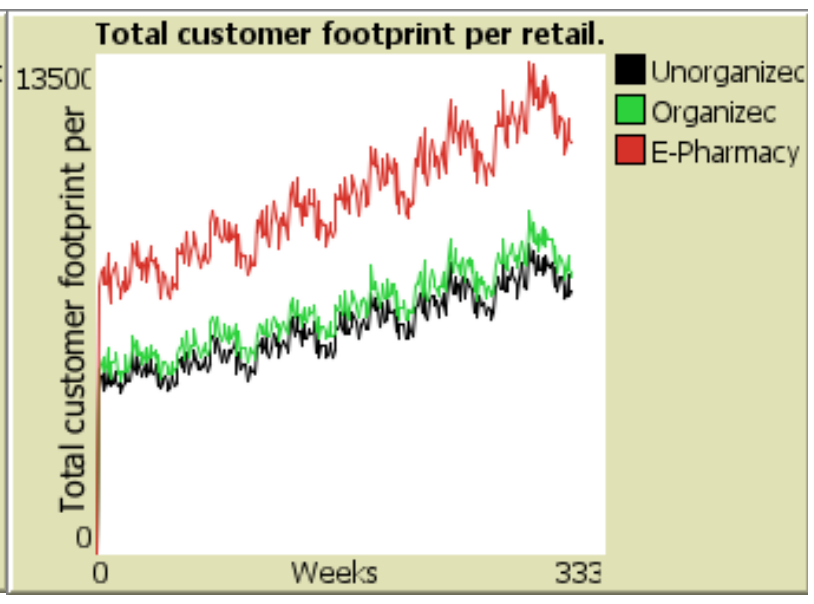

Experiment 15

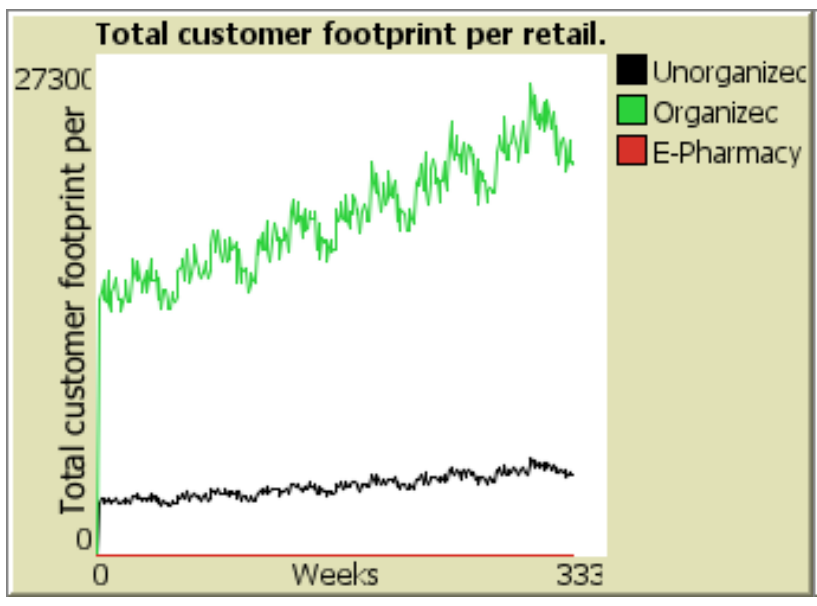

Experiment 16

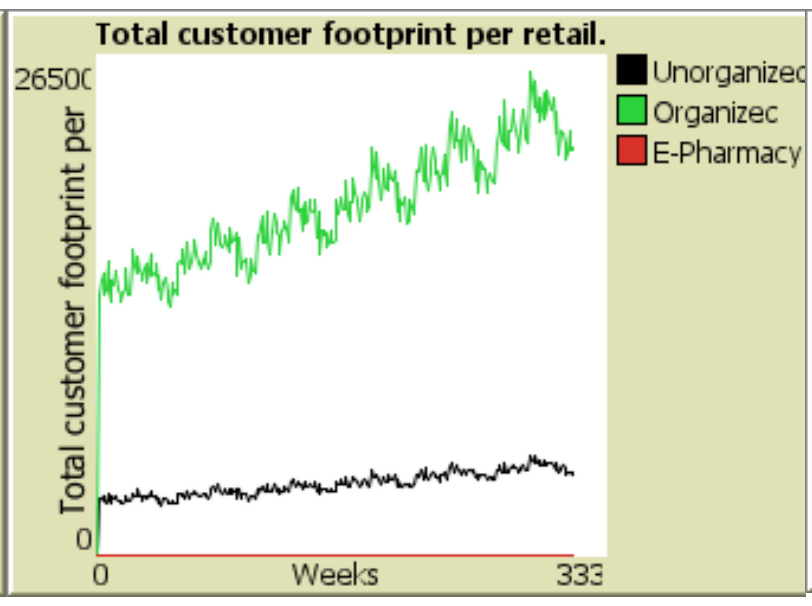

Experiment 17

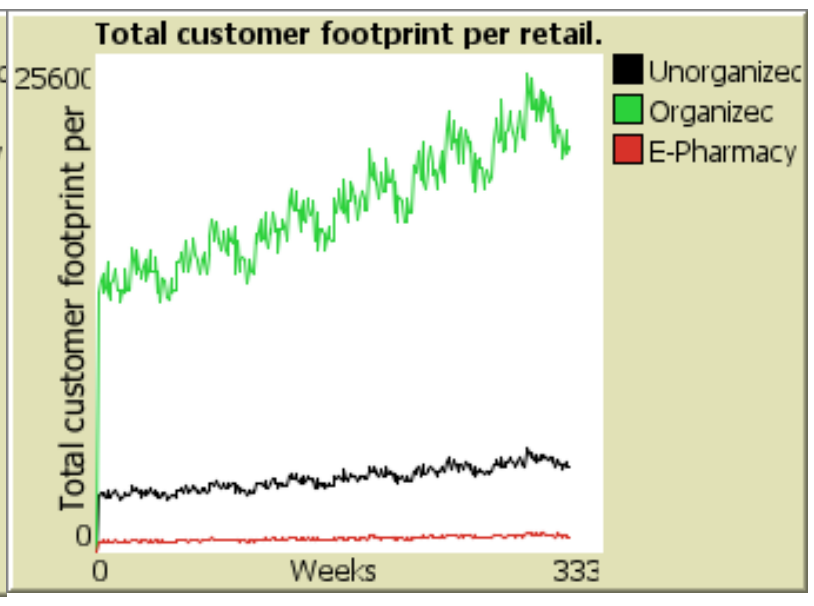

Experiment 18

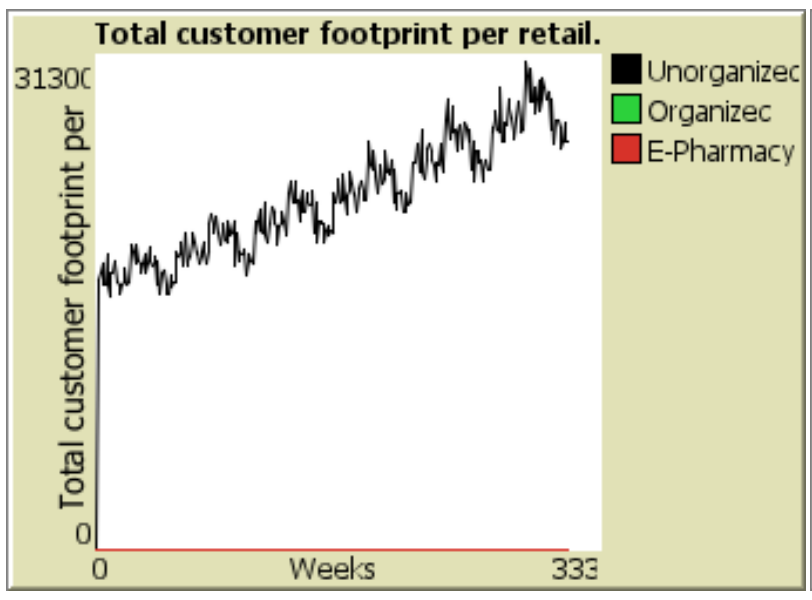

Experiment 19

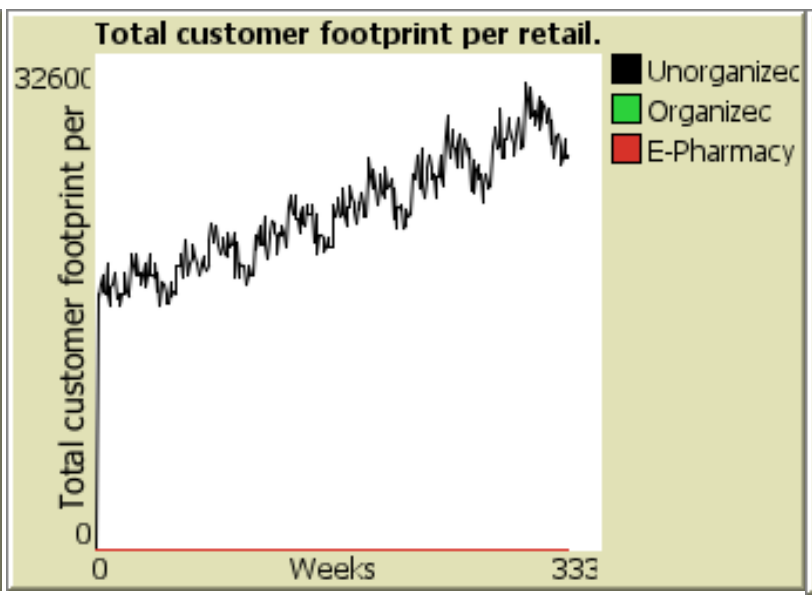

Experiment 20

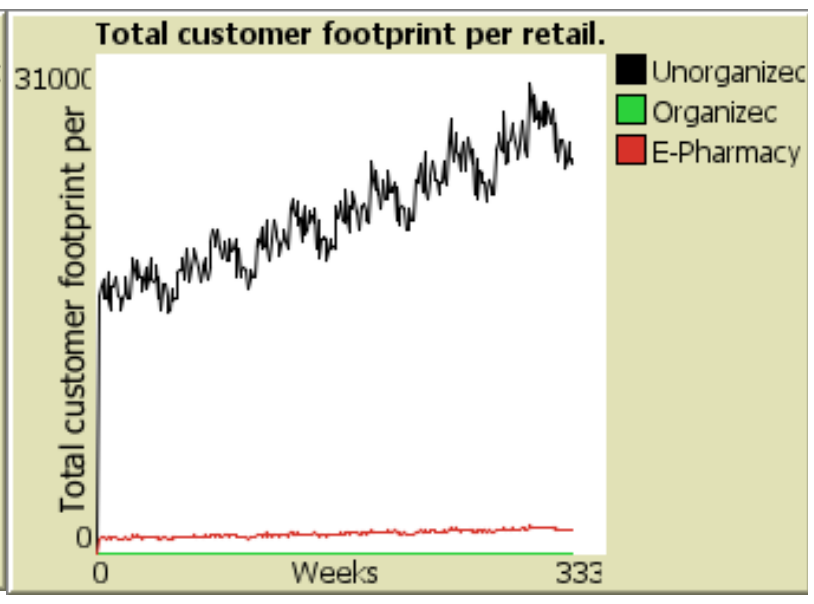

Experiment 21

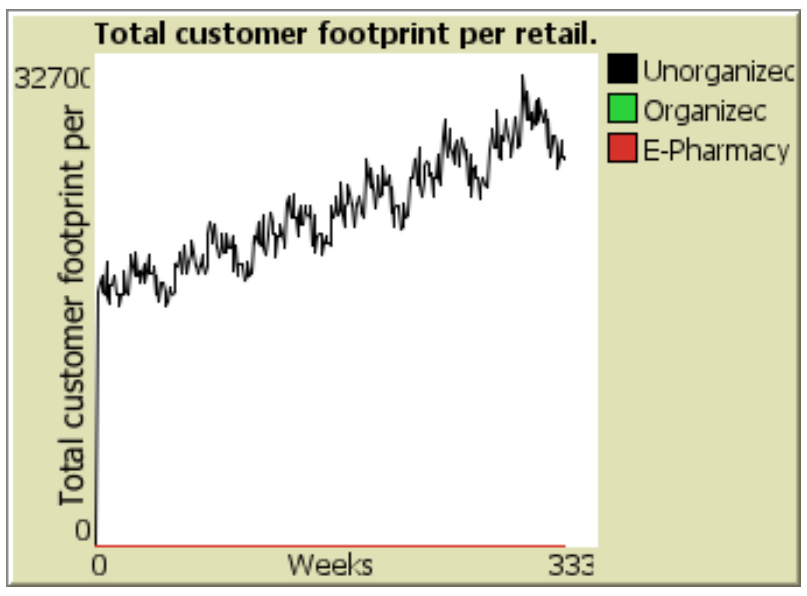

Experiment 22

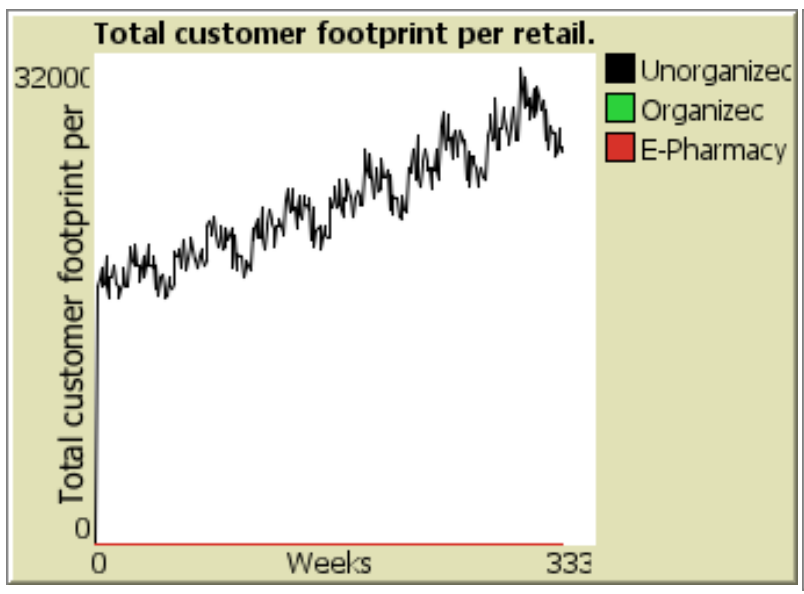

Experiment 23

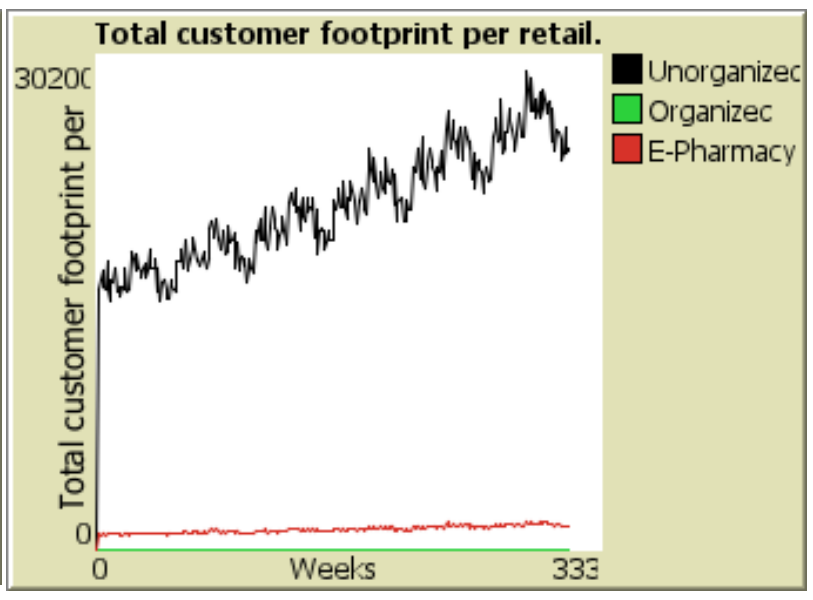

Experiment 24

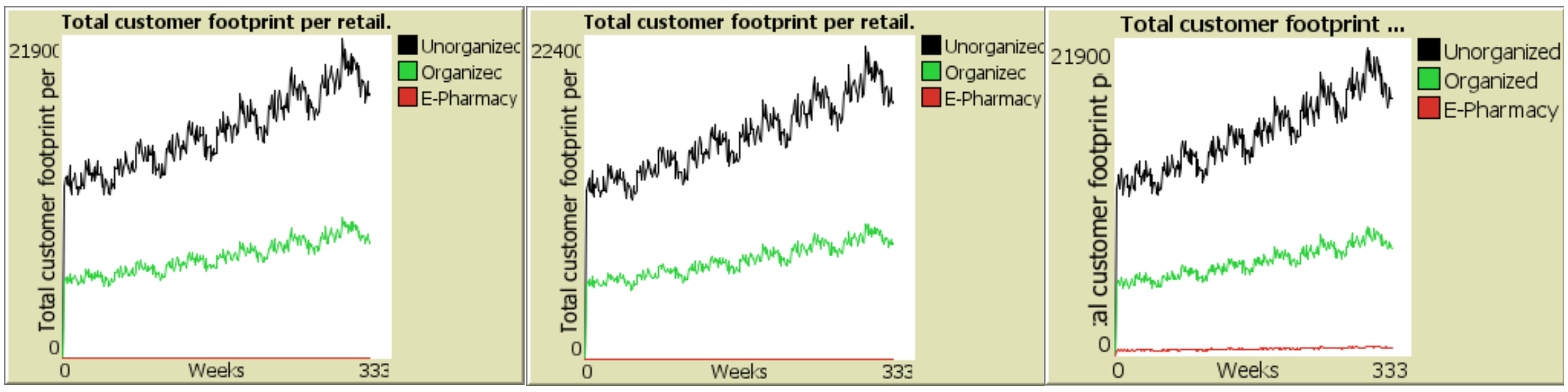

Experiment 25 Experiment 26 Experiment 27

FIGURE E7 Simulation results of weekly customer footprint as per experimental design for product quality.

**E2.3. Simulation Results of Market Share as per 27 Experimental Runs Utilising ANOVA for Quality of Product**

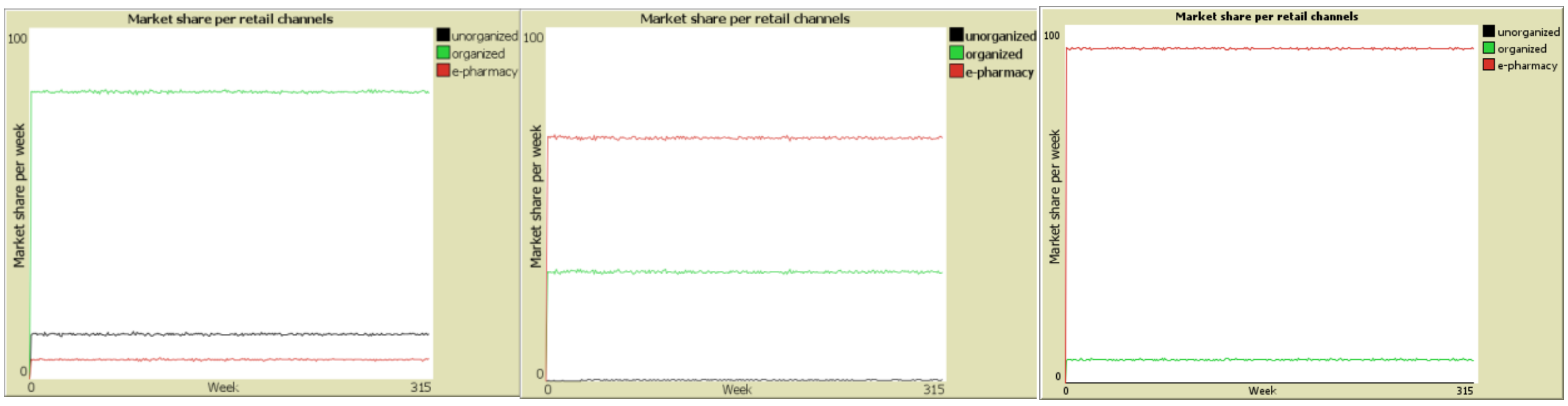

Experiment 1 Experiment 2 Experiment 3

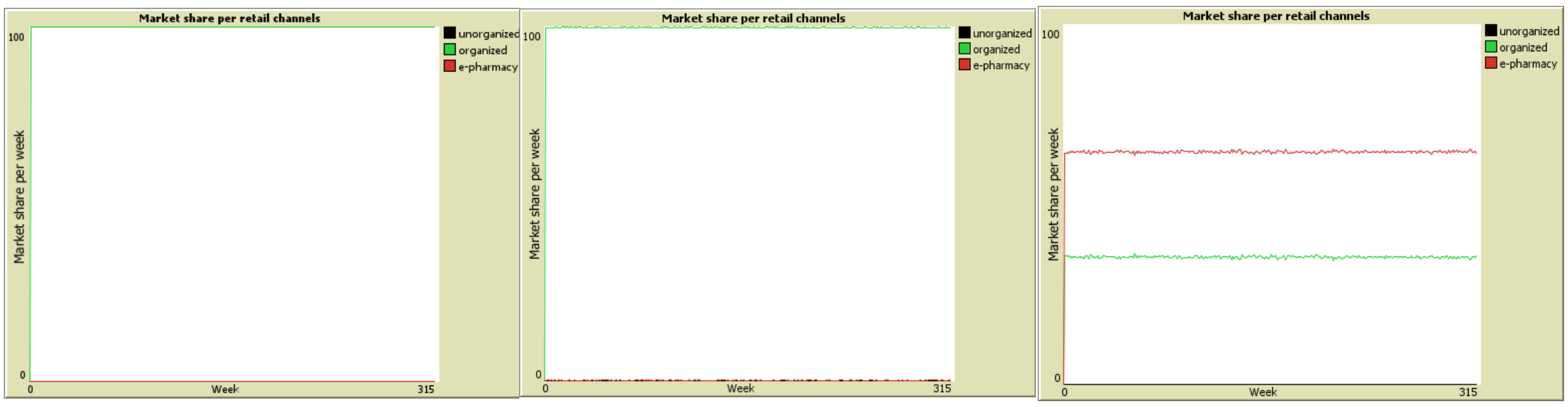

Experiment 4 Experiment 5 Experiment 6

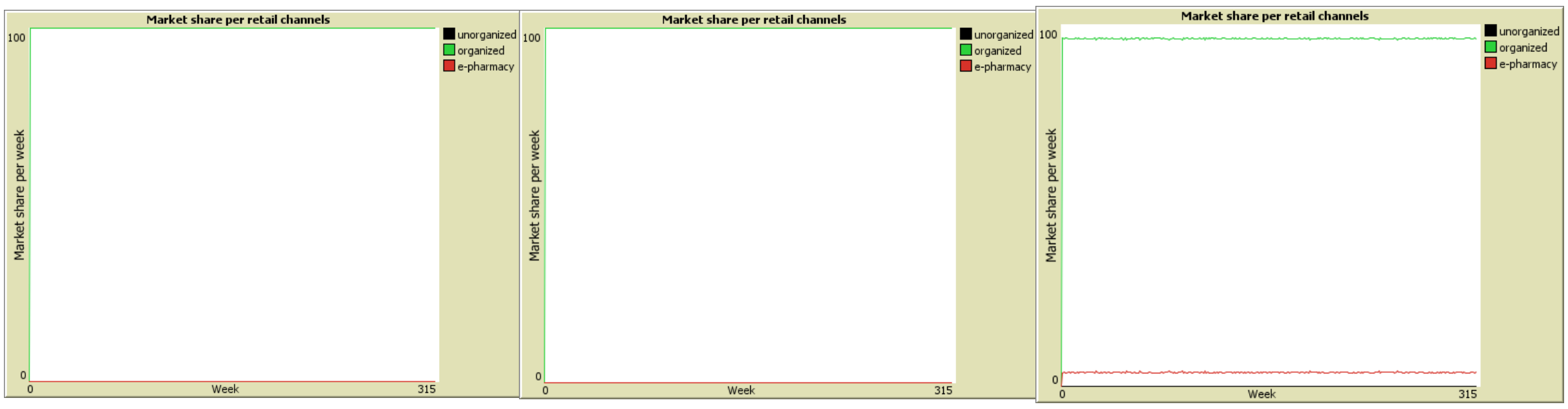

Experiment 7 Experiment 8 Experiment 9

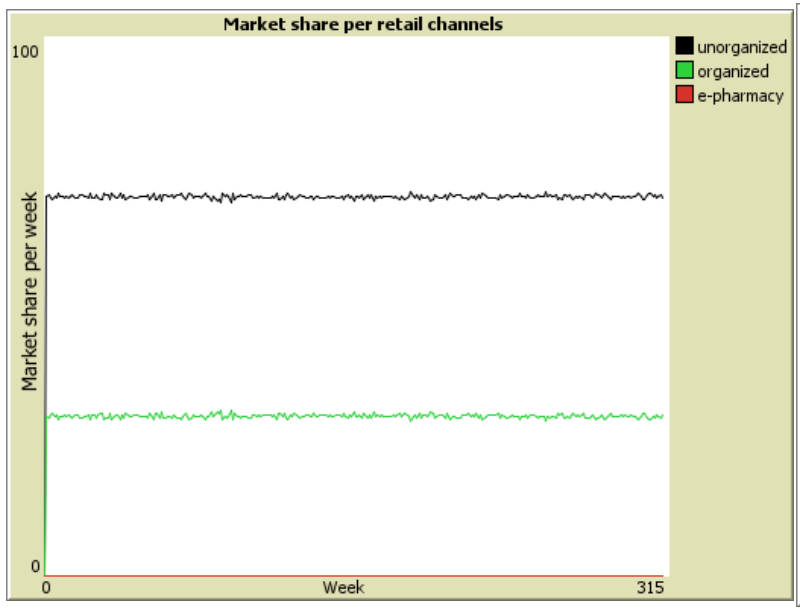

Experiment 10

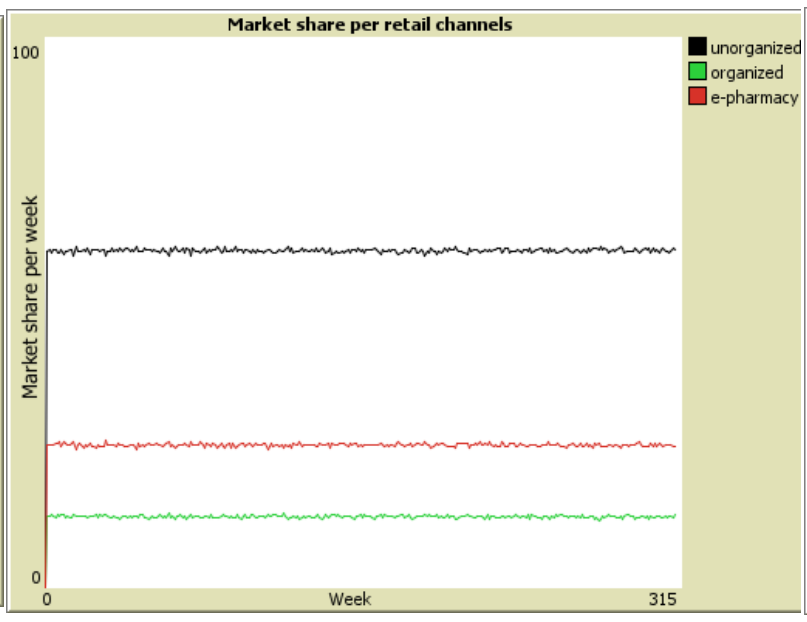

Experiment 11

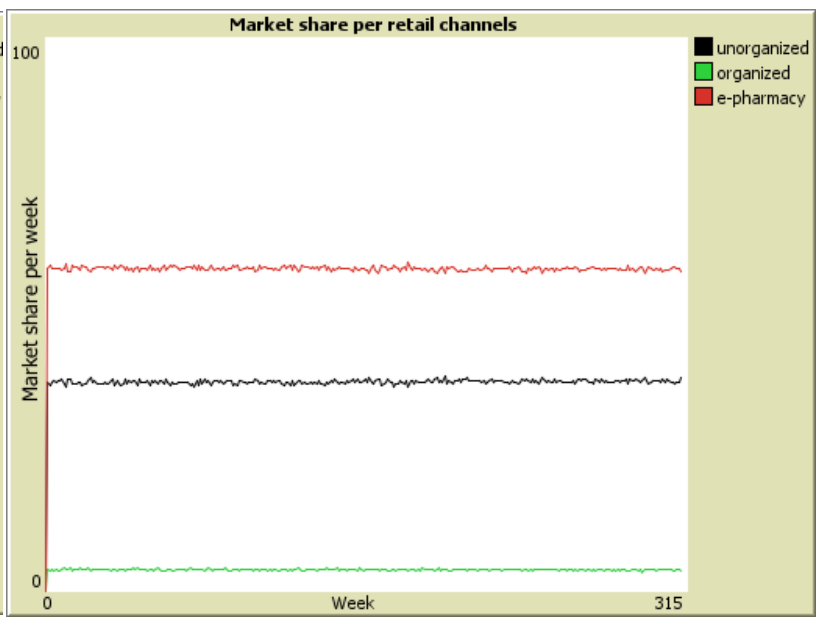

Experiment 12

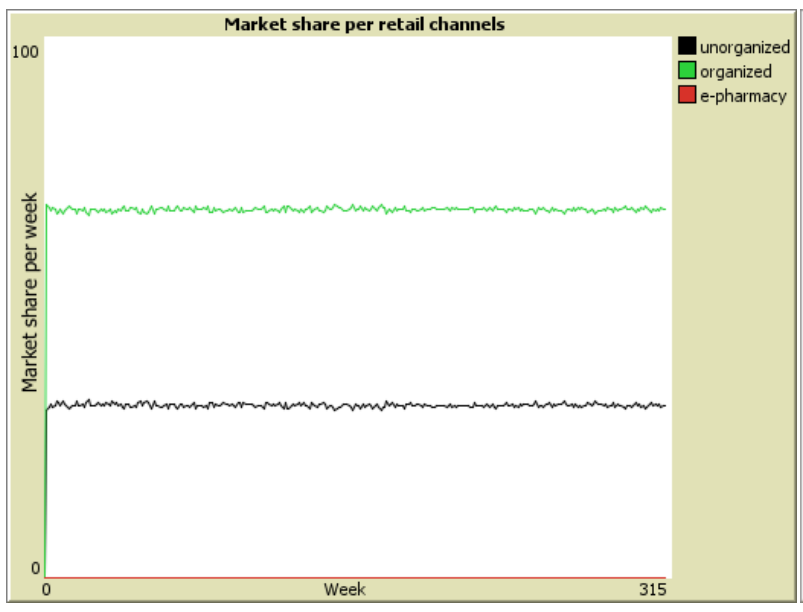

Experiment 13

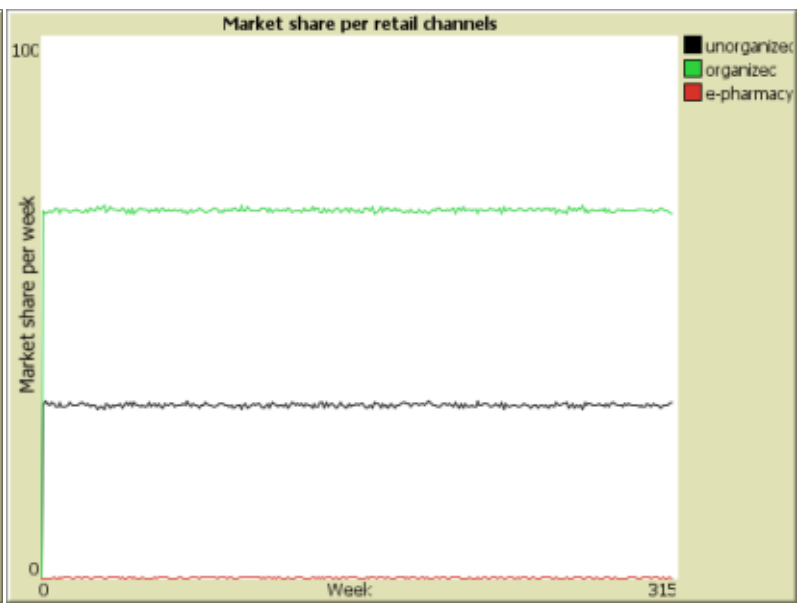

Experiment 14

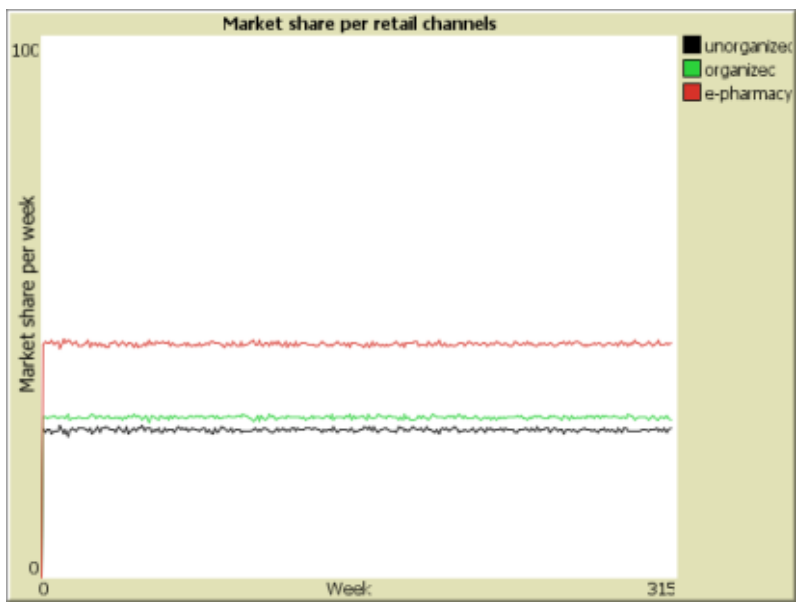

Experiment 15

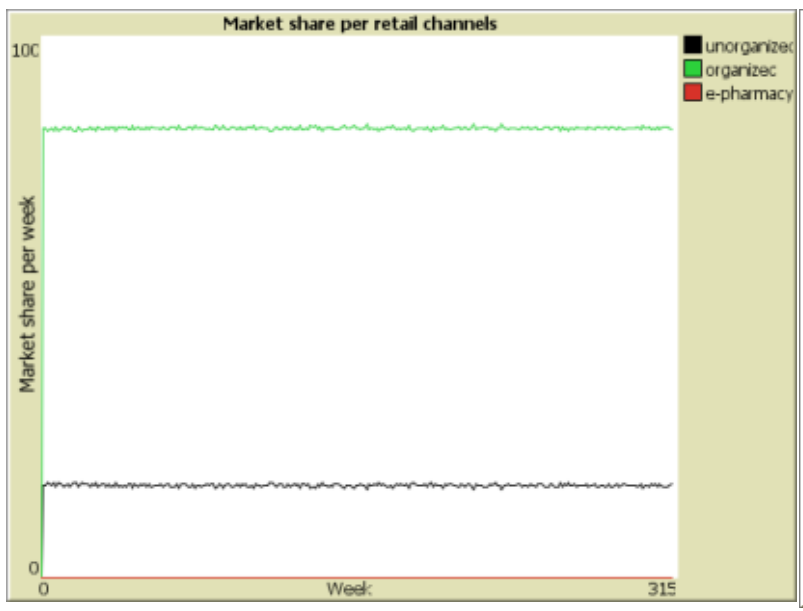

Experiment 16

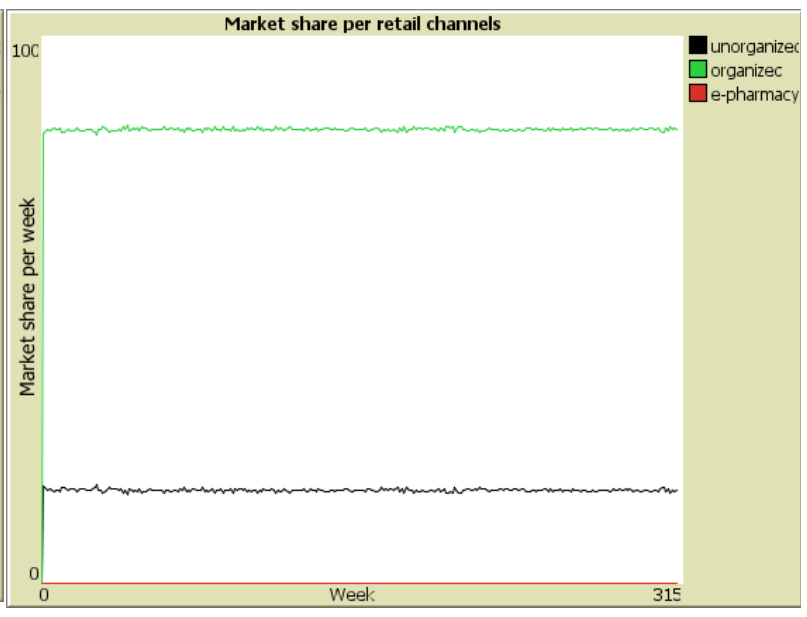

Experiment 17

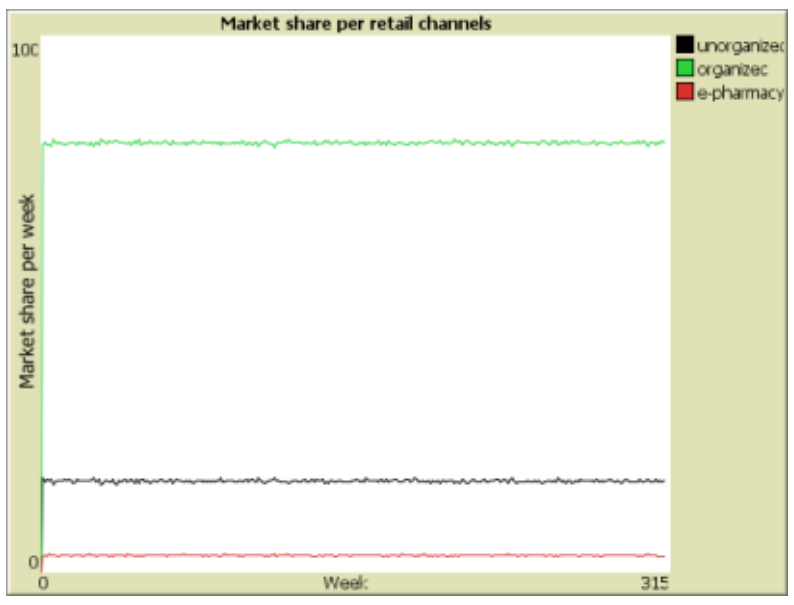

Experiment 18

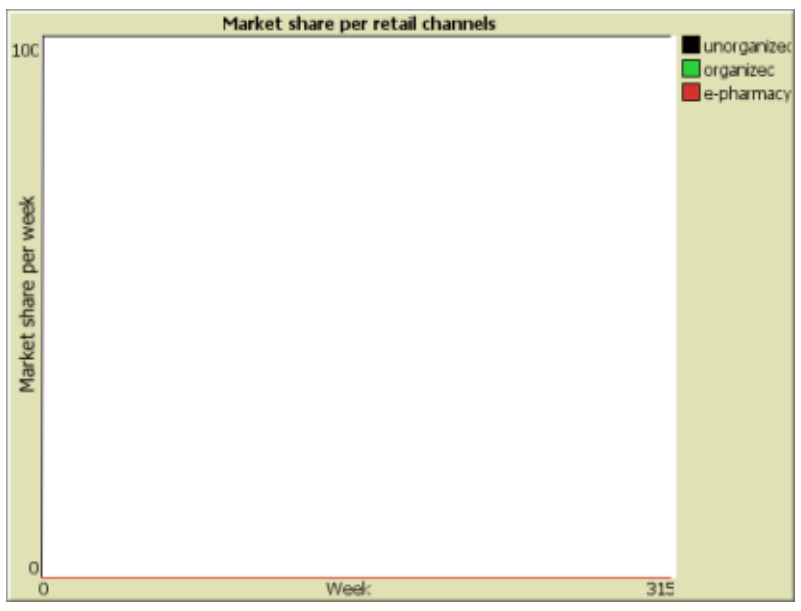

Experiment 19

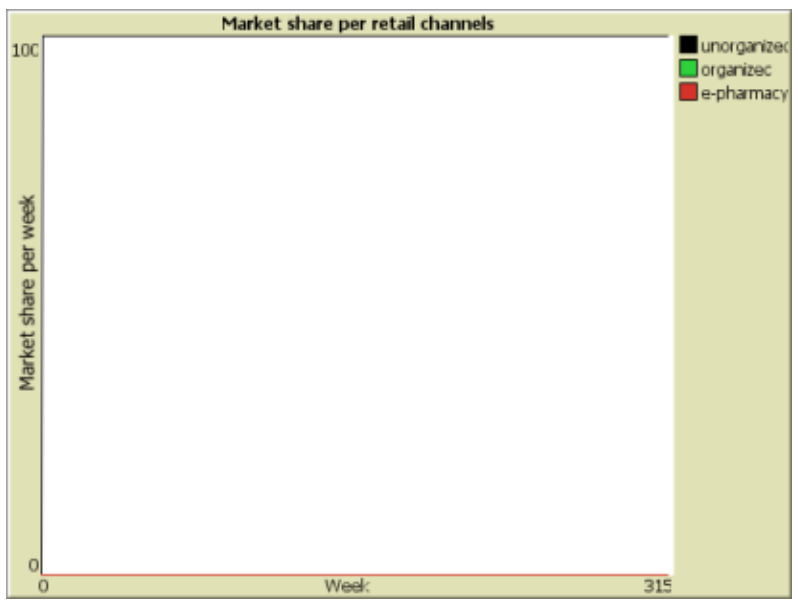

Experiment 20

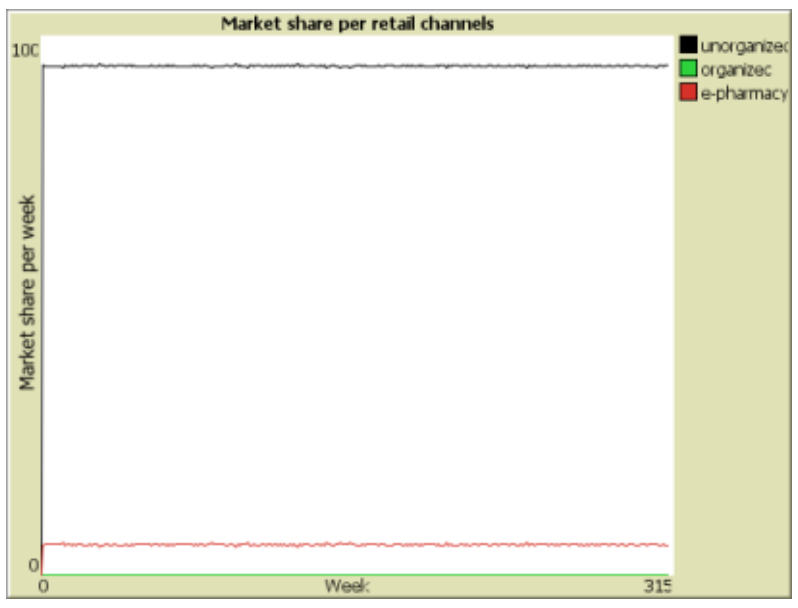

Experiment 21

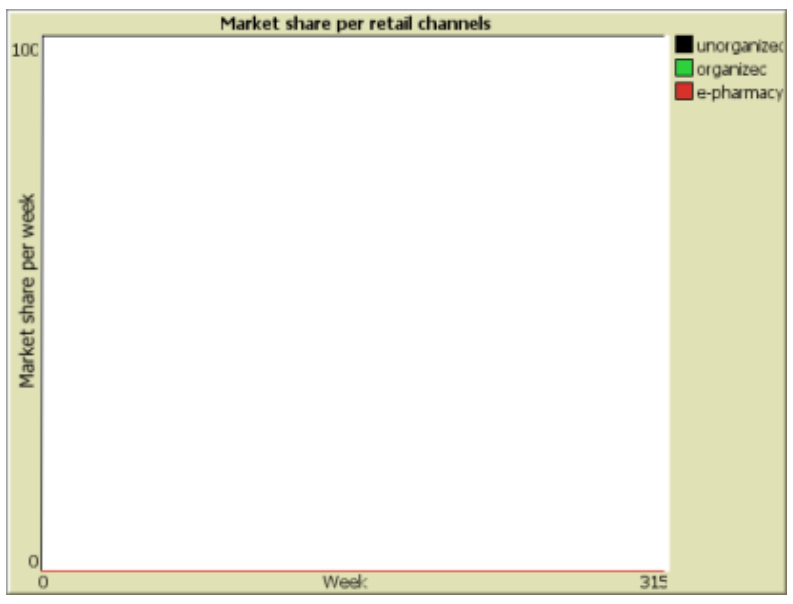

Experiment 22

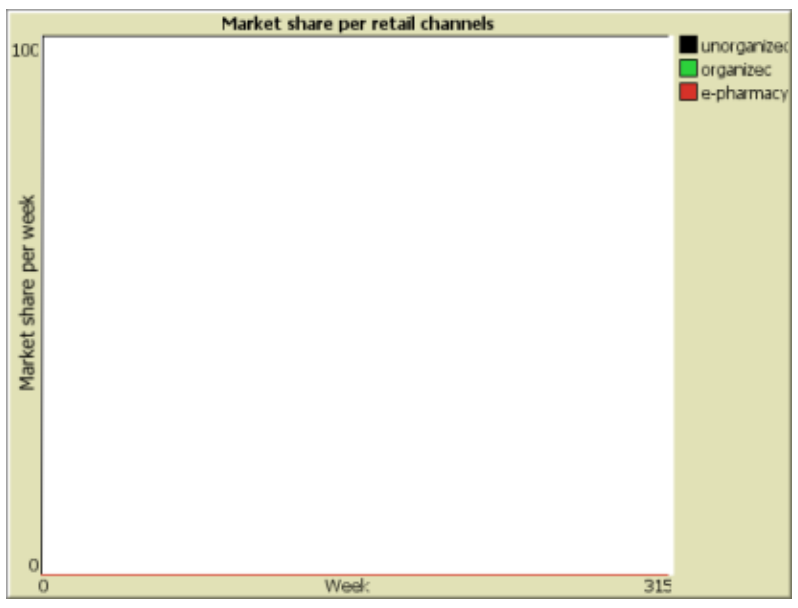

Experiment 23

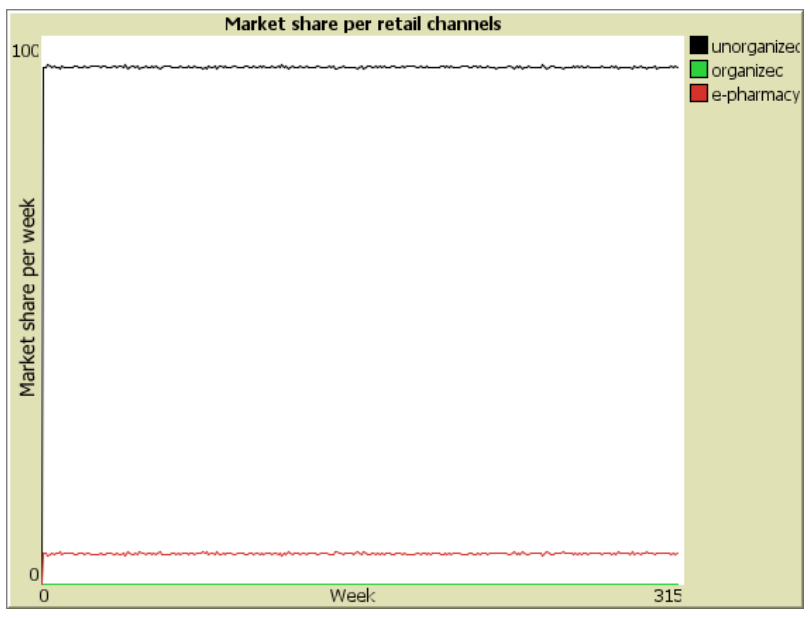

Experiment 24

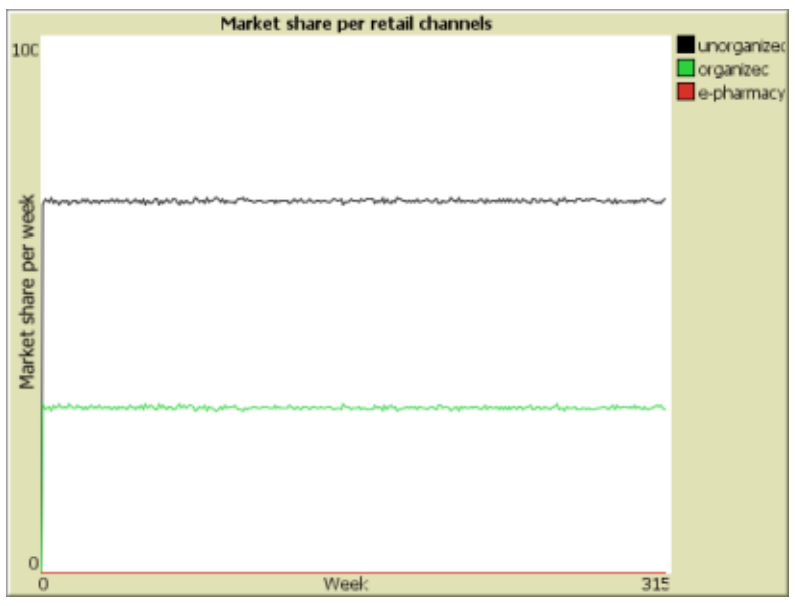

Experiment 25

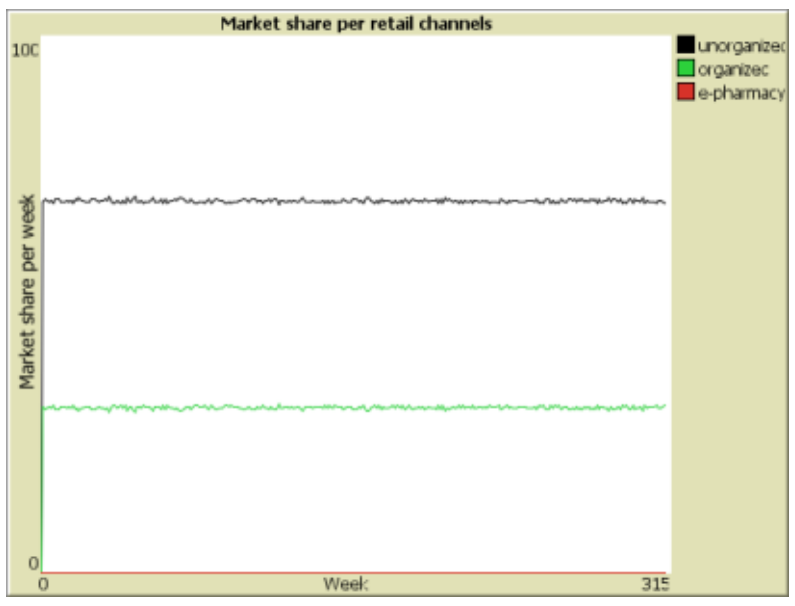

Experiment 26

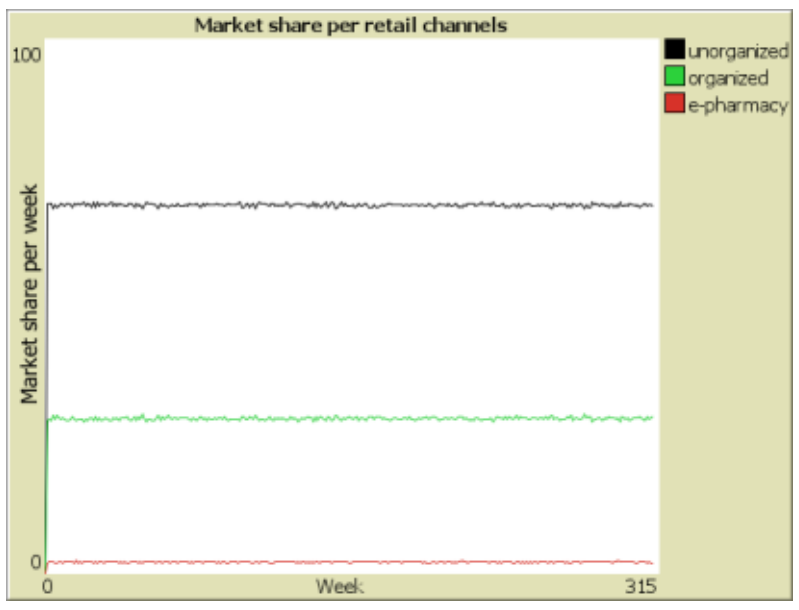

Experiment 27

FIGURE E8 Simulation results of market share as per experimental design for product quality.

### E2.4. Simulation Results of Unorganized Retails Present in the Market (No. of Stores Surviving After Shut Down) as per 27 Experimental Runs Utilising ANOVA for Quality of Product

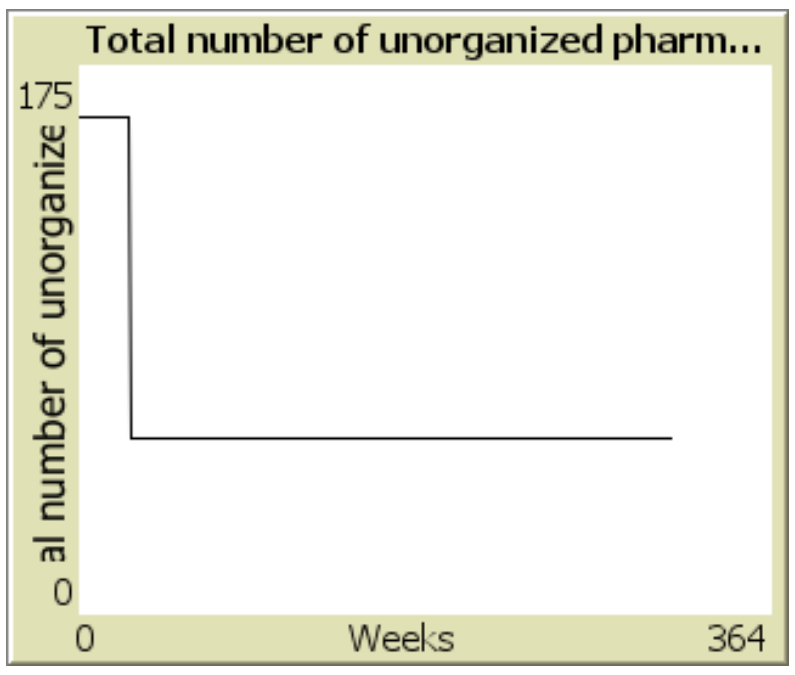

Experiment 1

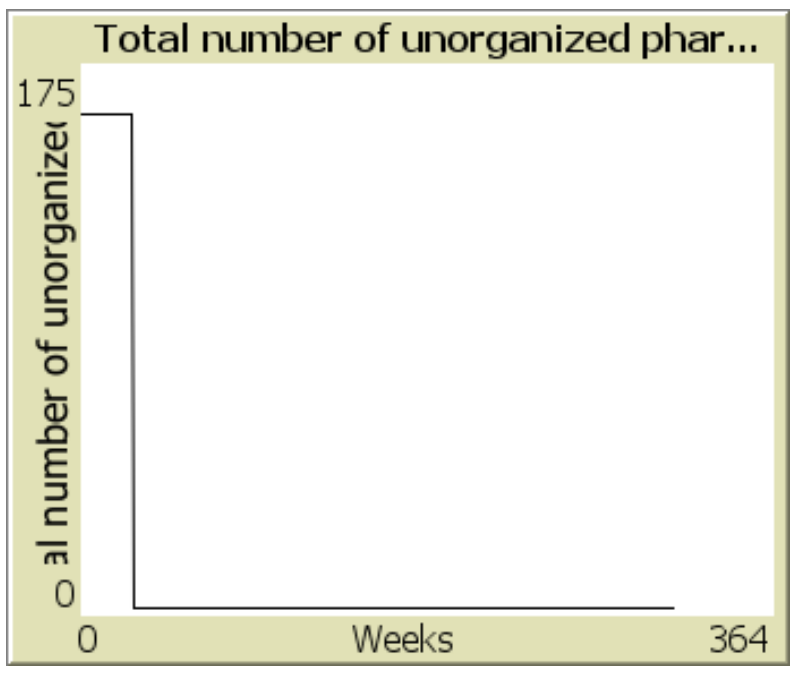

Experiment 2

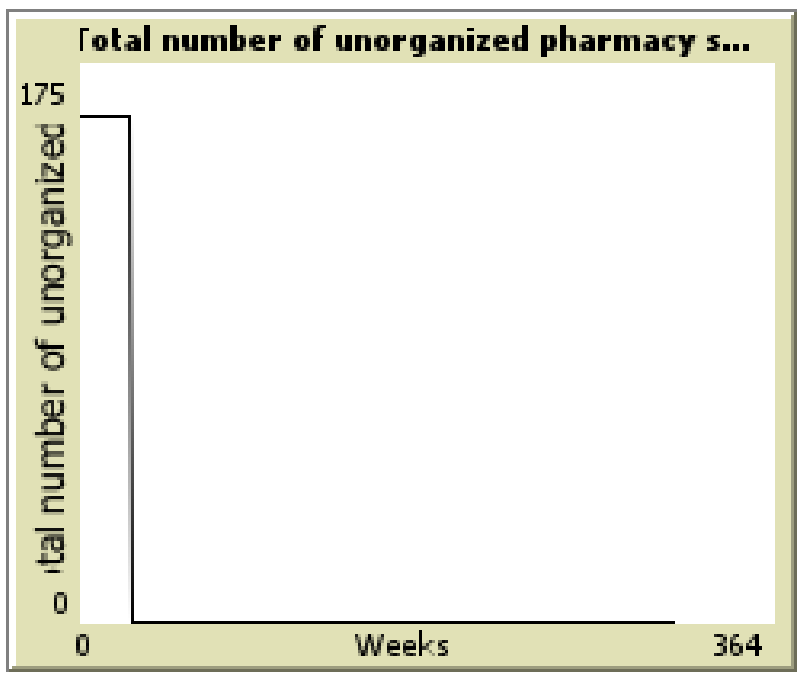

Experiment 3

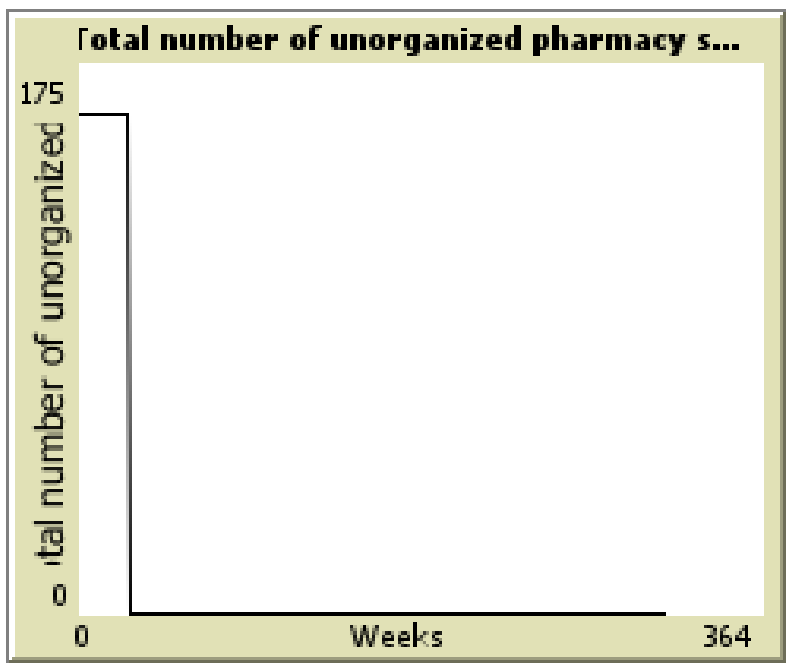

Experiment 4

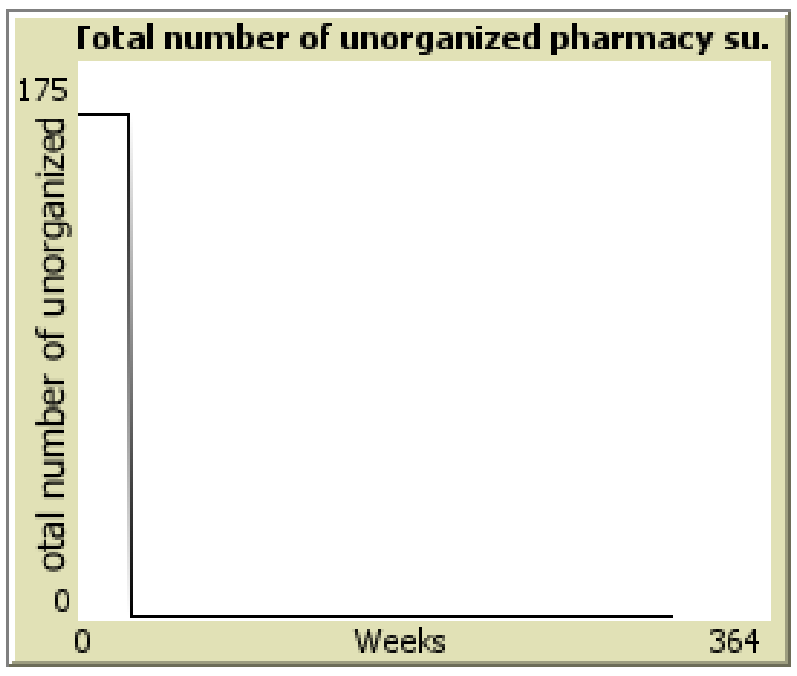

Experiment 5

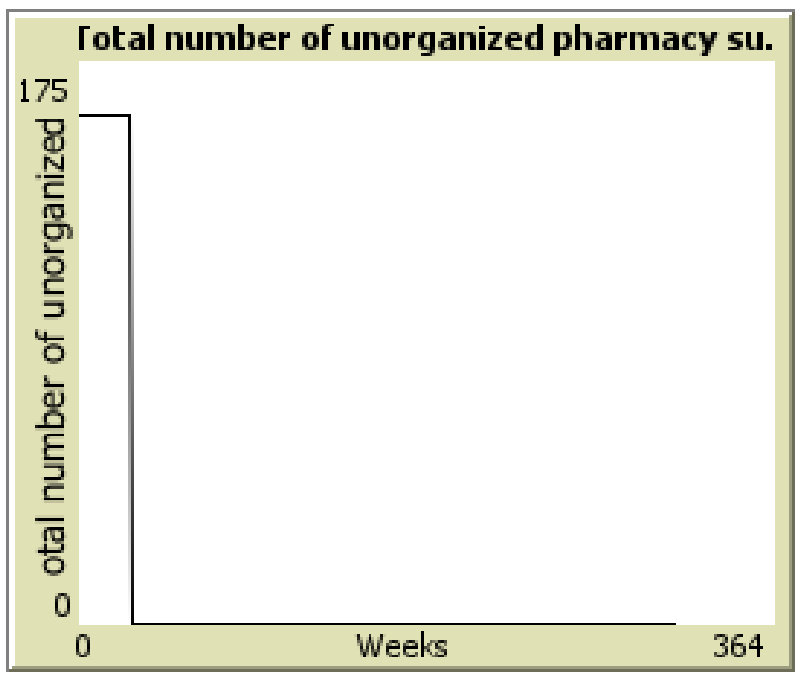

Experiment 6

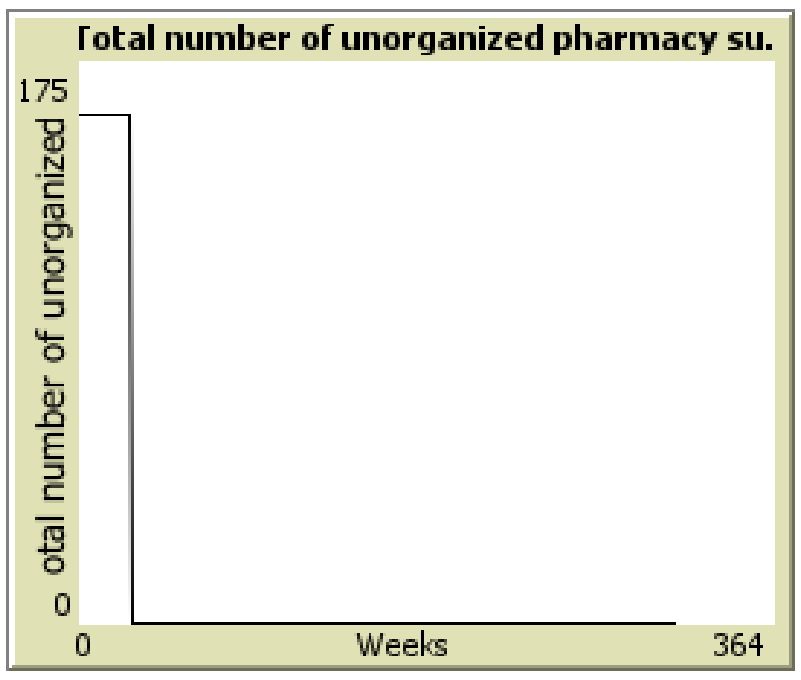

Experiment 7

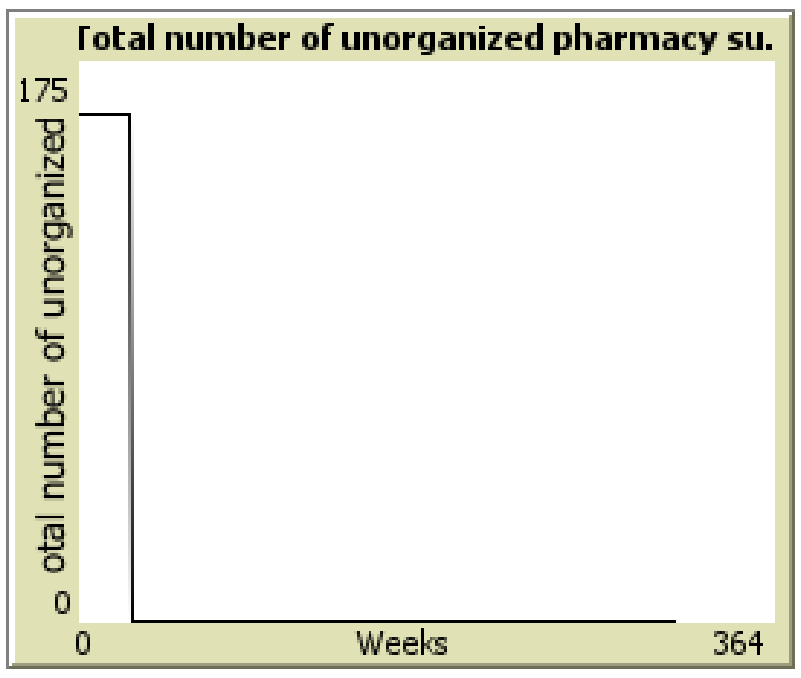

Experiment 8

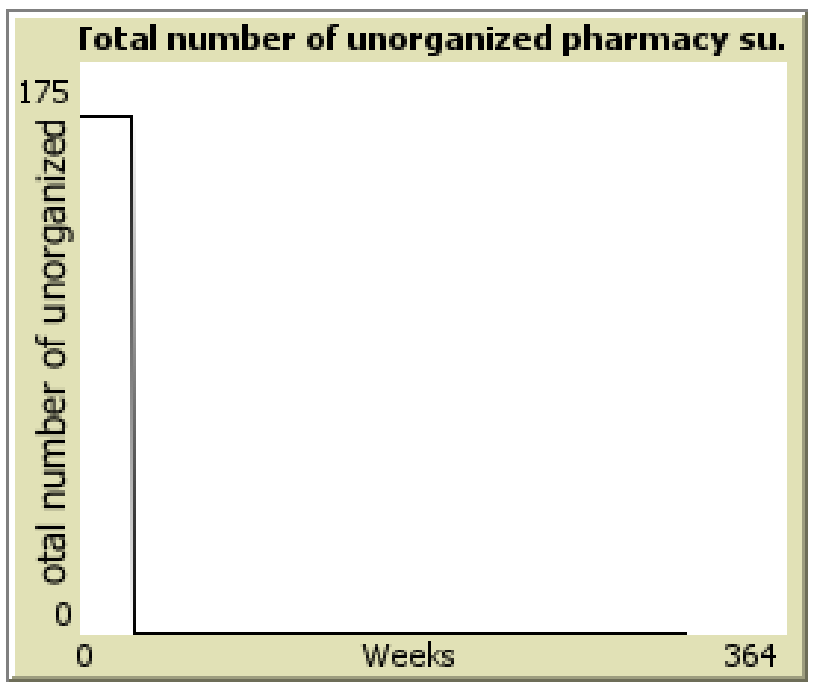

Experiment 9

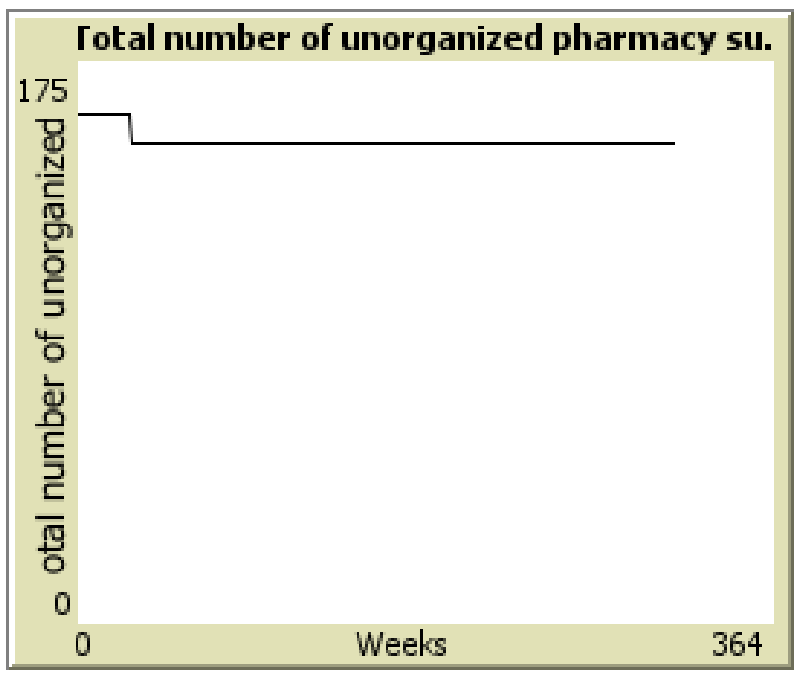

Experiment 10

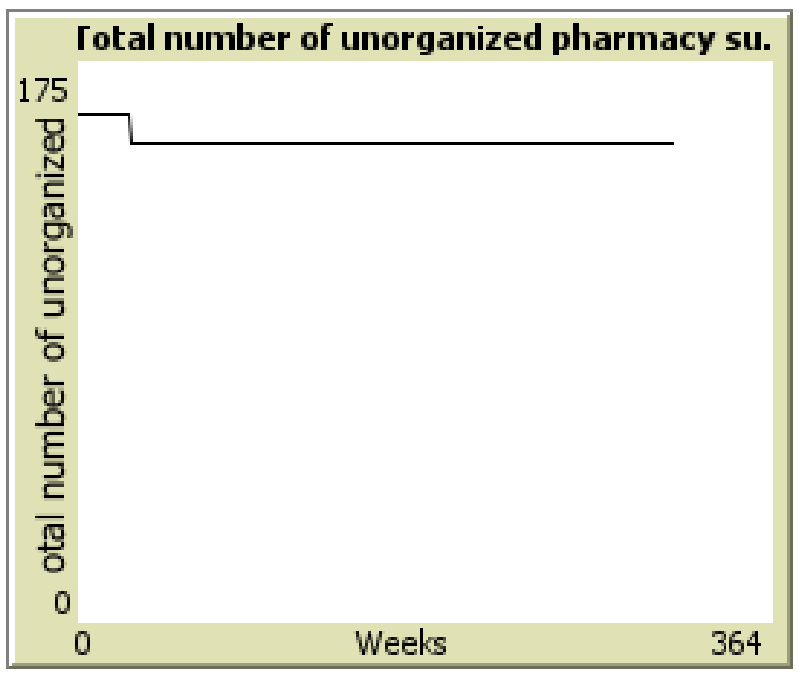

Experiment 11

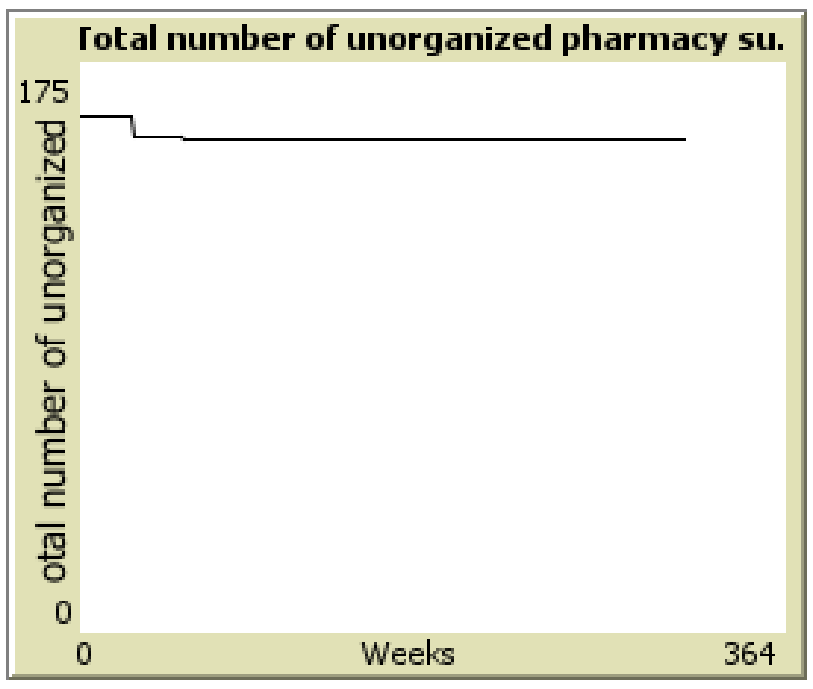

Experiment 12

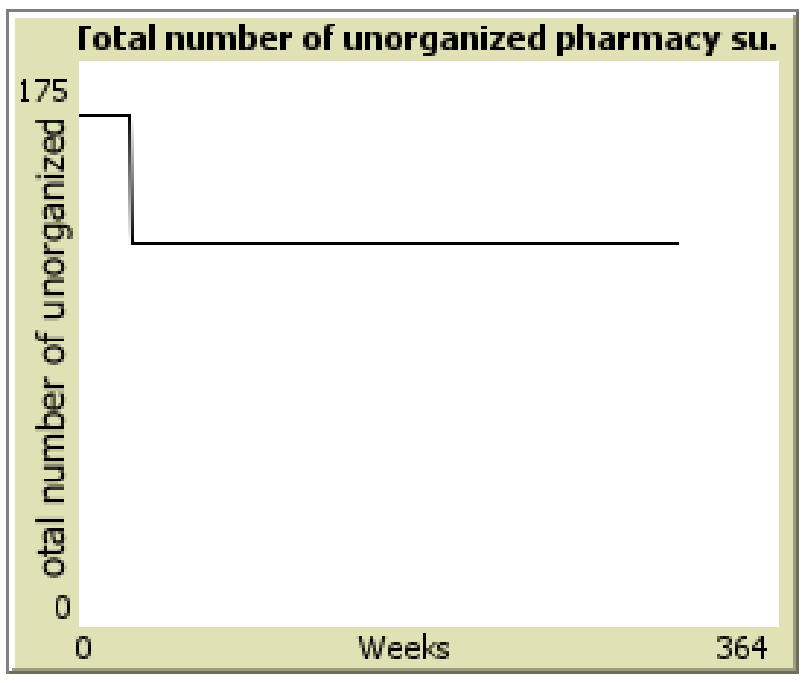

Experiment 13

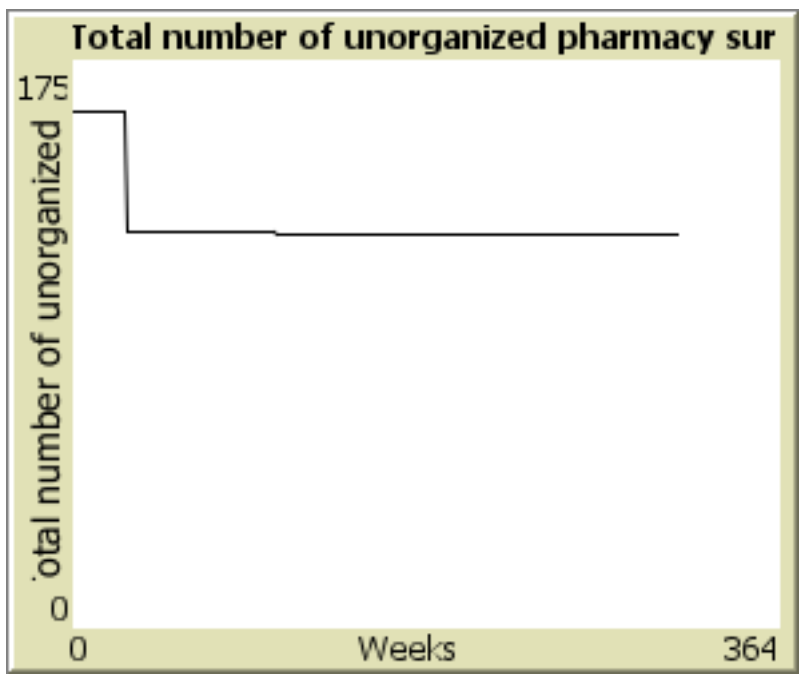

Experiment 14

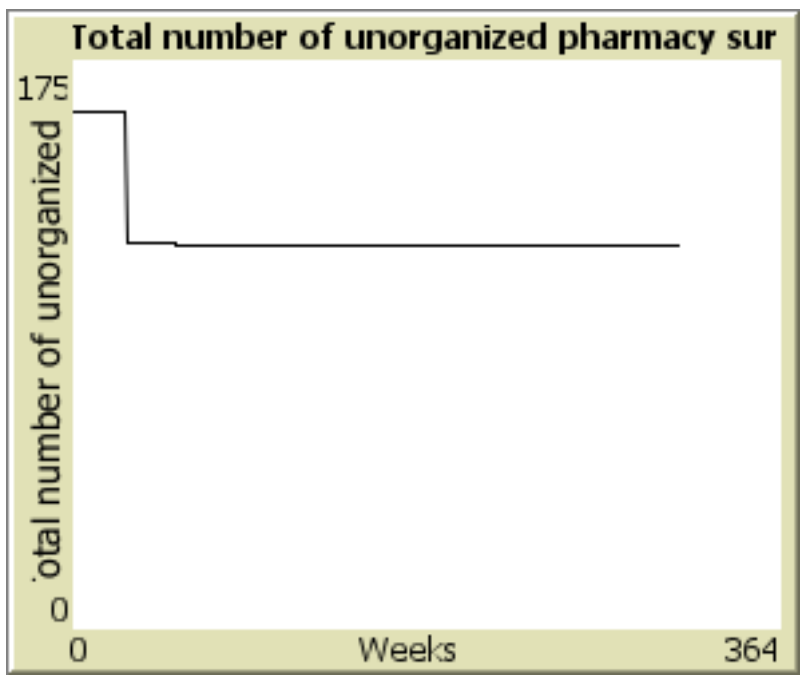

Experiment 15

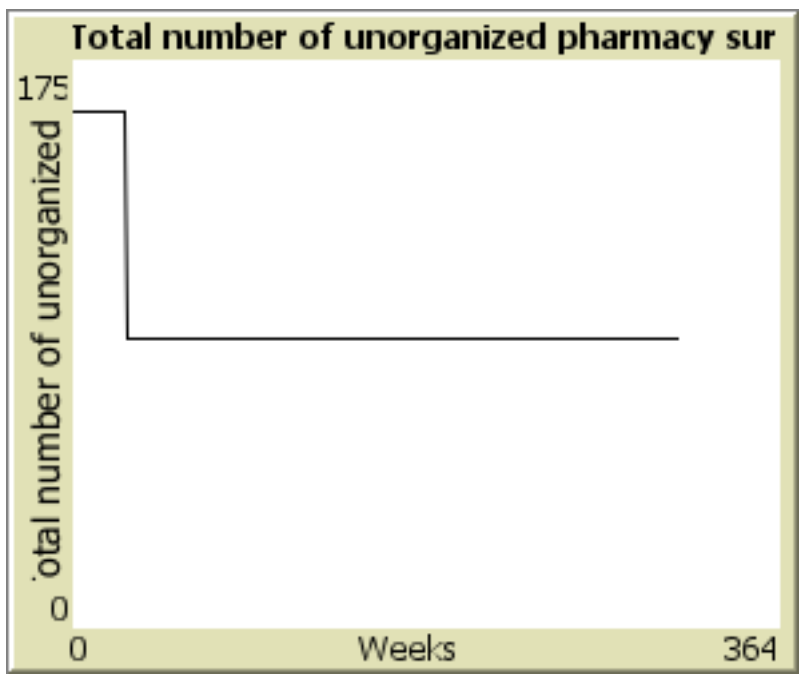

Experiment 16

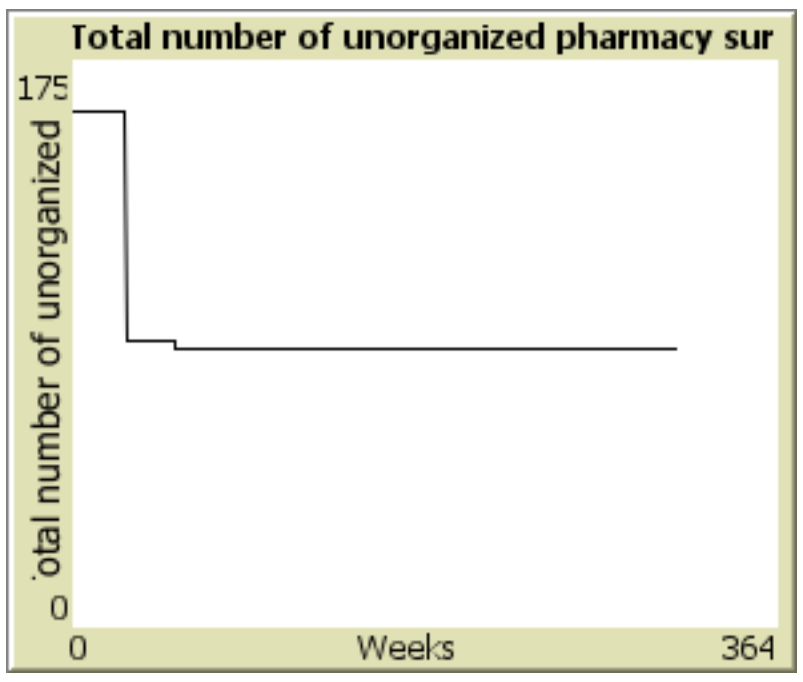

Experiment 17

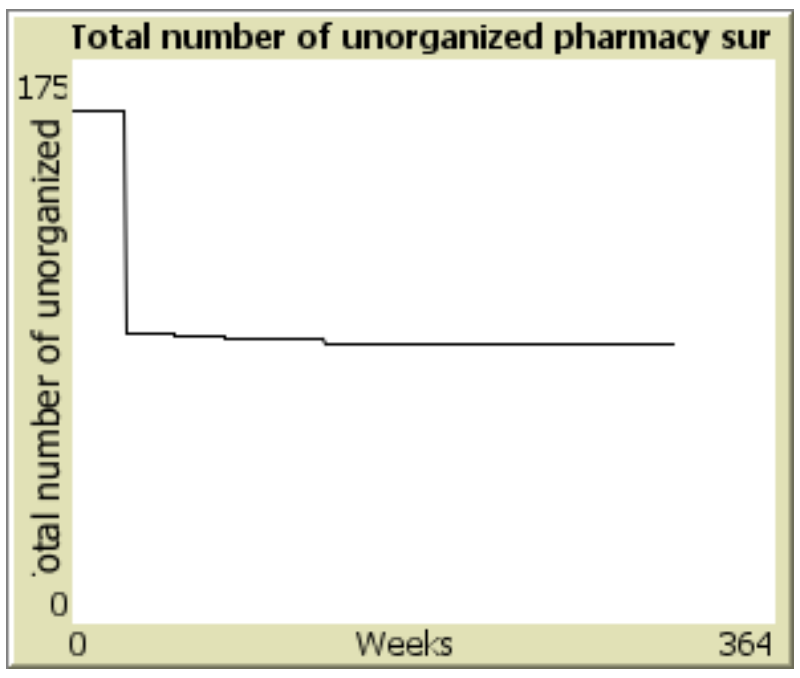

Experiment 18

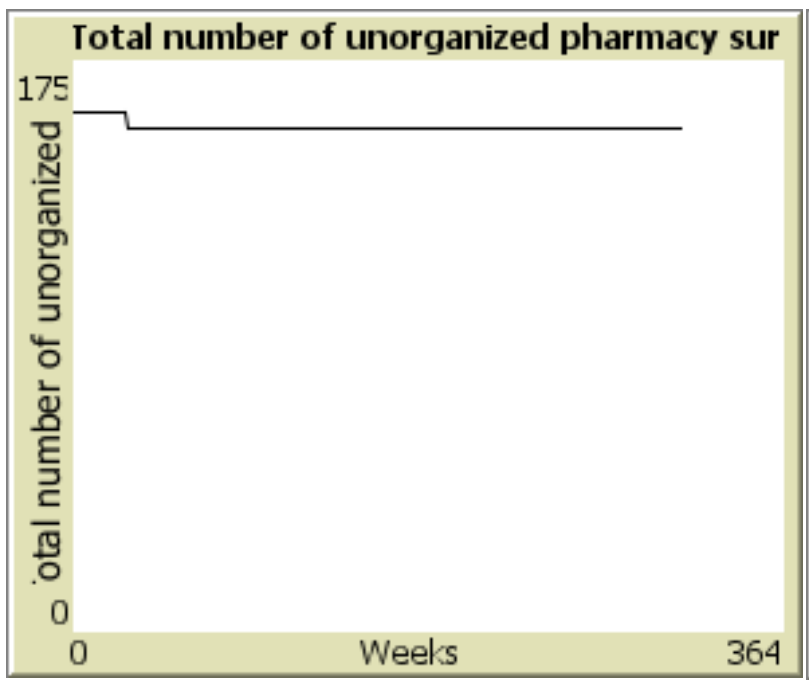

Experiment 19

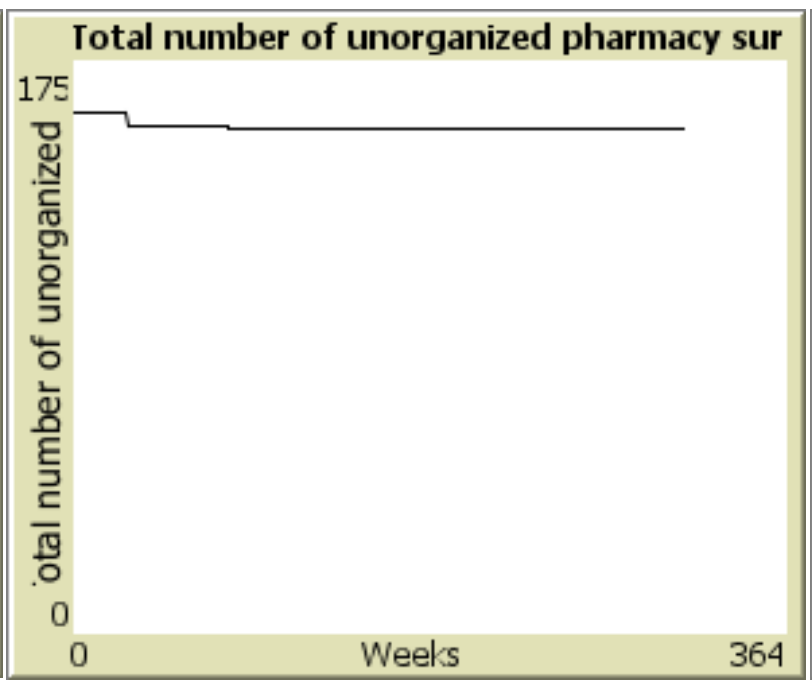

Experiment 20

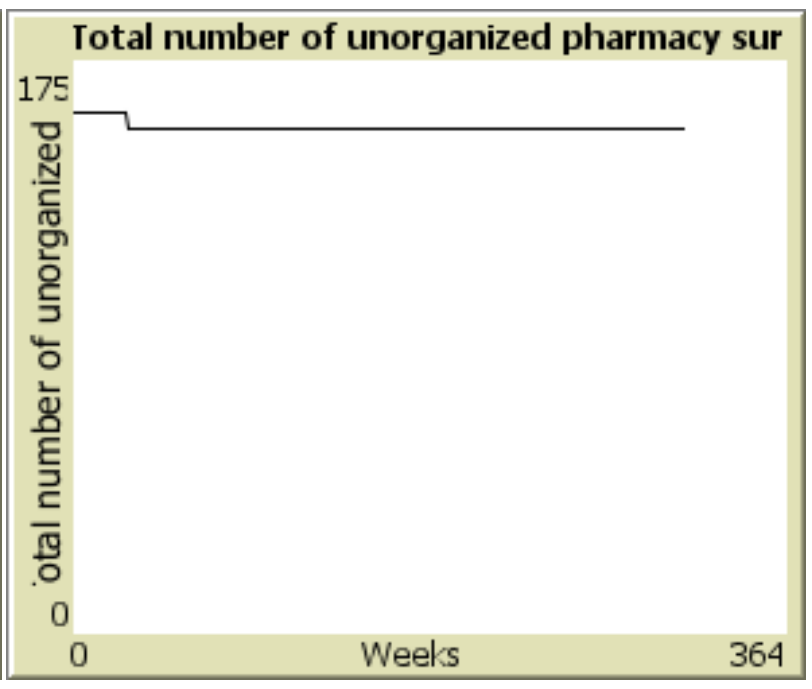

Experiment 21

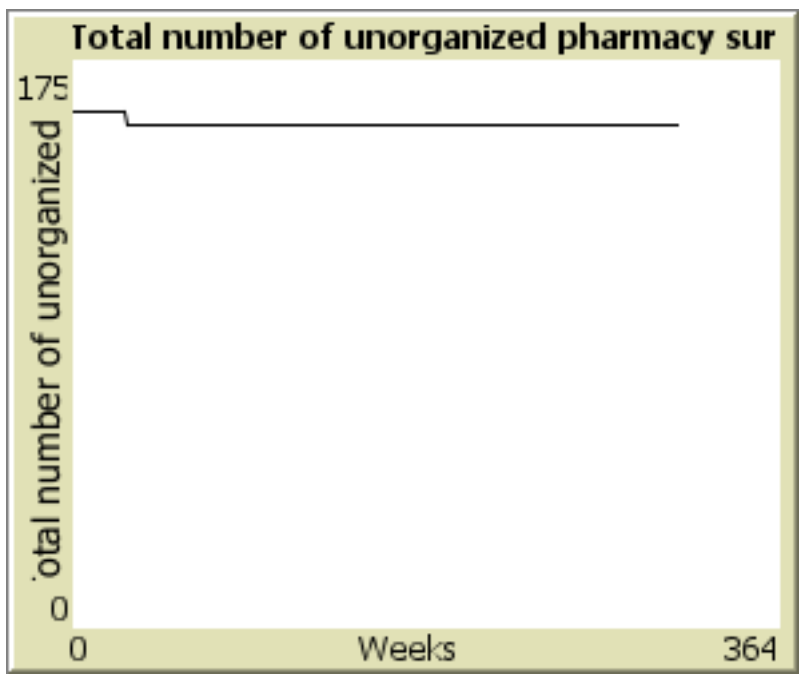

Experiment 22

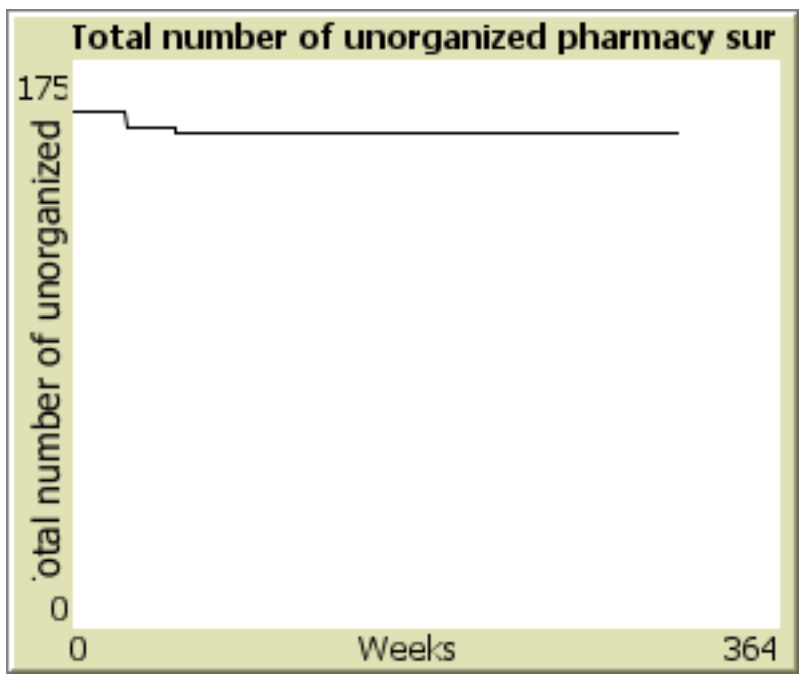

Experiment 23

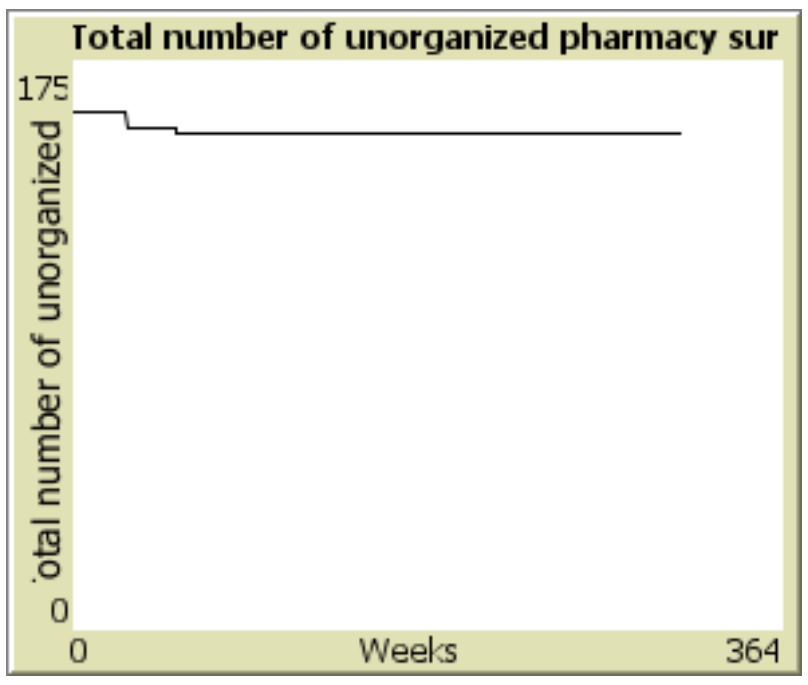

Experiment 24

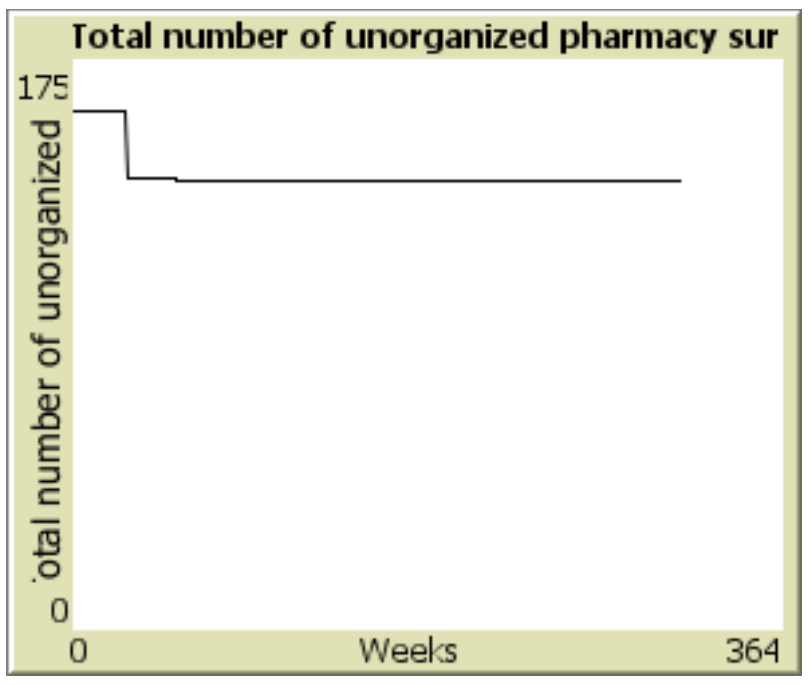

Experiment 25

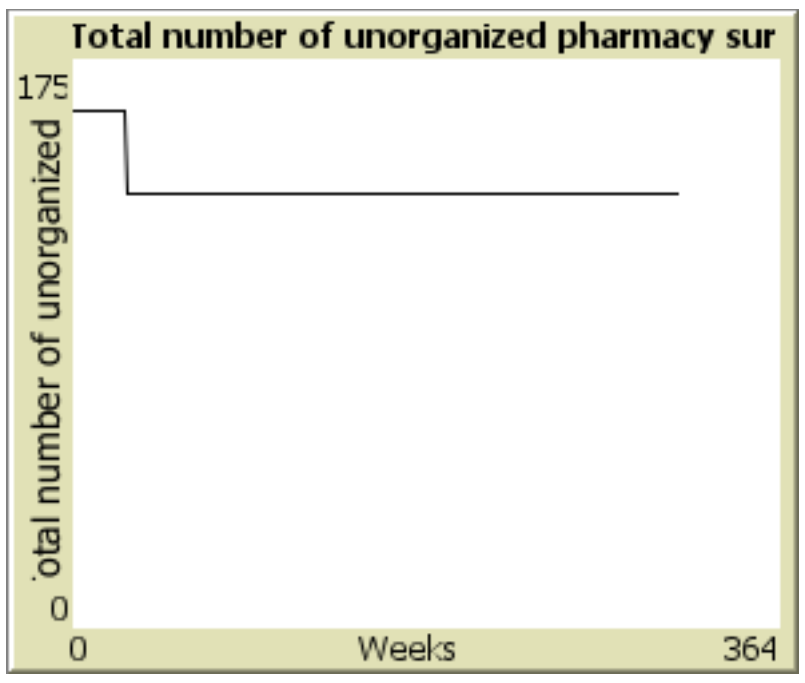

Experiment 26

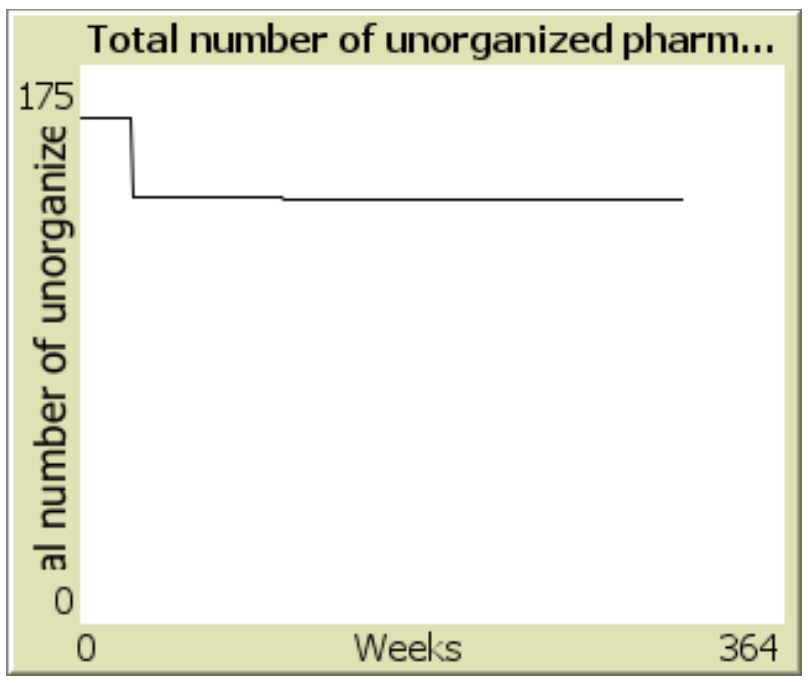

Experiment 27

FIGURE E9 Simulation results of unorganized retails present in the market (no. of stores) as per experimental design for product quality.

**E2.5. Simulation Results of Distance (km) a Customer Travelled as per 27 Experimental Runs Utilising ANOVA for Quality of Product**

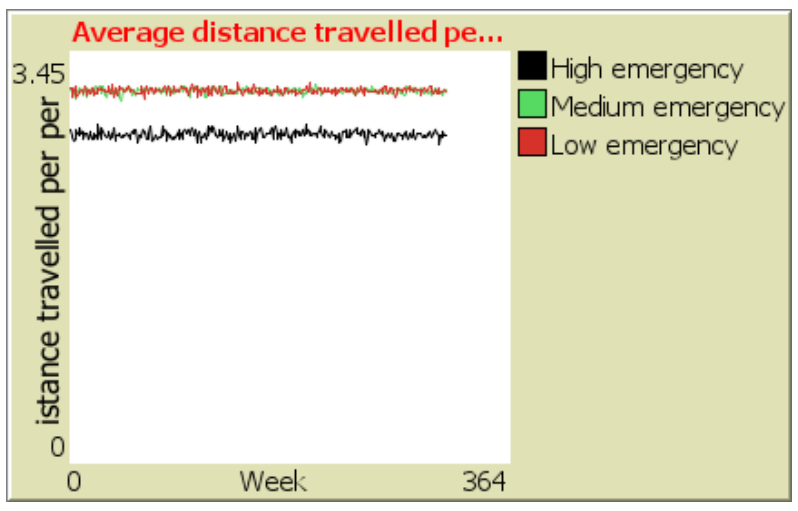

Experiment 1

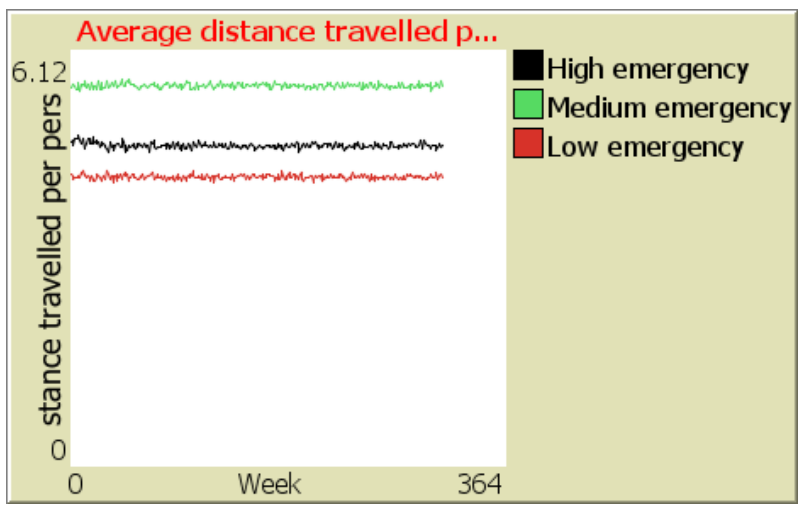

Experiment 2

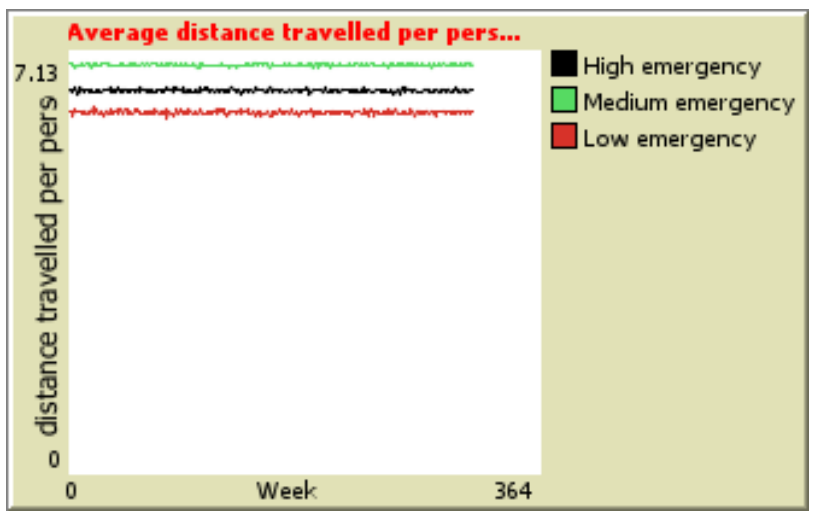

Experiment 3

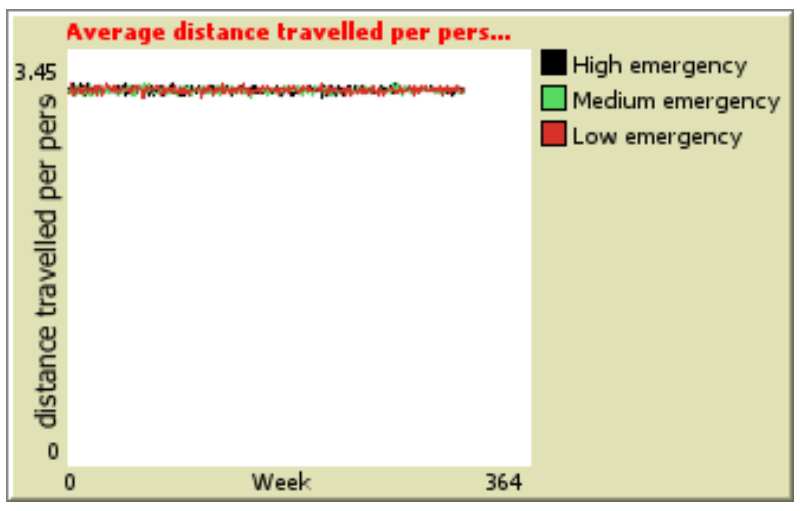

Experiment 4

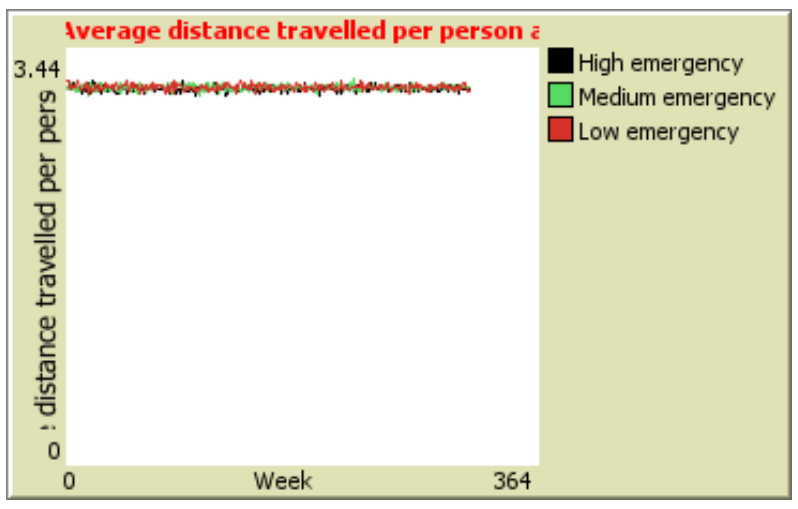

Experiment 5

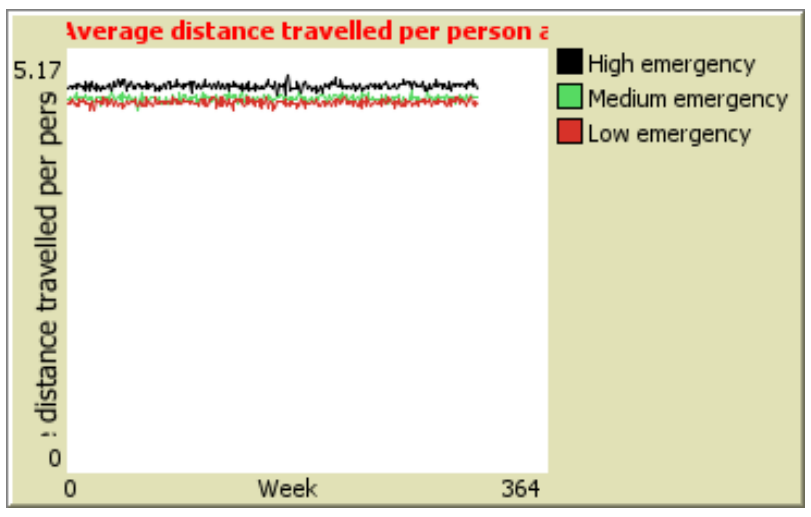

Experiment 6

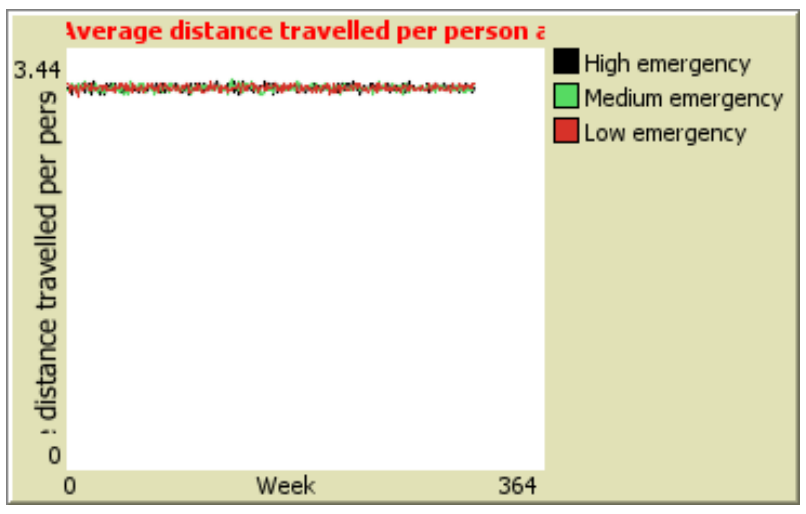

Experiment 7

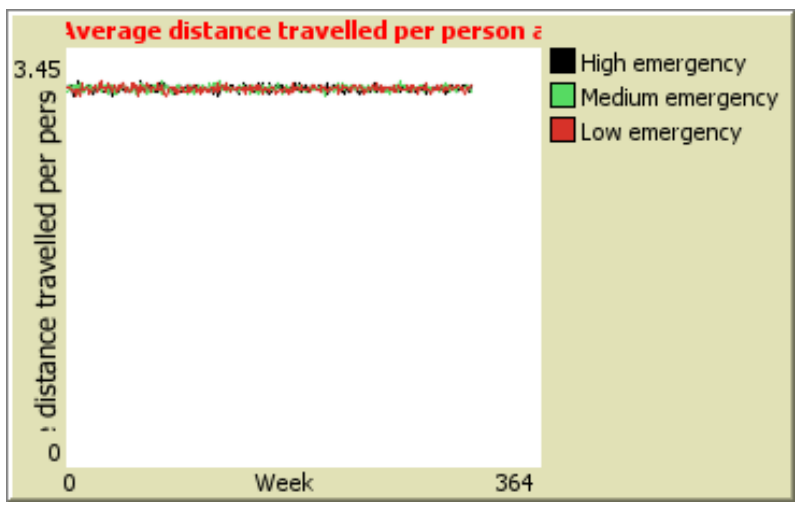

Experiment 8

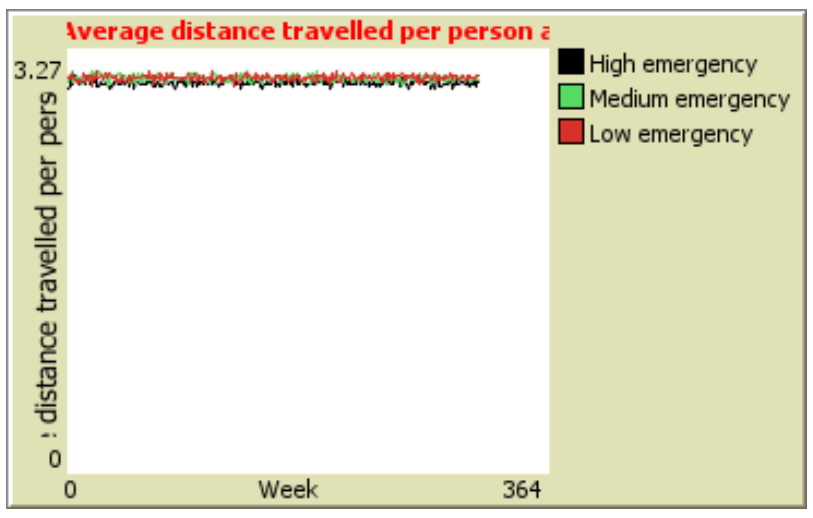

Experiment 9

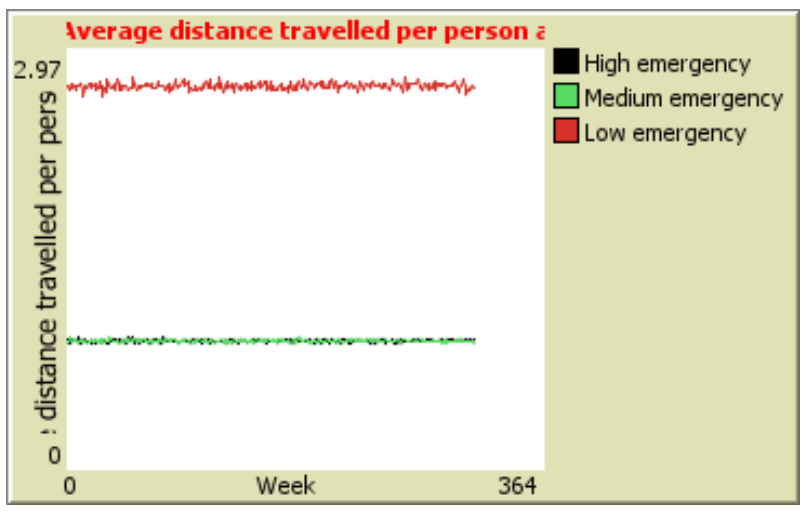

Experiment 10

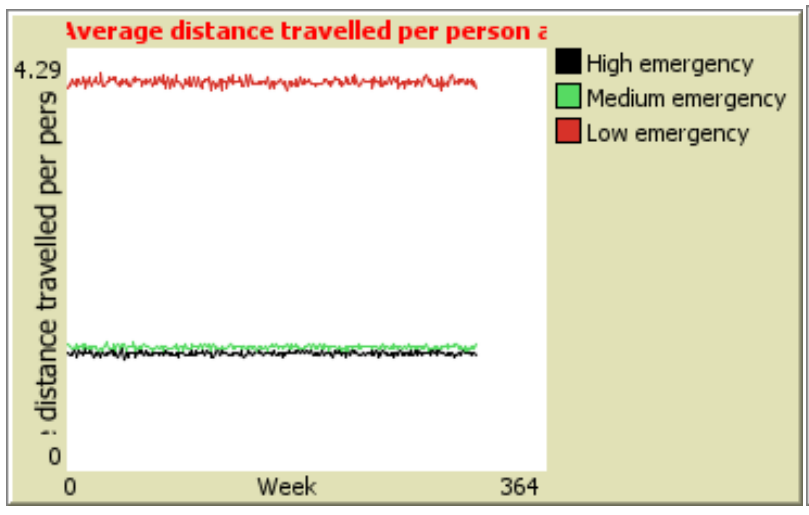

Experiment 11

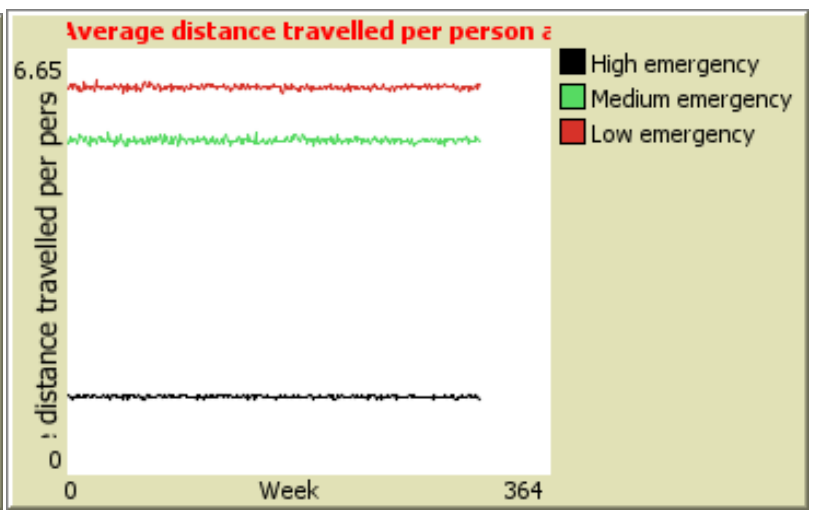

Experiment 12

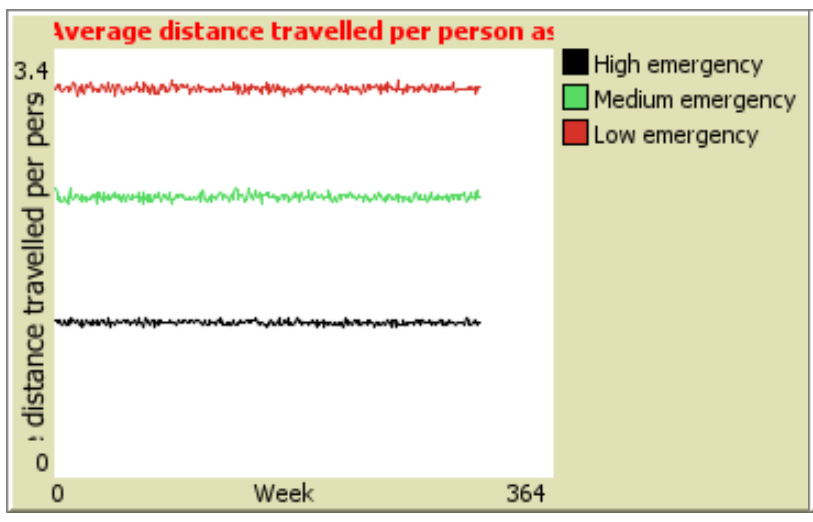

Experiment 13

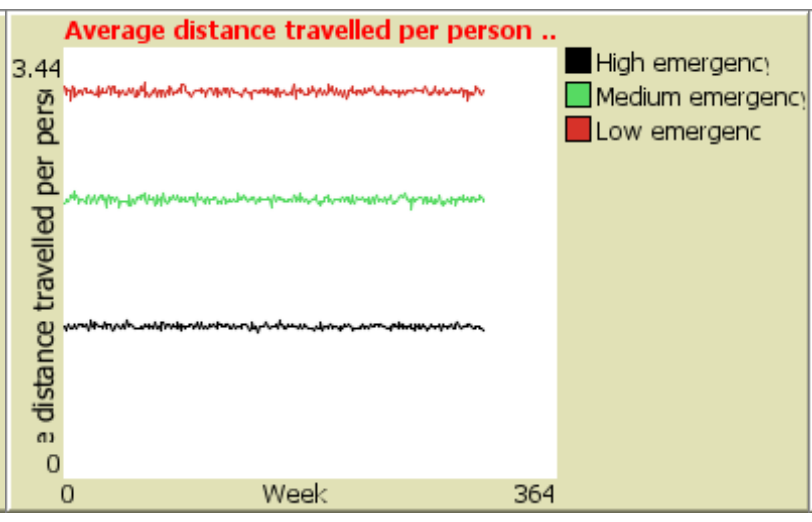

Experiment 14

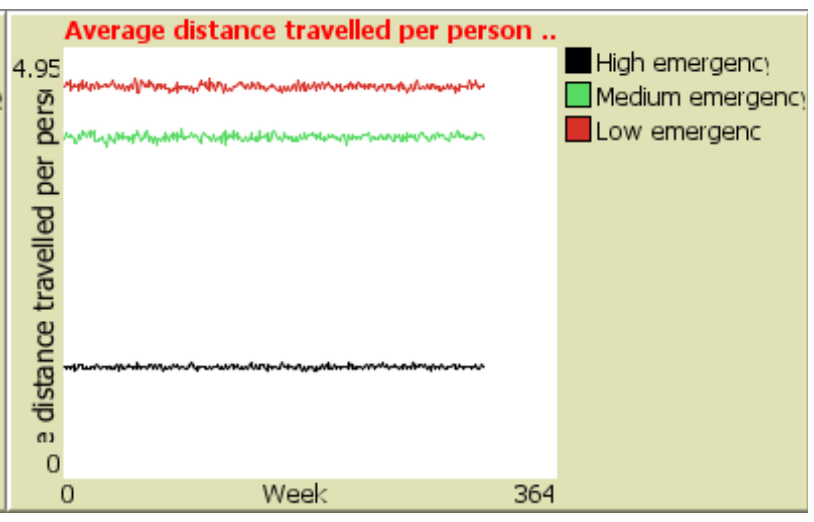

Experiment 15

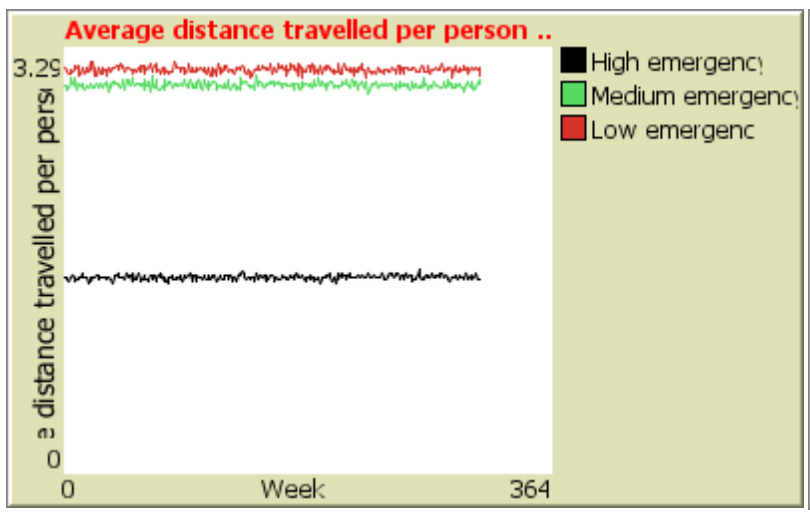

Experiment 16

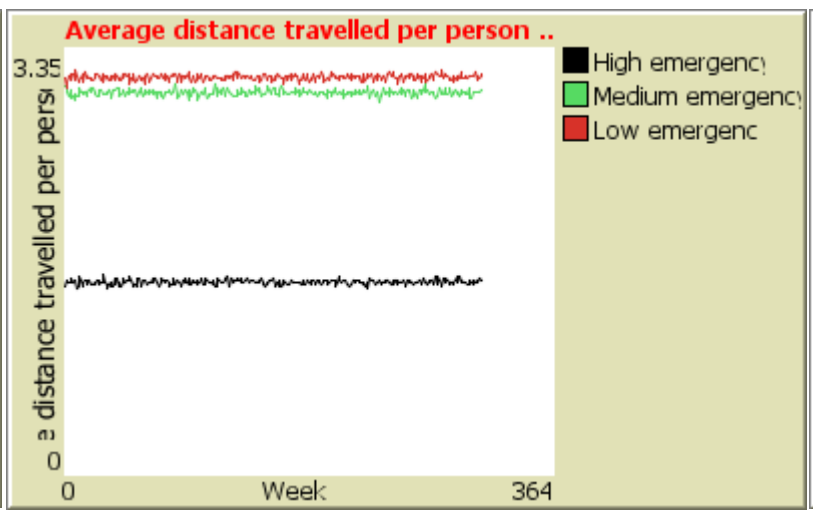

Experiment 17

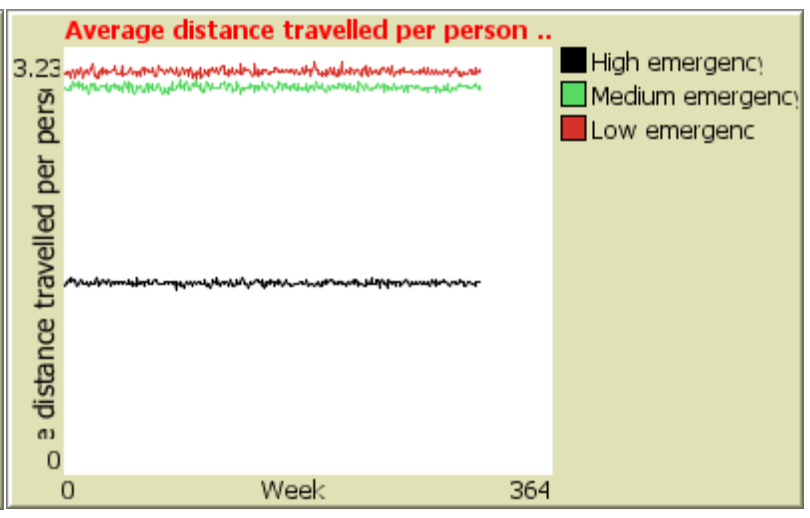

Experiment 18

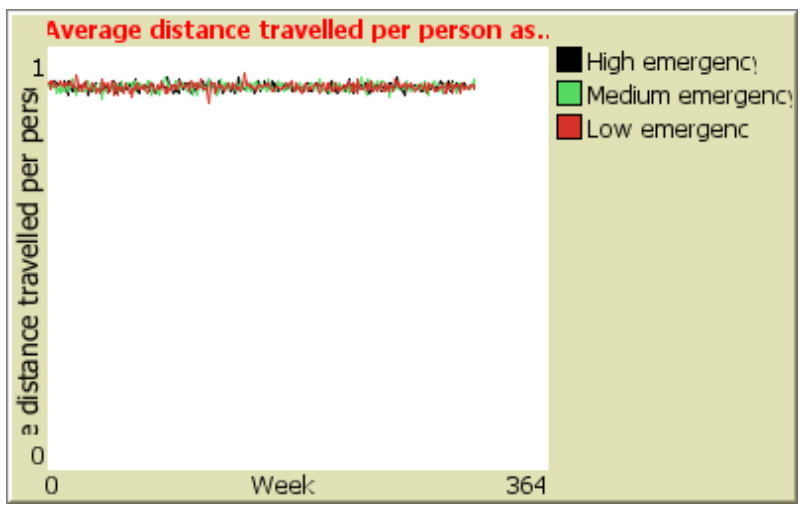

Experiment 19

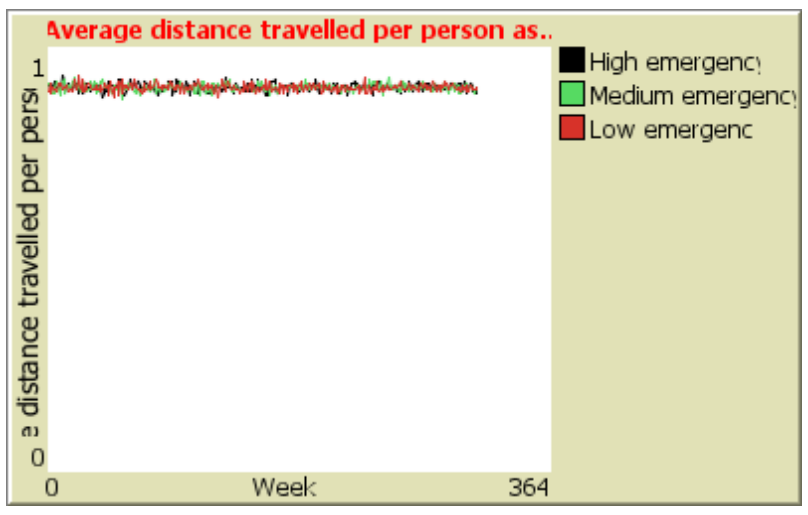

Experiment 20

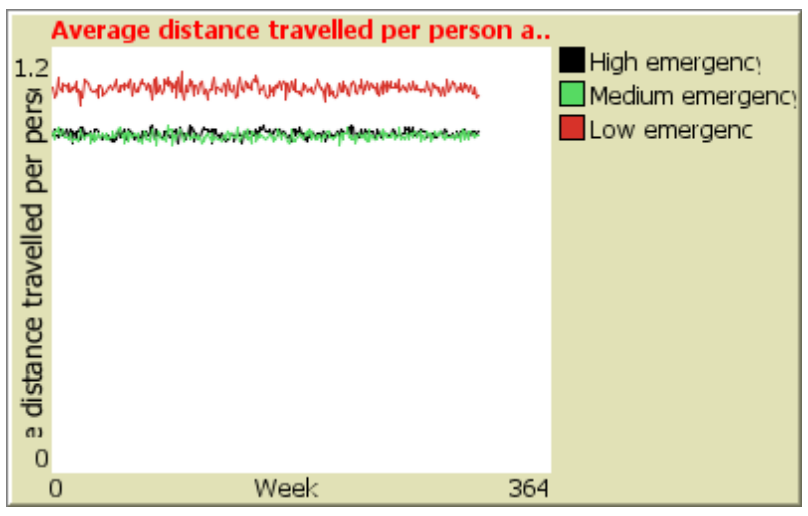

Experiment 21

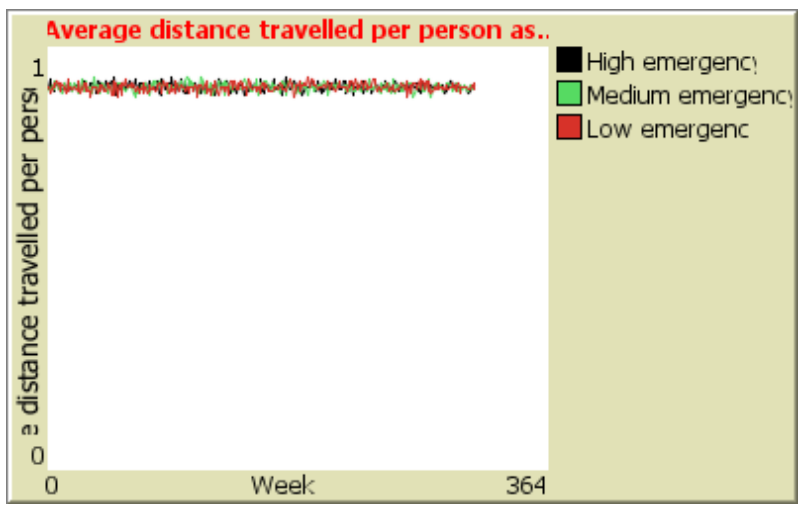

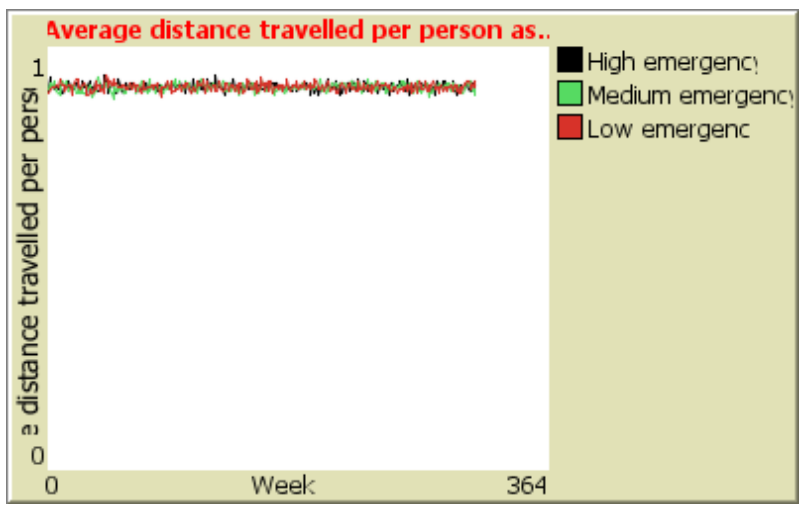

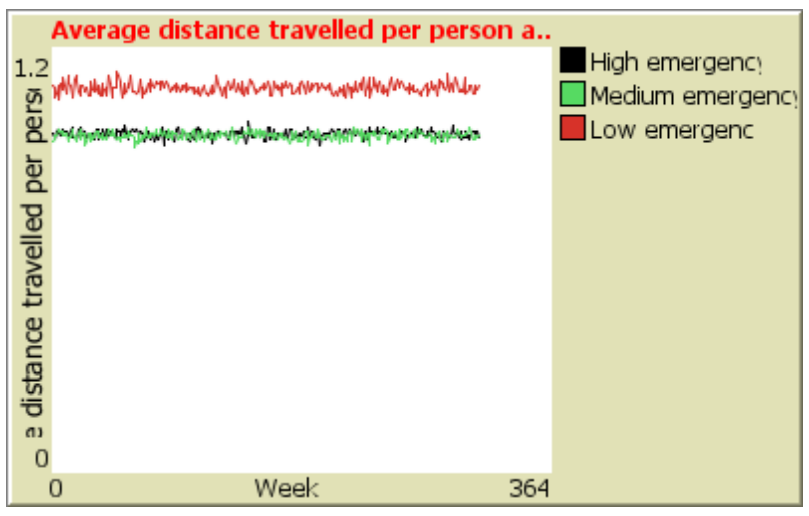

Experiment 22 Experiment 23 Experiment 24

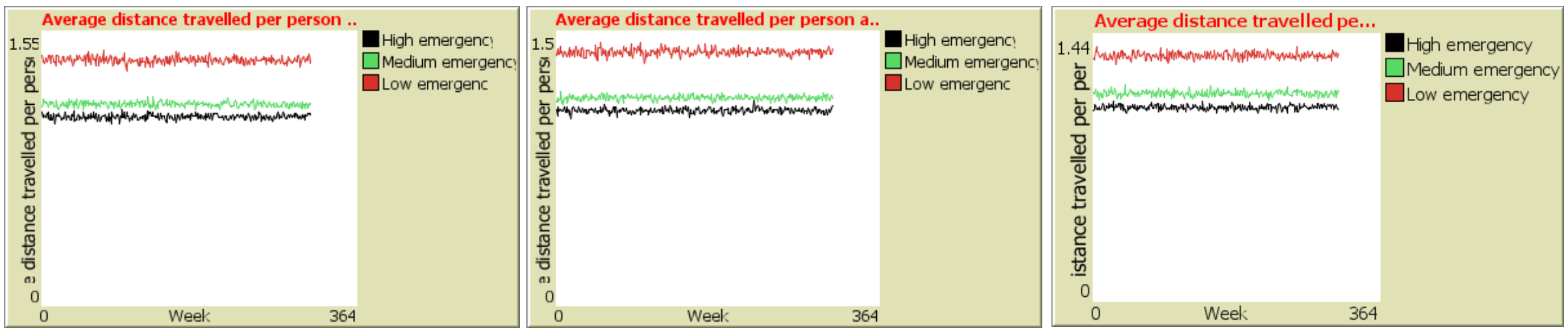

Experiment 25 Experiment 26 Experiment 27

FIGURE E10 Simulation results of distance (km) a customer travelled as per experimental design for product quality.